\documentclass[11pt]{book}

\usepackage{amsfonts}
\usepackage{amssymb}
\usepackage{amsmath}
\usepackage{psfig}
\usepackage{epsfig}
\usepackage{graphics}
\usepackage{dcolumn}

\renewcommand{\baselinestretch}{1.0} 
\newcommand{\be}{\begin{equation}}       
\newcommand{\ee}{\end{equation}}         
\newcommand{\bea}{\begin{eqnarray}}       
\newcommand{\eea}{\end{eqnarray}}

\newcommand{\clearemptydoublepage}{\newpage{\pagestyle{empty}\cleardoublepage}}

\begin{document}
\renewcommand\topfraction{.85}
\renewcommand\textfraction{.15}

\pagestyle{plain}


\thispagestyle{empty}

\begin{titlepage}
\begin{figure}[!h]
\includegraphics[width=1.8cm,clip]{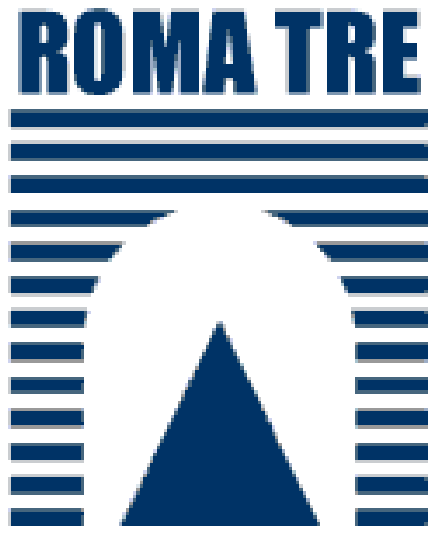}
\hspace{.3 cm}\parbox[b]{17cm} {\Large{UNIVERSIT\`A DEGLI STUDI
ROMA TRE} \\ \medskip
\Large{Dipartimento di Fisica ``Edoardo Amaldi''} \\
\Large{Dottorato di ricerca in fisica - XV ciclo}}
\end{figure}
\begin{center}
\vspace{3.0cm} {\bf \LARGE Complete Angular Distributions
\\ \medskip of the $\gamma d \rightarrow pn$ Reaction \\\medskip
in the Few GeV Region} \\
\vspace{1.0cm} {\LARGE Federico Ronchetti}
\end{center}
\vspace{5.0cm}
\begin{table}[!h]
\begin{center}
\begin{tabular}{ccc}
\large \it Coordinatore & \qquad \qquad \qquad \qquad
& \large \it Tutore \\ & &  \\
\large prof. Filippo Ceradini & & \large prof. Enzo De Sanctis
\\ & & \large \ 
\end{tabular}
\end{center}
\end{table}
\vspace{1.cm}
\begin{center}
{\large Aprile 2003}
\end{center}
\end{titlepage}

\pagenumbering{roman}  
\setcounter{page}{0}  
\tableofcontents 
\clearemptydoublepage
\listoffigures
\clearemptydoublepage
\listoftables
\clearemptydoublepage
\thispagestyle{empty}

\begin{center}
\section*{Abstract}
\end{center}
\indent
\par

The study of two body deuteron photo-disintegration is ideal for investigating
the transition region from meson exchange to the Quantum Chromo Dynamics (QCD) regime 
because of its simplicity and amenability to calculation \cite{HOLT}. 



In the past years, the theoretical efforts have been focused in 
two different directions: extending at higher energies the models based on 
meson exchange, and improving the models inspired by QCD principles to 
extrapolate their predictions at lower energies, in order 
to find out the approach that best describes the experimental 
data and in which energy region the transition between the two descriptions takes place.

Discriminating among the available models is still difficult because 
data are relatively scarce and mainly cover a limited angular range. 
\par

The possibilities for a complete experimental investigation of the  \mbox{$\gamma d \rightarrow p n$}
process are much improved since the availability of an intense \mbox{(200 $\mu$A)} electron beam of 
6 GeV energy at the Jefferson Lab Continuous Electron Beam Accelerator Facility (CEBAF).
In this new experimental context, a comprehensive study of the \mbox{$\gamma d \rightarrow p n$} 
reaction was started in recent years, using complementary equipments:
\\
\begin{itemize}
\item
very high energy, high-intensity polarized and unpolarized {\em bremsstrahlung} photon beams 
in conjunction with high resolution spectrometers
of moderate angular acceptance for measurements at few scattering angles  (JLab Halls ``A'' and ``C'');
\item 
high energy (unpolarized) tagged photon beam in conjunction with the 
large angular acceptance (nearly $4\pi$) and good momentum resolution of 
the CLAS detector (JLab Hall ``B''). 
\end{itemize}

\par
This thesis reports on the first, wide-ranging survey of the \mbox{$\gamma d \rightarrow p n$} 
differential cross section, for incident photon energies between 0.5 and 3.2 GeV,
characterized by the high statistics and large kinematic coverage 
obtained with the CEBAF Large Acceptance Spectrometer at Hall B (CLAS). 

The subject is introduced from the physical motivation for the 
measurement together with an overview of the theoretical efforts made
to describe the photo-disintegration process in the few GeV region.
Then, the CLAS detector characteristics are briefly described in order to
better illustrate the full data analysis procedure which,
together with the final results and their theoretical 
interpretation, constitute the very original part of this work.

\clearemptydoublepage
\pagenumbering{arabic}  

\setcounter{chapter}{0}
\setcounter{chapter} {0}   
\chapter{Physical Motivations}

\section{Introduction}
\indent
\label{sec:intro}
\par

The study of high-energy two body deuteron photo-disintegration has received 
a renewed interest in recent years in the attempt to evidence quark related 
effects in nuclei. In fact, this reaction is well suited for studying 
the interplay of nuclear and particle physics while 
allowing the best possible separation between the known 
electromagnetic reaction mechanism and the nuclear system structure. 

The physics interest deals with how and at what energy the transition takes place from the 
hadronic picture of the deuteron, established in the reaction at 
low energies (below 1~GeV), to the quark-gluon picture, 
which is expected to be correct for energies much higher then 1~GeV, 
where distances of the order of few tenth of Fermi play a role.

From the experimental point of view, the history of $\gamma d \rightarrow p n$ measurements
has been characterized by large discrepancies among different datasets.
The situation was complicated by the fact that the majority of the older experiments 
used {\em bremsstrahlung} photons and were subject to uncertainties 
due the poor knowledge of the photon flux and energy.
Moreover, large solid angle detectors were not widely available, 
leading to relatively small coverage in the proton scattering angle.

\section{The Low Energy Region}
\indent
\par
\label{sec:low-en}
The total photo-disintegration cross section data at incident photon energy below 1~GeV 
show a discrepancy up to 40\% in the $\Delta(1232)$ resonance region.
However, considering data from experiments performed with quasi 
mono-energetic photons or improved untagged {\em bremsstrahlung} 
techniques, a fair agreement (of order 8$\div$10\%) is found especially 
above 300 MeV of photon energy~\cite{PROS,JENK}.

In this energy region the cross section data can be described with a 
number of theoretical models all based on the ``traditional''
nuclear picture of meson exchanges such as:
the Impulse Approximation~\cite{ARE1}, the Coupled Channel Approach 
\cite{WILH}, the Diagrammatic Approach by~\cite{LAGE}, 
and the Modified Current Conservation~\cite{JAUS}.

The measured differential cross section is reproduced 
using a $4^{\rm{th}}$ order Legendre polynomial expansion:

\begin{equation} 
\label{multipole}
\sum_{l=0}^4 {A_l(E_\gamma)P_l \left( \cos{\theta_p^{\rm{CM}}} \right)  }
\end{equation}\\
and the total cross section can be computed using 
the relation:
\mbox{$\sigma_{tot}=4\pi A_0$}.

\begin{figure}[htbp]
\begin{center}
\leavevmode
\epsfig{file=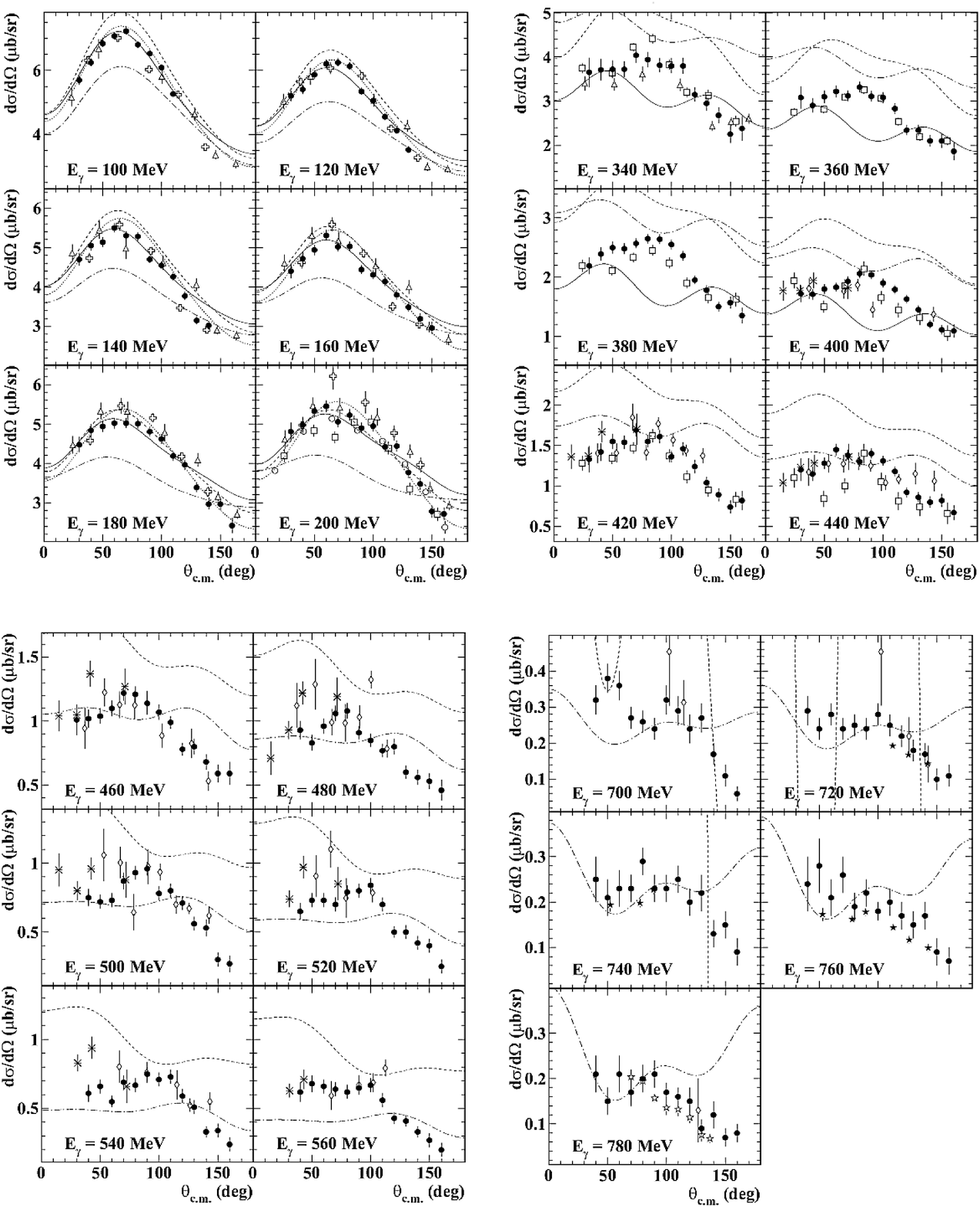,width=13cm,height=16cm,angle=0}
\caption{ \small Differential cross section angular distributions for the deuteron 
photo-disintegration process for selected incident photon energies below 0.8~GeV.
Data points are from the following experiments:
Bonn~\cite{AREND} (open squares), Frascati~\cite{EDSLEV} (open crosses), 
MIT~\cite{MATT} (open triangles), LEGS~\cite{LEGS} (open dots), and Mainz/DAPHNE~\cite{CRAW} (solid dots). 
The curves belong to the following theoretical models: the Impulse Approximation~\cite{ARE1} (dashed line), 
the Coupled Channel Approach~\cite{WILH} (up to 0.4~GeV) (continuous line), the Diagrammatic Approach ~\cite{LAGE} 
(dot dashed line), and the Modified Current Conservation~\cite{JAUS} (dotted line). 
As can be seen from plots, the predictions of the theoretical models have 
problems in reproducing the data, starting from relatively low incident photon energies.
(The plots are a courtesy of P. Pedroni - INFN, Pavia)}
\label{fig:mainz10}
 \end{center}
\end{figure}

A selection from the measured angular distributions 
for the \mbox{ $\gamma d \rightarrow pn$} cross section 
in the photon energy range
between 0.1 and 0.8~GeV 
is shown in Fig.~\ref{fig:mainz10} where the results are from 
the following experiments: Bonn~\cite{AREND}, Frascati~\cite{EDSLEV}, 
MIT~\cite{MATT}, LEGS~\cite{LEGS}, and Mainz/DAPHNE~\cite{CRAW}.
The overall dataset cover 
a relatively wide range in the proton scattering angle 
$\theta_p^{\rm{CM}}$, that is from $~25^\circ$ up to $~160^\circ$.

\begin{figure}[htbp]
\begin{center}
\leavevmode
\epsfig{file=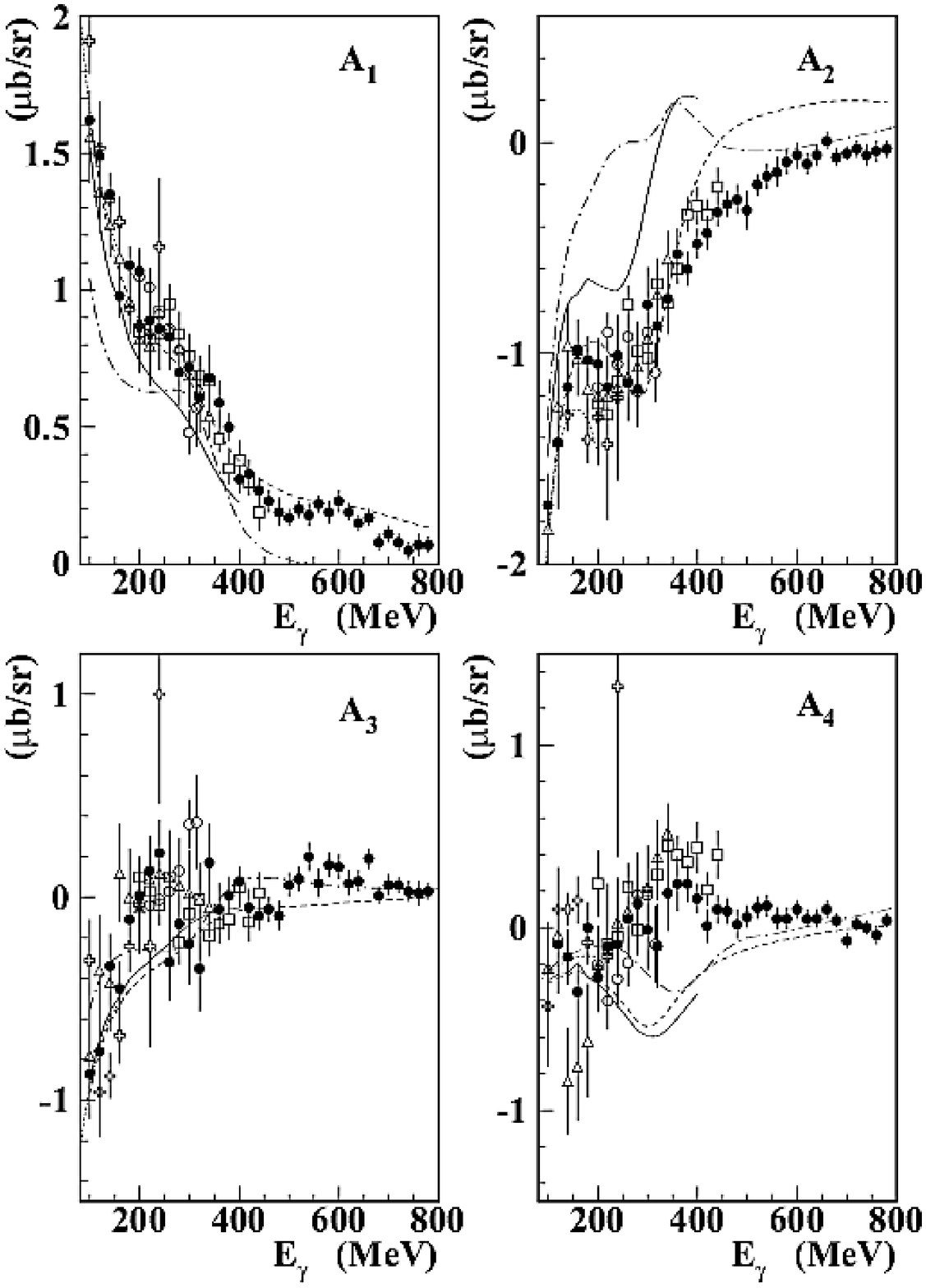,width=14cm,height=16cm}
\caption{ \small Legendre coefficients for the differential cross section.
At the highest photon energy reached at Mainz (around 0.8~GeV), the $A_1$ and $A_3$ coefficients
exhibit a small but significant deviation from the trend for the 
lowest energies. Data points are from the following experiments:
Bonn~\cite{AREND} (open squares), Frascati~\cite{EDSLEV} (open crosses), 
MIT~\cite{MATT} (open triangles), LEGS~\cite{LEGS} (open dots), and Mainz/DAPHNE~\cite{CRAW} (solid dots). 
The curves belong to the following theoretical models: the Impulse Approximation~\cite{ARE1} (dashed line), 
the Coupled Channel Approach~\cite{WILH} (up to 0.4~GeV) (continuous line), the Diagrammatic Approach ~\cite{LAGE} 
(dot dashed line), and the Modified Current Conservation~\cite{JAUS} (dotted line). 
(Courtesy of P. Pedroni - INFN, Pavia)}
 \label{fig:mainz12a}
 \end{center}
\end{figure}

The most significant contribution to the cross section data 
in this incident photon energy region 
is provided by the Mainz experiment, performed with a tagged photon beam 
in the energy range from 0.1 to 0.8~GeV in conjunction with the 
DAPHNE large-angle spectrometer. 
Results from this experiment are shown in Fig.~\ref{fig:mainz10} 
together with data points from Refs~\cite{BABA,CHIN,DOUG}. 
It can be seen that below 0.4~GeV the DAPHNE data better match 
the results of experiments carried out using a tagged photon 
beam such as Bonn and LEGS.
This peculiarity is also reflected by the Legendre coefficients 
$A_1-A_4$ plotted in the left panel of Fig.~\ref{fig:mainz12a}. 

\begin{figure}[htbp]
\begin{center}
\leavevmode
\epsfig{file=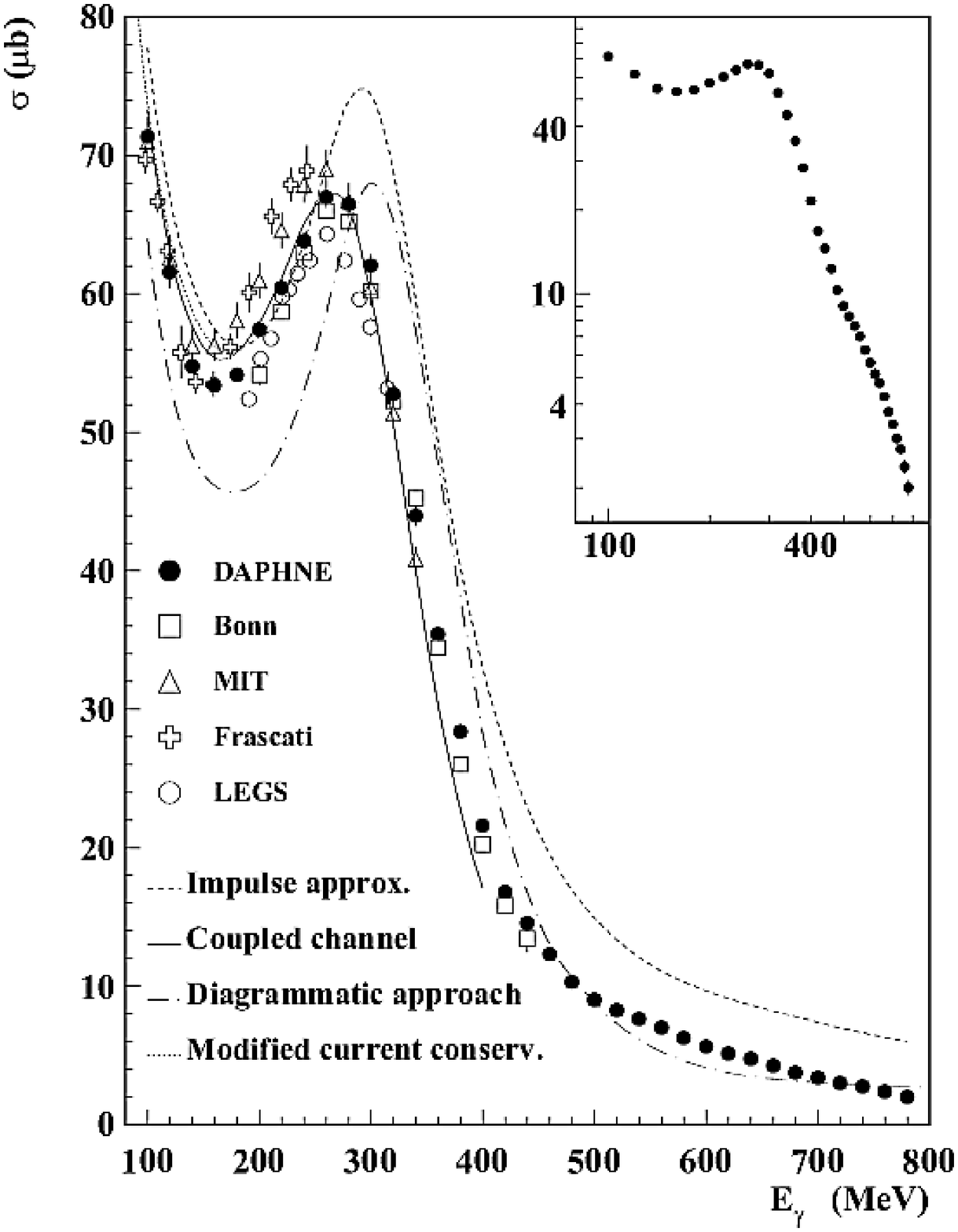,width=14cm,height=16cm}
\caption{ \small Total photo-disintegration cross section.
In the energy region around 0.8~GeV the total cross section behavior
is not a smooth decreasing function of the incident photon energy:
this could be explained by the intermediate excitation 
of baryon resonances in the second resonance
region which includes $P_{11}(1440), D_{13}(1520)$, and $S_{11}(1535)$. 
Data points are from the following experiments:
Bonn~\cite{AREND} (open squares), Frascati~\cite{EDSLEV} (open crosses), 
MIT~\cite{MATT} (open triangles), LEGS~\cite{LEGS} (open dots), and Mainz/DAPHNE~\cite{CRAW} (solid dots). 
The curves belong to the following theoretical models: the Impulse Approximation~\cite{ARE1} (dashed line), 
the Coupled Channel Approach~\cite{WILH} (up to 0.4~GeV) (continuous line), the Diagrammatic Approach ~\cite{LAGE} 
(dot dashed line), and the Modified Current Conservation~\cite{JAUS} (dotted line). 
(Courtesy of P. Pedroni - INFN, Pavia)}
 \label{fig:mainz12b}
 \end{center}
\end{figure}

The consistency between the angular distributions provided by 
different experiments is better than that of the absolute 
cross section shown in Fig.~\ref{fig:mainz12b}. 
The DAPHNE results show a noticeable change in slope of the cross section, 
moving toward higher photon energies. The effects starts around 0.5~GeV 
and it is reflected in the shape of the Legendre polynomial
coefficients shown in Fig.~\ref{fig:mainz12a}.

In this energy region, also the $A_1$ and $A_3$ coefficients
exhibit a small but significant deviation from the trend for the 
lowest energies: this could be explained by the intermediate excitation 
of baryon resonances in the second resonance
region which includes $P_{11}(1440), D_{13}(1520)$, and $S_{11}(1535)$.

\subsection{Theoretical Interpretation}
\label{sub:low-teo}
\indent
\par

Included in Figs \ref{fig:mainz10}, \ref{fig:mainz12a} and \ref{fig:mainz12b} are the results of four
theoretical models based on meson-nucleon degrees of freedom.

The Impulse Approximation and the Coupled Channel Approach 
consider the nucleon internal degrees of freedom explicitly
in the hamiltonian and the deuteron wave function is a superposition of different
intrinsic configurations ($NN$, $N\Delta$, $\Delta \Delta$) containing also excited nucleons.
These configurations are mutually coupled by non-diagonal transition potentials.
The two models need some kind of phenomenological input and
start on the same assumptions, but handle 
the resulting set of coupled equations differently: 
the Impulse Approximation is a perturbative solution in which only 
the $NN$ ground configuration is retained while the isobar 
degree of freedom are only implicitly taken into account via
the dispersive part of the $N-N$ potential. 
On the other hand, the Coupled Channel Approach gives 
the full solution for the $NN-\Delta\Delta$ coupled channel using 
custom models for the different potentials involved and for the $\Delta$ width. 
Its results includes only multipoles up to $L=4$
while for scattering waves with $j \geq 4$ it still relies on the Impulse Approximation. 

Figs \ref{fig:mainz10} and \ref{fig:mainz12a} show that the Impulse Approximation  
reproduces the angular distributions up to 360 MeV even though 
it  overestimates the total cross section.
On the other hand, the Coupled Channel Approach 
gives a good description of the total cross section, but has some problems 
describing the shape of the angular distributions, 
since a dip is found at $90^\circ$ which is not reflected by the data.

This feature appears also in the behavior of the $A_2$ coefficient 
shown in Fig.~\ref{fig:mainz12a} (upper left)
which determines the curvature of the angular distribution 
around $90^\circ$ and which is strongly overestimated above 160 MeV. 
This anomalous behavior is produced by the $N \Delta$ configurations
in higher partial waves~\cite{ARE1,WILH} with $j \geq 4$ which 
are important for the differential cross sections,
but negligible for the total cross section.

In the Laget Diagrammatical Approach, 
the amplitude is expanded in a series of relevant interaction mechanisms.
Each elementary transition amplitude is then determined starting from effective 
Lagrangians for the basic $N \pi$ and $\gamma N$ couplings. 
This model is again similar to the Impulse Approximation 
but predicts an incorrect energy for the $\Delta$ peak excitation.  
This discrepancy is due to dynamical effects on the $\Delta$ mass given 
by the corresponding $N-N$ and $N-\Delta$ channels that are not taken accounted for, 
so that the resulting energy for the $\Delta$ peak is shifted.
This effect would be compensated by the $N-\Delta$ re-scattering, 
which is expected to shift back in place the $\Delta$ peak.

The approach of Jaus, Bofinger and Woolcock, calculates the multipole 
amplitude from an effective current density which takes into account 
convection, spin, and meson exchange currents.
This procedure is based on an effective one-pion exchange current density 
which does not violate the current conservation law.
The model allows for two adjustable parameters: the $\pi NN$ coupling constant and the
pion cut-off mass, which can be used to fit the existing experimental data below 200 MeV.
This limit stems from the fact that only a limited number of multipole amplitudes have 
been considered and the finite $\Delta$ width has not been taken in to account.
The calculations predict a total cross section which is very similar to the Coupled Channel
result. Both methods agree with the Mainz cross section values up to about 120 MeV,
but then between 120 and 180 MeV they both overestimate the experimental results.

\clearpage
\section{The High Energy Region}
\indent
\par
\label{sec:high-en}

\begin{figure}[htbp]
\begin{center}
\leavevmode
\epsfig{file=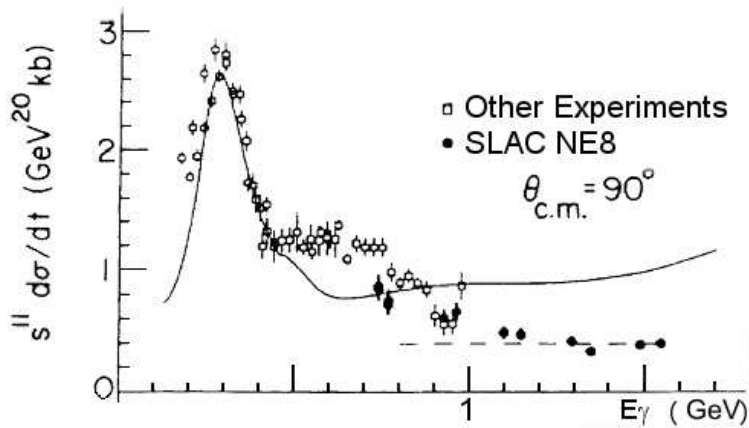,width=10cm,height=6cm}
\caption{\small Results form the NE8 experiment: the cross section shows scaling for incident
photon energies larger then 1~GeV. The solid line represents a meson-nucleon 
description which do not account for the data starting from 0.5~GeV of incident photon energy.}
\label{fig:ne8}
\end{center}
\end{figure}

For incident photon energies of the order of 1~GeV, 
the deuteron photo-disintegration differential cross section 
was measured for the first time by the SLAC NE8 experiment~\cite{FREE}.
Results from the NE8 experiment are shown Fig.~\ref{fig:ne8} 
for a proton scattering angle $\theta_p^{\rm{CM}}=90^\circ$.
The data are multiplied by the $s^{11}$ factor, where $s$ is the square of
the total energy in the usual Mandelstam notation.

Clearly, for  $E_\gamma \geq$ 1~GeV, the data shows a flat behavior.
This represents the first evidence for the possibility of
the so called {\em scaling} in the
deuteron photo-disintegration cross section, suggesting
that point-like constituents as 
quarks and gluons ({\em i.e.} the QCD degrees of freedom) 
could be the correct description for the process, 
in analogy to the Deep Inelastic Scattering (DIS).
On the contrary, the traditional meson exchange description, 
discussed in the previous section to interpret the 
low energy data, breaks down.

This striking result was supported and extended by another experiment 
subsequently performed at SLAC for photon energies up to 
2.8~GeV~\cite{BELZ}.
 
The SLAC results have renewed the interest for the study of 
the deuteron photo-disintegration at high energy and as a consequence 
a vast physics program has been approved at Jefferson Lab.
This program involves all the three JLab collaborations
in a full investigation of the physics topics related to 
the deuteron photo-disintegration process:

\begin{itemize}
\item Hall A Collaboration: angular distributions at three medium energies and investigation of the polarization observables~\cite{HALLA};
\item Hall B Collaboration: nearly complete angular distributions from 0.5 to 3.0~GeV of photon energy~\cite{HALLB};
\item Hall C Collaboration: differential cross section at few angles and very high energy \mbox{(up to 5.5~GeV)}~\cite{HALLC}.
\end{itemize}

\subsection{Theoretical Interpretation}
\indent
\par
\label{sub:high-teo}

\begin{figure}[htbp]
 \begin{center}
 \leavevmode
  \epsfig{file=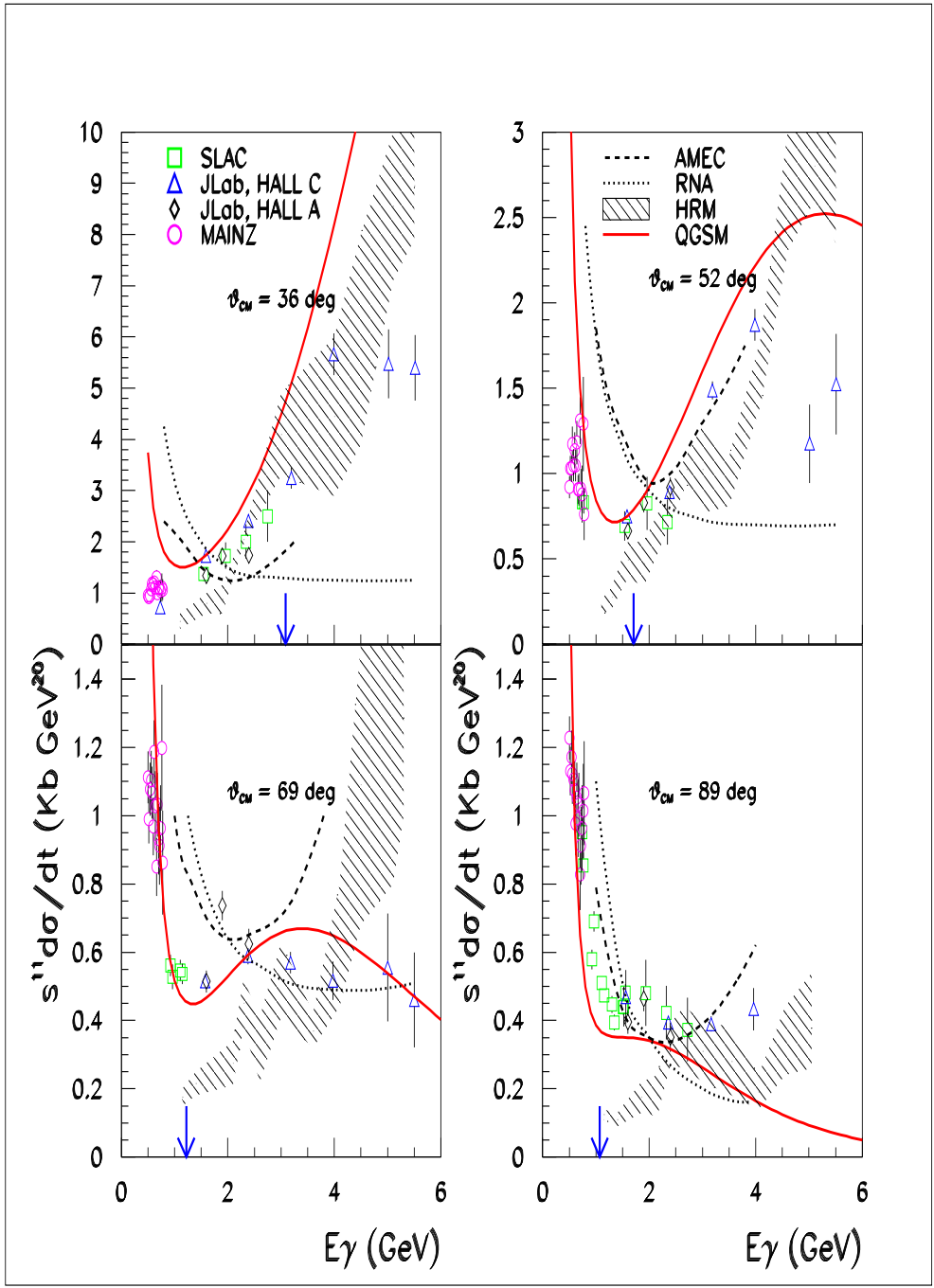,width=13cm,height=18cm}	 
\caption{\small Deuteron photo-disintegration differential cross section multiplied by $s^{11}$. 
Experimental data are from Mainz~\cite{CRAW} (open/magenta circles), SLAC~\cite{FREE,BELZ} 
(open/green squares), 
Jlab Hall C~\cite{BOCH,SHU1} (open/blue triangles), and Hall A~\cite{SHU2} (open/black diamonds). 
Points are multiplied by $s^{11}$, and arrows indicate the expected threshold for the onset 
of the CCR scaling, where the momentum transfer to the proton, $t$ is of the order of
1~GeV$^2$~\cite{BRO1}. Theoretical curves are described in the text.}
 \label{fig:fixed-angle}
 \end{center}
\end{figure}

The experimental data provided by SLAC, JLab Hall A, 
and Hall C are shown in Fig.~\ref{fig:fixed-angle} for the measured  
proton scattering angles. The cross section is again multiplied by $s^{11}$
in order to evidence the possible scaling behavior. 

In this energy region an asymptotic behavior could be expected 
on the basis of the simple model of the Constituent Counting Rules (CCR,~\cite{BRO1}).
This model considers the carriers of the currents within the hadrons as structureless and 
describable in terms of a renormalizable field theory, with the conditions of asymptotically 
scale invariance for the interactions among the constituents, and finiteness 
at the origin of the hadronic wave functions.
Accordingly, applying dimensional counting to the minimum quark field component of a hadron 
accounts for many experimental consequences of its compositeness.
The constituent counting rule prediction for exclusive scattering is:

\be
\frac{d\sigma}{dt} = s^{(2-n)}f\left(\frac{t}{s}\right)\ ,
\label{CCR}
\ee\\ 
where $s \rightarrow \infty$ and $t/s$ is fixed, where $s$, $t$ (and $u$) are the 
usual relativistic invariants in the Mandelstam notation.
Here $n$ is the number of elementary fields in the initial and final states.

Considering the \mbox{$\gamma d \rightarrow pn$} process it turns out that 
$n = 13$ since the initial state has one photon plus 6
quark fields while the final state has 6 quark fields which means that at 
at high incident photon energy and intermediate angles, conventionally $90^\circ$,
where both the $t$-dominance and $u$-dominance are suppressed 
the  $\gamma d \rightarrow pn$ differential cross section
is expected to scale as $s^{-11}$.
As shown in Fig.~\ref{fig:fixed-angle}, this prediction is well confirmed 
by the experimental data for $\theta_p^{\rm{CM}} = 69^{\circ}$ and $89^{\circ}$.

The overall data behavior is less clear and may suggest a deviation 
from the predictions of the simple CCR scaling indicating, in turn, 
that pQCD itself is not plainly applicable. 

In fact, the cross section at the forward angles $\theta_p^{\rm{CM}} = 52^{\circ}$ 
and $36^{\circ}$ falls off more slowly, with an $\simeq s^{-9}$ scaling at 
lower energies, until the onset of the $s^{-11}$ behavior at about 3 and 4~GeV
beam energy, respectively. However it results that an absolute pQCD calculation
would not correctly predict the data. Further observations may be made for 
other photo-reactions, and it remains to be seen how this 
behavior arises and if pQCD is a genuine explanation for it.

In this case, a different approach have to be used and it seems very 
reasonable that a soft description for the process will have to be introduced.

This can be done following opposite strategies, such as the inclusion 
of QCD degrees of freedom in the low energy description or the 
extrapolation of the conventional $N-\pi$ interaction mechanisms to the higher 
energy region. It is also possible to merge suitable parts of this 
two techniques to obtain an hybrid description 
(for a full review see Ref.~\cite{GILM}).
The curves shown in Fig.~\ref{fig:fixed-angle} represent the predictions of 
several theoretical models implementing the outlined strategies.
In the Reduced Nuclear Amplitude (RNA)  
model~\cite{RNA,RNAEL} the binding of the quarks inside the nucleons 
and the deuteron is taken into account with empirical form factors 
and the elementary cross section is computed assuming CCR scaling.
This approach is able to describe the $\gamma d \rightarrow p n$ 
cross section with appropriate normalization factors 
for $\theta_p^{\rm{CM}} = 89^{\circ}$ and $69^{\circ}$ with $E_{\gamma} >$ 2~GeV.

In the Hard quark Rescattering Model (HRM)~\cite{HRM}, 
the elementary interaction consists of a quark exchange between the 
two nucleons. The incoming photon is absorbed by a quark of one 
nucleon which then gives up its momentum via a hard gluon exchange 
with a quark of the other nucleon.
The model assumes that this rescattering mechanism is analogous,
with some approximation, to the wide angle $p-n$ scattering 
which is also dominated by quark exchange. 
The limits for the applicability of the model are $E_{\gamma}>$ 
2.5~GeV and momentum transfer $t>$2~GeV$^2$, but, under particular 
assumptions for the short distance $p-n$ interaction, they can be extended. 
However, the agreement with the data is poor, especially at higher 
energies where the uncertainty is large,
due to the limited knowledge of the $pn$ cross section for the actual 
kinematic conditions.

\begin{figure}[htbp]
 \begin{center}
 \leavevmode
\hspace*{-2.0cm}
\epsfig{file=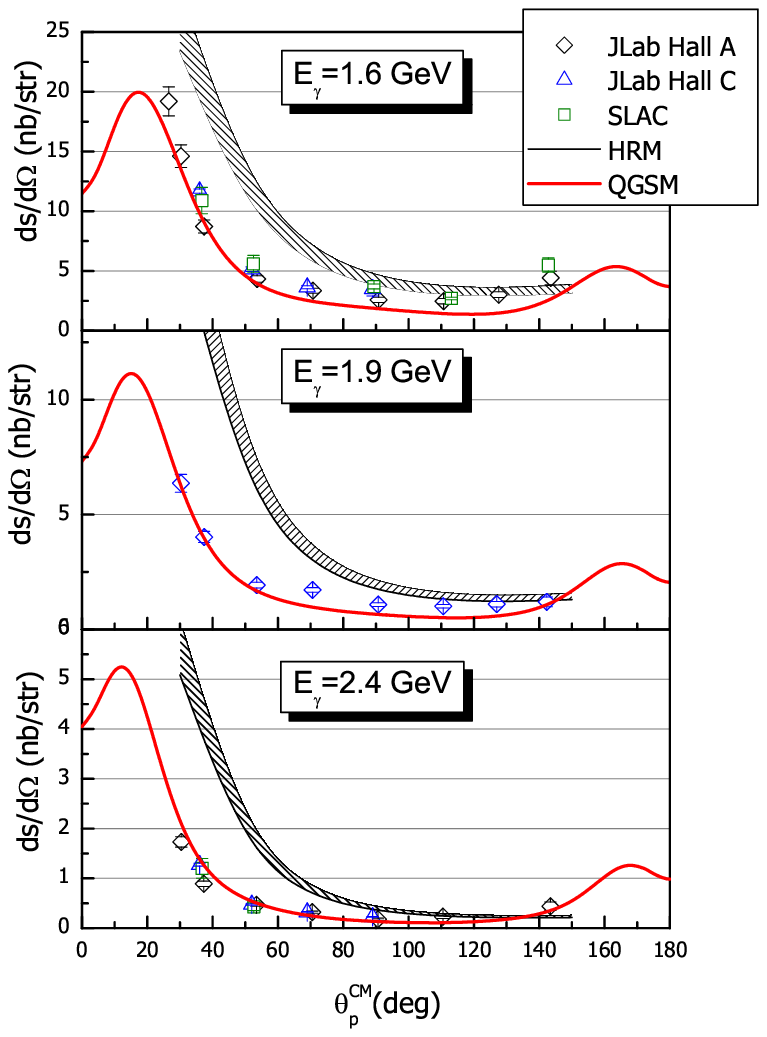,width=14cm}
 \caption{ \small Angular distributions for three incident photon energies: \mbox{1.6 (top plot)}, 
\mbox{1.9 (middle plot)}, and \mbox{2.4~GeV (bottom plot)}.
The experimental data are from Refs~\cite{BELZ,BOCH,SHU1,SHU2} while the curves represent 
the theoretical predictions of the QGSM~\cite{gris} and the HRM~\cite{SARG}. 
Prediction from other models are not reported since the were not provided or could not be calculated. From the plots it is seen that
the QGSM well describes the data. This model also predicts a forward to backward asymmetry.
More data especially at very forward and backward angles are necessary to verify this prediction.}
 \label{fig:ang1}
 \end{center}
\end{figure}

Even if traditional models based on the $N-\pi$ picture fail 
to reproduce the data at $E_{\gamma} >$ 1~GeV, this approach can be extended,
with the appropriate modifications, in the few~GeV region.
The Asymptotic Meson Exchange Current (AMEC) model~\cite{AMEC},
uses form factors to describe the $d-NN$ interaction 
vertex and an overall normalization factor fixed by fitting the experimental 
data at 1~GeV. The results reproduce the energy dependence of the 
cross section only for $\theta_p^{\rm{CM}} = 89^{\circ}$. 

It is worth noticing that this non QCD-based model provides a scaling law for
the cross section, with an exponent depending on the scattering angle.

A non perturbative description of the photo-disintegration
cross section data is given by the Quark Gluon String Model (QGSM) (Ref.~\cite{gris})
This model assumes that the scattering amplitude at very high
energy is dominated by the exchange of three valence 
quarks in the $t$-channel. The duality property of scattering
amplitudes~\cite{LEV,CLOS} allows its extension at intermediate energies.

As shown in Fig.~\ref{fig:fixed-angle} the predictions of the
QGSM are in good agreement with the available data in the
full energy range at $\theta_p^{\rm{CM}} = 89^{\circ}$, $69^{\circ}$. 
At $52^{\circ}$ the trend of the QGSM theoretical curve could be compatible
with the last two points at the highest energies since their errors are
quite large. At $\theta_p^{\rm{CM}} = 36^{\circ}$ the situation is
more delicate even if the QGSM curve has a better trend with respect to 
the other calculations.

A better insight and a more effective discrimination 
among the predictions of the competing theoretical models 
are obtained examining the angular distributions of the 
differential cross section~\cite{FEDRO}.

Experimental data recently obtained from~\cite{SHU2}
for three different incident photon energies: 
$E_\gamma =$ 1.6, 1.9, and 2.4~GeV are shown in Fig.~\ref{fig:ang1}.

These angular distribution were measured only for the indicated 
energies and are not complete since the very forward and backward regions
are not covered.
Additional data from Refs~\cite{BELZ,BOCH,SHU1} are also shown
in Fig.~\ref{fig:ang1} together with the curves representing 
theoretical predictions.
The shaded area is the HRM~\cite{SARG} 
calculation which has an uncertainty of 15-20\% while
the QGSM prediction is represented by a dashed line. 
Other calculations are not shown
since they were not provided (AMEC) or the model could not
be applied in this case (RNA) since the curves for different angles
have arbitrary normalizations.

Also in this case, the QGSM well reproduces the available distributions
while the HRM (apart from a possible overall shift) show a different behavior 
at small and large angles.

From the experimental side, it is clear that the knowledge of the full angular distributions,
including the points at very forward and backward angles and in a wider incident photon energy 
range will be extremely helpful to better
discriminate among the models and to further check the QGSM predictions.

\subsection{The Origin of Cross Section Scaling}
\indent
\par
\label{sub:scaling}

It has to be said that the applicability  
of the CCR at this incident photon energies is still an open question. 
In fact, this approach is based on the 
assumption that soft and hard contributions
factorize~\cite{hhc} even if this hypothesis 
(relying on pQCD) may be valid only for very high momentum
transfers which are not attained in the few~GeV energy region
where experimental data are available (see Refs~\cite{IOF,ISGU,RADI,AZNA}).
\begin{figure}[htbp]
 \begin{center}
 \leavevmode
\epsfig{file=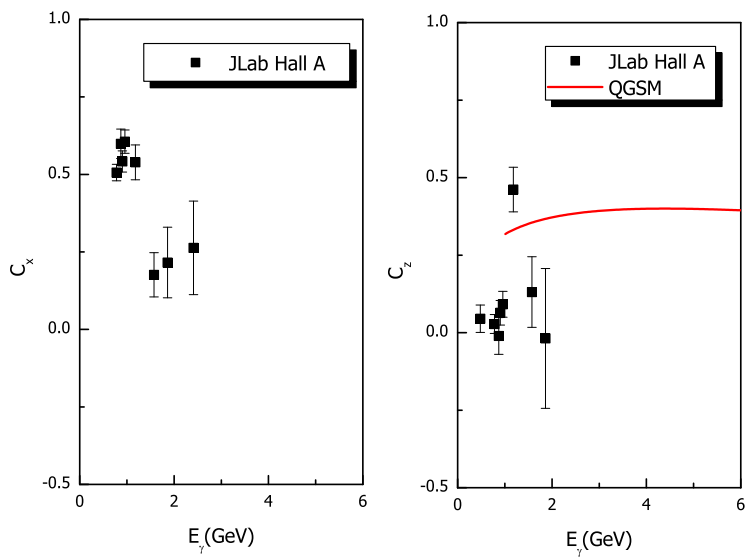,width=11cm}
 \caption{ \small Proton polarization transfers $C_x$ (transverse in-plane, left panel) and $C_z$ 
(longitudinal in-plane, right panel) in the deuteron photo-disintegration at $\theta_p^{\rm{CM}} = 90^{\circ}$.
The $y$-axis is perpendicular to the scattering plane, the $z$-axis is parallel to the scattered 
proton momentum and the $x$-axis is perpendicular to $y$ and $z$.  
Experimental data are from~\cite{pol} and have been corrected for spin rotation due to the LAB $\rightarrow$ CM
transformation. The solid/red curve represents the QGSM prediction for $C_z$~\cite{qnp-vera}. The QGSM 
calculation for $C_x$ is underway.
}
\label{fig:hhc}
\end{center}
\end{figure}
In order to clarify this point it is useful to consider
the complementary information to the differential cross section 
given by polarization observables.
Significant data on this topic were recently provided by the JLab 
Hall A Collaboration~\cite{pol} and are shown in Fig.~\ref{fig:hhc} 
which gives the proton in-plane polarization transfers and the 
photon asymmetry as measured 
in the $\vec{\gamma} d  \rightarrow \vec{p} n$  
reaction for the proton angle $\theta_p^{\rm{CM}} = 90^{\circ}$.
In this case, the signature of pQCD effects 
rests in the so called Hadron Helicity Conservation (HHC) 
\cite{hhc,hhcprot,landmecprot,landmecgam}.
HHC arises from the fact that vector interactions 
such as photon-quark or gluon-quark couplings 
conserve chirality, leading, in turn, to conservation of
the sums of the components of the hadronic spins
along their respective momentum directions.

If pQCD can be really invoked to explain the cross section scaling at
$\theta_p^{\rm{CM}} = 90^{\circ}$ it is expected that at the same
angle the polarized observables will satisfy the pQCD based prediction
of HHC of  $C_x = C_z = 0$. 

As shown in Fig.~\ref{fig:hhc}, for $E_\gamma \geq$ 1.5~GeV
the results on $C_x$ seem not in agreement with the HHC prediction
while the those on $C_z$ are more difficult to interpret but
the QGSM calculation for $C_z$ has a reasonable trend with respect 
to the JLab data. A similar calculation for $C_x$ is underway.

The problem with HHC lies in the limiting hypothesis of chirality 
conservation while the QGSM, being a non perturbative approach, 
is far from the chiral limit (characteristic of pQCD) and is able 
to introduce a chirality violation which may be reflected in the
data.

For photon energies below 1.5~GeV the proton polarization
transfers $C_x$ and $C_z$ show quasi-resonance structures 
at $\sqrt{s}=$2.7 and 2.9~GeV, respectively.
Such structures are conceptually 
out of the QGSM range of applicability~\cite{qnp-vera12}.

\begin{figure}[htbp]
 \begin{center}
 \leavevmode
\epsfig{file=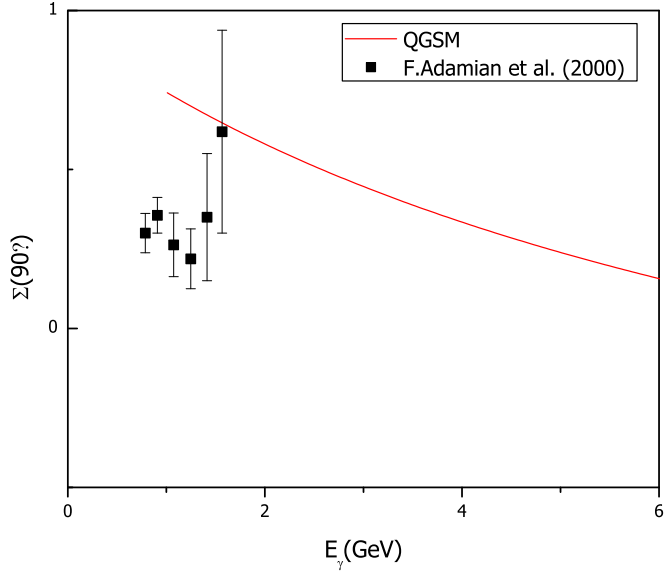,width=10cm}
 \caption{ \small Asymmetry $\Sigma$ for linearly polarized photons in the deuteron 
photo-disintegration at $\theta_p^{\rm{CM}} = 90^{\circ}$ as 
a function of the incident photon energy.
Experimental data are from Ref.~\cite{adam}. The solid/red curve is the prediction
of the QGSM~\cite{qnp-vera}. 
}
\label{fig:hhc2}
\end{center}
\end{figure}

Another observable carrying information on HHC predictions
is the photon asymmetry, defined as

\begin{equation}
\Sigma(\theta_p^{\rm{CM}})=\frac{d\sigma_\parallel -d\sigma_\perp}{d\sigma_\parallel+d\sigma_\perp}  \ ,
\label{eq:asym}
\end{equation}

for which the HHC prediction at $\theta_p^{\rm{CM}} = 90^{\circ}$ is 1 
(see Ref.~\cite{qnp-vera}).

As can be seen from Fig.~\ref{fig:hhc2} most of the experimental 
data are limited to the incident photon energy region below 1.5~GeV
where the perturbative regime may be not fully established,
in any case the trend of the data does not seem to indicate that
the HHC prediction is verified.
In the higher energy region where $E_\gamma \geq$ 1.5~GeV, 
the QGSM predicts a slow decrease of $\Sigma(90^\circ)$ 
Unfortunately the overlap with the available data 
is quite limited so definite conclusions cannot be drawn yet.

Summarizing: at least for $\theta_p^{\rm{CM}} = 90^{\circ}$ 
the cross section data and the in-plane polarization transfers 
results do not appear consistently described in a coherent perturbative picture.

This indicates that a non perturbative approach may be 
needed in order to fully describe the deuteron photo-disintegration 
process (polarized and unpolarized). 
The QGSM gives a non perturbative description of the
process which reproduces fairly well the
available data: for this reason the QGSM foundations
will be given in the next Section.

\newpage
\section{The Quark Gluon String Model}
\indent
\par
\label{sec:qgsm}

Quantum chromodynamics is the fundamental theory of strong interactions. 
The asymptotic freedom of QCD implies that its coupling constant 
$\alpha_s(Q^2)$ becomes small at small distances 
(large $Q^2$) and this gives the possibility to use perturbation 
theory in processes with large momentum transfer. 
Experimental data for these processes are in good agreement 
with pQCD predictions.
On the other hand, at large distances (of order $1/\Lambda_{QCD}$, 
where $\Lambda_{QCD} \simeq $ 0.3~GeV) 
the coupling constant is not small anymore and non perturbative effects, 
which are responsible for confinement, come into play. 
The large distance dynamics is necessary to understand processes with small 
momentum transfer, which give dominant contributions to high energy 
hadronic interactions. 

In this region a non perturbative approach should be used and  
there are not many ways to introduce non perturbative effects in QCD.
One possibility is to consider lattice calculations~\cite{CRE}, 
which are most directly related to the exact QCD 
results. This method, however, needs very large lattices to obtain accurate and reliable results
and it is difficult to apply to complex phenomena as exclusive reactions at high energies.
Another possibility is from the QCD sum rules~\cite{SHIF} which express physical quantities in terms of 
several parameters such as vacuum correlators of gluonic and quark fields. 
As well as lattice calculations, this method has been applied mainly to {\it static} 
properties of hadrons.   
A third method, which will be briefly introduced in this overview, is the $1/N$ 
(or topological) expansion in QCD~\cite{THOO,VEN,CIA}, 
where $N$ is the number of colors or flavors. Dynamical content for this model 
will be drawn from the string (or color tube) models~\cite{CAS1,CAS2,ART,AND1,AND2}, which are closely 
related to the space-time picture of interaction in $1/N$ expansion. 
The string-like configuration of gluonic fields is confirmed by lattice calculations.

These ingredients form the basis for the Quark Gluon String Model (QGSM,~\cite{KA1,K93}) which then accounts
for the confinement of quarks and can be considered a microscopic theory underlying the Regge phenomenology.
As a result many relations between parameters of reggeon theory (based on analyticity and unitarity) can be 
established from the QGSM~\cite{KA3}, since a quark-gluon string can be identified with a 
corresponding Regge trajectory.
The QGSM has been applied to many different problems of strong interactions: 
hadronic mass spectrum~\cite{KA5}, widths of resonances~\cite{KA6}, 
relations between total cross sections, residues of Regge poles~\cite{KA7}, 
behavior of hadronic form factors and, in recent years, 
to heavy ion collisions at high energy~\cite{ions} and 
to the deuteron two body photo-disintegration~\cite{gris}. 


\subsection{The Topological Expansion in QCD}
\indent
\par
\label{sub:topo}

Large distance dynamics of QCD defines what is called soft process, 
{\em i.e.} when the value of the coupling constant $\alpha_s \simeq 1$ 
and cannot be used as a good expansion parameter. 
Since QCD has not yet been solved exactly, it is necessary to find 
out another parameter, which is small enough to allow a non perturbative 
expansion of amplitudes. 

It is possible~\cite{THOO} to consider as a small parameter 
the quantity $1/N_c$, where $N_c$ is the number of colors 
with corrections to the main term of order $1/{N_c}^2 \thickapprox 0.1$.
Another suitable parameter is $1/N_f$ where $N_f$ is the number of flavors~\cite{VEN}. 
Soft processes involve only the light quark flavor $(u,d,s)$ contributions,
so that$N_f\thickapprox3$.
An intuitive justification for the use of these quantities, lies in the fact that 
these two parameters $(1/N_c, 1/N_f)$ are the only dimensionless quantities 
in massless QCD, being the only suitable quantities to be 
used as expansion parameters in that limiting case.
The formal limit of $N_c \rightarrow \infty$~\cite{THOO} 
$(N_f/N_c \rightarrow 0)$ shows many interesting 
properties and has been intensively studied since in this limiting case 
there is hope to obtain exact solution of the theory. 
However, this approximation is too far from the real world 
(for instance, all resonances are considered infinitely 
narrow $(\Gamma \thickapprox 1/N_c)$).
The case when the ratio  $(N_f/N_c \rightarrow 1)$ is fixed and the 
expansion of all amplitudes in $1/N_f$ or $1/N_c$ 
is carried out~\cite{VEN} is more realistic 
and is called topological expansion, since each term of the series
corresponds to an infinite sets of Feynmann diagrams with definite topology.
The $1/N$ expansion has a dynamical character which means that it 
can work better in some regions of the kinematical variables, 
while in some other regions its convergence can be worse. 
For example, in the case of a binary ($2\rightarrow 2$) amplitude, 
which is a function of two variables $(s,t)$, the $1/N$ expansion 
becomes more accurate as $t$ increases in the positive $t$ region~\cite{CHE}, 
while for high energies additional terms have to be considered. 

\begin{figure}[htbp]
 \begin{center}
 \leavevmode
 \epsfig{file=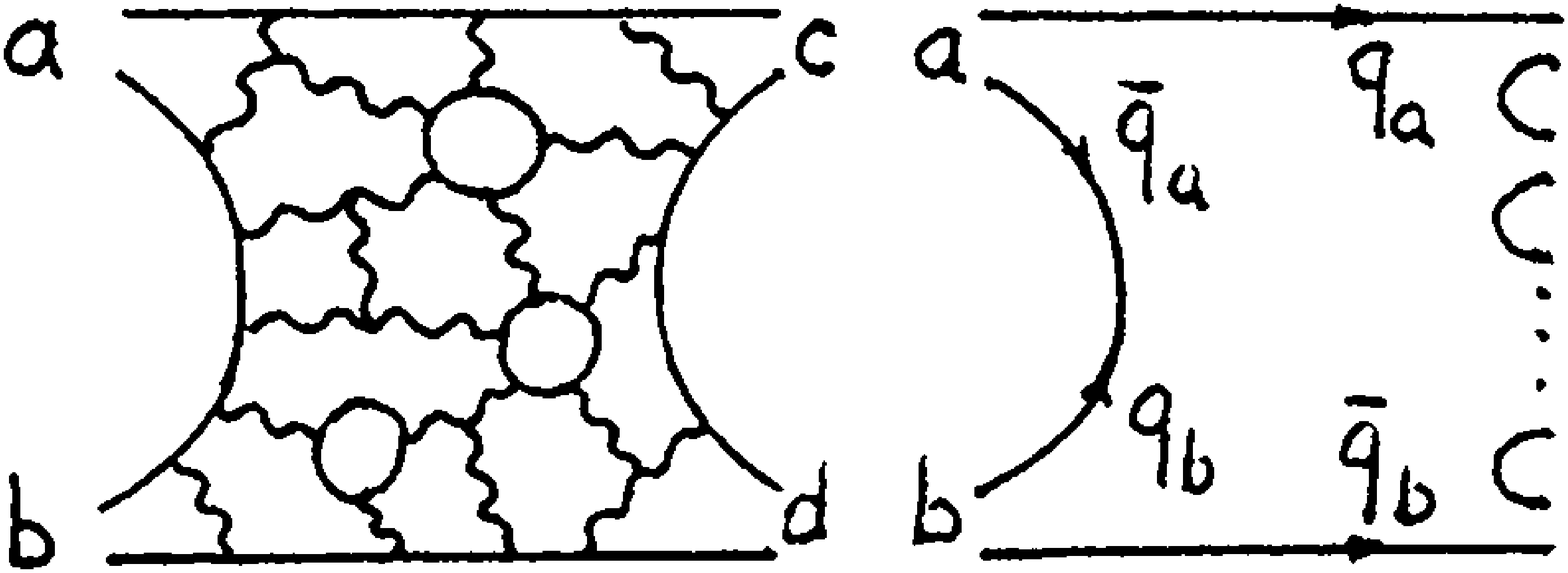,width=6cm}
 \caption{\small Left panel: planar diagram for the reaction $ab \rightarrow cd$. 
Right panel: the same for the reaction $ab \rightarrow X$. Full lines denotes quarks, wavy lines, gluons.}
 \label{fig:planar1}
 \end{center}
\end{figure}

The topological classification of diagrams, in this scheme, 
starts from the amplitude for the binary reaction $ab\rightarrow cd$. 
Here, the main contribution is given by the ``planar'' diagrams, 
shown in figure Fig.~\ref{fig:planar1} where planar means a class of 
diagrams where all lines lie on a plane without hard interactions 
inside the contour delimited by the external quark lines. 
These lines of valence quarks determine the {\it border} of the diagram.\\
\\
All planar diagrams give a contribution of the same order in $1/N$, taking  
into account that each color trace gives a factor $N_c$ and each fermion 
loop a factor $-N_f$, and considering the limit of large $N$ under 
the condition $g^2N\thickapprox 1$. 
It should be emphasized that the topological expansion can be 
applied only to colorless amplitudes.
The unitarity content of the planar diagrams can be reveled 
cutting such a diagram in the $s$-channel as shown in right panel of 
Fig.~\ref{fig:planar1}.
This operation gives the amplitude of multi-particle production, 
which is planar by  itself. 

\begin{figure}[htbp]
 \begin{center}
 \leavevmode
 \epsfig{file=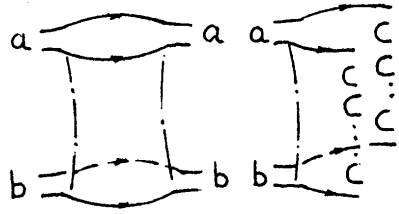,width=6cm}
 \caption{\small Left panel: Cylinder type diagram. Right panel: cutting of these diagrams in the $s$-channel.}

 \label{fig:planar2}
 \end{center}
\end{figure}

Cutting of the diagram on the left panel in Fig.~\ref{fig:planar1}, 
either in the $s$ or the $t$-channel, always contain a $q\bar{q}$ pair, 
which enters in the boundary of the diagram and determines 
the internal quantum numbers (charge, isospin, etc.) exchanged in the given channel. 
At high energies and small momentum transfers these planar diagrams are determined 
by exchange of Regge poles in the $t$-channel.
Diagrams for elastic scattering ({\em i.e.} reactions without 
quantum numbers exchange in the $t$-channel), 
are shown in \ref{fig:planar2} , where the valence quarks of the 
colliding hadrons are conserved in the process of interaction. 

\begin{figure}[hb]
 \begin{center}
 \leavevmode
 \epsfig{file=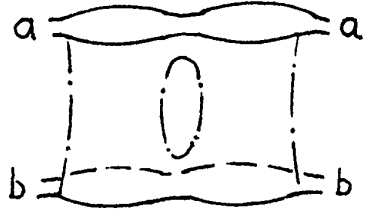,width=6cm}
 \caption{\small Topological diagram with one handle corresponding to the exchange of two 
Pomerons in the $t$-channel.}

 \label{fig:pom}
 \end{center}
\end{figure}

These diagrams are of cylinder type and their $s$-channel cutting 
corresponds to the multi-particle production configurations. 
This is shown by the picture on the right of Fig.~\ref{fig:planar2} which 
corresponds to the production of two chains of particles with the same 
structure as the one shown for the planar diagram cut of \ref{fig:planar1}.
From the $t$-channel point of view, the cylinder diagrams are due to gluon exchange 
in the $t$-channel, and they are usually assumed to give rise to a Pomeron pole. 
More complicated diagrams can be drawn, such the one shown in Fig.~\ref{fig:pom}, 
which has two Pomeron exchange contribution, and one hole in the surface.

Holes in a diagram defines ``handles'', so each topological class of surfaces 
is characterized by a given number of boundaries $(n_b)$ and handles $(n_h)$. 
The topological expansion allows a complete classification of diagrams based 
on the two parameters $(n_b)$ and $(n_h)$ and to determine their dependence 
on $1/N$. 

All the diagrams belonging to a given topological class have the following 
dependence on $1/N$:
\be
T\left(n_b,n_h\right)={\left(\frac{1}{N}\right)}^{n_b+2n_h}
\label{topo}
\ee
so that the contribution of planar diagrams $(n_b=1,n_h=0)$ to the scattering amplitude is $1/N$, 
the cylinder diagram $(n_b=2,n_h=0)$ contribution is $\sim{(1/N)}^2$ and that of the diagram of 
Fig.~\ref{fig:pom} $(n_b=2,n_h=1)$ is $(1/N)^4$. 

This equation is valid for 4-point amplitudes; amplitudes with a larger number of hadrons 
should take into account that each external hadronic state introduces a factor $1\sqrt{N}$, 
due to the normalization of its wave function.
The ratio of the cylinder to the planar diagram is $\sim 1/N$, however for amplitudes 
with vacuum quantum numbers in the $t$-channel, the cylinder type diagrams will dominate 
at high energies. 
This because its relative contribution increases as $s^{\alpha_{P(0)}-\alpha_{R(0)}}$ as 
energy increases as it is implied by the dynamical character of the $1/N$ expansion. 
In many cases, the type of the process (and the number of boundaries) is fixed by 
the quantum numbers in the $t$-channel. 
In this case the expansion parameter depends only by the number of handles and, 
according to Eq.~\ref{topo}, is $(1/N)^2$.

\subsection{Color Tube or Quark Gluon String}
\indent
\par

\begin{figure}[htbp]
 \begin{center}
 \leavevmode
 \epsfig{file=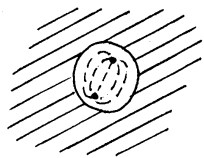,width=3cm}
 \caption{\small Hadron as a ``bubble'' in the vacuum.}

 \label{fig:bubble}
 \end{center}
\end{figure}

The topological expansion gives a useful classification of all QCD 
diagrams but needs a definite space time picture to become predictive.
This picture is based on some assumptions on the properties 
of confinement which are related to the properties of the QCD vacuum. 
Many reasons suggest that QCD vacuum has a complicated structure and 
contains important large scale fluctuations of gluon and quark fields. 
These fluctuations generate vacuum correlations of gluon and quark fields 
which are not described by perturbation theory.
In particular, the vacuum correlator of the square of gluonic field $G^{a}_{\mu \nu}$ 
has been found from the analysis of QCD sum rules~\cite{SHIF}:\\
\be
\frac{\alpha_s}{\pi} \langle 0| G^{a}_{\mu \nu} |0 \rangle \thickapprox 0.012 (\rm{GeV})^4\ ,
\label{corr}
\ee\\
and it can be interpreted as a vacuum condensate connected to the energy density of the vacuum.
\be
\varepsilon_\nu = -{\frac{b}{32}} \langle 0| {\frac{\alpha_s}{\pi}} G^{a}_{\mu \nu} G^{a}_{\mu \nu} 
|0 \rangle \mbox{ with } b={\frac{11}{3}} N_c - {\frac{2}{3}}N_f\ .
\label{cond}
\ee\\
Thus the energy density of the physical vacuum with the gluon condensate 
given by \ref{corr} is lower then for the "empty" vacuum. 
If the vacuum fluctuations will be broken in some region of space, 
then the energy of this region will be increased. 
Such a breaking of vacuum fluctuations in QCD 
(partial or complete) arises naturally inside hadrons, 
due to color fields induced by valence quarks. 
As a result, there will be two different states of vacuum condensate, 
one is the vacuum outside the 
hadrons and the other one inside the hadron. It is assumed 
that the chromo-electric field of quarks do not penetrate far 
into the ordinary vacuum medium since the gluons carry the color charge so that the 
flux lines attract each others. This behavior leads to the property of confinement, 
so that isolated states in the vacuum can exist only as ``white'' states with 
all color lines closed (see Fig.~\ref{fig:bubble}). 
In this picture a hadron can be considered as a bubble in a "vacuum liquid". 

\begin{figure}[hb]
 \begin{center}
 \leavevmode
 \epsfig{file=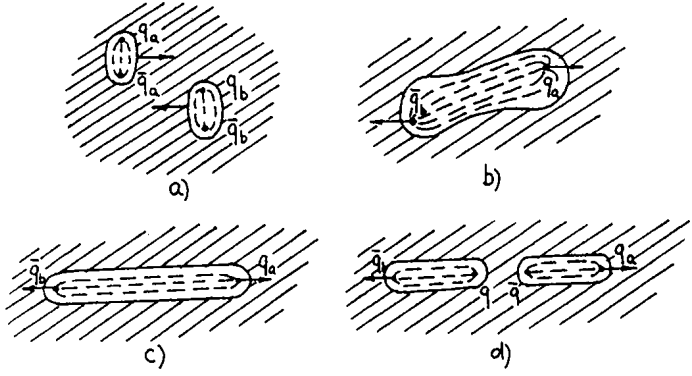,width=10cm}
 \caption{\small Interaction and decay of a color tube.}

 \label{fig:bubint}
 \end{center}
\end{figure}

The process of interaction of such bubbles at high energies (Fig.~\ref{fig:bubint}) 
leads to production of new objects: 
color tubes or strings (a string is a limiting case of a color tube with no transverse 
momentum distribution).

It is necessary to consider separately two cases.

\begin{itemize}
\item{The processes with the annihilation of valence quarks of the colliding hadrons, corresponding to 
the planar diagrams of the $1/N$ expansion.}
\item{The processes  of the diffraction type, where the valence quarks are conserved, connected with cylinder 
type diagrams.}
\end{itemize}

In the first case, after an interaction and annihilation of valence quarks the 
configuration shown in Fig.~\ref{fig:bubint} (b), where the color lines connect 
the spectator quarks moving in opposite directions (in the CM system), is realized. 
As they move apart the region where the vacuum fluctuations are broken 
enlarges (Fig.~\ref{fig:bubint} (c)) and the color tube or string is produced and this 
leads to an increase of the energy. 
At some point, the breaking of the color tube will be energetically favorable 
and the production of $q\bar{q}$  pairs from the vacuum (Fig.~\ref{fig:bubint} (d))
will occur. This process repeats until many white bubbles or hadrons will be produced. 
The time needed for production of a hadron with momentum $p$ and mass $m$ is 
$\tau \sim \frac{p}{m^2}$, so in the CM system the fastest hadrons, 
which contain the spectator quarks $q_a$ and $\bar{q}_b$, will be produced last.
Each produced $q$ and $\bar{q}$ pair has a small relative 
momentum in their rest system which means a small rapidity difference. 
As a result of the Lorentz invariance, finally produced hadrons at high energy 
will be uniformly distributed in the rapidity and will have limited transverse momenta.


\subsection{Non Perturbative Approach in the QGSM}
\indent
\par

The QGSM describes rather well the experimental data on 
exclusive and inclusive hadronic cross sections 
at high energy~\cite{KA5,THOO,VEN,KA9}  and gives new 
important insights into the Regge phenomenology. 
It can explain the quark-gluon content of different Reggeons, 
relate their residues, predict inclusive spectra of
different particles in different kinematical regions, 
find the relations between exclusive and multiple 
production amplitudes through the unitary conditions.
Concerning this last topic, multiple production at high energy, 
the QGSM is very similar to the dual parton model~\cite{CAP}. 

\begin{figure}[htbp]
 \begin{center}
 \leavevmode
 \epsfig{file=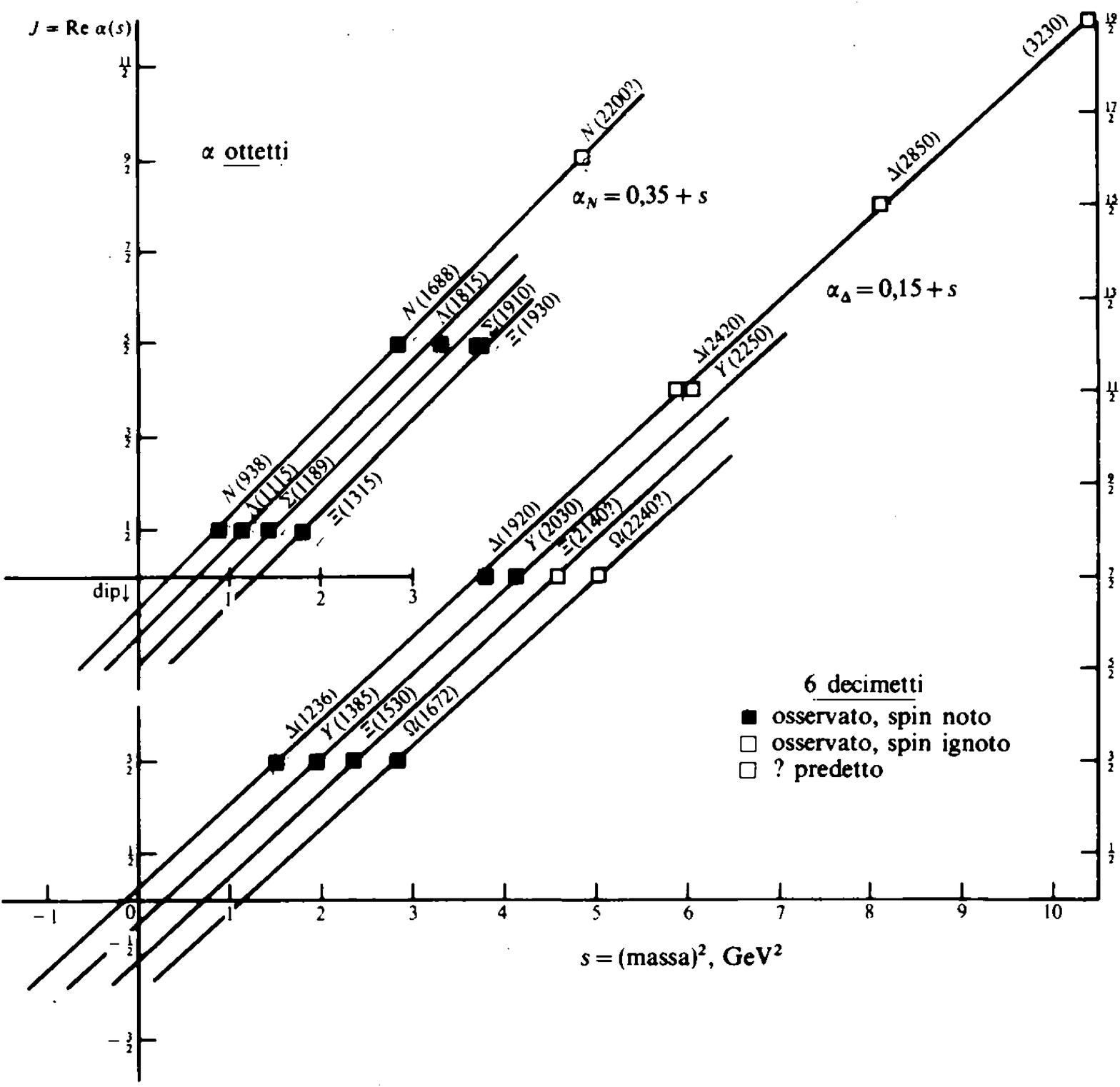,width=10cm}
 \caption{\small Simple linear baryon Regge trajectories separated by $\Delta J=2$~\cite{SEGR}.}

 \label{fig:regge}
 \end{center}
\end{figure}

The Regge pole analysis of the exclusive hadronic reactions~\cite{COLL} was 
formulated much before QCD and for this reason is very often seen as a 
dormant topic being so strongly identified with the pre-quark 
era of the $S$-matrix and dispersion relations approach to strong interactions.  
This point of view is not correct~\cite{BJO} since the description 
of high energy behavior in terms of singularities in the complex angular 
momentum plane of two body scattering amplitudes  
is completely general being based on the general principles of unitarity, 
analiticity and crossing symmetry. 
In recent years it has been resurrected by all the theories 
which aim to a string formulation for hadrons 
and the conjecture of Regge that high energy behavior of two 
body scattering amplitudes might be describable in 
terms of moving poles in the $J$-plane is {\it experimentally verified } 
and all hadrons do lie on trajectories 
which connects the particle mass with 
its total angular momentum (Fig.~\ref{fig:regge}).

Despite the very broad applications of such approaches to the different hadronic reactions and to the 
inclusive hadron-nucleus collisions at rather high energy they were poorly used in the description of exclusive 
reactions with nuclear targets. One of the main reason is apparently related to the fact that those approaches 
were originally introduced to describe high energy data, but data on exclusive reactions with nuclear targets 
have been limited, up to now, mainly to the intermediate energy region.
Let us say something more on the region of validity in $s$ and $t$ of this approach. From the quantum mechanical 
point of view, high energy means that the wavelength of the photon is much smaller than the radius of the 
target nucleon which in QCD can be taken as the radius of confinement $R_0$, which is about 1 Fm.
Therefore, at $E_\gamma \geq$1~GeV, the necessary condition $E_{\gamma}R_0 \gg 1$ can be considered as fairly 
satisfied. In the Regge approach, high energy means $s \gg t_{char}$ where $t_{char}$ is some characteristic 
value of $t$ which is usually taken as $\simeq 1$ (GeV)$^2$.
As the region of validity in $t$ or $u$ is concerned, there is a common convention that this is limited by $|t|$ 
or $|u|$ $\leq 1$ (GeV/c)$^2$. This convention is based on the consideration that with increasing of $|t|$ or 
$|u|$ the secondary singularities in the complex angular momentum plane as daughter Regge trajectories 
or branch points~\cite{LEON} may also become important. However, this considerations cannot be formulated in a 
quantitative way because the residues of Regge poles and discontinuities of Regge cuts are usually 
parameterized and are different for different reactions. 
This means that at the phenomenological level, the description of data include contributions of some effective 
Regge trajectories which in turn may include the effects of secondary singularities.

A different philosophy is used in the QGSM since each graph in this model is classified
according to its topology and the corresponding Regge singularity is fixed by the quark-gluon 
content of the graph itself. 
Therefore the corresponding amplitude can be considered for all values 
of $s$ and $t$ as an analytic functions of those variables.
The QGSM gives the possibility to calculate parameters that before were considered 
only at the phenomenological level in the Regge theory and to establish a bridge between QCD
and the Regge approach to strong interactions.

\subsection{QCD Content of Regge Trajectories}
\indent
\par

There is experimental evidence and theoretical reasons that both meson and baryon states should 
be connected, on the $J$ vs $M^2$ plane by an essentially non linear relation (trajectory) instead of the 
simple scheme shown in Fig.~\ref{fig:regge}.
In fact, the starting point for Regge's work was the partial wave expansion of the scattering amplitudes 
\cite{LEON} but here we would like to reinterpret the original idea of Regge pole 
and trajectory using the modern QCD concepts and language
to make evident that the Regge trajectories must contain some non linearity.

\begin{figure}[htbp]
 \begin{center}
 \leavevmode
 \epsfig{file=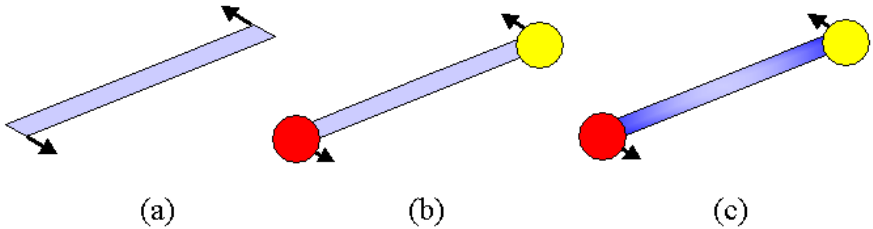,width=12cm}
 \caption{\small Color strings: (a) massless string, (b) homogeneous string connecting a $q\bar{q}$ couple, 
(c) inhomogeneous string with variable tension due to color vacuum polarization.}

 \label{fig:string}
 \end{center}
\end{figure}

Let us consider a meson state composed of a $q\bar{q}$ pair connected by the color flux 
tube (Fig.~\ref{fig:string}).

In a simple potential description the energy can be written as:

\be
H=\sqrt{\overrightarrow{p_1}^2+ m_1^2} + \sqrt{ \overrightarrow{p_2}^2+ m_2^2} + V(r)\ , 
\ee\\
where $V(r) = \sigma r$ is a linear potential and $\sigma \simeq 400$ (MeV)$^2$. 

For a sake of simplicity, (Fig.~\ref{fig:string} (a)) massless quarks will be
considered. In the CM system the energy will be  $H=2p+\sigma r$ 
where the orbital angular momentum $\overrightarrow{\ell}$ will be 
$\overrightarrow{r} \times \overrightarrow{p}$ so that:\\ 
\be
H(r)=2\frac{\ell}{r} + \sigma r\ .
\ee\\
From the relation $\frac{\partial H}{\partial r}= - \frac{2\ell}{r^2}+ \sigma =0$ 
the minimum of $H$ is found at $r_0=\sqrt{\frac{2\ell}{\sigma}}$. If this position is 
chosen as the ground state of the system, 
the energy can be expressed as: $H(r_0)=E=2 \sqrt{2} \sqrt{\sigma \ell}$ or 
$E^2=8\sigma \ell $.

This relation gives the dependence of the angular momentum on the energy of the system: 
$\ell=\frac{1}{8\sigma}E^2$. Performing the substitutions
$\alpha \Rightarrow \ell$ and $E^2 \Rightarrow t$:
\be
\alpha(t)=\frac{1}{8\sigma}t\ ,
\ee\\
so that the expression for a simple linear Regge trajectory can be readily recognized.

The choice of a massless $q\bar{q}$ system gives rise to a linear relation in the Mandelstam variable $t$. 
The exact calculation yields  $\alpha(t)=\frac{1}{2\pi\sigma}t=\alpha^\prime t $ 
for our string with massless ends where $\alpha^\prime(0) \sim 1$~GeV$^2$.\\
This scheme can be generalized to a string connecting two massive quarks (Fig.~\ref{fig:string} (b)). 
In this case a static tension is introduced on the color string, and
an additional term is collected in the Regge trajectory:

\be
\alpha(t)=\alpha(0)+\alpha^\prime(0)t\ .
\ee\\
Another correction term can be obtained considering the screening effects 
introduced by the vacuum polarization in QCD which complicates the string 
structure.
A distance dependent tension (Fig.~\ref{fig:string} (c)) can account for this 
effects since the formation of $q\bar{q}$ pairs from the vacuum weakens the flux tube
opening ``holes'' in the string. 
A possible form for such a potential due to  $q\bar{q}$ loops comes from 
lattice QCD calculations (see Ref.~\cite{BORN}) and can 
be parametrized in terms of the screening length $\mu^{-1}$:

\be
\label{lattice}
V(r) = \left(-\frac{\alpha}{r}+\sigma r\right)\frac{1-e^{-\mu r}}{\mu r}\ ,
\ee\\
where $\alpha=0.21\pm 0.01$ Fm. 

This potential has the correct Coulomb behavior at small distances and approaches a constant at large
distances. The ratio $\frac{\sigma}{\mu}$ represents the splitting energy of the $q\bar{q}$ pair, with 
$\mu^{-1}=0.90 \pm 0.20$ Fm.

As was shown in the simple model described above, it is expected that
composite hadronic systems such mesons and baryons give rise to (slightly) non linear
Regge trajectories. 
In the QGSM, the non linearity of baryon Regge trajectories in incorporated 
according to Refs~\cite{LYU1,BRA,INO,CHI}.
For small momentum transfers, the non linear part is negligible, 
but when $t\geq 1 $(GeV/c)$^2$ effects introduced by the non linear
pert of the Regge trajectories become critical and must be considered.

\section{Quark Hadron Reactions in the QGSM}
\indent
\par

The QGSM was originally developed to describe hadron-hadron 
collisions at high energy. In order to better define its extension to photo-reactions
a brief review of its application and predictions 
for hadronic collisions will be given.

Reactions as $pp \rightarrow d\pi^+$ and  $\bar{p}d \rightarrow p\pi^-$ 
are dominated by diagrams with three valence quark exchanges 
in the $t$-channel~\cite{KA10}  
in analogy to the deuteron photo-disintegration case (see Ref.~\cite{K93} and Refs~\cite{KA11,KKG}). 

\begin{figure}[htb]
 \begin{center}
 \leavevmode
 \epsfig{file=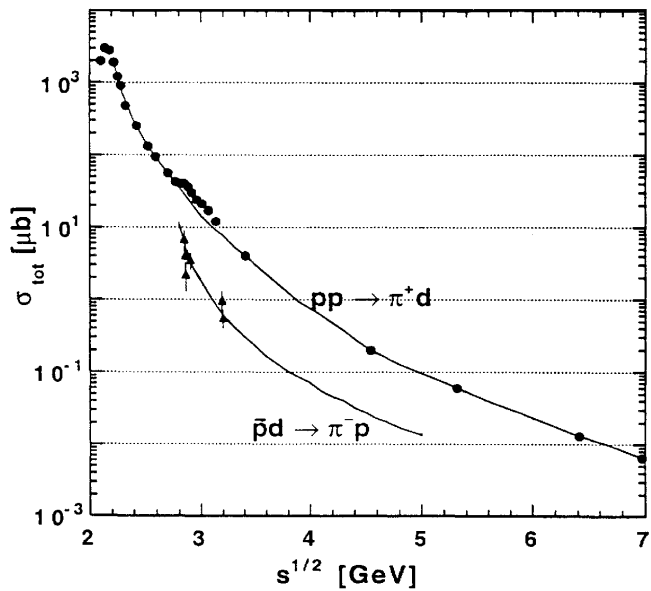,width=9cm}
\caption{\small Total cross sections of the reactions  $pp \rightarrow d\pi^+$ and 
$\bar{p}d \rightarrow p\pi^-$ as a function of the CM energy, and predictions of
the QGS model.}
 \label{fig:ppqgsm}
 \end{center}
\end{figure}

Fig.~\ref{fig:ppqgsm} shows the total cross section data for 
$pp \rightarrow d\pi^+$~\cite{ALLA}.
It is seen that the QGSM very well reproduces the data 
with the single exception of the near threshold region. 
The QGSM describes quite well also the data from the $\bar{p}d \rightarrow p\pi^-$ 
reaction which have been obtained from the reversed channel $\pi^-p\rightarrow \bar{p}d $ 
\cite{BAG,BIZZ}.

\subsection{Transition Probabilities and Transition Amplitudes}
\indent
\par

\begin{figure}[htb]
 \begin{center}
 \leavevmode
 \epsfig{file=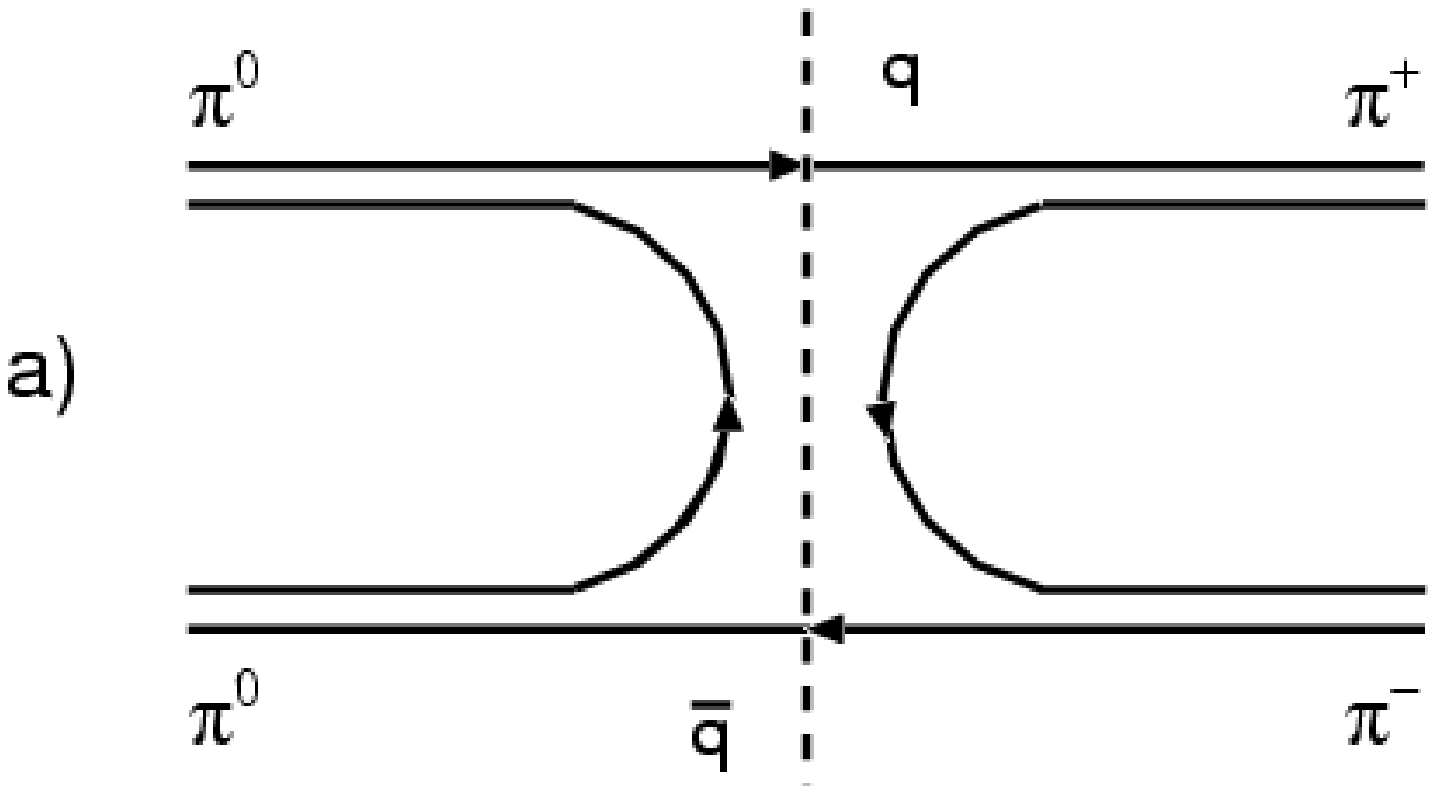,width=4.5cm}
\epsfig{file=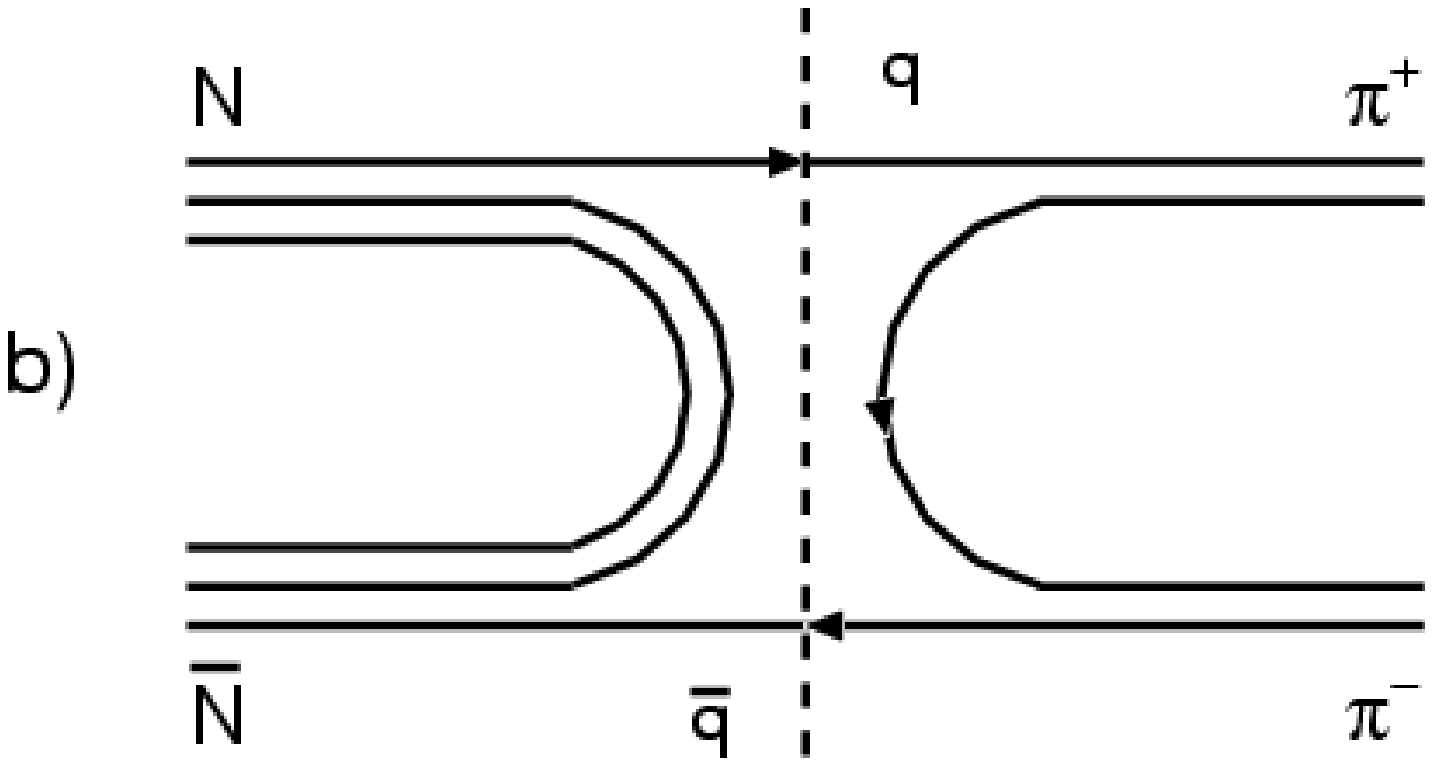,width=4.5cm}
\epsfig{file=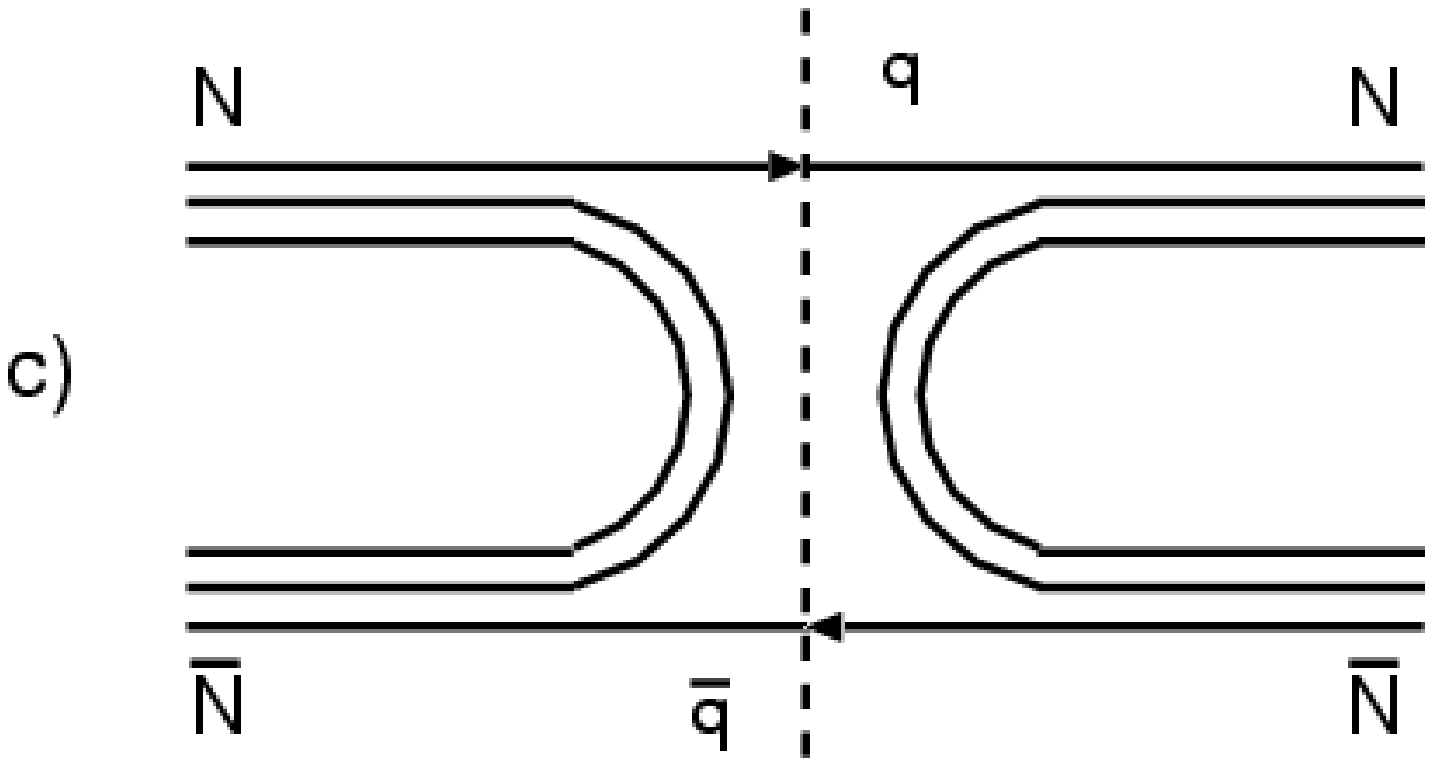,width=4.5cm}
 \caption{\small Planar diagrams describing the binary reactions 
\mbox{$\pi^0\pi^0\rightarrow \pi^+\pi^-$} (a: left panel), \mbox{$N\bar{N} \rightarrow \pi^+\pi^- $} (b: central panel), and 
\mbox{$N\bar{N} \rightarrow N\bar{N}$} (c: right panel).
}
 \label{fig:binary}
 \end{center}
\end{figure}

Each diagram of the topological expansion has a simple interpretation within
the framework of the space-time pattern formulated in terms of a color string 
(see Refs~\cite{KA9,BBG}).
For instance, at high energy $\sqrt{s}$, 
the binary reaction shown in Fig.~\ref{fig:binary} (a)
occurs due to a specific quark configuration in each pion, 
where (in the CM system) one quark (or anti-quark) takes almost
the entire hadron momentum and plays the role of a spectator, 
while the valence anti-quark or quark is rather slow. 

The difference in the rapidities $\Delta y$ between the quark 
$q$ and the anti-quark $\bar{q}$ in each pion is

\be
\Delta y =y_q - y_{\bar{q}} \simeq \frac{1}{2}\ln{\frac{s}{s_0}}\ ,
\label{rapdiff}
\ee\\
with the scale $s_0 \simeq 1 $(GeV)$^2$. 

When the two slow valence particles $q$ and $\bar{q}$ from the two initial 
$\pi^0$s annihilate, the fast spectator quark and anti-quark continue to move in the 
previous directions and form a color string in the intermediate state. 
After that, the string breaks to produce  
$q\bar{q}$-pairs from the vacuum and the formation of the $\pi^+\pi^-$ system 
takes place in the final state.
The same space-time pattern holds also for the situation depicted in Fig.~\ref{fig:binary} (b)
with the only difference, 
that the string is formed after the annihilation of a diquark anti-diquark pair from the 
$N\bar{N}$ system in the initial state.
Correspondingly, the graph of Fig.~\ref{fig:binary} (c) shows the formation of 
the $q\bar{q}$ string due to annihilation of the 
valence diquark anti-diquark pair in the initial state and the production of a diquark anti-diquark pair 
as a consequence of the breaking of the string.
The annihilation of the initial  $q\bar{q}$  pair takes place, when a gap in rapidity
of the valence $q$ and $\bar{q}$  is small and both interacting partons are almost
at rest in CM system so that the relative impact parameter ${\bf b}_\perp -{\bf b}_{0\perp}$ 
is less than their interaction radius.
It is possible to prove that the probability to find a valence quark with a rapidity $y_q$ at impact
parameter ${\bf b}_\perp$ inside a hadron can be written as (\cite{KA1,KA5,BBG})

\begin{equation}
\label{probability} w\left(y_q-y_0,{\bf b}_{\perp}-{\bf
b}_{0\perp}\right)= \frac c{4\pi R^2(s)}\exp \left[ -\beta
(y_q-y_0)-\frac{({\bf b}_{\perp}-{\bf
b}_{0\perp})^2}{4R^2(s)}\right] \ ,
\end{equation}\\
where $c$ is a normalization constant, $y_0$ is the average rapidity, ${\bf b}_{0\perp}$ 
is the transverse coordinate in the CM system in the impact parameter representation. 
Furthermore, it is possible to relate the parameter 
$\beta$ and the effective interaction radius squared $R^2(s)$ in \ref{probability} 
that specify the quark distribution
inside a hadron, to the phenomenological parameters of a Regge trajectory 
$\alpha_i(t)$ which gives the dominant 
contribution to the amplitude for the considered planar graph. In this case one gets

\begin{equation}
\label{r2} R^2(s)=R_0^2+\alpha ^{\prime }_i (y_q-y_0)\ ,\quad
\beta =1-\alpha_i(0) \ ,
\end{equation}\\
where $\alpha^{\prime}_i=\alpha^{\prime}_i(0)$ is the slope of the dominant Regge trajectory.

Due to the creation of a string in the intermediate state the amplitude of a binary reaction 
$ab \rightarrow cd$ has the $s$-channel factorization property, {\em i.e.} the probability for the
string to produce different hadrons in the final state does not depend on the type of the
annihilated quarks and it is only determined by the flavors of the produced quarks.
The same independence also holds for the production of the color string in the intermediate
state from the initial hadron configuration: it depends only on the type of the annihilated quarks.
This $s$-channel factorization has been formulated in Refs~\cite{KA1,KA5,BBG} in terms 
of transition probabilities
as defined in Eq.~(\ref{probability}).

An example of this can be seen for the reactions shown in Fig.~\ref{fig:binary} 
as well as for the crossed reactions
\mbox{$\pi^+\pi^-\rightarrow \pi^0\pi^0$}, \mbox{$\pi^+\pi^-  \rightarrow N\bar{N}$}, and 
\mbox{$N\bar{N} \rightarrow N\bar{N}$}.

For large $s$ values and finite values of the 4-momentum transfer 
squared $t$, these channels are described by planar diagrams with $t$-channel 
valence-quark exchanges as in Fig.~\ref{fig:binary}, where the single and double 
solid lines correspond to valence quarks and diquarks, respectively
(soft gluon exchanges are not shown).
Introducing the hadron($\gamma$)-quark and quark-hadron transition amplitudes,
analytic expressions for the form factors have been found in the time-like region
which admit continuation in the space-like region.

This approach can be generalized~\cite{KAKO}
introducing the amplitudes $\widetilde{T}^{ab\to q\bar q} (s,{\bf
b}_{\perp})$ and $\widetilde{T}^{q\bar q\to cd}(s,{\bf b}_{\perp
})$, that describe the formation and the fission of an
intermediate color string.
The amplitude for the reaction $ab \to cd$ described by the planar graph   
can be written, using the $s$-channel factorization
property, in the form of a convolution of two amplitudes:

\begin{equation}\label{ImpRepr}
A^{ab\rightarrow cd}\left(
s,{\bf q}_{\perp }\right) = \frac i{8\pi ^2s}\int d^2{\bf k}_{\perp } \
T^{ab\rightarrow q\overline{q}} \left( s,{\bf k }_{\perp }\right)
T^{q\overline{q}\rightarrow cd}\left( s,{\bf  q }_{\perp }-{\bf
k}_{\perp }\right)
\end{equation}\\
in momentum representation, or as the product

\begin{equation}\label{factorization}
\widetilde{A}^{ab\rightarrow cd}(s,{\bf b}_{\perp })=\frac
i{2s}\ \widetilde{T}^{ab\to q\bar q}(s,{\bf b}_{\perp })\ \widetilde{T}
^{q\bar q\to cd}(s,{\bf b}_{\perp })\ ,
\end{equation}\\
in the impact-parameter representation.\\
The solution for the quark-hadron transition amplitudes $T^{ q
\overline{q} \rightarrow \pi \overline{\pi } }\left( s,{\bf
k}_{\perp }\right)$ and $T^{q \overline{q}\rightarrow
N\overline{N}}\left( s,{\bf k}_{\perp }\right) $ at large
invariant energy $\sqrt{s}$ can be found using single Regge-pole
parameterizations  of the binary hadronic amplitudes $A^{\pi
^{0}\pi ^{0}\rightarrow \pi^{+} \pi^{-}}$,
$A^{N\overline{N}\rightarrow \pi \overline{\pi}}$ and
$A^{N\overline{N}\rightarrow N\overline{N}}$
\begin{equation}
 \label{MBDReggeAmplitude}
\begin{array}{c}
\displaystyle A^{\pi ^{0} \pi ^{0}\rightarrow \pi
^{+}\pi ^{-}}\left( s,t\right) =N_M\ \left( -\frac
s{m_0^2}\right) ^{\alpha _M\left( t\right) }\exp \left( R_{0M}^2t\right), \\
\displaystyle A^{ N
\overline{N}\rightarrow \pi \overline{\pi } }
\left( s,t\right) =N_B\ \left( -\frac
s{m_0^2}\right) ^{\alpha _B\left( t\right) }\exp \left( R_{0B}^2t\right), \\ \\
\displaystyle A^{N\overline{N}\rightarrow
N\overline{ N }}\left( s,t\right) =N_D\ \left( -\frac
s{m_0^2}\right) ^{\alpha _D\left( t\right) }\exp \left( R_{0D}^2t\right).
\end{array}
\end{equation}\\
Here $\alpha _M\left( t\right) $, $\alpha _B\left( t\right) $ and
$\alpha_D\left( t\right) $ are the dominant meson, baryon and
diquark-antidiquark trajectories while $N_M$, $N_M$ and $N_D$ are
normalization constants; \mbox{$m_{0}^2=s_0$} and \mbox{$R_{0 i}$} is the
interaction radius for the $i$-th trajectory. The following
intercepts and slopes are found for the dominant Regge trajectories
\begin{equation}
\label{ReggePoles}\alpha _M\left( 0\right) \simeq 0.5,\quad \alpha _B\left(
0\right) \simeq -0.5,\quad \alpha _D\left( 0\right) \simeq -1.5
\end{equation}
and
\begin{equation}\label{ReggePoles2}
\alpha _M^{\prime }\left( 0\right) \simeq \alpha _B^{\prime }\left( 0\right)
\simeq \alpha _D^{\prime }\left( 0\right) \simeq 1.0\ \rm{GeV}^{-2}   .
\end{equation}\\
Using equations (\ref{factorization}) and (\ref{MBDReggeAmplitude}) 
the amplitudes $\widetilde{T}^{q\overline{q}
\rightarrow \pi \overline{\pi}}\left( s,{\bf b}_{\perp }\right) $
and $\widetilde{T}^{q \overline{q}\rightarrow N\overline{N}}\left(
s,{\bf b} _{\perp }\right)$ can be written as
\begin{equation} \label{PionNucleonFragmentation}
\begin{array}{c}
\displaystyle \widetilde{T}^{q\overline{q} \rightarrow \pi \overline{\pi
}}(s,  {\bf b}_{\perp})=N_M^{1/2}\frac 1{2\sqrt{\pi }R_M\left( s\right)
}\left( -\frac s{m_0^2}\right) ^{\left( \alpha _M\left( 0\right) +1\right)
/2}\exp \left( -  \frac{{\bf b}_{\perp }^2}{8R_M^2\left( s\right) }\right), \\
\\
\displaystyle \widetilde{T}^{q\overline{q}\rightarrow N \overline{N}}(s,{\bf
b}_{\perp})= N_D^{1/2}\frac 1{2\sqrt{\pi }R_D\left(s\right) }
\left( -\frac s{m_0^2}\right)^{\left( \alpha _D\left( 0\right)+1\right) /2}
\exp \left( -\frac{{\bf b}_{\perp }^2}{8R_D^2\left( s\right)}\right),
\end{array}
\end{equation}\\
where $R_M\left( s\right) $ and $R_D\left( s\right) $ are the
effective interaction radii given by
\begin{equation}\label{Radii}
\begin{array}{c}
\displaystyle R_M^2\left( s\right) =R_{0M}^2+\alpha _M^{\prime }\left( 0\right) \ln \left(
-\frac s{m_0^2}\right),  \\ \\
\displaystyle R_D^2\left( s\right) =R_{0D}^2+\alpha _D^{\prime }\left( 0\right) \ln \left(
-\frac s{m_0^2}\right).
\end{array}
\end{equation}\\
The substitution of the amplitudes (\ref{PionNucleonFragmentation})
into the factorization formula (\ref{factorization}) gives:
\begin{equation}\label{CrossAmplitude}
\begin{array}{c}
\displaystyle \widetilde{A}^{N\overline{N}\rightarrow \pi \overline{\pi }}
(s,{\bf b}_{\perp })= \\ \\
\displaystyle  \left( N_MN_D\right) ^{1/2}
\frac 1{4\pi R_D\left( s\right) R_M\left(  s\right)
}\left( -\frac s{m_0^2}\right) ^{\frac 12\left( \alpha _D\left(
0\right) + \alpha _M\left( 0\right) \right) }
\exp \left[ -{\bf b}_{\perp }^2 \left(
\frac 1{8R_M^2\left( s\right) }+\frac 1{8R_D^2\left( s\right) }\right) \right].
\end{array}
\end{equation}\\
Consistency of Eqs. (\ref{CrossAmplitude}) and
(\ref{MBDReggeAmplitude}) requires the following
relations between the Regge parameters and normalization constants
(\cite{KA5,KAKO,KA1}):
\begin{equation}\label{OurPlanar} \begin{array}{c}
\displaystyle 2\frac 1{R_B^2\left( s\right) }=\frac 1{R_M^2\left( s\right)
}+\frac 1{R_D^2\left( s\right) } , \\ \\
\displaystyle 2\alpha \left( 0\right) _B=\alpha _D\left( 0\right) +\alpha _M\left(
0\right) ,
\end{array}
\end{equation}
\begin{equation}
\label{NormCnst}
\left( N_M N_D\right) ^{1/2}\frac 1{R_D\left( s\right)
R_M\left( s\right) }=N_B\frac 1{R_B^2\left( s\right) } .
\end{equation}\\
If only light $u$, $d$ quarks are involved it can be assumed that
(\cite{KA5,KA11,KAKO})
\begin{equation}\label{UDPlanar}
\begin{array}{c}
\displaystyle \alpha _M^{\prime }\left( 0\right) =\alpha _B^{\prime }\left(
0\right) =\alpha _D^{\prime }\left( 0\right)\equiv  \alpha ^{\prime }\left(
0\right),\\ \\
\displaystyle R_{0M}^2\left( 0\right) =R_{0B}^2\left( 0\right) =R_{0D}^2\left(
0\right)  \equiv  R_{0}^2\left( 0\right),\\ \\
\left( N_MN_D\right) ^{1/2}=N_B .
\end{array}
\end{equation}\\
Then the relations (\ref{OurPlanar}) and (\ref{NormCnst}) can be
fulfilled at all $s$. Otherwise, they can only be satisfied at
sufficiently large $s$ (Ref.~\cite{HITO}).

\section{Deuteron Photo-disintegration in the QGSM}
\label{spin}
\indent
\par

It is convenient start the description of the photo-disintegration reaction considering
the amplitudes for spin-less constituents before going over to the case in which 
the photon spin is included.

\begin{figure}[htbp]
\begin{center}
\leavevmode
\epsfig{file=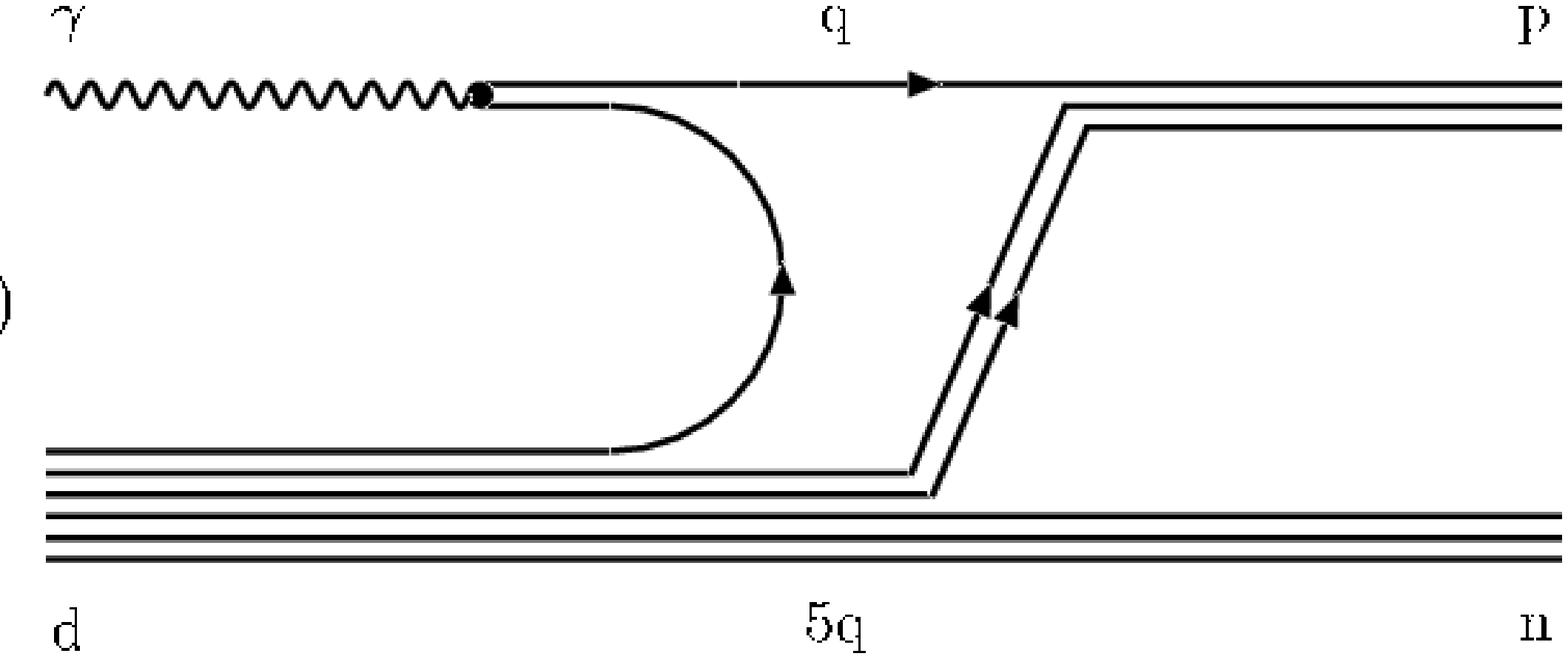,width=5cm}
\epsfig{file=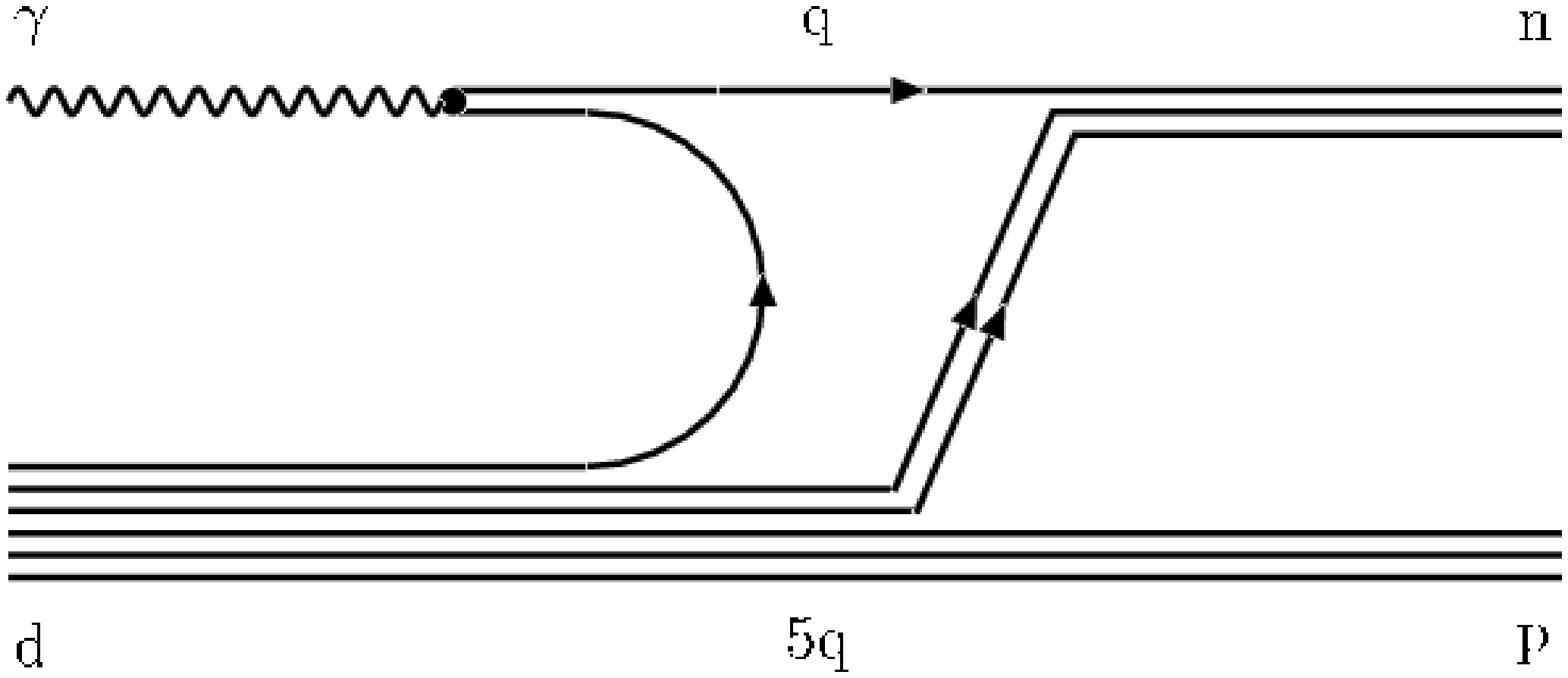,width=5cm}	
\caption{ \small Diagrams describing three valence quark exchanges in the $t$-channel 
(left panel) and $u$-channel (right panel).}
 \label{fig:vg-gdpn}
 \end{center}
\end{figure}

From Eq.~\ref{factorization} the transition 
amplitude corresponding to each quark diagram of 
Fig.~\ref{fig:vg-gdpn} can be written as:

\begin{equation}\label{factorizgamd}
\widetilde{A}^{\gamma d\to pn}(s,{\bf b}_{\perp })=\frac
i{2s}\ \widetilde{T}^{\gamma d\to q (5q)}(s,{\bf b}_{\perp })\ \widetilde{T}
^{q (5q)\to pn}(s,{\bf b}_{\perp }),
\end{equation}\\
where the amplitudes  $\widetilde{T}^{\gamma d\to q (5q)}(s,{\bf
b}_{\perp })$ and $\widetilde{T}^{q (5q)\to pn}(s,{\bf b}_{\perp
})$ are given by (cf. (\ref{PionNucleonFragmentation}))

\begin{equation} \label{deutFragmentation}
\begin{array}{c}
\displaystyle \widetilde{T}^{\gamma d \rightarrow q(5q)}
(s,  {\bf b}_{\perp})=N_{M(6q)}^{1/2}\frac 1{2\sqrt{\pi }R_{M(6q)}
\left( s\right)
}\left( -\frac s{m_0^2}\right) ^{\left( \alpha _M\left( 0\right) +1\right)
/2}\exp \left( -  \frac{{\bf b}_{\perp }^2}{8R_{M(6q)}^2\left( s\right) }
\right), \\
\\
\displaystyle \widetilde{T}^{q(5q)\rightarrow pn}(s,{\bf
b}_{\perp})= N_{D(6q)}^{1/2}\frac 1{2\sqrt{\pi }
R_{D(6q)}\left(s\right) }
\left( -\frac s{m_0^2}\right)^{\left( \alpha _D\left( 0\right)+1\right) /2}
\exp \left( -\frac{{\bf b}_{\perp }^2}{8R_{D(6q)}^2\left( s\right)}\right).
\end{array}
\end{equation}\\
Here the effective interaction radii $R_{M(6q)}\left( s\right)$ 
and $R_{D(6q)}\left( s\right)$ are defined as

\begin{equation}\label{Radii1}
\begin{array}{c}
\displaystyle R_{M(6q)}^2\left( s\right)
=R_{0M(6q)}^2+\alpha _M^{\prime }\left( 0\right) \ln \left(
-\frac s{m_0^2}\right),  \\ \\
\displaystyle R_{D(6q)}^2\left( s\right) =
R_{0D(6q)}^2+\alpha _D^{\prime }\left( 0\right) \ln \left(
-\frac s{m_0^2}\right),
\end{array}
\end{equation}
where $R_{0M(6q)}^2$ and $R_{0D(6q)}^2$ are, in general, different
from $R_{0M}^2$ and $R_{0D}^2$ in \mbox{Eq.~(\ref{Radii})}.\\

Since the photon possess an hadronic structure, 
a photo-reaction such as $\gamma d \rightarrow pn$ 
could be assimilated to a hadron-hadron collision.
In this case, the photon spin must be taken into account 
since its transversal polarization leads to a non-trivial angular dependence 
of the residue of the amplitude~\cite{K93}.
Spin effects in the QGSM where first introduced using the $s$-channel 
factorization property of amplitudes, developed in Ref.~\cite{KAKO} with respect to 
the description of the electromagnetic nucleon form factors $F_1$ and $F_2$.

The resulting form factors for nucleons and pions 
comply well with the experimental data for both positive 
and negative values of $q^2$ (see Refs~\cite{ARNO,ARMS,BISE,BEBE,BISE1,MILA}.

In the case of the deuteron photo-disintegration, the
amplitude can be written in the form:

\begin{eqnarray}\label{ImpReprspin}
\lefteqn{ \langle p_3, \lambda_{p}; p_4,
\lambda_{n} | \hat{T}\left(s,{\bf p}_{3 \perp}\right)| p_2,
\lambda_{d}; p_1, \lambda_{\gamma}\rangle =}  \nonumber \\
&&\frac i{8\pi ^2s}\int d^2{\bf k}_{\perp } \
\langle \lambda_{p};
\lambda_{n} | \hat{T}^{q(5q)\rightarrow pn}\left(s,{\bf k}_{\perp}\right)|
\lambda_{q};\lambda_{(5q)}\rangle\nonumber \\
&& \langle \lambda_{q}; \lambda_{(5q)} | \hat{T}^{\gamma d\rightarrow q(5q)}
\left(s,{\bf p }_{ 3 \perp}-{\bf k }_{\perp}\right)|
\lambda_{d};\lambda_{\gamma}\rangle \ ,
\label{convolutionspin}
\end{eqnarray}\\
where $p_1$, $p_2$, $p_3$, and $p_4$ are the 4-momenta of the
photon, deuteron, proton, and neutron, respectively, while
$\lambda_i$ is the $s$ channel helicity of the $i$-th particle.
Furthermore, making the simplifying assumption that the spin of the
$(5q)$ state is $1/2$, the amplitude $\hat
{T}^{\gamma d\rightarrow q(5q)}$ can be written as

\begin{eqnarray}
\lefteqn{ \langle \lambda_{q};
\lambda_{(5q)} | \hat{T}\left(s,{\bf k}_{\perp}\right)|
\lambda_{d};\lambda_{\gamma}\rangle =}  \nonumber \\
&& \bar u_{\lambda_q}(p_q) \hat {\epsilon}_{\lambda_{\gamma}}
\left(\frac{-\hat k+m_q}{k^2-m_q^2}\right)
\hat {\epsilon}_{\lambda_d} v_{\lambda_{(5q)}}(p_{(5q)}) \
D^{\gamma d\rightarrow q(5q)}(s,{\bf k}_{\perp})\ ,
\label{gamdq5qAmpl}
\end{eqnarray}\\
where ${\epsilon}_{\lambda_{d}}$ and
${\epsilon}_{\lambda_{\gamma}}$ are the deuteron and photon
polarization vectors, $D^{\gamma d\rightarrow q(5q)}(s,{\bf
k}_{\perp})$  is the scalar amplitude and  $m_q$ is the quark
mass. In analogy to $q\overline{q}\rightarrow N\overline{N}$,
which was analyzed in Ref.~\cite{KAKO}, we can describe the spin
structure of the amplitude $\hat {T}^{q(5q)\rightarrow pn}$ in
terms of eight invariant amplitudes

\begin{eqnarray}
\lefteqn{ \langle \lambda_{p};
\lambda_{n} | \hat{T}^{q+(5q)\rightarrow
pn}\left(s,{\bf k}_{\perp}\right)|
\lambda_{q};\lambda_{(5q)}\rangle =}  \nonumber \\
&&D_1(s,{\bf k}_{\perp})\ \delta_{\lambda_p \, \lambda_q}
\delta_{\lambda_n \, \lambda_{(5q)}}+
D_2(s,{\bf k}_{\perp})\ (\sigma_y)_{\lambda_p \, \lambda_q}
\delta_{\lambda_n \, \lambda_{(5q)}}+\nonumber\\
&&D_3(s,{\bf k}_{\perp})\ \delta_{\lambda_p \, \lambda_q}
(\sigma_y)_{\lambda_n \, \lambda_{(5q)}}+
D_4(s,{\bf k}_{\perp})\ (\sigma_x)_{\lambda_p \, \lambda_q}
(\sigma_x)_{\lambda_n \, \lambda_{(5q)}}+\nonumber\\
&&D_5(s,{\bf k}_{\perp})\ (\sigma_y)_{\lambda_p \, \lambda_q}
(\sigma_y)_{\lambda_n \, \lambda_{(5q)}}+
D_6(s,{\bf k}_{\perp})\ (\sigma_z)_{\lambda_p \, \lambda_q}
(\sigma_z)_{\lambda_n \, \lambda_{(5q)}}+\nonumber\\
&&D_7(s,{\bf k}_{\perp})\ (\sigma_x)_{\lambda_p \, \lambda_q}
(\sigma_z)_{\lambda_n \, \lambda_{(5q)}}+
D_8(s,{\bf k}_{\perp})\ (\sigma_z)_{\lambda_p \, \lambda_q}
(\sigma_x)_{\lambda_n \, \lambda_{(5q)}} \ ,
\label{q5qpnAmpl}
\end{eqnarray}
where the $z$- and $x$-axes are directed along the photon
momentum and the momentum transfer ${\bf k}_{\perp}$,
respectively, and the $y$-axis is orthogonal to the scattering
plane.\\
Now the experimental data on the proton form factor are in
agreement with the assumption that the dominant contribution
stems from the amplitude corresponding to the conservation of the
$s$-channel helicities (\cite{KAKO}). Here 
the same assumption is used, taking into account only the amplitude
$D_1(s,{\bf k}_{\perp})$. Thus:
\begin{eqnarray}
\lefteqn{ \langle \lambda_{p};
\lambda_{n} | \hat{T}\left(s,{\bf p}_{3 \perp}\right)|
\lambda_{d};\lambda_{\gamma}\rangle =}  \nonumber \\
&&\frac i{8\pi ^2s}\int \! d^2{\bf k}_{\perp } \
 \bar u_{\lambda_p}(p_3) \hat {\epsilon}_{ \lambda_{\gamma}}
\left(\frac{-\hat k+m_q}{k^2-m_q^2}\right)
\hat {\epsilon}_{\lambda_{d}} v_{\lambda_n}(p_4) \nonumber \\
&& \times D^{\gamma d\rightarrow q(5q)}(s,{\bf k}_{\perp})
\ D_{1}(s,{\bf p}_{3 \perp}-{\bf k}_{\perp}).
\label{convolutionspin1}
\end{eqnarray}
Furthermore, taking into account that at high energy
$p_{\gamma}\gg \sqrt{s_0}$ and finite momentum transfer $t\simeq
|{\bf p}_{3\,\perp}|^2\simeq s_0$ the momentum $k$ is almost
transversal $k=(k_0,{\bf k}_{\perp},k_z)$, where $\displaystyle
k_0\simeq k_z\simeq O\left(\frac{s_0}{2 p_{\gamma}}\right)$ and
$\int \! d^2{\bf k}_{\perp}\ {\bf k }_{\perp}(...) \sim {\bf
p}_{3\,\perp}$, it is found the following representation for the spin
structure of the $\gamma d\rightarrow pn$ amplitude:
\begin{eqnarray}
\lefteqn{ \langle \lambda_{p};
\lambda_{n} | \hat{T}\left(s,{\bf p}_{3 \perp}\right)|
\lambda_{d};\lambda_{\gamma}\rangle =}  \nonumber \\
&& \bar u_{\lambda_p}(p_3) \hat {\epsilon}_{\lambda_{\gamma}}
\left({-A(s,t){\bf p}_{3\,\perp}\cdot \mbox{\boldmath $\gamma$}
+B(s,t) m}\right)
\hat {\epsilon}_{\lambda_{d}} v_{\lambda_n}(p_4)\ ,
\label{spin1}
\end{eqnarray}
where
\begin{eqnarray}
&&A(s,t)=\frac i{8\pi ^2s}\int \! d^2{\bf k}_{\perp } \
\frac{{\bf k}_{\perp}\cdot{\bf p}_{3\,\perp}}
{|{\bf p}_{3\,\perp}|^2}\
\frac{1}{k^2-m_q^2}  \nonumber \\
&&\times D^{\gamma d\rightarrow q(5q)}(s,{\bf k}_{\perp})
\ D_{1}^{q(5q)\rightarrow pn}(s,{\bf p}_{3\, \perp}-{\bf k}_{\perp})\ ,
\label{convolutionscalarA}\\
&&B(s,t)=\frac i{8\pi ^2s} \frac{m_q}{m}\int \! d^2{\bf k}_{\perp } \
\frac{1}{k^2-m_q^2}  \nonumber \\
&&\times D^{\gamma d\rightarrow q(5q)}(s,{\bf k}_{\perp})
\ D_{1}^{q(5q)\rightarrow pn}(s,{\bf p}_{3\, \perp}-{\bf k}_{\perp})\  ,
\label{convolutionscalarB}
\end{eqnarray}
and $m$ is the nucleon mass. In the case of a Gaussian parametrization
for  $ D^{\gamma d\rightarrow q(5q)}(s,{\bf k}_{\perp})$ and $D_1$
(in Eqs. (24,25)) the ratio $R={A(s,t)}/{B(s,t)}$
is a smooth function of $t$. Further on, it will be assumed that it is a
constant and will be considered as a free parameter.
The differential cross section for the reaction $\gamma
d\rightarrow pn$ is then 
\begin{eqnarray}
&\displaystyle \frac{d\sigma^{I}_{\gamma d\to pn}}{d t}& =
\frac{1}{64\,\pi s}\ \frac{1}{(p_{\gamma}^{\mathrm{cm}})^2}\
\left[S_{t}\ |B(s,t)|^2+S_u\ |B(s,u)|^2 \right. \nonumber\\
&&\left. +(-1)^{I+1}\  2S_{tu}\
{\mathrm Re}(B(s,t)B(s,u)\right],
\label{eq:sigt}
\end{eqnarray}
where~$I$ is the isospin of the reaction, {\em i.e.} $I=1$ (or $0$) for
isovector (or isoscalar) photons. The kinematical functions $S_t$,
$S_u$, $S_{tu}$ in (\ref{eq:sigt}) are given by
\begin{eqnarray}
S_t=\frac{1}{6}\  \sum _{\lambda_{\gamma},\ \lambda_{d}}
{\mathrm {Tr}}\ \left[ \hat {\epsilon}_{\lambda_{\gamma}}
\left(R (\hat{p}_3-\hat{p}_1)
+ m \right)
\hat {\epsilon}_{\lambda_{d}}
\left(\hat {p}_4- m  \right) \right.  \nonumber \\
\times \left. \hat {\epsilon}^{\star}_{\lambda_{d}}
\left(R (\hat{p}_3-\hat{p}_1)
+  m \right)
 \hat {\epsilon}^{\star}_{\lambda_{\gamma}}
\left(\hat {p}_3+ m  \right) \right] \ ,\nonumber \\
S_u=\frac{1}{6}\  \sum _{\lambda_{\gamma},\ \lambda_{d}}
{\mathrm {Tr}}\  \left[ \hat {\epsilon}_{\lambda_{d}}
\left(R (\hat{p}_3-\hat{p}_1)
+  m \right)
 \hat {\epsilon}_{\lambda_{\gamma}}
\left(\hat {p}_4- m  \right) \right. \nonumber \\ \times \left.
\hat {\epsilon}^{\star}_{\lambda_{\gamma}} \left(R
(\hat{p}_3-\hat{p}_1)
+  m \right) \hat
{\epsilon}^{\star}_{\lambda_{d}} \left(\hat {p}_3+ m
\right)\right] \ ,\nonumber\\ S_{tu}=\frac{1}{6}\  \sum
_{\lambda_{\gamma},\ \lambda_{d}} {\mathrm {Tr}}\ \left[ \hat
{\epsilon}_{\lambda_{\gamma}} \left(R (\hat{p}_3-\hat{p}_1)
+ m \right) \hat
{\epsilon}_{\lambda_{d}} \left(\hat {p}_4- m  \right)\right.
\nonumber \\ \times \left. \hat
{\epsilon}^{\star}_{\lambda_{\gamma}} \left(R (\hat{p}_3-\hat{p}_1)
+ m \right) \hat
{\epsilon}^{\star}_{\lambda_{d}} \left(\hat {p}_3+ m
\right)\right] \ .
\end{eqnarray}
In order to fix the energy dependence of the amplitude $B(s,t)$ it is
required that \begin{equation}  \displaystyle \left.\frac{{\mathrm
d}\sigma}{{\mathrm d }t} \right|_{\; \theta^{{\mathrm
CM}}=0}\sim \left(\frac{s}{s_0}\right)^{2\alpha_N(0)-2} \ .
\end{equation}\\ 
Taking into account that $S_{t}\sim s$ for $s\gg s_0$ it is
found that \begin{equation}
 \displaystyle
B(s,t)\sim\left(\frac{s}{s_0}\right)^ {\alpha_N(0)-1/2} \ .
\end{equation} Moreover, a good approximation for the energy dependence
of $S_t(\theta^{{\mathrm CM}}=0)$ in the region $p_{\gamma}=
1\div 7.5$~GeV is \begin{equation} \left.S_t \right|_{\;
\theta^{{\mathrm CM}}=0}\approx C p_{\gamma}^2 \end{equation}
with $C=(36\pm 3)~$GeV$^2$. Using this approximation 
$B(s,t)$ can be related to the Regge-pole exchange amplitude as
\begin{equation}
|B(s,t)|^{2}=\frac{1}{C p_{\gamma}^2}\
|\mathcal{M}_{{\mathrm Regge }}(s,t)|^2\ ,
\label{Bst}
\end{equation}

where
\begin{equation}
\mathcal{M}_{{\mathrm Regge}}(s,t)= F(t)
\left(\frac{s}{s_0}\right)^{\alpha_{N}(t)} \exp{\left[
      -i\ \frac{\pi}{2}\left(\alpha_{N}(t) -
        \frac{1}{2}\right)\right]}\  .
\label{eq:Mregge}
\end{equation}
Here $\alpha_N(t)$ is the trajectory of the nucleon Regge pole and
$s_0 =4~\mathrm{GeV}^2 \simeq m_d^2$.

\subsection{Nonlinear Nucleon Regge Trajectories}
\indent
\label{sub:qcd-mot}

According to the data on $\pi N$ backward scattering (see {\em e.g.} the
review~\cite{LYU1}) the nucleon Regge trajectory has a
nonlinearity of the form:

\begin{equation}
  \alpha_{N}(t)= \alpha_{N}(0) +\alpha'_{N}(0)\, t +
  \frac{1}{2}\, \alpha''_{N}(0)\, t^2 \, K(t),
\label{eq:alpha1}
\end{equation}\\
where $\alpha_{N}(0)=-0.5$, $\alpha'_{N}(0)=0.9\
\mathrm{GeV}^{-2}$ are the intercept and slope of the Regge
trajectory, and $\alpha''_{N}(0)= 0.20 \div 0.25\  \mathrm{GeV}^{-4}$ is
the coefficient of the nonlinear term. In (\ref{eq:alpha1}) it is
introduced also a cut-off function $K(t)$. Assuming that $K(t) =1$ 
the amplitude will grow very fast with $s$ at large $t$ which would violate
unitarity.
To prevent this fast growth the exponential form
\begin{equation}
K(t)=\exp\left({-\beta t^2}\right) \end{equation}
is chosen with $\beta =0.008\, \mathrm{GeV}^{-4}$. The small value of
$\beta$ does not destroy the parameterization of $\alpha(t)$ for
$-t \leq 1.6\ \mathrm{GeV}^2$ derived from Ref.~\cite{LYU1}. 
Note also that the phenomenological Regge trajectory (\ref{eq:alpha1})
with a power-like or exponential cut-off is nonlinear only for
moderate values of $t$; at large $t$ the quadratic term becomes
small and the trajectory becomes essentially linear again.

On the other hand, the QCD motivated Regge trajectories as
suggested by Brisudov\'{a}, Burakovsky and Goldman (BBG)
\cite{BBG} show a different behavior at large $t$. 
As shown in the previous Section the screened quark-antiquark potential (Eq.~\ref{lattice})
leads to nonlinear meson Regge trajectories. 
These trajectories can be parametrized on the whole physical sheet as
\begin{equation}
  \alpha(t) = \alpha(0)+ \gamma \left[ T^{\nu} - \left(T-t\right)^{\nu}
\right] \label{eq:nonlinBBG}
\end{equation}
with $0\leq \nu \leq 1/2$. The limiting cases $\nu=1/2$ and $\nu
\to 0$ ($\gamma \nu =$ const) correspond to the square-root
trajectory
\begin{equation}
  \alpha(t) = \alpha(0)+ \gamma \left[ \sqrt{T} - \sqrt{T-t}
\right],
\label{eq:sqroot}
\end{equation}
and the logarithmic trajectory
\begin{equation}
  \alpha(t) = \alpha(0)- (\gamma \nu) \ln\left(1 - \frac{t}{T}
\right) \label{eq:log}
\end{equation}\\
respectively. Such trajectories arise not only for heavy
quarkonia, but also for light-flavor hadrons.

In order to find the possible forms of nonlinear Regge trajectories for
mesons composed of light quarks, the analytical model 
shown in subsection \ref{sub:qcd-mot} has to be considered.  
In this picture, the color flux tube is
stretched between quark and antiquark at the tube ends. 
The varying string tension is introduced to
simulate dynamical effects such as the weakening of the flux tube
due to pair ($q \bar{q}$) creation. Within this framework
it is possible to recover the form of the underlying potential 
for a given Regge trajectory. Potentials leading to
``square-root'' and ``logarithmic'' Regge trajectories have been taken
from Ref.~\cite{BBG} where it was demonstrated that their effect is very 
similar to what is obtained by a screened potential in the unquenched lattice QCD.
In addition, this approach is able to reproduce with very good accuracy 
all the available meson spectra.

From experimental data it is known that the slopes of meson and 
baryon Regge trajectories are almost the same $\alpha^{\prime}_N \simeq
\alpha^{\prime}_{\rho} \simeq 0.9-1 $~GeV$^{-2}$ (see Ref.~\cite{KA9}). 
Since the slope is determined by the string
tension which, in turn, depends on the color charges at the string ends,
a baryon Regge trajectory can be described using 
the form suggested by the analytical string model for the meson.
In the baryon case one of the quarks at the ends is substituted by a diquark (\cite{KOBZ}).

\section{\mbox{Predictions for the $\gamma d \rightarrow pn$}\\ Differential Cross Section}
\label{sec:results}

The dependence of the residue $F(t)$ on $t$ can be written 
in the form (see~\cite{KA11,KKG}):

\begin{equation}
  F(t) = B {\left[\frac{1}{m^2 - t}\ \exp{(R_1^2t)} + C\, \exp{(R_2^2
        t)} \right]}\ ,
\label{eq:resid1}
\end{equation}\\
where the first term in the square brackets contains the nucleon
pole and the second term accounts for the contribution of
non-nucleonic degrees of freedom in the deuteron such as the {\bf D} wave 
components and isobar contributions.

The set of parameter for the Regge trajectory
is the same used in the phenomenological nonlinear trajectory~\cite{gris}:

\begin{eqnarray}
&&\mathrm{}\ 
B=4.01\cdot10^{-4}\, \mathrm{kb}^{1/2}\cdot\mathrm{GeV} ,\;
C = 0.7\ \mathrm{GeV}^{-2}  , \nonumber\\
&&  R_1^2 = 2\ \mathrm{GeV}^{-2} , \;
R_2^2 = 0.03\  \mathrm{GeV}^{-2},\; 
\alpha''_{N(0)}= 0.25\, \mathrm{GeV}^{-4}
\:
\end{eqnarray}\\
the ratio $R=A(s,t)/B(s,t)$ being equal to 1. 

The QCD motivated logarithmic trajectory differs only for 
an overall normalization factor $B$ taken as $1.8 \cdot 10^{-4}$ kb$^{1/2}
\cdot $~GeV and for the ratio $R=A(s,t)/B(s,t)$ being equal to 2.

The free parameter $T$ is fixed by comparison with the available experimental data.
The QGSM predictions, based on the QCD motivated ``logarithmic'' trajectory 
of Eq.~\ref{eq:log}  were shown in Fig.~\ref{fig:ang1} for $T=$1.7~GeV$^2$
taking into account the interference of the isovector and isoscalar components of the
scattering amplitude. The forward-backward asymmetry arises 
from the interference of two amplitudes describing the contribution of
isovector ($\rho$ like) and isoscalar ($\omega $ like) photons,
so that in this case the differential cross section can be written as
\begin{eqnarray}
&\displaystyle \frac{d\sigma^{\rho+\omega}_{\gamma d\to pn}}{d t}& =
\frac{1}{64\,\pi s}\ \frac{1}{(p_{\gamma}^{\mathrm{cm}})^2}\
\left[S_{t}\ |B^{\rho}(s,t)+B^{\omega}(s,t)|^2+ \right. \nonumber\\
&&S_u\
|B^{\rho}(s,u)-B^{\omega}(s,u)|^2  \nonumber\\
&&\left. +  2S_{tu}\
{\mathrm Re}\left(B^{\rho}(s,t)+B^{\omega}(s,t)\right)^{\star}
\left(B^{\rho}(s,u)-B^{\omega}(s,u)\right)\right] \ ,
\label{eq:sigtas}
\end{eqnarray}
having used the vector dominance model in the form:
\begin{equation}
B^{\omega}(s,t)=B^{\rho}(s,t)/\sqrt{8}, \hspace{1cm}
B^{\omega}(s,u)=B^{\rho}(s,u)/\sqrt{8}. \end{equation} 

\section{Summary}
\label{sec:summ}

The study of the deuteron photo-disintegration at photon energies 
higher then 1~GeV can shed light on the reaction mechanisms 
at work in the transition from the nuclear to the QCD picture of the deuteron.

To this aim, the world cross section data on the deuteron
photo-disintegration have been reviewed, both at low ($E_\gamma \le$ 1~GeV) and high 
(1 $\leq E_\gamma \leq$ 5.5~GeV) incident photon energies.  
In the high energy region a new phenomenon emerges: the cross section scaling.
According to a simple interpretation it could be a consequence 
of perturbative QCD.
The scaling appears at different energies for different angles
corresponding to a perpendicular momentum transfer of approximately 1~GeV$^2$.
From a closer look at the photo-disintegration data for the differential cross section 
emerges that the pQCD description is not plainly applicable especially at
intermediate momentum transfers.

The complementary information derived from polarization observables
suggests that non perturbative physics could be at work 
even at relatively high momentum transfers so that  
soft mechanism could mimic the scaling behavior.
Some non perturbative quark models well describe the experimental data.
Among the various models, the Quark Gluon String model
stands out since it gives a good description
of both cross section and proton in-plane polarization transfers data.

For this reason, the theoretical foundations and predictions 
of the QGSM have been fully illustrated.

From the experimental point of view, the possibility to 
further check the theoretical predictions and to discriminate among the
different models requires the knowledge of the differential cross section 
in a wide range of photon energies and for very forward and backward angles.

\setcounter{chapter} {1}     
\pagestyle{plain}
\chapter{Experimental Apparatus}


\section{The Jefferson Lab Accelerator Facility }
\indent
\par

\begin{figure}[htbp]
 \begin{center}
 \leavevmode
 \epsfig{file=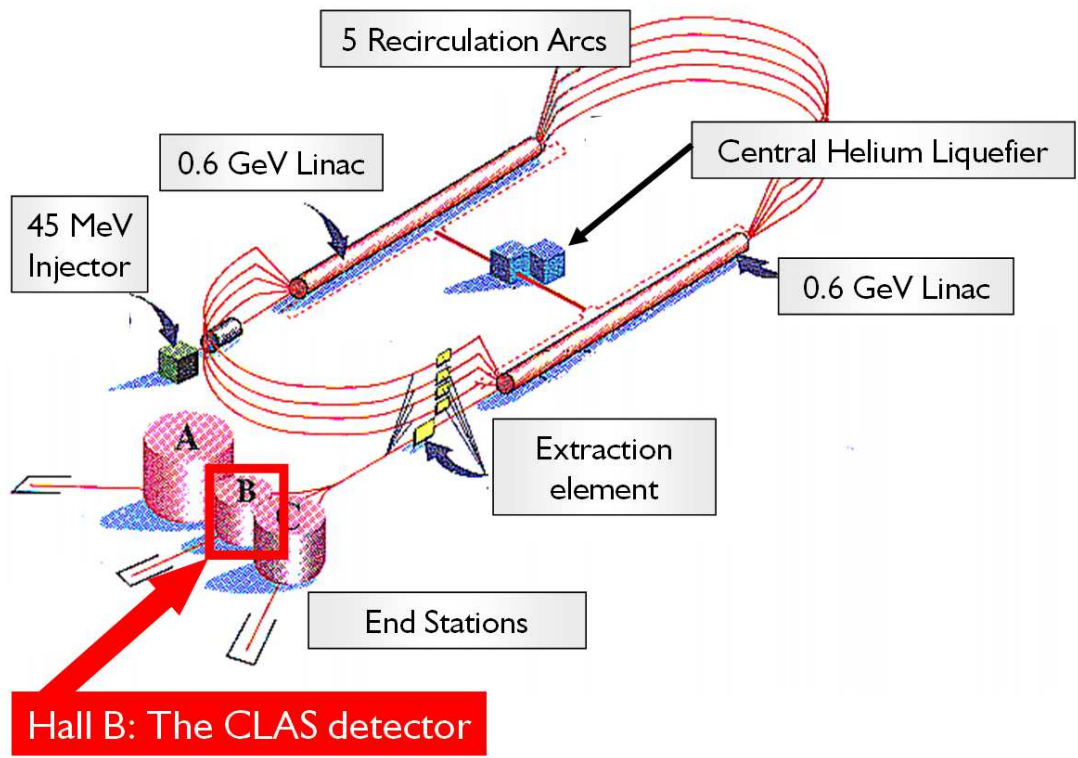,width=15cm}
 \caption{ \small Schematic layout of the JLab Continuous Electron Beam Accelerator.}
 \label{fig:acc}
 \end{center}
\end{figure}

The Continuous Electron Beam Accelerator (CEBAF) at JLab uses a high power  
electron beam with energies up to 6~GeV and 100\% duty cycle. 
A schematic layout of the accelerator and experimental halls is shown in Fig.~\ref{fig:acc}. 
The machine design, based on a race-track configuration, has the advantage 
of limiting the length and cost of the accelerating sections by recirculating
the beam several times into the linacs while keeping low the  
energy loss due to {\em bremsstrahlung} in the curvature parts of the tracks. 
The injection system consists of a 45 MeV linac inserting electrons
into the main accelerator. 
The different beams can coexist in the linacs, but must occupy different arcs.
At the heart of the machine are the 8 five-cell niobium superconducting radio frequency 
(RF) cavities per cryo-module having an high average gradient 
(see Tab. \ref{tab:cells}) and operating at 1.5~GHz. 
\begin{table}[htbp]
\begin{center}
\begin{tabular}{|l|c|c|} \hline
 & Gradient & Quality Factor  \\ \hline
Design specs & 5.0 MeV/m& $2.4\cdot 10^9$ \\
Average performance & 9.56 MeV/m & $7.0 \cdot 10^9$ \\
Best performance & 18.0 MeV/m &    $9.0 \cdot 10^9$      \\ \hline
\end{tabular}\\
\caption{\small Performance figures for the CEBAF RF super-conducting cells.}
\end{center}
\label{tab:cells}
\end{table}
Once into the main accelerator, the beam enters the north linac 
and gains an additional boost of 600~MeV of kinetic energy and 
then is bent through $180^\circ$ and sent to the south linac where 
it is boosted once again by 600~MeV. 
At this point, the beam could be sent to an experimental hall or recirculated 
through the north linac. The beam can be further recirculated up to 4~times, 
picking 1200~MeV each time is sent around the accelerator. 

The primary electron beam delivered by the accelerator can be separated 
to be sent to the experimental areas, Halls A, B, and C, 
for simultaneous experiments. 
The electron packet structure has a frequency of 1.5~GHz and can reach a bunch-to-bunch 
intensity modulated with a periodicity of 3, in order to deliver beam to the 
different Halls with variable intensity and energy so that 
beams in the three Halls can have either the same energy or energies 
multiples of 1/5 of the end-point energy.

The experimental equipment in the Halls 
is complementary, addressing a wide range of physics issues. 
Hall A has two identical focusing high-resolution spectrometers with a 
maximum momentum of 4~GeV/c~\cite{N6}. Hall C has two symmetric focusing 
spectrometers: one featuring acceptance of high-momentum particles, the other a 
short path length for the detection of decaying particles~\cite{N7}. 

Hall B houses the CEBAF Large Acceptance Spectrometer, CLAS,  
designed for operation with both electron and tagged-photon beams 
and described in more detail in the next section.


\section{The CEBAF Hall B}
\indent
\par

The Hall B facility is devoted to experimental studies 
of electro and photo-reactions requiring 
the detection of several only loosely correlated particles in the hadronic 
final state and measurements at limited (for the JLab standards) luminosities.
In order to achieve high detection efficiency for multi-particle final states
a large-acceptance detection is required. 
In addition, a large-acceptance detector may compensate for various restrictions 
limiting the experimental operation to moderate luminosities. 
For example, in experiments using a tagged {\em bremsstrahlung} photon beam, 
a lower luminosity is required to keep down accidental coincidences.
 
\begin{figure}[htpb]
 \begin{center}
 \leavevmode
 \epsfig{file=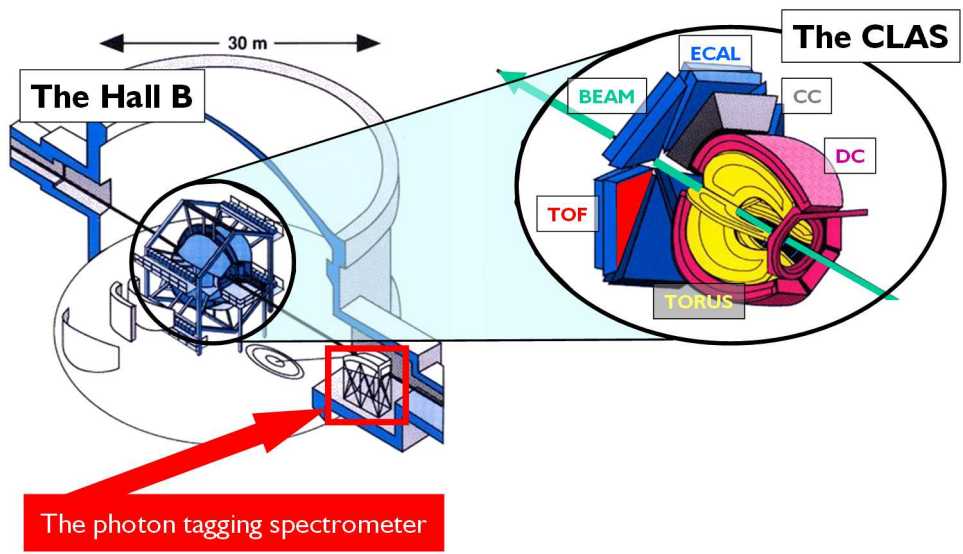,width=13cm}
 \caption{ \small Left panel: Schematic view of the Hall B enclosure, including the CLAS
metal supporting structures. The tagging spectrometer is shown at the entrance on the Hall,
upstream whit respect to the electron beam direction. Right panel: The CLAS detector magnified
view showing the schematics for the detector subsystems.}
 \label{fig:hallb-c}
 \end{center}
\end{figure}

\begin{figure}[htbp]
 \begin{center}
 \leavevmode
 \epsfig{file=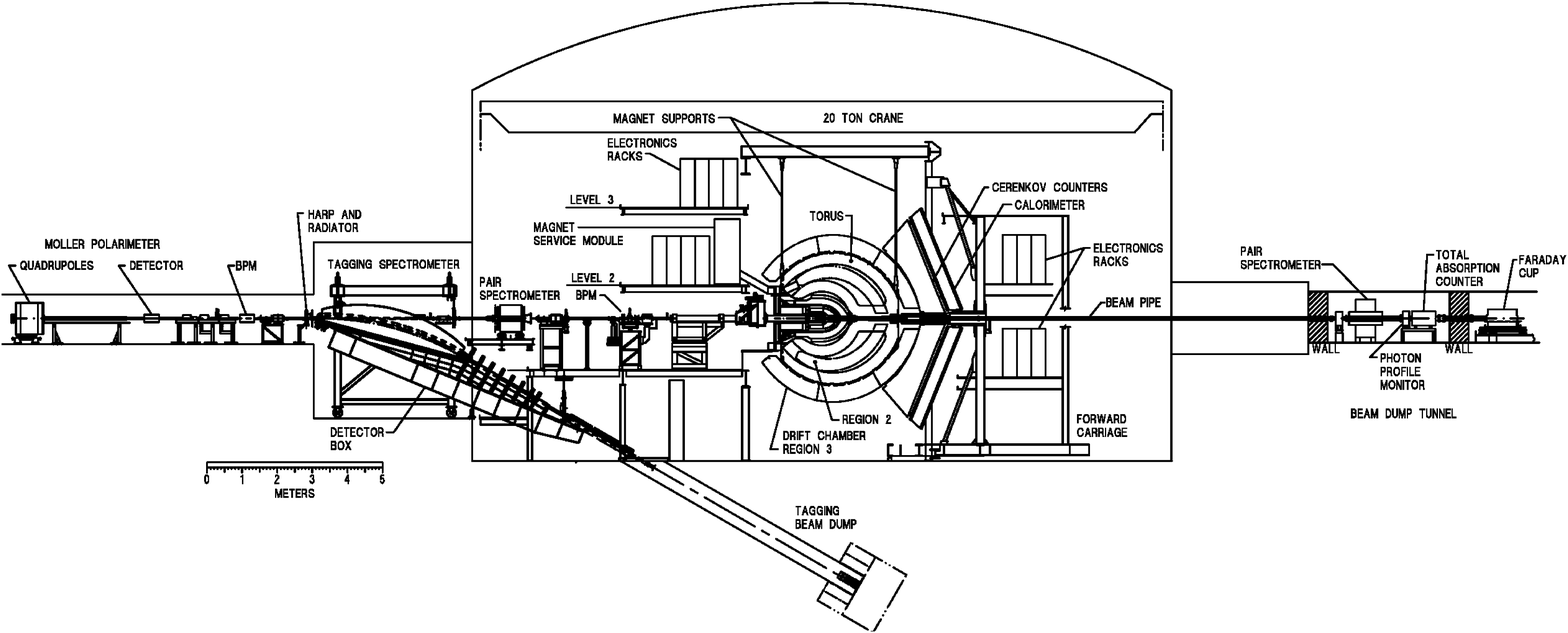,width=20cm,height=10cm,angle=90}
 \caption{ \small Hall B beam-line, the CLAS detector, and associated equipment.}
 \label{fig:hallb-d}
 \end{center}
\end{figure}

A view of the Hall B layout is shown in Figs \ref{fig:hallb-d} and \ref{fig:hallb-c} where 
the electron-photon beam line, the tagging spectrometer for photon beam
operations, the large angle spectrometer CLAS and its associated equipments 
are represented.
Also, solid-state polarized targets can only be operated at lower beam currents, 
corresponding to low luminosity. A high detection efficiency for multi-particle events and a useful event 
rate at limited luminosity both require a detection system with a large acceptance.
The instantaneous intensity of a tagged photon beam produced by the
Hall B tagging spectrometer is limited to approximately $10^7$ $\rm{tagged}\ \gamma/\rm{s}$ 
by accidental coincidences, typically resulting in relatively low background 
rates in a large-acceptance detector.
\vspace{4cm}

\clearpage
\subsection{Hall B Photon Production and Tagging}
\indent
\par

Electrons from the CEBAF accelerator strike a thin target (the "radiator") 
just upstream from a magnetic spectrometer~\cite{N37} (the "tagger" ). 
Photons produced in the radiator continue toward
the CLAS target following the same beam-line through the magnet yoke that 
is traversed by the electron beam when the tagger is not in use.
The {\em  bremsstrahlung} tagging technique for direct measurement of the 
incident photon energy in photo-nuclear interactions is well established~\cite{ADON}.
The JLab system is the first photon tagger in the multi~GeV energy range to
combine high resolution ($\simeq 10^{-3}E_0$) with a broad tagging range 
\mbox{($20\%-95\%$)} of $E_0$.

There are three modes of operation for photon beams in Hall B: a normal non-
polarized mode, a circularly polarized mode, and a linearly polarized mode. 
These are governed by the polarization state of the incident electron beam, 
and by the nature of the photon radiator.

\begin{figure}[htbp]
 \begin{center}
 \leavevmode
 \epsfig{file=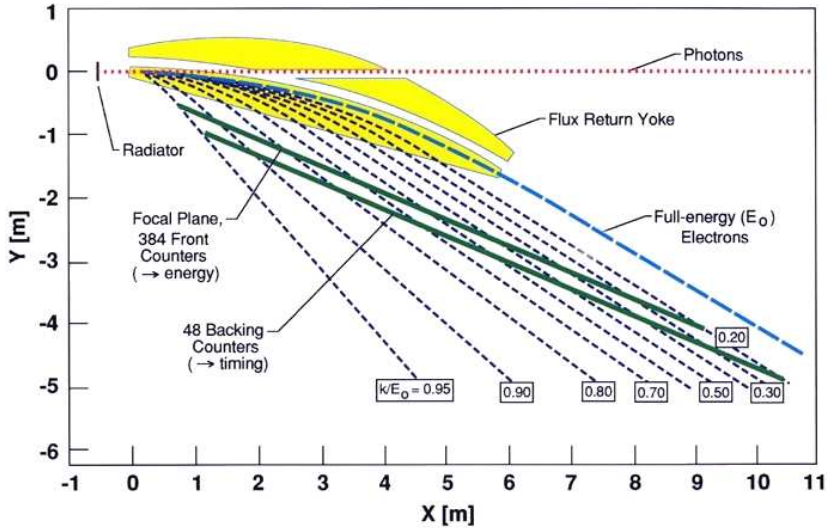,width=11cm}
 \caption{ \small Hall B photon tagging system.
Photons produced in the radiator continue toward
the CLAS target following the same beam-line through the magnet yoke that 
is traversed by the electron beam when the tagger is not in use.
The photon-tagging system can tag photons with energies between 20\% and 
95\% of the incident electron energy, and is capable of operating with 
electron beam energies up to 6.1~GeV.
}
 \label{fig:tagger}
 \end{center}
\end{figure}

\subsection{Photon Beam Position and Profile}
\indent
\par

The electron beam used in photon experiments is monitored up to a point 
just before the radiator by RF cavities providing a good determination
of both the production point and initial direction of 
the photon beam. Then, downstream of the tagging magnet, the 
photon beam passes through a collimator.

In the present experiment, the collimator has a 0.861~cm diameter hole in a 
cylindrical nickel block 25~cm long. This collimator is located approximately 
14~m from the radiator, and restricts the diameter of the photon beam at the 
CLAS target to less than 3~cm. A magnet placed just downstream from the 
collimator sweeps aside low energy secondary charged particles created in the 
collimator.

The photon beam position and size are monitored by a 
fixed array of crossed scintillator fibers located 20~m behind the CLAS target. 
The scintillators respond to the electron-positron pairs produced by photons 
in the CLAS target, in the atmosphere between the target and the hodoscope, 
or within the scintillating fibers themselves.

\subsection{Photon Energy and Timing}
\indent
\par
The photon-tagging system can tag photons with energies between 20\% and 
95\% of the incident electron energy $E_0$, and is capable of operating with 
electron beam energies up to 6.1~GeV. The field setting of the tagger magnet 
is matched to the incident beam energy so that those electrons that do not 
radiate will follow a circular arc just inside the edge of the pole face, 
and will be directed into a secondary shielded beam dump below the floor 
of Hall B with a maximum capacity of 800~W.

An electron that radiates a {\em bremsstrahlung} photon has lower momentum, 
and consequently smaller radius of curvature in the tagger dipole field, so 
that it emerges from the magnet along the open edge of the pole gap. A 
scintillator hodoscope along the flat focal plane downstream from this straight 
edge detects this electron, and thereby allows for the determination of the 
energy and timing of the radiated photon.

The overall geometry of this arrangement may be seen in Fig.~\ref{fig:tagger}. 
The tagger dipole is topologically a C-magnet with a full-energy radius of 
curvature of 11.80~m, a full-energy deflection angle of 30$^\circ$, and a gap 
width of 5.7~cm. The required magnetic field in the gap is 1.13~T 
for a beam energy of 4~GeV.

The focal plane hodoscope consists of two separate planes of scintillator 
detectors. The first detector plane (called the E-plane for energy) is composed of 
384 plastic scintillators 20-cm long and 4-mm thick. Their widths (along the 
dispersion direction) range from 6 to 18 mm in order to subtend approximately 
constant momentum intervals of 0.003$\cdot E_0$. Each counter optically overlaps its 
adjacent neighbors by one third of their respective widths, thus creating 767 
separate photon energy bins that provide an energy resolution of 
$0.001\cdot E_0$.

The second detector plane (called T-plane for timing) lies 20~cm downstream 
of the E-plane and contains 61 counters, 2-cm thick, which are read out using 
PMTs attached by solid light guides at both ends (transverse to the particle
direction) of each scintillator. The RMS timing resolution of these counters is 
110~ps.

\begin{figure}[htpb]
 \begin{center}
 \leavevmode

 \epsfig{file=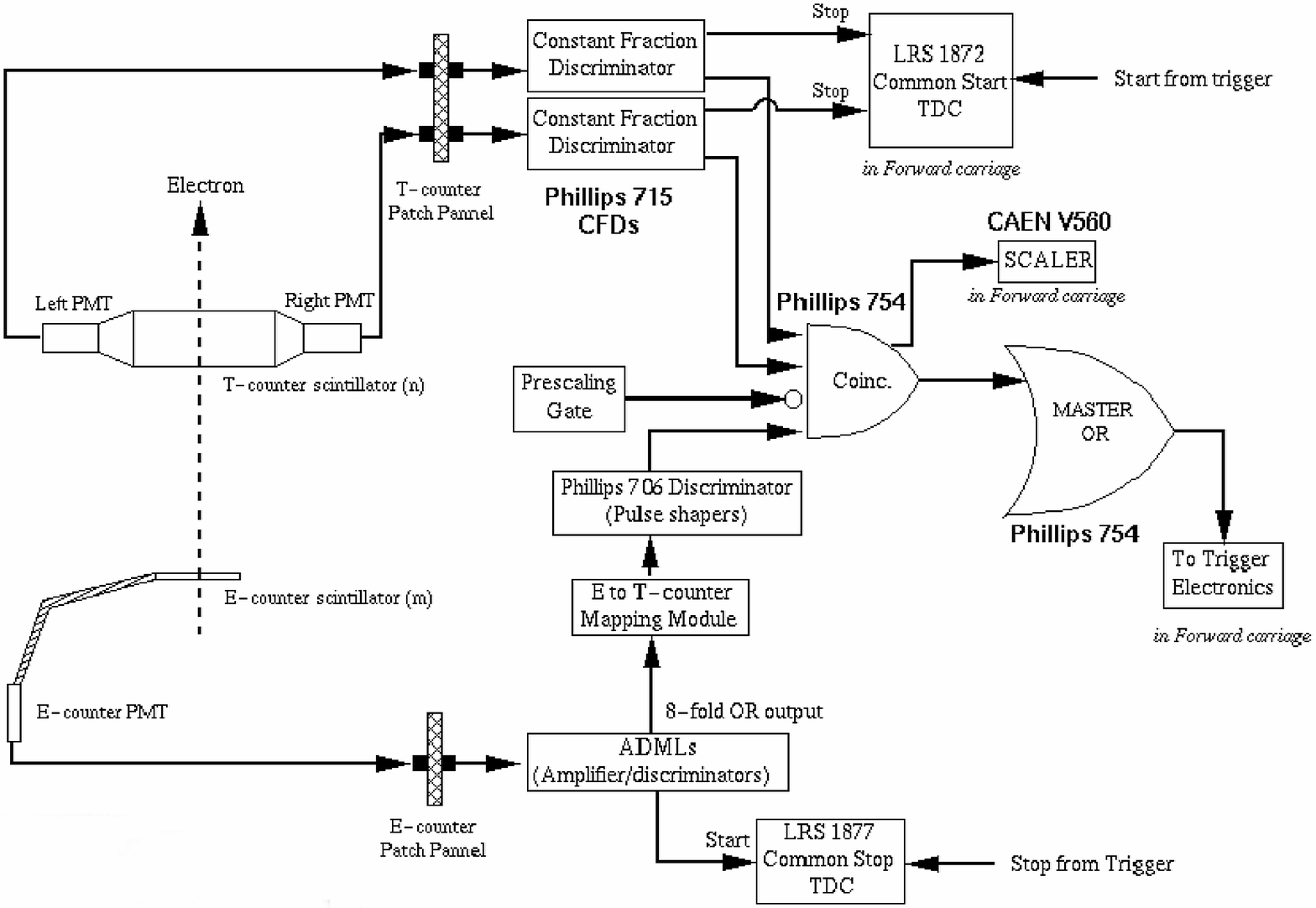,width=11cm,height=15cm}
 \caption{ \small Tagger spectrometer electronics scheme. See the text for details. }
 \label{fig:tagger_el}
 \end{center}
\end{figure}

A schematic of the electronics of the tagger focal plane can be seen in 
Fig.~\ref{fig:tagger_el}. 
A scattered electron first encounters an E-counter and then passes a 
T-counter producing signals in both devices that are processed in separate 
electronic chains and participate in the construction of the 
tagger ``event''. 

The signal coming from the E-counter single PMT is sent to an ADML board, 
amplified and discriminated to produce an ECL logic pulse. 
In turn, this pulse is then sent to an E-T mapping module 
which performs appropriate coincidences according to E and T geometries 
in order to reduce the production of accidental triggers. 
The same ECL pulse acts as a start for the E-counter 
TDC which stop will be given by the CLAS trigger.
To produce a tagger event, an ECL signal from the hit 
T-counter must be in coincidence with the E-counter signal so
that both signals can be interpreted in term of a definite electron  
trajectory. To this aim, the output of the E-T mapping module is
elongated using a discriminator so that the discriminated 
T-counter signal determines the timing.
Thus, a three-fold coincidence is required between the ``left'' and ``right'' 
signals from the two sides of a T-counter scintillator bar and
a geometrically matched E-counter signal.
For the timing measurement, the output of each T-counter 
discriminator is sent to one multi-hit FASTBUS TDC, 
in common start mode, with 50~ps per channel resolution. 
The signal for the TDC start comes from the CLAS trigger while
the stop from the T-counter.
Since the tagger has 61 T-counters, there are 61 possible pulses
which are sent through a cascade of logic modules set in OR mode 
(referred as ``Master OR'', MOR).

\subsection{Photon Beam Flux}
\indent
\par
The absolute photon flux is determined at very low flux rates by inserting a 
large lead-glass total absorption shower counter (TAC) into the photon beam. 
\begin{table}[htbp]
\begin{center}
\begin{tabular}{|l|c|} \hline
Property & Specs  \\ \hline
Glass composition & $55\%\rm{Pb} + 45\%\rm{SiO}_2$\\
Effective Z & 48.9\\
Radiation Length & 2.36 cm\\
Moliere Radius & 3.1 cm\\
Refraction index & 1.67\\
Cerenkov angle & 34.3$^\circ$\\
Density & 4.08 g/cm$^3$\\
Effective length & 17 RL\\
Energy resolution & 16\% at $E_\gamma=$300 MeV\\
Timing resolution & 4 ns\\
Efficiency & 99\%  \\ \hline 
\end{tabular}\\
\end{center}
\caption{\small Characteristics of the total absorption counter (from Ref.~\cite{TIER}).}
\label{tab:tac}
\end{table}
The TAC is essentially 100\% efficient (see Tab. \ref{tab:tac}), allowing
for the efficiency of the different elements of the tagging hodoscope to be determined. 
Due to counting pile-up problems, the TAC can only be operated at beam currents 
up to 100~pA, and must be retracted from the beam-line under normal running 
conditions. 
\begin{figure}[htbp]
 \begin{center}
 \leavevmode
 \epsfig{file=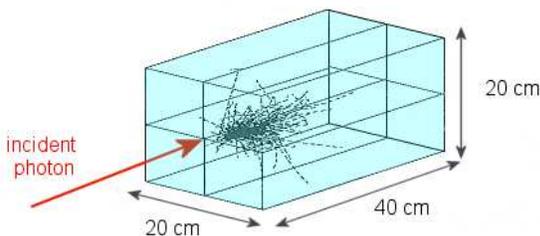,width=8cm}
 \caption{ \small Hall B Total Absorption Counter (TAC).
The TAC is essentially 100\% efficient allowing for the efficiency of 
the different elements of the tagger hodoscope to be determined. Due to 
counting pile-up problems, the TAC can only be operated at beam currents 
up to 100~pA, and must be retracted from the beam-line under normal running 
conditions. }
 \label{fig:tac}
 \end{center}
\end{figure}
Thus, a secondary monitor, linear in flux over a wide range, 
is cross-calibrated against the TAC at low rates and then used to monitor 
the flux at higher intensities. More details on the photon flux normalization
will be given in the next Chapter.
\clearpage
\section{The CEBAF Large Angle Spectrometer (CLAS)}
\indent
\par
\begin{figure}[htbp]
 \begin{center}
 \leavevmode
 \epsfig{file=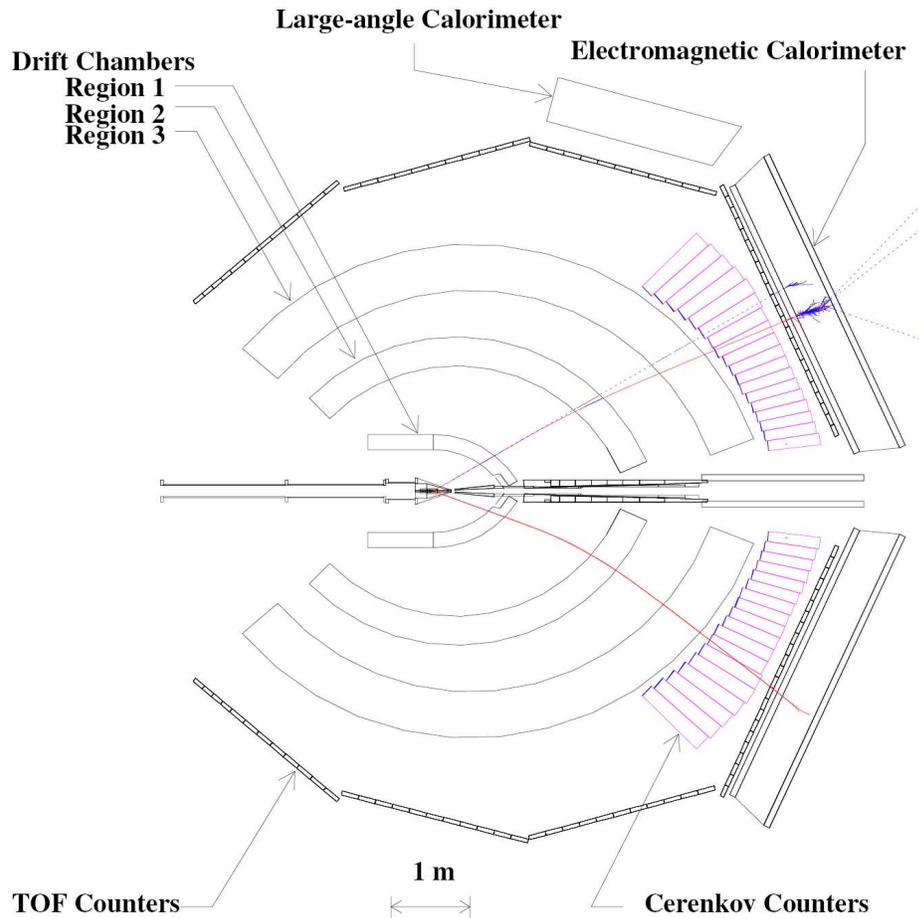,width=13cm}
 \caption{ \small A schematic top view of the CLAS detector cut along the beam line.
Typical photon, electron and proton tracks (from top to bottom) from an interaction
in the target are shown (the photon track through the CLAS drift chamber system
is drawn only to guide the eye).
}
 \label{fig:hallb}
 \end{center}
\end{figure}

The CLAS is nearly $4\pi$ spectrometer based on a toroidal magnetic field. 
The primary requirements driving this choice were the ability 
to measure charged particles with good momentum resolution, 
provide geometrical coverage of charged particles 
to large angles in the laboratory, and keep a magnetic-field-free region around 
the target to allow the use of dynamically polarized targets. 
A view of the CLAS detector and its ancillary equipment (such as the
photon tagging spectrometer and the downstream photon flux normalization devices) 
is shown in Fig.~\ref{fig:hallb}.
A spherical coordinate system is used in all of the descriptions.
The $z$-axis is taken to lie along the beam direction, with B as the polar 
(scattering) angle, and $\phi$ as the azimuthal angle. 
The $x$ and $y$ directions are then, respectively, horizontal and vertical 
in the plane normal to the beam.
The CLAS magnetic field is generated by six superconducting coils arranged 
around the beam line to produce a field pointing primarily in the $\phi$-direction. 
The particle detection system consists of drift chambers (DC)~\cite{N8,N9,N10} to 
determine the trajectories of charged particles, 
gas Cerenkov counters (CC)~\cite{N11} for electron identification, scintillation counters 
\cite{N12} for measuring time of flight (TOF), and electromagnetic 
calorimeters (EC)~\cite{N13,N14} to detect showering particles (electrons and photons) 
and neutrons by induced ionization. 

The six segments are individually instrumented to form six 
independent magnetic spectrometers with a common target, trigger and data
acquisition system (DAQ).
A two-level trigger system is used to begin data conversion and readout. 
The Level-1 trigger can make use of the fast information from 
the time-of-flight counters, the Cerenkov counters, and the electromagnetic 
calorimeters. A Level-2 trigger adds crude track finding using hit patterns in 
the drift chambers. The data acquisition system collects the digitized data 
and stores the information for off-line analysis.

In the following sections, each of the CLAS subsystems and the Hall B 
ancillary equipment are briefly described. 

\subsection{Target}
\indent
\par
\begin{figure}[htpb]
 \begin{center}
 \leavevmode
 \epsfig{file=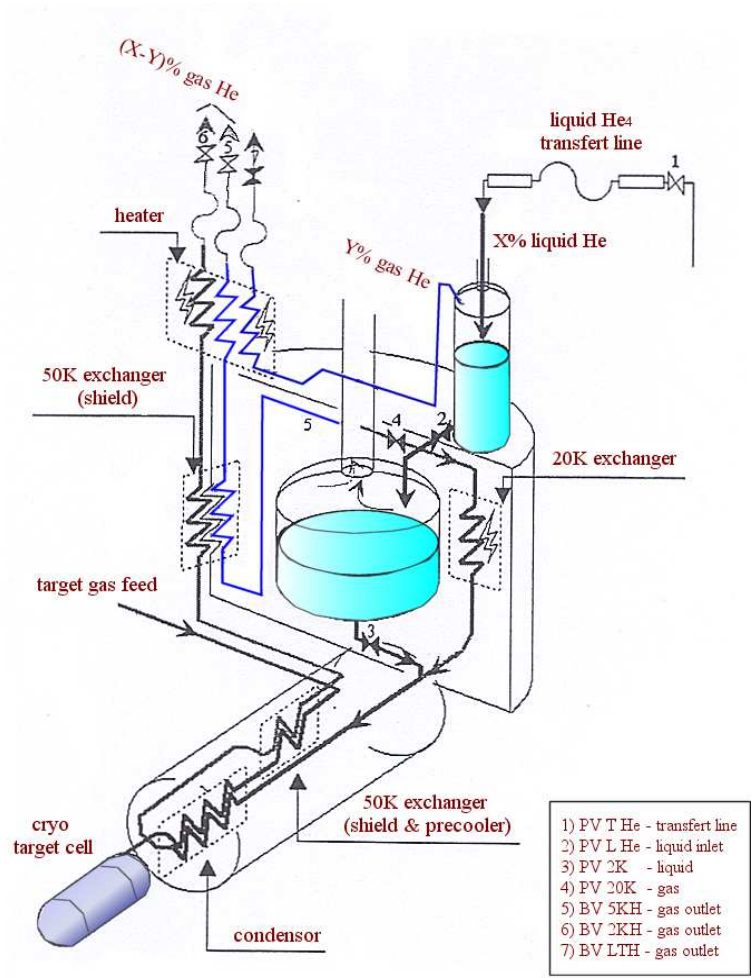,width=10cm,height=13cm}
 \caption{ \small Unpolarized cryogenic target cooling system (from Ref.~\cite{TIER}).}
 \label{fig:tgt}
 \end{center}
\end{figure}

A variety of targets have been used to date, with dimensions 
adapted to the particular needs of either electron or photon running. The 
most common target used has been liquid $\rm{H}_2$. However, reactions have also 
been studied using liquid $\rm{D}_2$ (as in the present case), $^3\rm{He}$, and $^4\rm{He}$; 
solid $^{12}\rm{C}$, Al, Fe, Pb, and $\rm{CH}_2$; 
and polarized $\rm{NH}_3$ and $\rm{ND}_3$ targets. 
All targets are positioned inside CLAS using support 
structures, inserted from the upstream end, and are 
independent of the detector itself (Fig.~\ref{fig:tgt}).
Unpolarized targets have commonly been positioned inside CLAS 
at the center of curvature of the inner toroidal coil. 

\clearpage
\subsection{Start Counter}
\indent
\par
The start counter (ST, see Ref.~\cite{N15}) is a set of scintillators surrounding the target 
with the same geometric acceptance of the CLAS returning the time at which 
a charged particle has left the target region.
This information, combined with time-of-flight information, 
determines the time needed to a charged particle to pass across the CLAS
in tagged-{\em bremsstrahlung} experiments. 
\begin{figure}[htbp]
 \begin{center}
 \leavevmode
 \epsfig{file=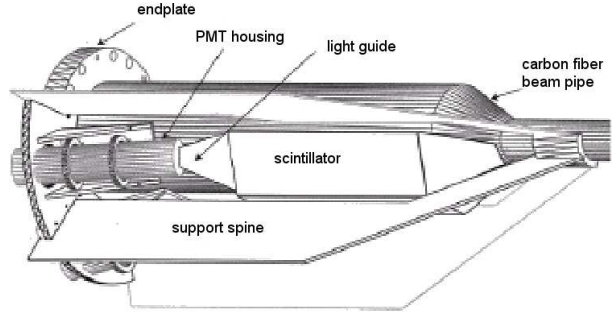,width=8cm}
 \caption{ \small The start counter assembly is a set of 
scintillators surrounding the target 
with the same geometric acceptance of the CLAS returning the time at which 
a charged particle has left the target region.
}
 \label{fig:st}
 \end{center}
\end{figure}
The timing resolution relies on the determination of the charged 
particle impact point along the start counter scintillator panels. 
For this reason, light readout is done at both ends of the panels. 
To save space, in the forward direction two sectors of the start counter 
panels are merged in one scintillator.
This scintillator is bent in the forward polar angle such that both ends 
of the panel are in the upstream direction allowing for a placement of a single
photomultiplier tube. The start counter timing resolution is of 260~ps.

\subsection{Torus Magnet}
\indent
\par
The magnetic field for the momentum analysis of charged particles is generated 
by six superconducting coils arranged in a toroidal geometry around the 
electron beam line. There is no iron in the system, so the magnetic field is 
calculated directly from the current in the coils. 
\begin{figure}[htbp]
 \begin{center}
 \leavevmode
 \epsfig{file=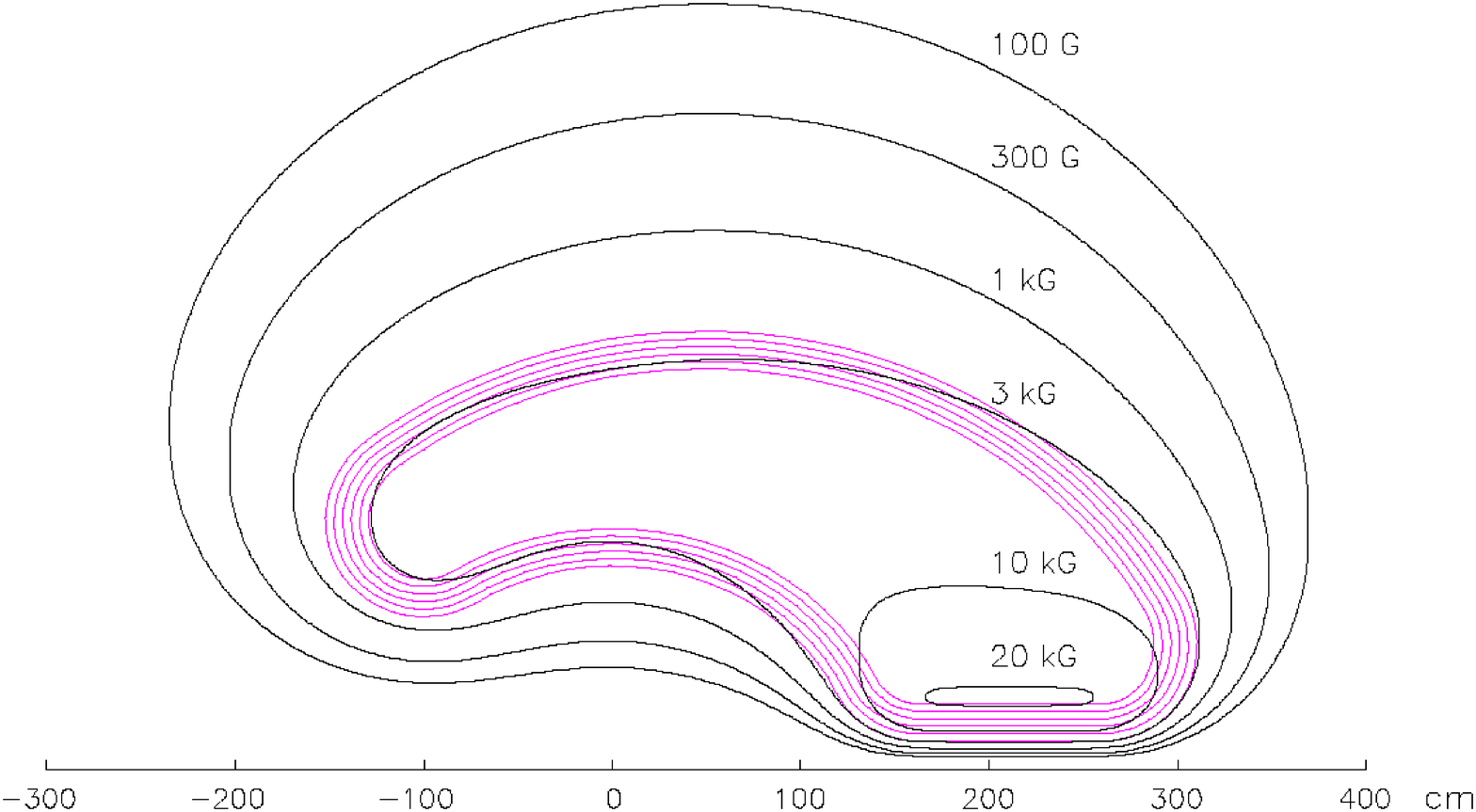,width=6.5cm}
 \epsfig{file=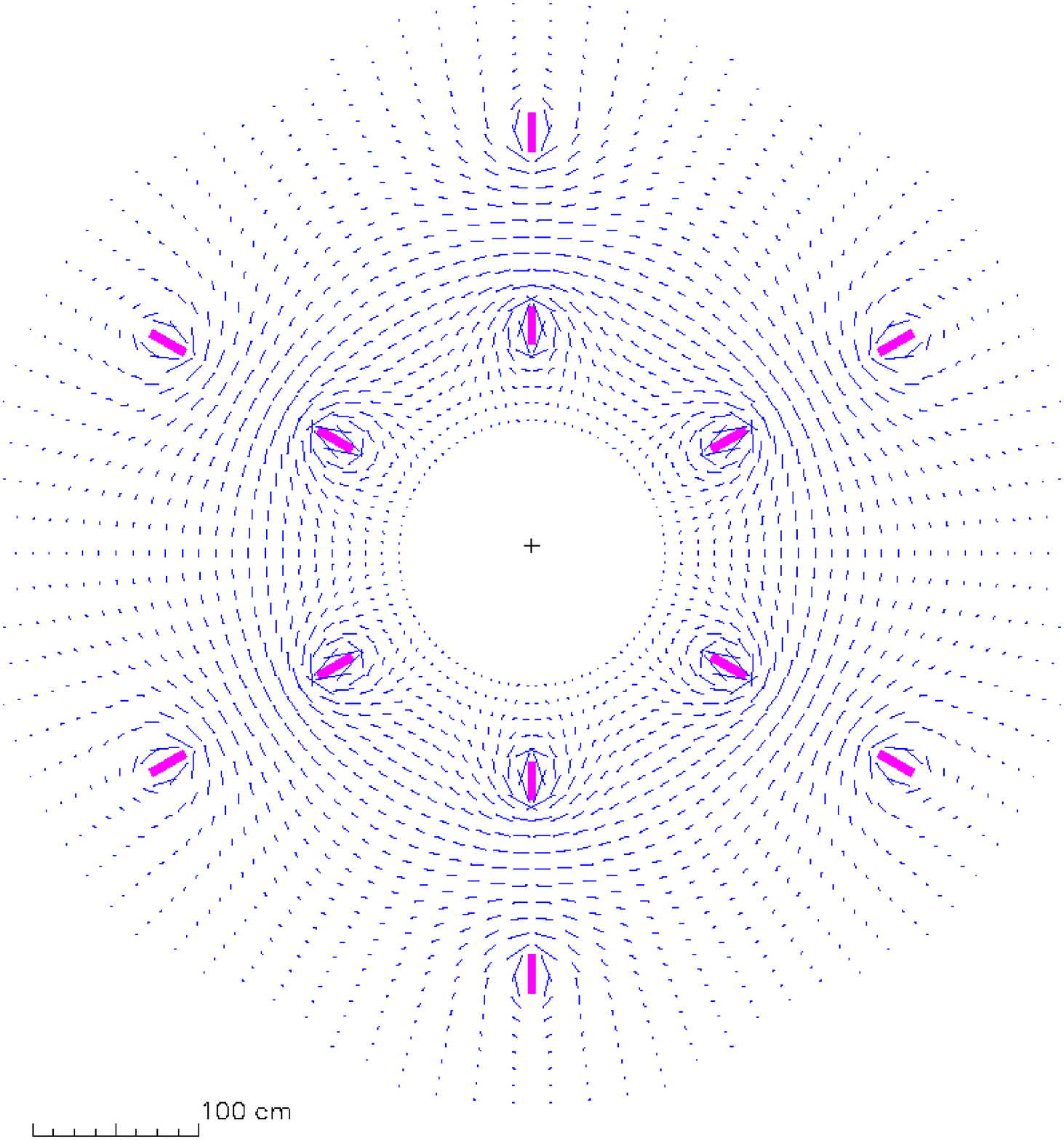,width=4cm}
\caption{ \small Left panel: Contours of constant magnetic field for the main CLAS
toroid in the mid-plane between two coils (see also Fig.~\ref{fig:hallb}). 
Right panel: Magnetic field vectors for the CLAS toroid transverse to the 
beam direction in a plane centered on the target. The six coils are shown in cross-section.}
 \label{fig:field}
 \end{center}
\end{figure}

The kidney-shape of the coils results in a high field integral for forward-going 
particles (typically high momentum), and a lower field integral for particles 
emitted at larger angles. At the same time, this coil geometry preserves a 
central field-free volume for the operation of a polarized target. 

At the maximum design current of 3860~A, the total number of amp-turns is 
$5 \cdot 10^6$. At this current the integral magnetic field 
\mbox{$\int{\overrightarrow{B} \cdot d\overrightarrow{l}}$} reaches 
2.5~\mbox{$\rm{T} \rm{m}$} in the forward direction, dropping to 
0.6~\mbox{$\rm{T} \rm{m}$} at a scattering angle of  $90^\circ$.
The magnetic field is reversible: when magnets are turned on in the normal 
condition, positive particles are bent outward from the beam axis.

The main field component is in the $\phi$-direction, however, there are 
significant deviations from a pure $\phi$-field close to the coils. 
The effect of these deviations on the particle trajectories is 
minimized by the circular inner shape of the coil: particles coming from 
the target do not experience a significant deflection in $\phi$ when crossing 
the inner boundary of the coil.

Routine operation has been limited to 88\% (3376~A) 
of the maximum current to keep internal mechanical stresses 
within conservative limits. 

\subsection{Drift Chambers}
\indent
\par
The CLAS toroidal magnetic field bends charged particles toward or away from the 
beam axis but leaves the azimuthal angle essentially unchanged. 
The magnet coils naturally define the detector into six independent 
tracking areas or "sectors". 
To simplify detector design and construction, 18 separate drift 
chambers were built and located at three radial positions in each of the six
sectors. 
These radial locations are referred to as "Regions". The six "Region 
One" chambers surround the target in an area of low magnetic field, 
the six "Region Two" chambers are situated between the magnet coils 
in an area of high field near the point of maximum track sagitta, while the 
six "Region Three" chambers are located outside of the magnet coils~\cite{N8,N9,N10}.
\begin{figure}[htbp]
 \begin{center}
 \leavevmode
 \epsfig{file=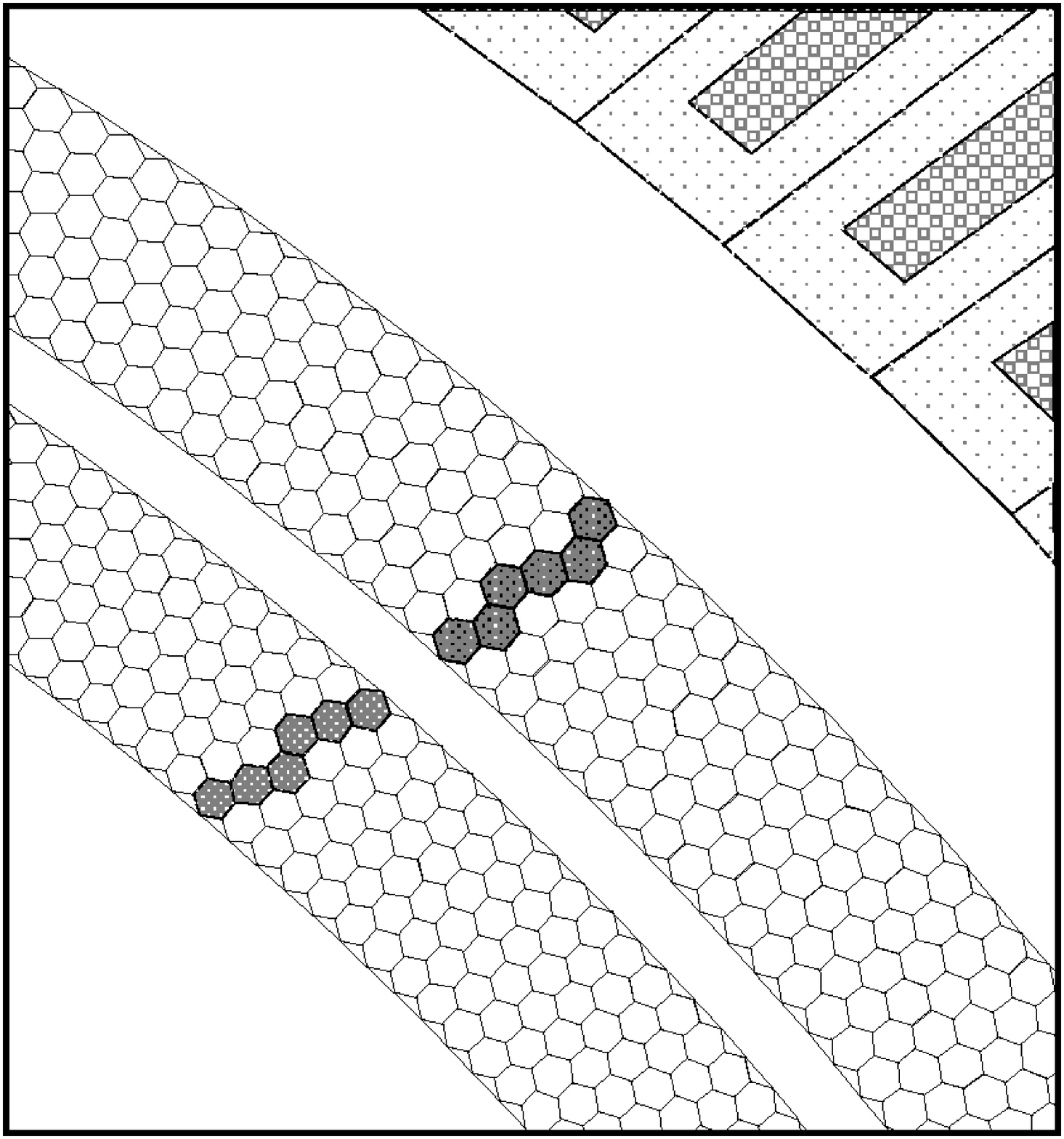,width=4cm}
\epsfig{file=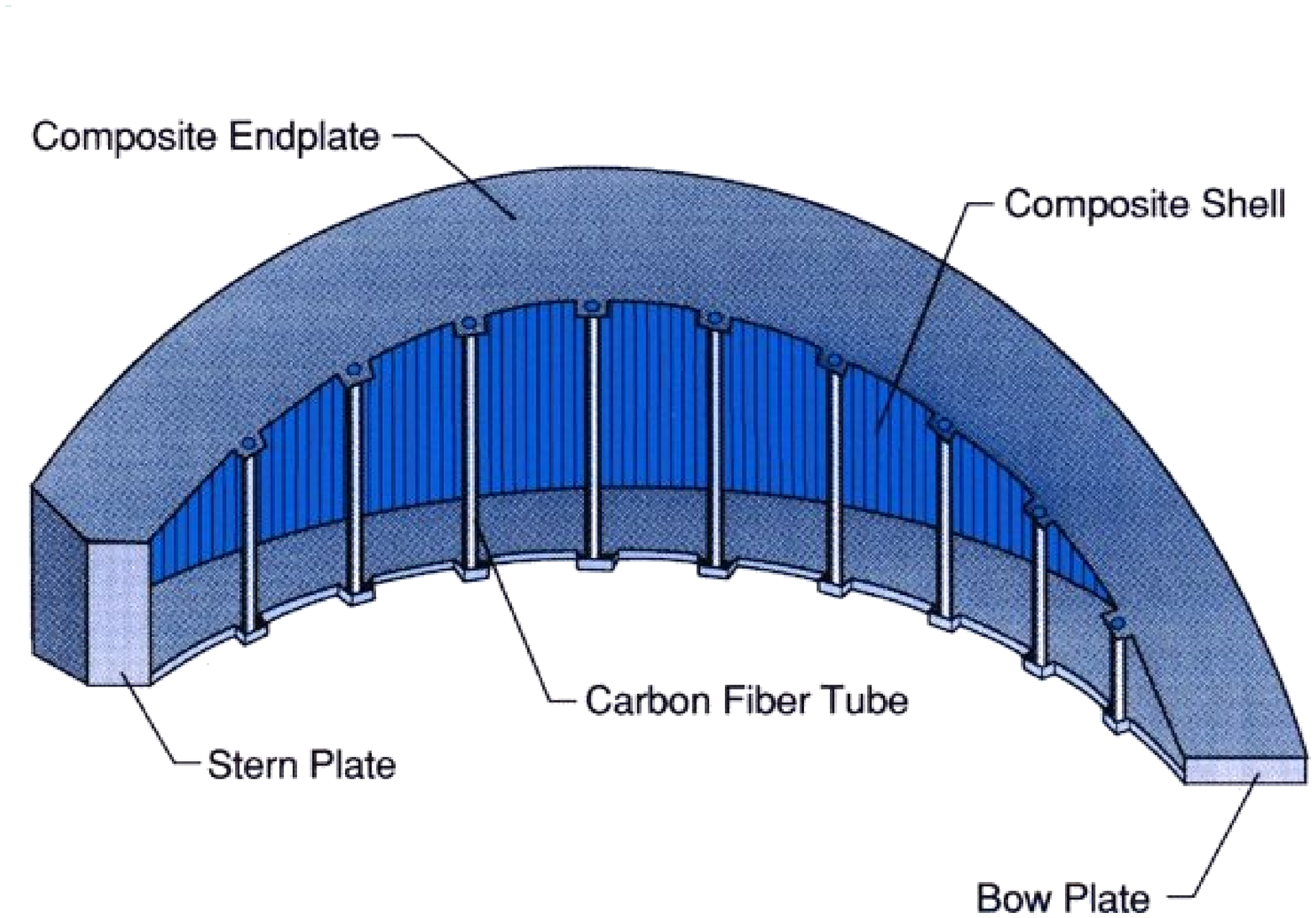,width=7cm}
 \caption{ \small Left panel: Portion of a region three drift chamber, showing
the layout of its two super-layers. The high-lighted drift cells have fired at
a charged particle passing by. Right panel: Section of the whole region 3 drift chamber.}
 \label{fig:dc}
 \end{center}
\end{figure}

To optimally fill the sector volume, the chamber bodies were 
designed to support wires stretched between two end-plates, each parallel to 
its neighboring coil plane, and thus tilted at 60$^\circ$ with respect to each other. 
This design provides maximum sensitivity to the track curvature since the 
wire direction is approximately perpendicular to the bend plane. The wire 
midpoints are arranged in "layers" of concentric (partial) circles, with the wire 
positions shifted by half the nominal wire spacing in successive layers. This 
pattern of wires in neighboring layers, with a repeating pattern of two 
field-wire layers and one sense-wire layer, results in a quasi-hexagonal pattern with 
six field wires surrounding one sense wire. The cell size increases uniformly 
with increasing radial distance from the target. 

For pattern recognition and tracking redundancy, the wire layers in each 
chamber are grouped into two "super-layers" of six wire layers each, one axial to the 
magnetic field, and the other tilted at a 6$^\circ$ stereo angle to provide azimuthal 
information. The stereo super-layer of Region one is an exception to this rule, consisting 
of only four wire layers due to space constraints. A detail of the wire layout is 
shown in Fig.~\ref{fig:dc} The total number of sense wires in the drift chamber 
system is 35,148.

A high voltage system maintains the sense wires at a positive potential and 
the field wires at a negative potential whose absolute value is half that of the 
sense wires. A layer of guard wires surrounds the perimeter of each super-layer, 
with the high voltage potential adjusted to approximate the electric field 
configuration of an infinite grid. This three-voltage scheme minimizes the effects 
of nearby grounded surfaces such as the end-plate.

\subsection{Cerenkov counters}
\indent
\par

The Cerenkov Counter (CC)~\cite{N11} performs the dual function of triggering on 
electrons and separating electrons from pions. The design of the Cerenkov 
detector aims at maximizing the solid-angle coverage in each of the six sectors 
out to an angle B = 45$^\circ$ with the least possible amount of material (to prevent 
degradation of the energy resolution). This is achieved by placing the light 
collecting cones and photomultiplier tubes (PMTs) in the regions of $\phi$ that 
are already obscured by the magnet coils, and covering as much of the available 
space as possible with mirrors. 
Since charged-particle trajectories lie approximately in planes 
of constant $\phi$, the placement of the PMTs in the shadows of the 
magnet coils does not affect the angular coverage. 
The light collection optics was designed to focus the light only in the 
$\phi$ direction, which preserves information on the electron polar angle. 
The Cerenkov radiator gas used in the detector is perfluorobutane 
($\rm{C}_4\rm{F}_{10}$) having an index of refraction of 1.00153. 
This results in a high photon yield and a pion momentum threshold of 2.5~GeV/c. 


\subsection{Time of Flight Scintillators}
\indent
\par
\begin{figure}[htbp]
 \begin{center}
 \leavevmode
 \epsfig{file=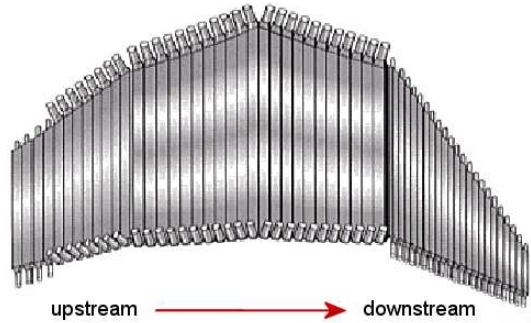,width=7cm}
 \caption{ \small Coverage of one CLAS sector by the time-of-flight scintillators.}
 \label{fig:tof_size}
 \end{center}
\end{figure}
The time-of-flight (TOF) counters~\cite{N12} cover the polar angular range between 
8$^\circ$ and 142$^\circ$ and the entire active range in azimuthal angle $\phi$. 
The scintillators are located radially outside the tracking system and the Cerenkov counters 
but in front of the calorimeters. Their alignment and relative positioning with 
respect to the beam-line are shown in Fig.~\ref{fig:tof_size}. 
\begin{figure}[htbp]
 \begin{center}
 \leavevmode
 \epsfig{file=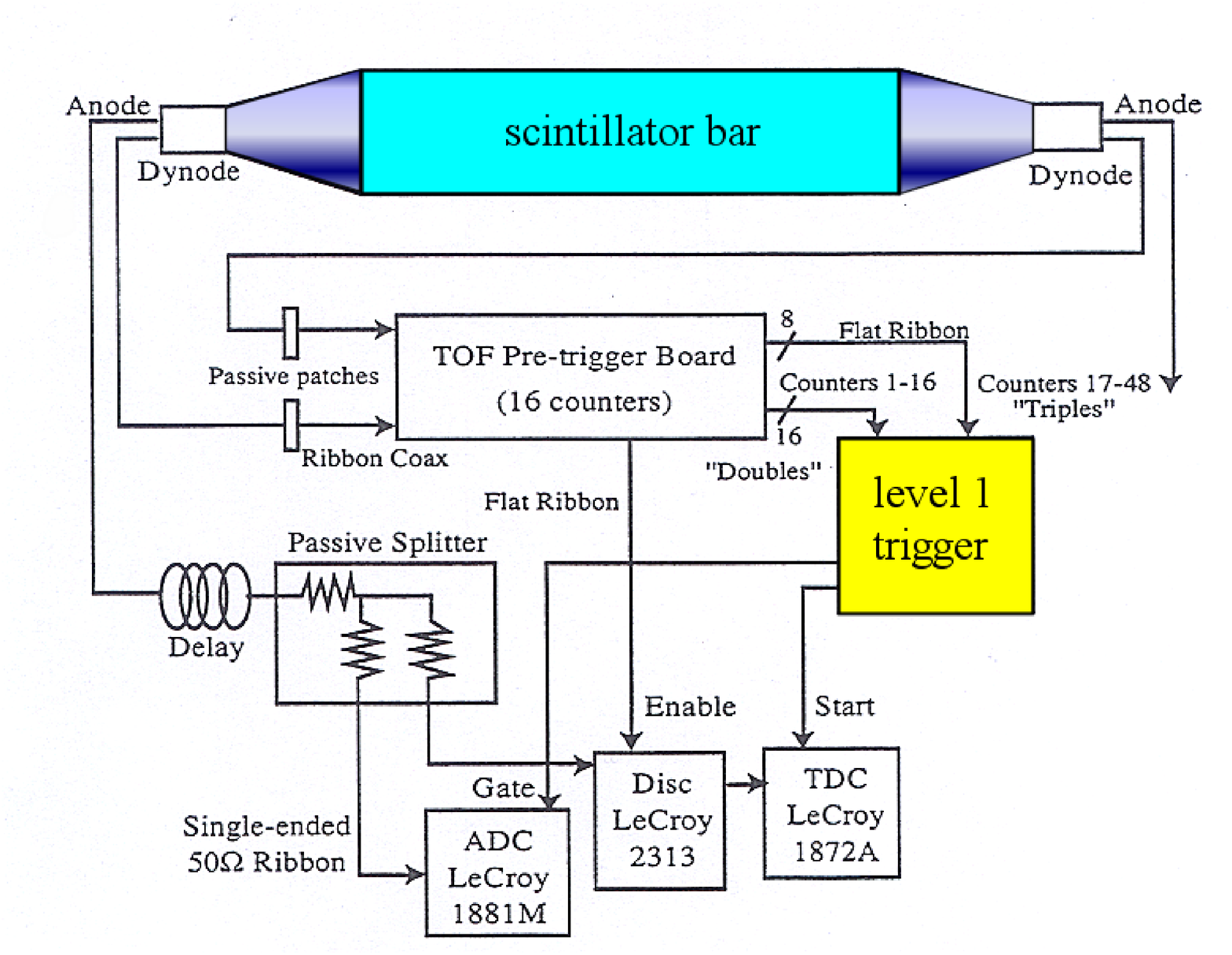,width=10cm}
 \caption{ \small Scheme of the electronic setup for a TOF counter.
Each counter consists of single scintillator bar with a PMT at 
each end. The timing resolution of the scintillators varies from 
80~ps for the shorter counters to 160~ps for the longer ones. }
 \label{fig:tof_el}
 \end{center}
\end{figure}

The scintillator thickness of 5.08 cm is uniform throughout, 
chosen to give a large signal for traversing minimum-ionizing particles. 
Each scintillator is positioned such that it is perpendicular to the average 
local particle trajectory. The width of each counter subtends about 
1.5$^\circ$ of scattering angle. 
The forward counters (at polar angles $\theta$ less than 45$^\circ$) are 15-cm wide, 
and the large-angle counters are 22-cm wide, in the polar angle direction.

Each TOF counter consists of a scintillator with a PMT at 
each end. The timing resolution of the scintillators varies from 
80~ps for the shorter counters to 160~ps for the longer ones.
The electronic readout of a single counter is shown in Fig.~\ref{fig:tof_el}.

\subsection{Electromagnetic Calorimeters}
\indent
\par
\begin{figure}[htbp]
 \begin{center}
 \leavevmode
 \epsfig{file=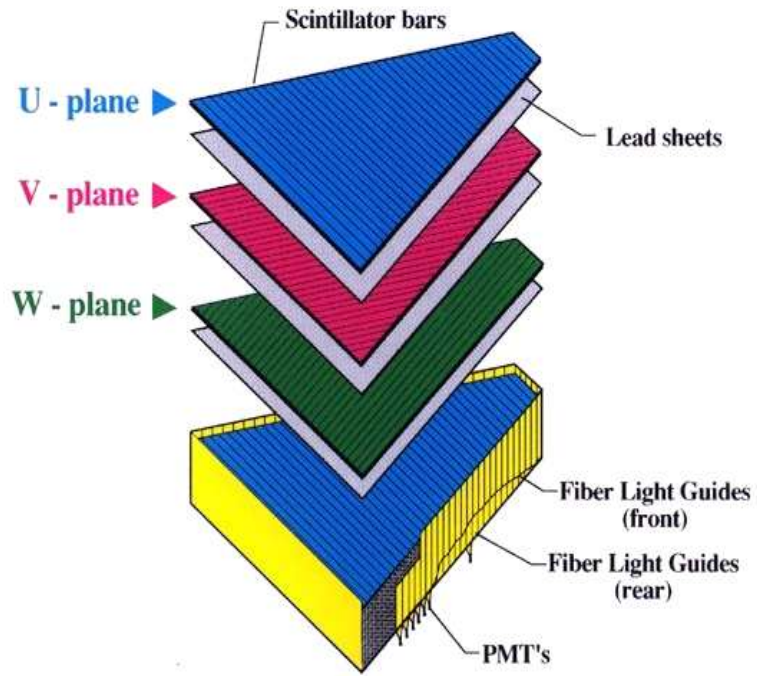,width=12cm,height=10cm}
 \epsfig{file=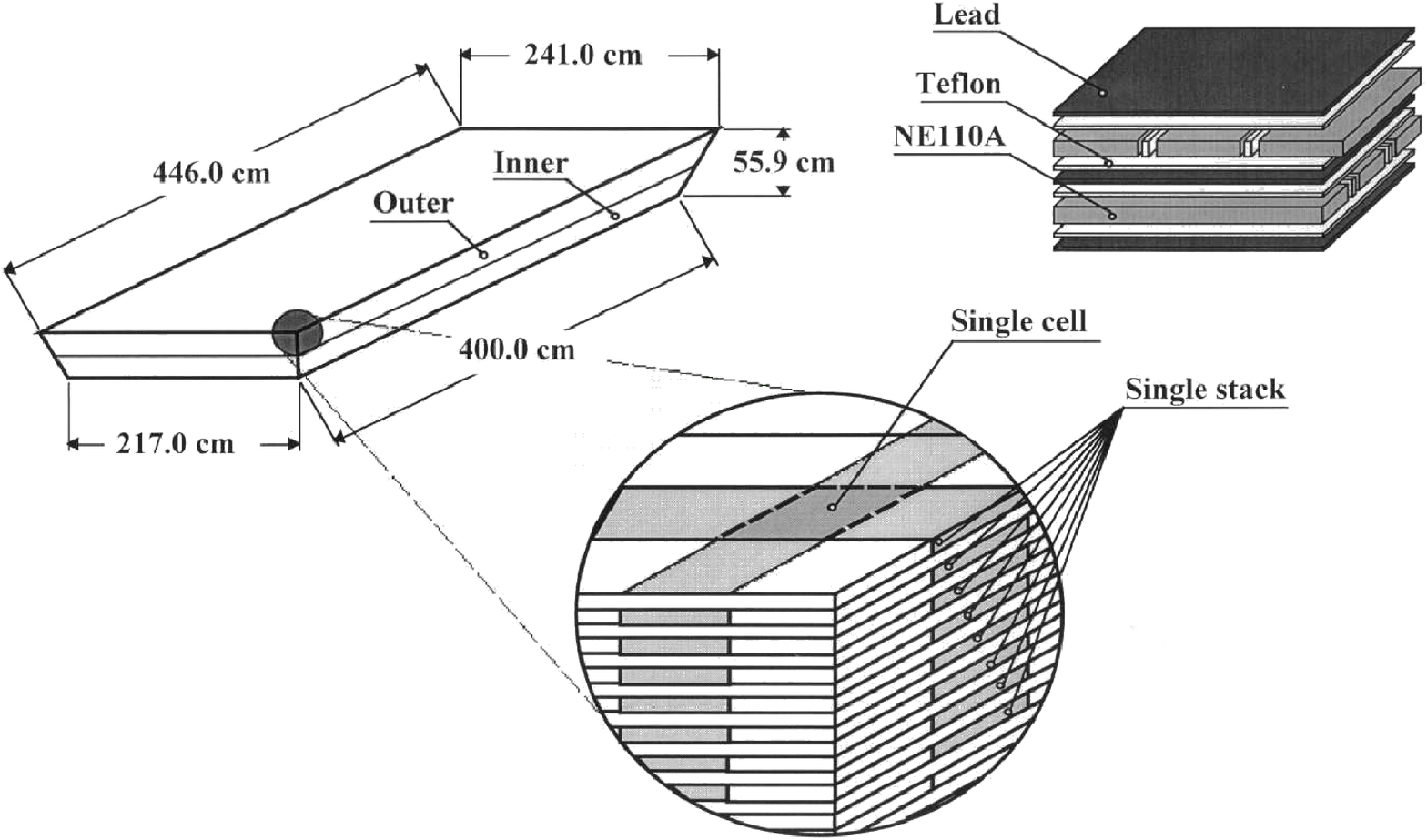,width=10cm,height=10cm}
\caption{ \small Upper panel: schematic view of one  module of the forward electromagnetic calorimeter.
The expanded view shows the three layers of lead and scintillator (conventionally referred as $u$, $v$, and $w$)
making up the calorimeter sandwich structure. Lower panel: the large angle calorimeter
tapered geometry (up-left), its lead-scintillator sandwich structure (up-right), and granularity 
(bottom-center).}
 \label{fig:ecals}
 \end{center}
\end{figure}

The forward region of the CLAS (polar angles from $8^\circ$ to  $45^\circ$)
is equipped with an electromagnetic calorimeter segmented in six independent 
modules of triangular shape. Each module is composed by 13 units, 
and each unit consists of 3 layers of lead sheets 2~mm thick 
interleaved with scintillator bars of 10~cm length and thickness
(top part of Fig.~\ref{fig:ecals}).
Such a configuration provides the best compromise between energy 
resolution and neutron detection efficiency.

Each layer of scintillator is rotated by $120^\circ$ with respect 
to the following layer to allow the stereo readout needed to resolve multiple 
hits into the calorimeter. The five units closest to the target 
are referred as the {\em inner} part of the calorimeter while the remaining 
units form the {\em outer} part. Inner and outer parts have independent
electronic readout to sample the longitudinal shape of the showers developed 
within the calorimeter. The total thickness in terms of radiation lengths is 14.8
and the energy resolution of the forward calorimeter is $\sigma_E/E=10\%\sqrt{E(GeV)}$.
The sampling fraction is approximately 0.3 for 
electrons of 3~GeV and greater, and for smaller energies, there is a monotonic 
decrease to about 0.25 for electrons of 0.5~GeV. The average RMS position 
resolution is 2.3~cm for electron showers with more than 0.5~GeV of energy 
deposited in the scintillator. 
The timing resolution for electrons averages 200~ps over the entire detector.

For two CLAS sectors the coverage is extended to polar angles of $75^\circ$
by the Large Angle Calorimeter (LAC). 
The LAC is also a sampling calorimeter composed of two identical modules with a 
multi-layer structure of lead sheets and scintillator bars, 
similar to the forward calorimeter. A single LAC module consists of 33 layers, each composed of a 
0.20~cm thick lead foil and 1.5~cm thick plastic scintillator bar~\cite{N18}. 

The calorimeter geometry is projective, with scintillators of 10-cm average width. 
To avoid optical coupling, 0.2~mm thick Teflon sheets separate 
neighboring scintillators. Scintillator bars in consecutive layers are rotated by
90$^\circ$ to form a \mbox{$40 \times 24$} matrix of roughly 
\mbox{$10 \times 10$}~cm$^2$ cells (bottom part of Fig.~\ref{fig:ecals}). 

To improve the $e^-/\pi^-$ discrimination, the LAC modules are longitudinally 
divided into inner (17~layers) and outer (16~layers) regions with individual light 
readouts.

Each module is equipped with 256 photo-multipliers tubes~\cite{N20}. 
The gain stability of each LAC photomultiplier is monitored with 
a precision of higher than 0.5\% using a radioactive source of $\alpha$-particles 
($^{241}\rm{Am}$) sealed within a YAP scintillator crystal~\cite{N21}.
Each LAC module corresponds to 12.9 radiation lengths and 1.0 hadronic absorption 
lengths and has an energy resolution of ${\sigma(E)}/{E} = {7.5\%}/{\sqrt{E(GeV)}}$.

\clearpage
\section{Trigger System and Data Acquisition}
\indent
\par

Many of the CLAS detector subsystems use the same type of electronic 
modules for readout to minimize the complexity of the system and to simplify 
maintenance and integration. Initially, all analog signals were 
digitized using commercial FASTBUS modules. 
Recently, however, more electronic modules are available in VME, and are being 
integrated into the system. 

The wire signals from the most complex CLAS subsystem, the
drift chambers, are read out using LeCroy 1877 pipeline time-to-digital 
converters (TDCs) operated in common stop mode~\cite{N8}
with a least count of 0.5~ns. 

For the PMT-based detectors, the signal times are digitized using LeCroy FASTBUS 1872A TDCs, 
and the pulse heights are converted using LeCroy FASTBUS 1881M analog-to-digital converters (ADCs). 
Most TDCs are fed from leading-edge LeCroy 2313 CAMAC discriminators set at a low threshold, 
typically 20~mV. 
The high-resolution TDCs, set to a nominal least count of 50~ps, are needed to achieve 
the required timing resolution for time-of-flight measurements. 
The ADC information is used to determine the energy deposition in each detector, 
as well as to correct to the TDC measurements for time walk.

The inputs to the Level 1 trigger are generated using custom electronics 
to sum appropriate groupings of PMT signals that can be used to generate
fast signals with high efficiency. 
In order to study the trigger in software, these inputs are also recorded 
in the data stream with LeCroy 1877 FASTBUS pipeline TDCs along with all other trigger inputs.

\subsection{Trigger System}
\indent
\par
To acquire events of interest while minimizing the dead-time, 
a two-level hierarchical trigger system was designed~\cite{N24}.
The Level 1 trigger is dead-timeless, processing all prompt PMT 
signals through a pipelined memory lookup within 90~ns. 
The resulting signals are sent to a trigger supervisor module, 
where they are used to gate the front-end electronics~\cite{N25}.
This custom board~\cite{N30} takes the Level~1 (and Level~2) 
inputs from the trigger system and produces all common signals, 
busy gates, and resets required by the detector electronics. 
The Level~2 trigger uses the fast-clear 
capability by clearing events that satisfied Level~1, 
but which have no tracks in the drift chambers. 

\subsection{Data Acquisition}
\indent
\par

\begin{figure}[htbp]
\begin{center}
\leavevmode
\epsfig{file=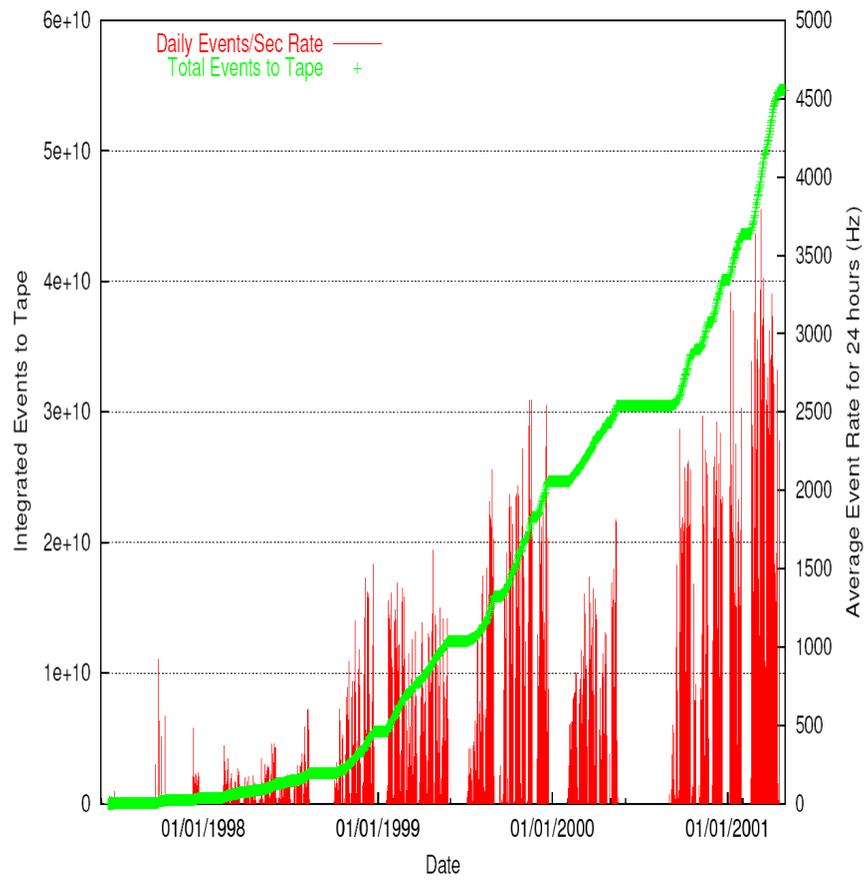,width=12cm,height=12cm}
\caption{\small 
 Event rate and integrated events accumulated by the CLAS data system vs. time for a three year period.}
\label{fig:daq}
\end{center}
\end{figure}

\begin{figure}[htbp]
\begin{center}
\leavevmode
\epsfig{file=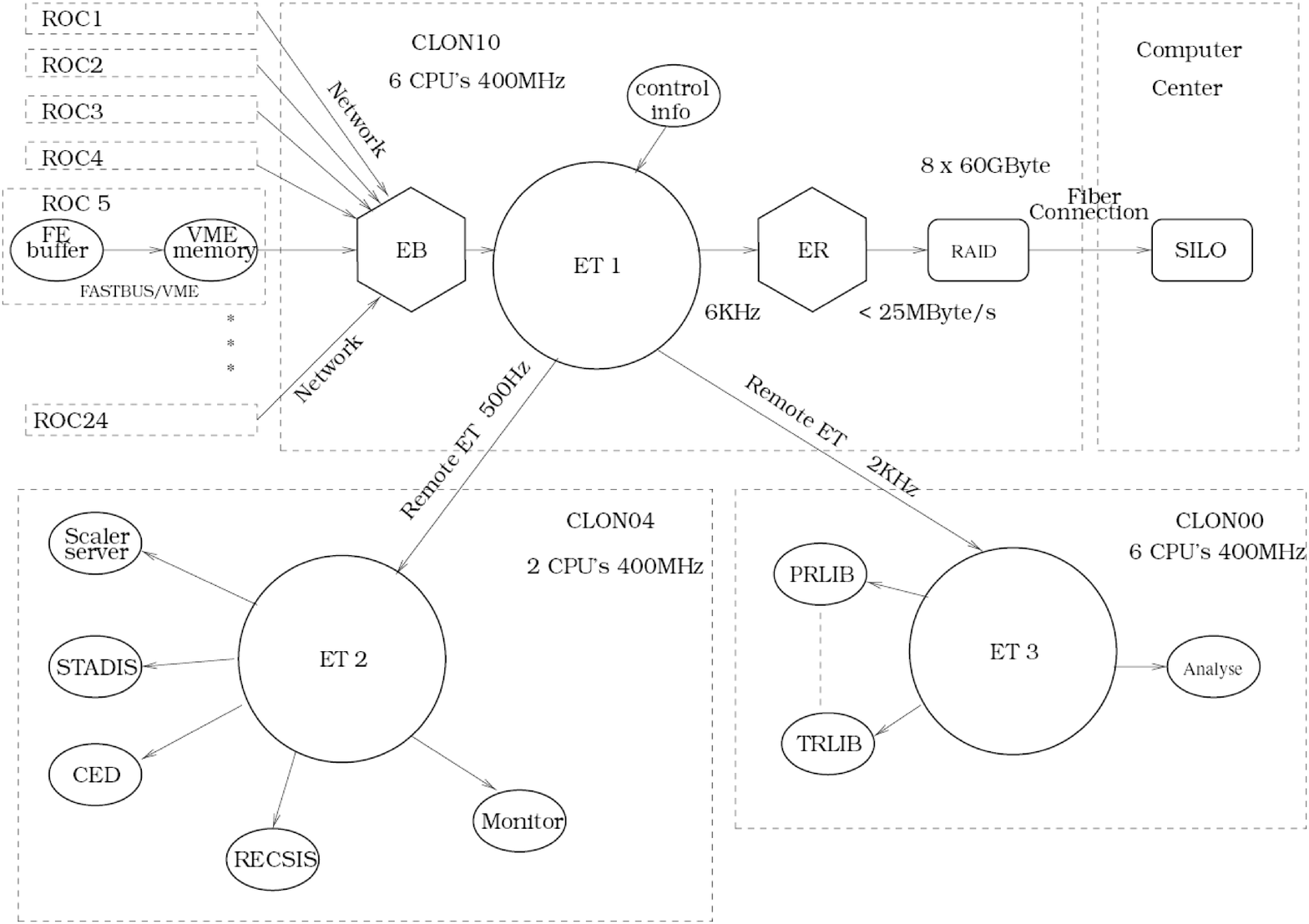,width=14cm,height=18cm}
\caption{\small Data flow schematic for CLAS, with the primary stream using the ET1 shared memory. 
Also shown are a variety of monitoring processes accessing data on other shared memory (ET2, ET3) managed by the ET system.
See text for additional details.}
\label{fig:flow}
\end{center}
\end{figure}

The CLAS data acquisition (DAQ) system was designed for an event rate of 2~kHz. 
Continued development of the DAQ over five years resulted in routine operation 
at event rates between 3 and 4~kHz for the 2000/01 running period
as can be seen from Fig.~\ref{fig:daq}.  
The data from the various detector components are digitized in 24 FASTBUS and 
VME crates within the experimental hall and collected by the 24 VME Readout 
Controllers in these crates. 
These data arrays, or event fragments, are buffered and then transfered via fast Ethernet 
lines to the CLAS on-line acquisition computer in the control room. 
Three primary processes, the Event Builder (EB), Event Transport (ET), 
and Event Recorder (ER), comprise the main data-flow elements in the acquisition computer. 
The EB assembles the incoming fragments into complete events.
The completed event is then labeled by a run and event number, an event type, 
and the trigger bits configuration.
At this stage the event has the final format it needs for off-line analysis.
The data flow is shown in Fig.~\ref{fig:flow}:
the EB passes the completed events to shared memory~\cite{DD} on the CLAS on-line computer. 
The ET system manages this shared memory, allowing access by various event producer 
and consumer processes on the same or remote processor systems. The ER picks up all 
events for permanent storage. 

Some events are transferred to remote ET systems, 
for raw data checks such as hit maps, status, and event displays, 
and for on-line reconstruction, analysis, and monitoring.
The ER writes the data in a single stream to magnetic media. The output files are 
striped across an array of local RAID disks. 
 
The CLAS data flow requires communication among some 100~processes running on a system 
of processors distributed around the detector in the experimental hall and linked to two 
Symmetric Multi-Processor (SMP) machines and a number of workstations in the control room. 
The framework for this organization is the CEBAF 
On-line Data Acquisition (CODA) system~\cite{N31,N32}.
The data runs are set up and performed by the CODA ``Run Control'' sequence, 
where a single run typically lasts for one or two hours. 
\clearpage
\section{Summary}
\indent
\par
The Large Acceptance Spectrometer (CLAS) allows the study 
photo- and electro-induced nuclear and hadronic reactions 
by providing efficient detection of neutral and charged particles over 
a good fraction of the full solid angle. 
The detector capabilities (summarized in Tab. \ref{tab:summ})
are being used in a broad experimental program 
to study the structure and interactions of mesons, nucleons, and nuclei using 
polarized and unpolarized electron and photon beams and targets. 
\begin{table} [htbp]
\begin{center}
\begin{tabular}{|l l c|} \hline
Capability    & Quantity  &  Range   \\ \hline
Coverage	& charged particle angle    	& 8$^\circ$ $\leq \theta  \leq$ 140$^\circ$\\
& charged particle momentum & $p \geq 0.2 $~GeV/c\\
& photon angle (4 sectors)  & 8$^\circ$ $\leq \theta \leq$ 45$^\circ$\\
& photon angle (2 sectors) 	& 8$^\circ$ $\leq \theta \leq$ 75$^\circ$\\
& photon energy			& $E_\gamma \geq$ 0.1~GeV\\
Resolution	& momentum ($\theta \lesssim 30^\circ$)	& $\sigma_p/p \approx$ 0.5\%\\
& momentum ($\theta \gtrsim 30^\circ$) & $\sigma_p/p \approx$ (1-2)\%\\
&polar angle			& $\sigma_\theta/\theta \approx$ 1 mrad\\
&azimuthal angle		& $\sigma_\phi / \phi \approx$ 4 mrad\\
&time (charged particles)	& $\sigma_t \approx$ (100-250) ps\\
&photon energy			& $\sigma_E/E \approx $10\%/$\sqrt{E}$\\
Particle ID	&$\pi/\rm{K}$ separation		& p$\leq$ 2~GeV/c\\
&$\pi^+/\rm{p}$ separation		& p$\leq$ 3.5~GeV/c\\
&$\pi^-$ misidentified as $e^-$	& $\leq$ $10^{-3}$\\
Luminosity	&electron beam			& $L \approx 10^{34}$ nucleon $\rm{cm}^{-2}\rm{s}^{-1}$\\
&photon beam			& $L \approx 5 \cdot 10^{31} $ nucleon  $\rm{cm}^{-2}\rm{s}^{-1}$\\
&data acquisition	&event rate		 4 kHz\\
&data rate & 25 MB/s \\
Polarized target &magnetic field & $B_{{\rm{max}}}=$ 5 T \\ \hline
\end{tabular} 
\end{center}
\caption { \small Summary of the CLAS detector characteristics.}
\label{tab:summ}
\end{table}

The CLAS, in conjunction with the associated photon tagging spectrometer 
represents an ideal experimental setup to measure the deuteron 
photo-disintegration differential cross section. 
In fact, it provides both large coverage in the proton scattering angle 
(from $\theta_p^{\rm{LAB}}=8^\circ$ to $140^\circ$) and an high energy 
(up to 3~GeV) tagged photon beam.

\setcounter{chapter} {2}     
\pagestyle{plain}
\chapter{Event Reconstruction\\ and Data Analysis}

\section{Introduction}
\indent
\par

The \mbox{$\gamma d \rightarrow pn$} differential 
cross section has been measured using the Hall B tagged photon beam in the energy 
range between $0.5-2.95$~GeV and the CLAS large acceptance spectrometer. 
The various CLAS production runs are organized by ``run periods''
where experiments sharing common running conditions take data simultaneously.
In this experiment (JLab E93-017, a part of the
g2 run period experiments, where g2 stands for ``$\gamma$ on a deuteron target'')
a high-statistics dataset has been collected using a loose trigger requiring only one charged 
hadron in the final state in coincidence with a tagger counter.
Four weeks of beam time have been used to accumulate 
approximately 2.4~billion valid triggers in two different
periods during the August (18 days) and September (14 days) 
months of the year 1999.

\section{Running Conditions}
\indent
\par
The E93-017 experiment is characterized as follows: 
\begin{figure}[htbp]
\begin{center}
\leavevmode
\epsfig{file=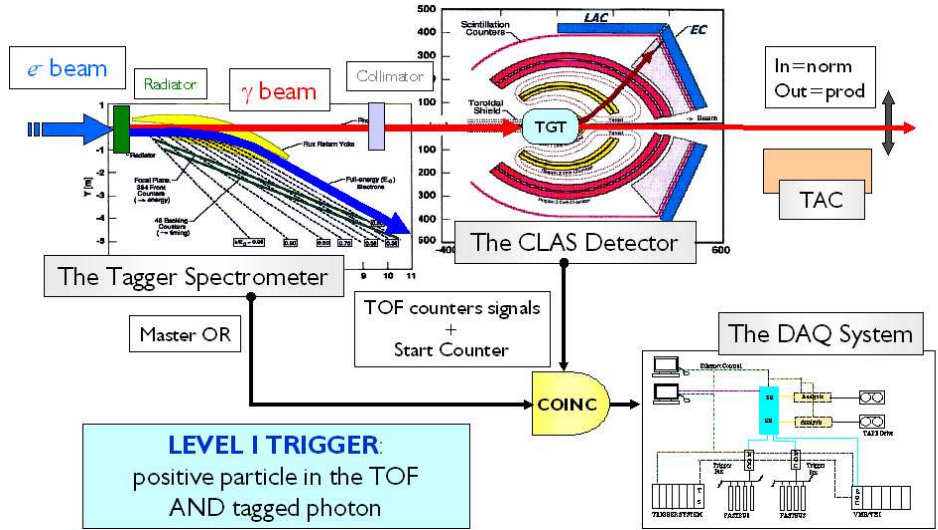,width=15cm,height=7cm}
\caption{\small Top pictures: Schematic view of the experimental setup used during 
the g2 run period. 
From left to right: electron beam, radiator, the tagger spectrometer, the photon beam-line, 
the deuteron target (TGT), the CLAS detector, and the down-stream photon flux normalization components.
For the present experiment, the hadronic events generated in the target volume
are detected in CLAS using the time-of-flight system, in coincidence 
with a tagged photon trigger and a start counter signal. 
Bottom pictures: the Level~1 trigger and the data acquisition system. 
Only the charged trigger scheme is indicated since the neutral trigger 
is not used in the current analysis.}
\label{fig:blt}
\end{center}
\end{figure}
\begin{itemize}
\item {Energy of the incident electron beam: 
\mbox{$E_0$=2.5 GeV} (``August'') and \mbox{$E_0$=3.1 GeV} (``September'').
The electron beam was impinging upon a gold radiator of \mbox{10$^{-4}$ radiation lengths};}
\item {Tagger spectrometer settings: 
full focal plane \mbox{$(0.2-0.95)E_0$}~GeV, or {\em prescaling OFF} configuration
for about 70\% of the running time;\\ high energy part of the focal plane \mbox{$(0.35-0.95)E_0$}~GeV
({\em prescaling ON} configuration) for the remaining time;}
\item {Photon energy: $0.5-2.38$~GeV (``August'') and $0.6-2.95$~GeV (``September'');}
\item {Current intensity: $10-13$ nA};
\item {Photon intensity on the single tagger counter: $\simeq 7 \cdot 10^6$ tagged $\gamma/s$ 
for {\em production} runs 
(standard current intensity reported above) and about $10^5$ tagged 
$\gamma/s$ for {\em normalization} runs 
(low current intensity runs $\simeq100$~pA);}
\item {Cryogenic target: mylar cylinder 4~cm 
in diameter and 10 cm long, filled with unpolarized liquid $D_2$ at 4~K of temperature;}
\item {CLAS superconducting torus magnetic field: 3375~A 
(88\% of the full field intensity) with
positive particles out-bending 
(so called ``normal'' field configuration)};
\item {Level 1 Trigger: coincidence between a signal in the tagger 
(Master OR) and either one charged track 
in the CLAS (80\%) or 2 neutral hits (20\%) 
in the electromagnetic calorimeters (the neutral trigger 
configuration is not used in the present analysis);}
\end{itemize}
\begin{table}[ht]
\begin{center}
\begin{tabular}{|c|} \hline
CHARGED TRIGGER RUNS \\ \hline
20086$\rightarrow$20150   \\
20188$\rightarrow$20350 \\
20584$\rightarrow$20705 \\
20741$\rightarrow$20760  \\ \hline
\end{tabular}
\end{center}
\caption{\small Runs collected with the charged trigger used in the 
\mbox{$\gamma d \rightarrow pn$} analysis.}
\label{tab:trg}
\end{table}

The experimental setup is schematically shown in Fig.~\ref{fig:blt}.
Each data acquisition session is labeled 
by a {\em run number} so that, for the present experiment, 
run numbers go from 19,948 to 20,760.
In addition to the basic dataset of 2.4 billion triggers, 
additional 22 million empty target triggers have been taken 
mainly for background studies together with 178 million 
triggers at very low current intensity.
These last data have been collected to measure 
the absolute photon flux and to study the 
photon tagger characteristics. 
In the present analysis, only the charged trigger runs have been
used (shown in Tab.~\ref{tab:trg}) for a total of 1,171 million of events.

\section{Event Reconstruction}
\indent
\par

\subsection{Physics Data}
\indent
\par
\begin{figure}[htbp]
\begin{center}
\leavevmode
\epsfig{file=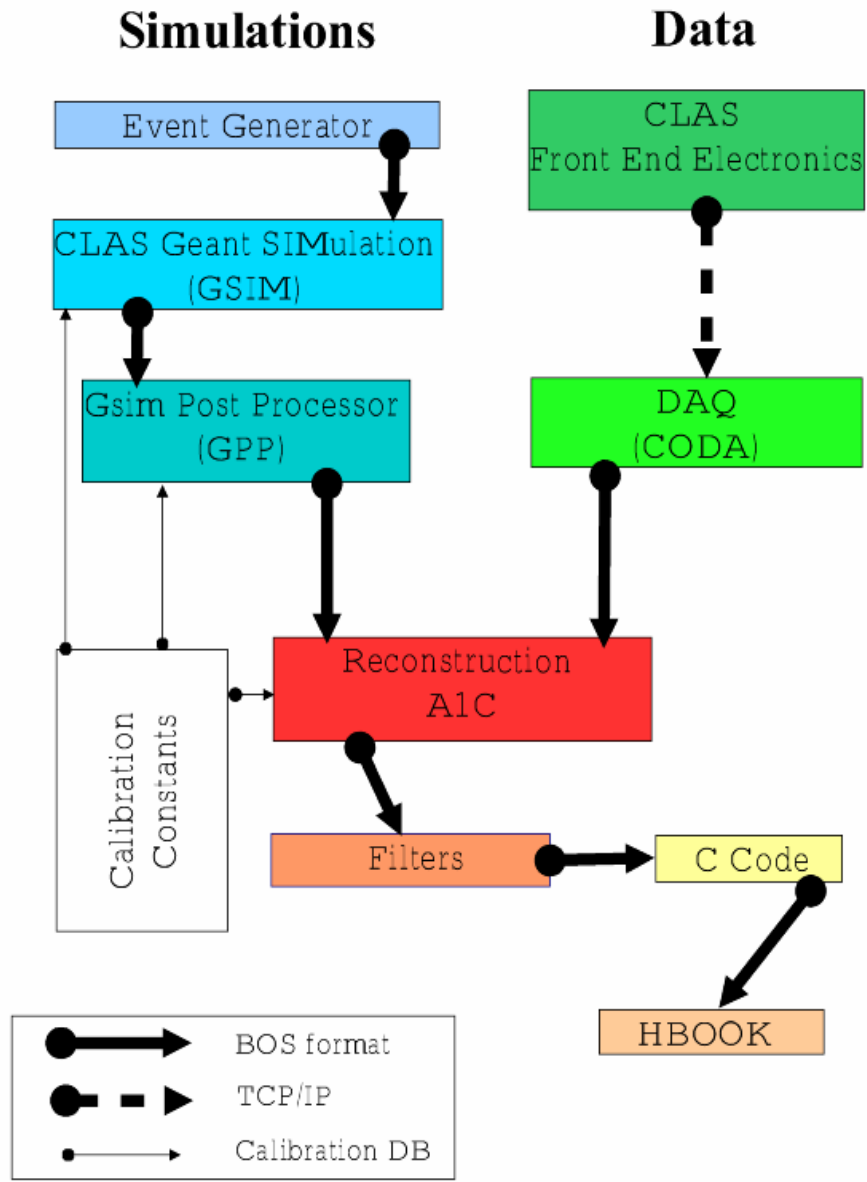,width=12cm,height=18cm}
\caption{\small  Block diagram of the off-line data processing. The raw data files 
(simulated -left panel- or measured -right panel-) are converted into 
reconstructed files and processed by CLAS applications or user written code to obtain
standard n-tuple files.}
\label{fig:offline}
\end{center}
\end{figure}

The event files collected during the on-line 
data acquisition (CODA) sessions are stored on tape silo system, 
to be retrieved in the off-line phase.
These files contain information on the detector ADC/TDC readouts 
at the time the event was digitized in CLAS and are called {\em raw} data files. 
The final codification, in the form of standard n-tuples, of the physical quantities of interest 
requires a multi step procedure performed using both official 
software packages (developed by the CLAS collaboration) 
and custom (user-written) code. The logic flow chart
of this procedure is shown in Fig.~\ref{fig:offline} (right side):
files collected during the experimental run are processed by the 
reconstruction code (called ``A1C'').
The first step of the reconstruction process is the conversion of 
raw ADC/TDC hits from the detector readout electronics 
into physical quantities such as energy and time.
This task is accomplished using
appropriate constants (also called calibration constants),
defined for each detector readout subsystem and run period.
The whole set of calibration constants is stored into the CLAS database system 
which can be accessed from the off-line software packages and
provides the information used for the reconstruction  
in the form of custom files (called {\em map} files) 
or via an on-line SQL-database server. 

Using the calibration constants, 
the charged particle 4-momentum is determined from the 
time of flight and the track radius of curvature 
measured by the time-of-flight and drift chamber systems, respectively.
Neutral particle momenta are basically reconstructed using 
the information coming from the electromagnetic calorimeters. 

The procedure for the determination of the calibration
constants is iterative:
a preliminary reconstruction pass is performed on a 
relatively small fraction of the total collected raw files
uniformly spaced through all the data taking period.
During the preliminary pass, up-to-date pedestals constants 
for the various CLAS subsystems are used together with
a ``test'' set of calibrations constants 
usually taken from a previous run period with similar 
experimental conditions. 
Then, the quality of the reconstructed data 
is monitored: if the result is not satisfying 
the constants are recalculated and successive passes of reconstruction 
are performed until the values of monitored quantities
are acceptable. At this point the set of calibration constants
is ``frozen'' into the database
so that massive data reconstruction can be started 
(this process is referred as ``cooking''). 
The output of the cooking process, is automatically screened by several 
data monitoring and filtering software utilities in order to check for 
possible problems or failures.
For instance, diagnostic programs produce the 
status files containing the drift chambers occupancy and time-of-flight 
paddles status for the run period of interest.

In order to optimize  the disk space requirements for the off-line analysis,
a filtering package is run on the entire reconstructed dataset
so that a set of reduced  files ({\em skimmed} files) 
are produced for each (group of) physics channel(s) of interest. 
These reduced files only contain events belonging the particular channel being 
analyzed, and only the banks strictly needed by the off-line analysis code. 

The physical quantities characterizing the reconstructed events are 
organized in a sequence of numerical fields colloquially called {\em banks}. 
Banks are accessed and managed using a FORTRAN-based memory allocation 
system known as BOS (Basic Object System) \cite{BOS}. 
Bank objects are addressed by capitalized 4-letters 
pointers.

\subsection{Simulated Data}
\indent
\par
A similar reconstruction chain is used for data files 
resulting from the GEANT Simulation (GSim) of the CLAS detector.
As shown on the left side of Fig.~\ref{fig:offline}, 
the output of the GSim tool is processed by the 
same reconstruction program used for real data files,
with the exception that a calibration procedure is not needed.
In this case, the database system provides the ``reverse'' 
calibration constants needed by the GSim tool to produce 
the appropriate ADC/TDC hits for the generated particles in their
simulated interaction with the CLAS.
The complete simulation of the CLAS detector is performed in 
two steps: GEANT simulation and Post Processing.

In the first step, raw data files are created from the
GSim tool considering the CLAS as an ``ideal'' detector
where all the subsystems are working properly and
the timing and spatial resolutions values are identical to the nominal ones.

On the contrary, the post processing phase introduces finite effects for
the timing and spatial resolutions and takes into account 
possible problems affecting the actual detector setup.
In fact, during the run period of interest, 
some detector subsystems may have out-of-work electronic channels 
and the drift-chambers may also have out of work wires.

An software package called GSim Post Processor (GPP) 
applies the necessary corrections to the GSim files
using the information on the detector status 
(obtained from the diagnostic programs run after
the reconstruction process) and takes into account
the timing and spatial resolutions finite effects.

The introduction of the finite resolution effects is done defining 
a certain number of {\em smearing} factors for the time-of-flight 
and drift chambers subsystems.

The smearing factor $f$ for the time-of-flight 
timing resolution is defined as follows:
\begin{equation}
\sigma_i = 2fR_i
\end{equation}\\
where $R_i$ is the ideal resolution of the TDC coupled to the time-of-flight
scintillator paddle $i$.

\begin{figure}[ht]
\begin{center}
 \leavevmode
\epsfig{file=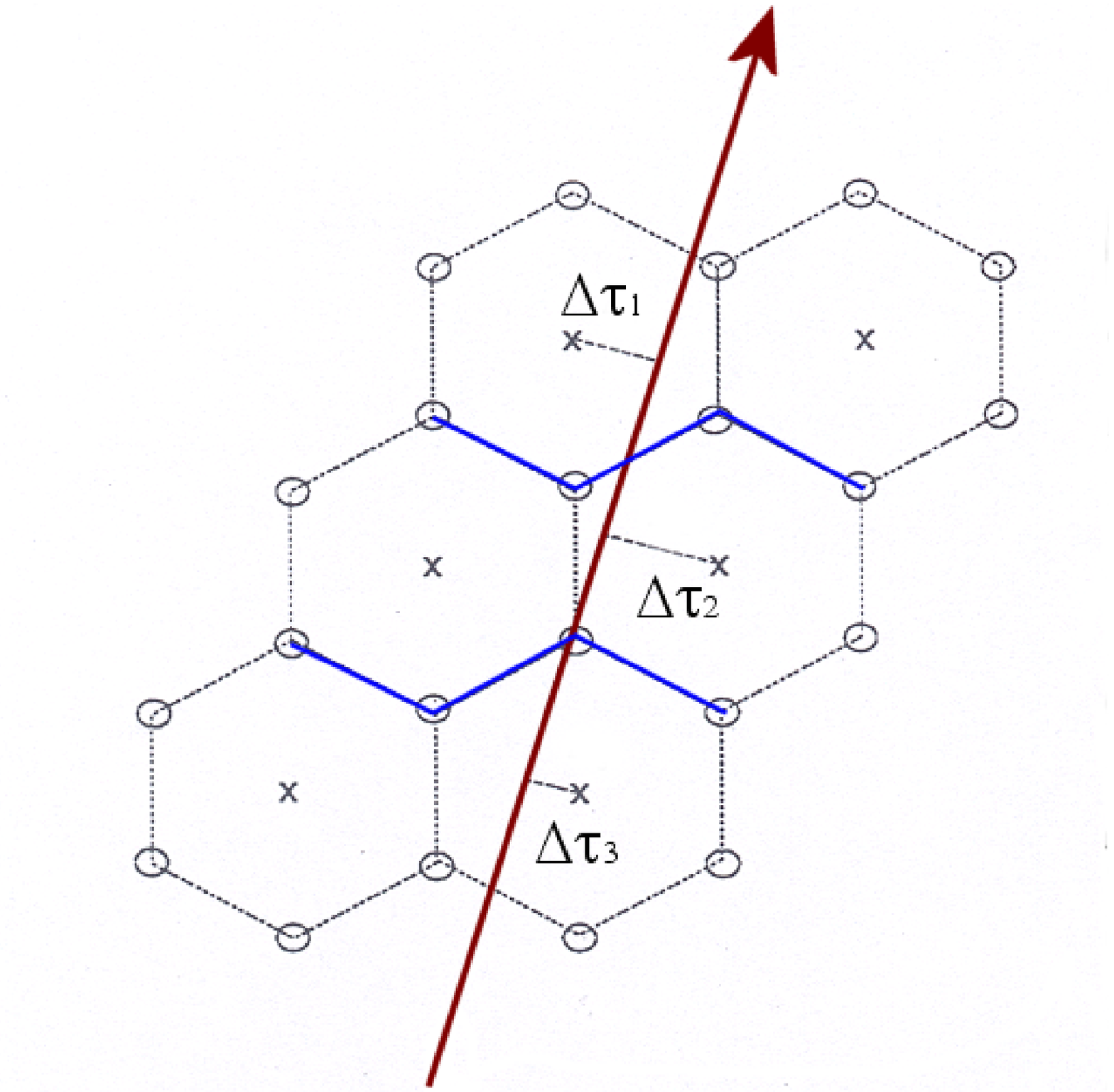,width=6cm}
 \epsfig{file=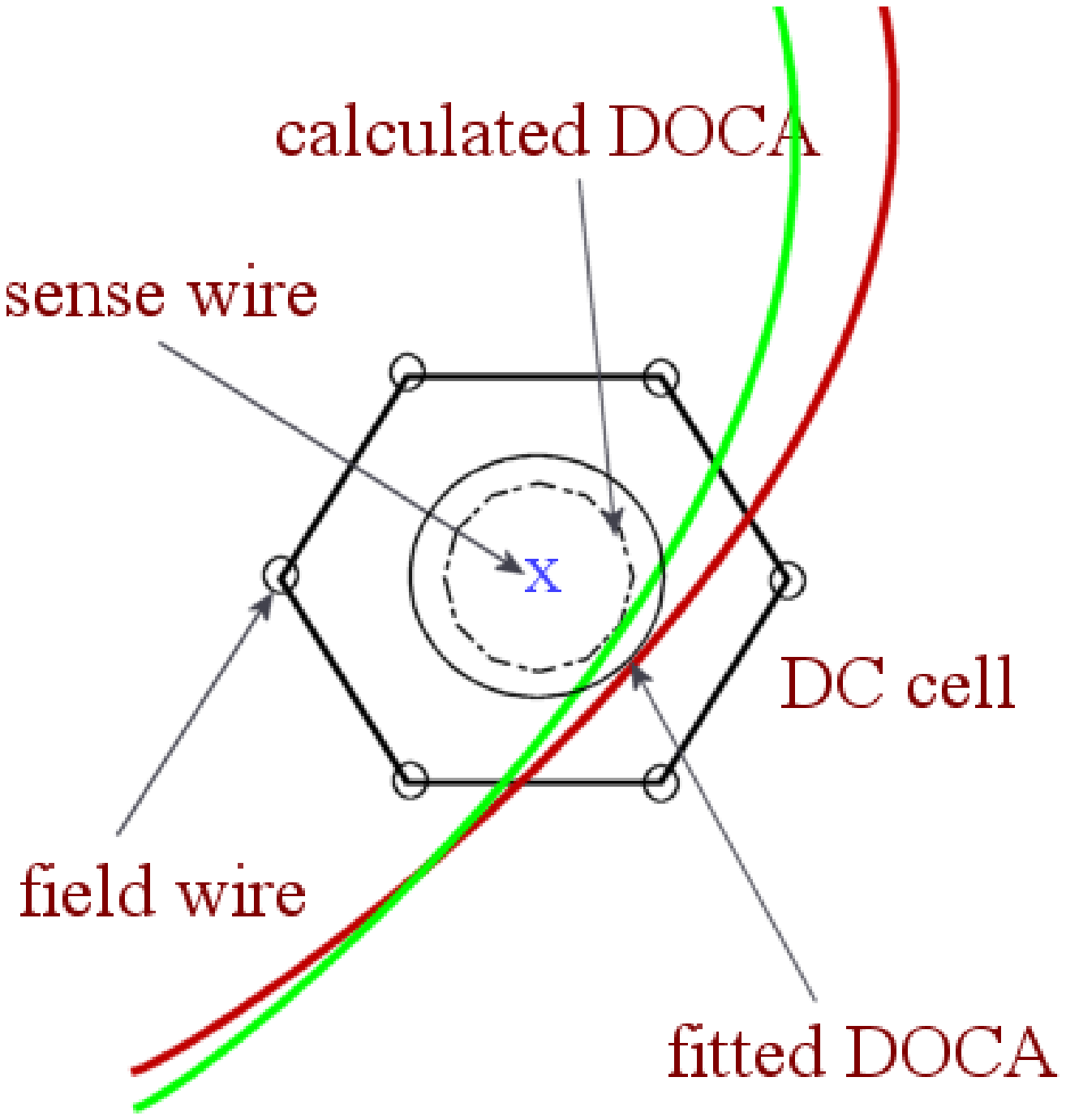,width=6cm}
 \caption{\small The tracking procedure in the CLAS drift chambers system. Left panel: a charged
particle passing through a super-layer. The field wires are indicated by the open dots
while crosses indicate the sensitive wires which collect the released charge. The drift times
are associated to distances of closest approach (DOCAs) represented by the dotted lines. Right panel: 
two different values for the distance of closest approach of the particle from the 
sense wire can be defined, giving origin to a loss of spatial resolution. See text for details.}
 \label{fig:dc-gpp}
\end{center}
\end{figure}

For what concerns the spatial resolution of the drift chambers,
the physical mechanism exploited to convert the drift
time into distance is shown in the left panel of Fig.~\ref{fig:dc-gpp}.
A charged particle passing through a DC cell, 
releases a charge deposition on the sense wire 
that yields a drift time which 
is converted into a distance
(DOCA, Distance Of Closest Approach) using pre-calculated functions. 
 
On the other hand, the overall track, as defined by the reconstruction
software, takes into account the time-distance information coming 
from all the cells crossed by the particle by means of a best fit procedure.
In this case, the distance of closest approach from the sense 
wire under consideration will be close to the 
calculated distance, but not exactly identical (right panel of Fig  \ref{fig:dc-gpp}). 

Inside each DC region the spatial residual is 
defined by the difference: 

\begin{equation}
| \rm{DOCA}_{\rm{calc}}^{(j)} | - | \rm{DOCA}_{\rm{fit}}^{(j)} |
\label{sr}
\end{equation}\\
where $j=1,2,3$ runs over the three DC regions.
It turns out that spatial residual as a function of $\rm{DOCA}_{\rm{calc}}$ 
is almost constant. 
The resulting distribution can be fitted using a 5$^{th}$-order 
polynomial:
\begin{equation}
S_j = \sum_{i=0}^{5}{ p_{i}^{(j)}  \rm{DOCA}_{\rm{calc}}^{i,(j)} }
\end{equation}
so that:
\begin{equation}
	\sigma_j = \rm{const} \cdot f_j  \cdot S_j 
\end{equation}\\
where the quantities $f_{(1,2,3)}$ are the three spatial resolution smearing
factors.
%
%

In order to introduce in the simulation the finite effects 
for the timing and spatial resolutions and the actual detector status, 
the GPP package is ran on the GSim output files before they are processed by the 
reconstruction program.

\section{Event Selection}
\indent
\par
\label{sec:id}

In order to constrain the two body deuteron photo-disintegration events,
the missing mass spectra have been calculated from the relation~\ref{eq:gdpn}: 
\begin{equation}
M_X^2 = \left( P_\gamma + P_d - P_p \right) ^ 2 
\label{eq:gdpn}
\end{equation}\\
where $P_\gamma$, $P_d$, and $P_p$ are the \mbox{4-momenta} obtained by the
reconstruction software for the identified particles.

\begin{figure}[htbp]
\begin{center}
\leavevmode
\epsfig{file=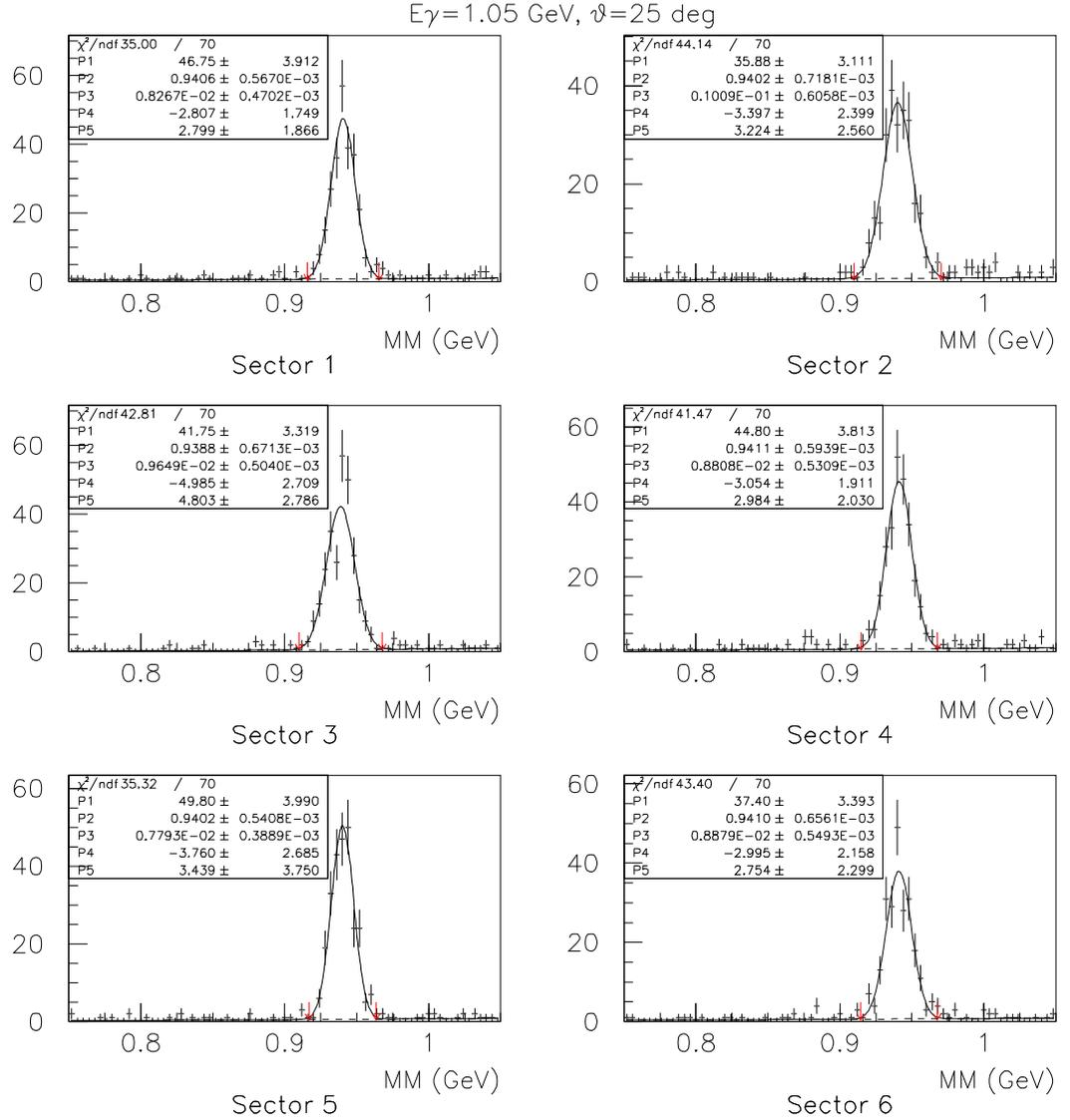,width=14cm}
\caption{\small 
Typical missing mass spectra calculated for the six CLAS sectors
from Eq.~\ref{eq:gdpn} for an incident photon energy of $E_\gamma=$1.05~GeV
and a forward proton scattering angle $\theta_p^{\rm{LAB}}=25^\circ$.
The peak corresponding to the neutron rest mass can be clearly identified, and the shape 
of the distributions is well reproduced by a Gaussian plus an exponential fitting function.
The red/dotted arrows shown in the
panels represent the $(-3\sigma,3\sigma)$
interval defining the width of the neutron peak, being
$\sigma$ the Gaussian function standard deviation.
}
\label{fig:id1}
\end{center}
\end{figure}

\begin{figure}[htbp]
\begin{center}
\leavevmode
\epsfig{file=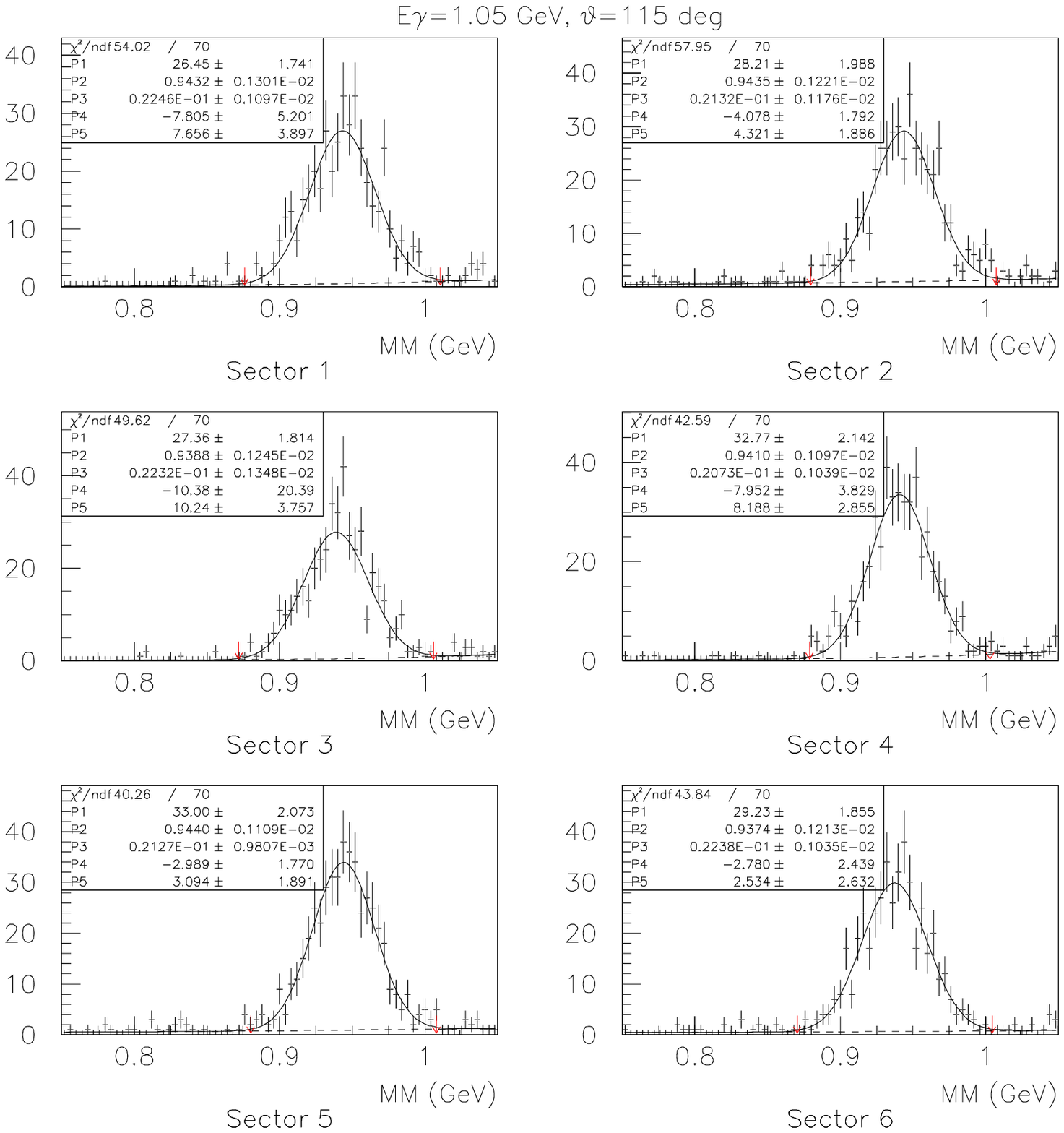,width=14cm}
\caption{\small
Example of missing mass spectra calculated for the six CLAS sectors
from Eq.~\ref{eq:gdpn} for an incident photon energy of $E_\gamma=$1.05~GeV
and a backward proton scattering angle $\theta_p^{\rm{LAB}}=115^\circ$.
The distribution is less pronounced compared to the one obtained at forward angles
since the momentum resolution of the CLAS is worse in the backward direction. 
Nevertheless the neutron signal is clearly evident and 
the shape of the distributions is well reproduced by a Gaussian 
plus an exponential fitting function.
Again, the red/dotted arrows represent the $(-3\sigma,3\sigma)$
interval defining the width of the neutron peak, being
$\sigma$ the Gaussian function standard deviation.
}
\label{fig:id2}
\end{center}
\end{figure}

The resulting distributions exhibit a clear peak corresponding 
to the neutron rest mass. An example is shown in Figs~\ref{fig:id1} and Fig.~\ref{fig:id2}
where the missing mass distributions
are calculated for the intermediate incident photon energy of
$E_\gamma=$1.05~GeV and for proton scattering angles of
$\theta_p^{\rm{LAB}}=25^\circ$ and $\theta_p^{\rm{LAB}}=115^\circ$, respectively.
The different plots represent the contribution to the missing mass spectra from 
the six CLAS sectors.
At forward angles, the neutron peak is quite sharp, the typical signal to noise ratio being 
of the order of $40\div60$ in this case.
The shapes of the missing mass distributions are well reproduced using
a Gaussian plus exponential fitting functions. The red/dotted arrows shown in the
panels of Fig.~\ref{fig:id1} represent the $(-3\sigma,3\sigma)$
interval (where $\sigma$ is the standard deviation of the Gaussian fitting function)
defining the width of the neutron peak.
At backward proton scattering angles, 
the distributions peaks are less pronounced compared 
to those obtained at forward angles, since the momentum resolution
of the CLAS is worse in the backward direction.
Nevertheless, the neutron peak is still clearly evident. 

In order to exclude  \mbox{$\gamma d \rightarrow p X$} 
events yielding mass values 
too far from the neutron rest mass the width $\sigma$ of the 
Gaussian function is used to define ($-3\sigma$, $3\sigma$) cuts 
around the distribution peak.

\begin{figure}[htbp]
\begin{center}
\leavevmode
\epsfig{file=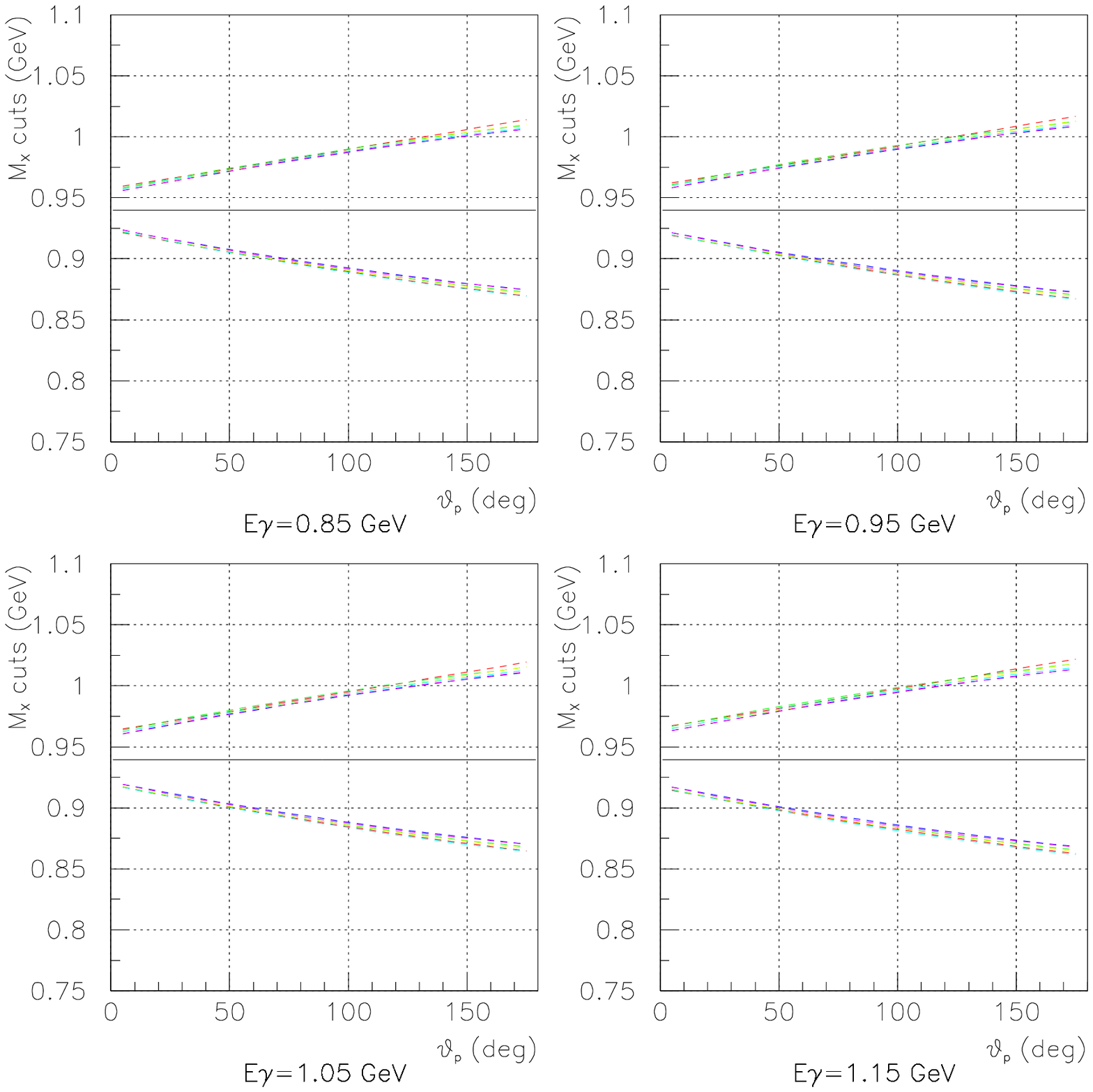,width=12cm}
\caption{\small Behavior of the missing mass cuts as a function of the 
proton scattering angle $\theta_p^{\rm{CM}}$ for four incident photon energies
around 1~GeV. The solid horizontal line represents the value of the neutron rest mass, 
while the different dotted lines represents the missing mass cuts for the six CLAS sectors}
\label{fig:bkg-cuts}
\end{center}
\end{figure}

These cuts are calculated, for each CLAS sector, 
as a function of the incident photon energy $E_\gamma$ 
and of the proton scattering angle $\theta_p^{\rm{CM}}$.
They have been used to define which proton events 
have to be associated to a two body photo-disintegration final state
using also the information
on the proton detection efficiency and detector fiducial cuts
as explained in Sec.~\ref{sec:eff}.
An example of the behavior of the missing mass cuts as a function of 
$\theta_p^{\rm{CM}}$
is shown in Fig.~\ref{fig:bkg-cuts} for four
incident photon energies around 1~GeV:
the different dotted lines correspond to cuts in the six CLAS sectors.
The missing mass cuts become wider at large scattering angles
reflecting the behavior of the missing mass distributions
shown in Fig.~\ref{fig:id1} and  Fig.~\ref{fig:id2}.

At higher incident photon energies the neutron peaks in the 
missing mass distributions
become less defined since the resolution of the spectrometer 
decreases. In fact, 
the reconstruction of the momentum of straighter tracks left by
protons of high momentum is less accurate 
and moreover the number of events is smaller since the 
cross section decreases very rapidly as 
a function of the incident photon energy.

\begin{figure}[htbp]
\begin{center}
\leavevmode
\epsfig{file=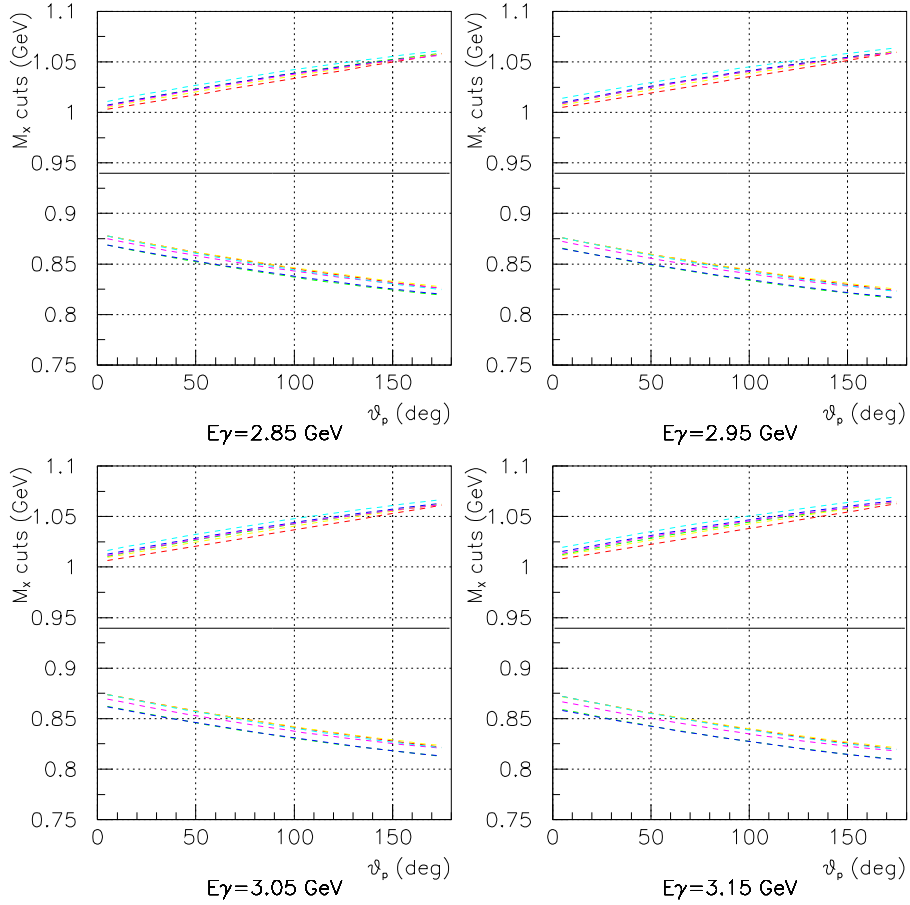,width=12cm}
\caption{\small Behavior of the missing mass cuts as a function of the 
proton scattering angle $\theta_p^{\rm{CM}}$ for four incident photon energies
around 2.95~GeV. The solid horizontal line represents the value of the neutron rest mass, 
while the different dotted lines represents the missing mass cuts for the six CLAS sectors.
In this case,
the cuts are wider than those calculated for lower photon energies nevertheless 
the missing mass distributions are still peaked enough for the fitting procedure
to be applied.
}
\label{fig:bkg-cuts1}
\end{center}
\end{figure}

An example of the behavior of the missing mass cuts 
as a function of the 
proton scattering angle $\theta_p^{\rm{CM}}$
at incident photon energies around 2.9~GeV is shown in Fig.~\ref{fig:bkg-cuts1}. 
In this case, for the reasons explained above, 
the cuts are wider than those calculated for lower photon energies even if 
the missing mass distributions are still peaked enough for the fitting procedure
to be applied.

\clearpage
\subsection{Interaction Vertex}
\indent
\par
\label{sec:vertex}
The interaction vertex of the incident photon with a deuteron 
in the target volume is calculated from events with more 
than one charged particle in the final state using 
the intersection between the different tracks.



\begin{figure}[htbp]
\begin{center}
\leavevmode
\epsfig{file=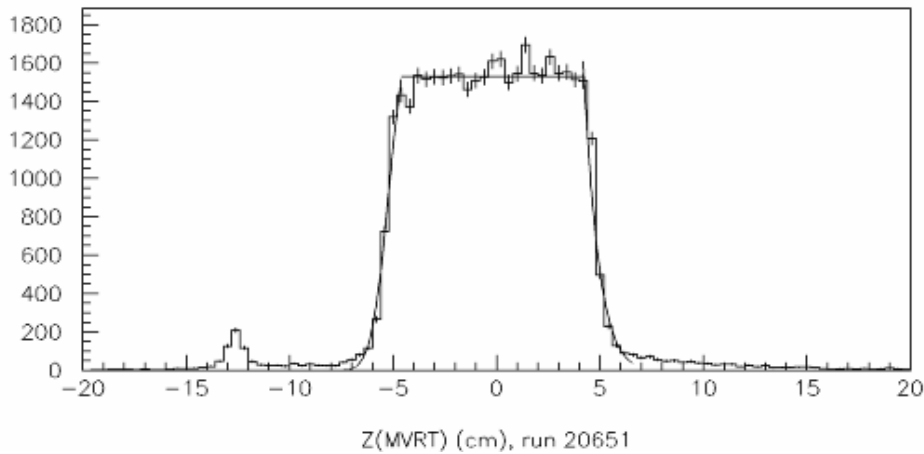,width=13cm}
\caption{\small Distribution of the vertex $z$-component for proton events originating 
in the target volume. The distribution shape reflects the target dimensions 
(10 cm length). The small peak on the left is the contribution of a target support 
structure and is eliminated by a cut on the vertex $z$-coordinate.
The information on the interaction vertex and the $x$ and $y$ positions 
of the beam is contained in the MVRT bank in the case of multi-track events.
The solid line represents a fit to the distribution defining the
effective target length.
}
\label{fig:vertex}
\end{center}
\end{figure}

The distribution of the vertex $z$ variable is shown in Fig.~\ref{fig:vertex},
where $z$ is the direction of the incoming photon beam.
As can be seen, the events are uniformly distributed along 
the 10 cm length target with a small contribution coming from 
a support structure located about 12 cm on the left of the target center.
The solid line represents a fit to the distribution defining the
effective target length.

In order to select events originating from the target volume 
containing liquid $D_2$ the following vertex cut is applied along the
$z$-axis:
\begin{equation}
-5.3 \leq z \leq 4.7 \ \rm{(cm)}\ ,
\label{eq:vz}
\end{equation}\\
where a displacement of $-0.3$~cm has been introduced to compensate
for the shift of the center of the target from the origin of the $z$ axis.  

\subsection{Energy Loss}
\indent
\par
The charged particles emitted in the target volume lose energy
from ionization processes  while traversing the liquid deuterium. 
The same effect occurs outside the target region when the charged hadron 
traverse  the start counter scintillators. 
For these reasons a momentum correction has been introduced (Ref.~\cite{ELOSS}) 
in the reconstruction procedure to compensate for the energy lost by the
outgoing charged particles in the target and start-counter regions.

\clearpage
\section{Photon Flux Normalization}
\indent
\par
\begin{figure}[htbp]
\begin{center}
\leavevmode
\epsfig{file=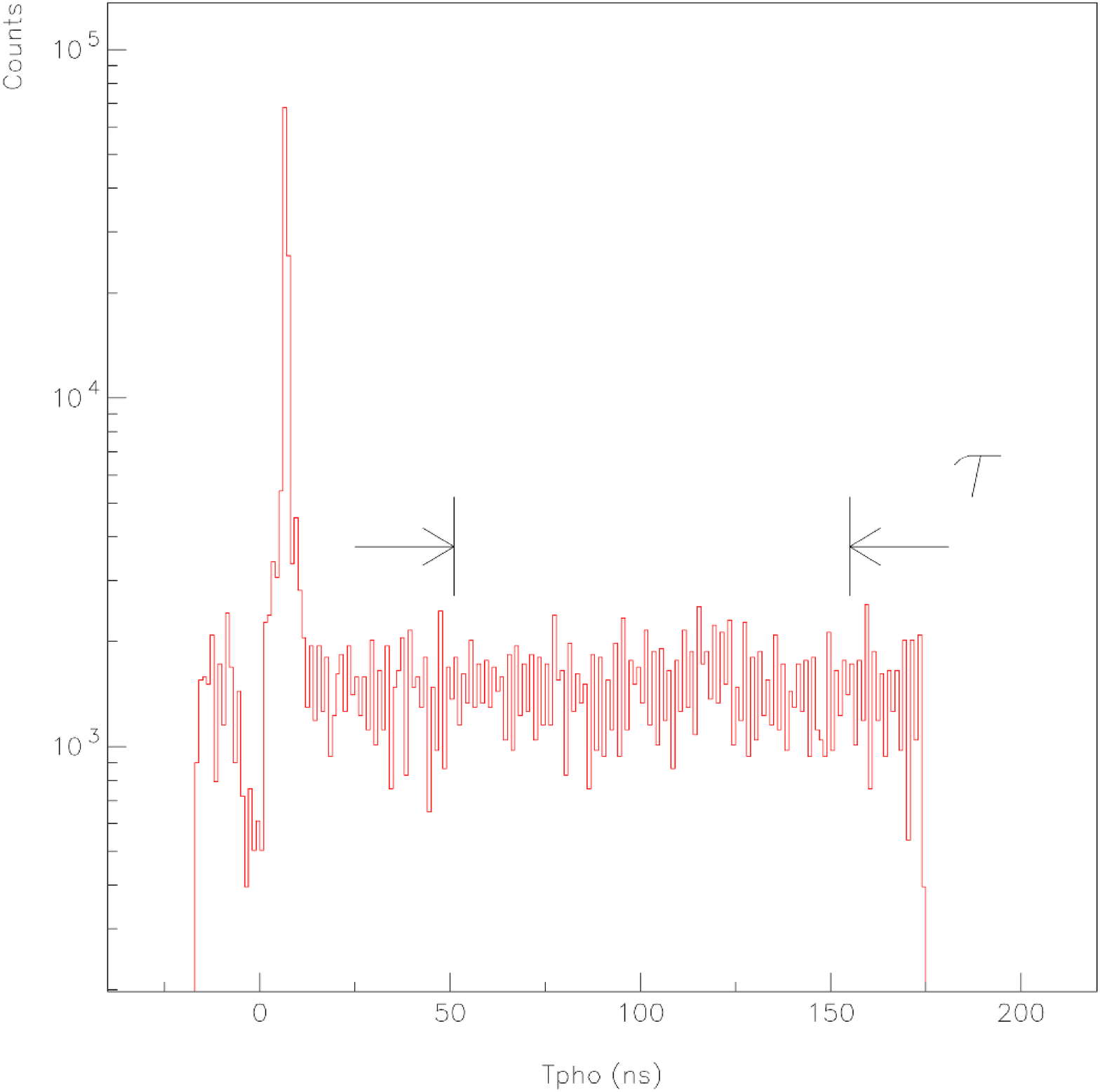,width=10cm,height=8cm}
\caption{\small TDC timing spectrum of a single \mbox{T-counter}.
The peak which caused the trigger and the accidental background can be clearly
identified.
Utilizing the accidental hits in the tagger, the normalization 
method determines the number of tagged photons from each \mbox{T-counter}. 
This is done by measuring the number of accidental hits of each \mbox{T-counter} in a given 
time interval $\tau$, which is called the out-of-time window. 
}
\label{fig:out-time}
\end{center}
\end{figure}

\label{sec:norm}
In order to calculate the cross section, 
the incident photon flux at the CLAS target must be known 
and the goal of the normalization procedure \cite{NORM}
is to determine it as a function of the energy.
This is done using the information of the out-of-time
photons by measuring the number of accidental hits of
each \mbox{T-counter} in a given time interval $\tau$,
which is called the out of time window.
Fig.~\ref{fig:out-time} shows the histogram of the TDC timing spectrum for a single 
\mbox{T-counter}. 
The peak which caused the trigger and the accidental background can be clearly
identified. Since the tagger TDCs operates in ``common start'' mode this peak implies that it 
was this \mbox{T-counter} which started all the other TDCs from the CLAS trigger
associated to the hadronic event generated by this particular photon (electron). 
Most of the time the TDCs which were not involved with the trigger 
will just overflow. But other times an accidental hit occurs in a \mbox{T-counter} 
which did not produce the trigger. This occurs with probability $1-\exp{(-N_{e}^{(i)}(s) \cdot \tau)}$ 
where $N_{e}^{(i)}(s)$ is the number of times the TDCs for each 
\mbox{T-counter} stop counting in the out of time window $\tau$. 

This measured electron rate is calculated ``instantaneously'' 
between each scaler dump in the DAQ, which is approximately every 10 sec. 
This procedure is based on the tagger \mbox{T-counters} and TAC TDCs
information only (scaler-less).

The total number of measured electrons (or out-of-time photons) 
giving the photon flux $N_\gamma(E_\gamma^{(i)})$ at a the photon energy 
$E_\gamma^{(i)}$ is then determined from the sum: 
\begin{equation}
N_\gamma(E_\gamma^{(i)}) =  \epsilon^{\rm{norm}}_{i} \cdot \sum_s N_e^{(i)}(s)
\label{eq:ngamma}
\end{equation}
where:
\begin{itemize}
\item {$i$  is the \mbox{T-counter} number ($1\div61$);}
\item {$N_e^{(i)}(s)$ is the number of measured electrons per \mbox{T-counter} per scaler interval 
corresponding to tagged photons not associated with an hadronic event in CLAS;}
\item {$\epsilon^{\rm{norm}}_{i}$ is the tagging efficiency for the $i^{\rm{th}}$-tagger channel measured during a normalization run.}
\end{itemize}

The live-time of the DAQ in this time interval, is 
multiplied to the measured electron rate to give the number 
of measured electrons on each \mbox{T-counter}. 
So the number $N_e^{(i)}(s)$ can be re-written as: 
\begin{equation}
N_e^{(i)}(s)= R^{(i)}(s) \cdot {K(s) \cdot \rm{LT}(s)} 
\end{equation}\\
where the DAQ live-time, indicated by LT$(s)$, is calculated on an event basis as:
\begin{equation}
\rm{LT}(s)= \frac{\rm{Accepted\ Triggers}}{\rm{Total\ Triggers}}
\end{equation}\\
where
\begin{equation}
K(s)= \frac{\rm{TRGS}[0] \cdot {\rm{Clock}_{\rm{RG}}}(s)} {10 KHz} 
\end{equation}\\
is the gated clock for the run, and
where $R^{(i)}(s)$ is the electron rate for each \mbox{T-counter} which 
can be written as:
\begin{equation}
R^{(i)}(s)= \frac{ N^ {(i)}_{\rm{TDC}}(s)} {\tau \cdot [ N^{(i)}_{\rm{TAGR}}(s) - N^{(i)}_{\rm{EP}}(s)]}
\label{eq:rate}
\end{equation}\\

In Eq.~\ref{eq:rate} ${N^{(i)}_{\rm{TDC}}}(s)$ is the number of 
TDC hits per \mbox{T-counter} and per scaler time interval, 
${N^{(i)}_{\rm{TAGR}}}(s)$ is 
the total number of events with successful tagger reconstruction
from the TAGR bank and ${N^{(i)}_{\rm{EP}}}(s)$
is the number of times per scaler interval where there is a TDC hit 
before the out-of-time window $\tau$.

In the present calculation scheme, ${N^{(i)}_{\rm{TAGR}}}(s) \cdot \tau$ 
represents the total time, during 10 s (scaler time interval), 
in which the TDC was open while ${N^{(i)}_{\rm{EP}}}(s) \cdot \tau$  
represents a dead time.  

In order to calculate the photon flux, 
the tagging efficiency, which
is defined as the number of tagged photons per energy interval divided
by the total number of counts observed in the tagging counter defining that
interval, must be measured.
The knowledge of this quantity is very important because it
directly affects the determination of the photo-reaction cross sections.
The tagging efficiency strongly depends on the beam setup.
For instance, it will be reduced whenever tight collimation is required 
on the photon beam.

\begin{figure}[htbp]
\begin{center}
\leavevmode
\epsfig{file=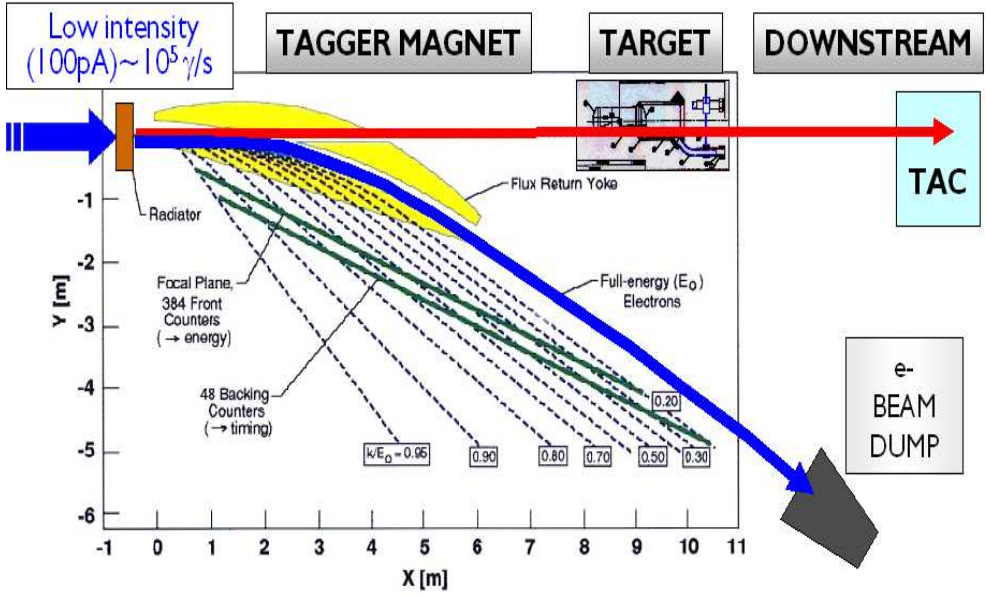,width=13cm,height=8cm}
\caption{\small Schematic representation of a normalization
run: the electron beam current is lowered to 100~pA
and the total absorption counter is inserted into the beam-line, downstream of the
CLAS. The trigger is formed only by the tagger master OR (MOR).
This setup allows for a direct measurement of the photon flux
from the total absorption counter (composed by a set of four lead-glass 
blocks with a detection efficiency close to 100\%).
}
\label{fig:nline}
\end{center}
\end{figure}

The measurement of the tagging efficiency is 
preformed during low intensity ``normalization'' runs 
(schematically shown if Fig.~\ref{fig:nline}) at 
approximately 10\% of the production beam current with 
the total absorption counter placed in the beam-line.
This setup allows for a direct measurement of the photon flux
from the total absorption counter (composed by a set of four lead-glass 
blocks with a detection efficiency close to 100\%).
Since the total absorption counter cannot handle photon beam rates 
higher than $10^5$ Hz without noticeable degradation it could not be inserted 
into the beam-line during the production runs to measure the photon flux 
directly.

During the data taking, normalization runs were performed 
at the end of each data acquisition session characterized by 
a group of production runs with common experimental conditions, 
such as trigger configuration and tagger prescaling settings.

Using this procedure, a total of 178 million normalization events 
have been collected. 
The relation giving the tagging efficiency for each tagger \mbox{T-counter} in a
normalization run is:
\begin{equation}
 \epsilon^{\rm{norm}}_{i} =\frac{N_\gamma^{\rm{TGT}}} {N^{\rm{TAGR}}_{e^-}}
= \frac{[T_i \cdot TAC]} {[T{_i}^{\rm{RAW}}]_{\rm{norm}}} 
\label{eq:tageff}
\end{equation}\\
where $[T{_i}^{\rm{raw}}]_{\rm{norm}}$ counts the number 
of coincidences $T_{\rm{left}} \cdot T_{\rm{right}}$  
while $[T_i \cdot TAC]$ is the number of those 
hits detected in coincidence with the TAC.
A small correction is done ($\simeq 2-3\%$) to the tagging 
efficiency to account for the loss of photons from the target to the 
TAC \cite{ATT}. 

\begin{figure}[htbp]
 \begin{center}
 \leavevmode
 \epsfig{file=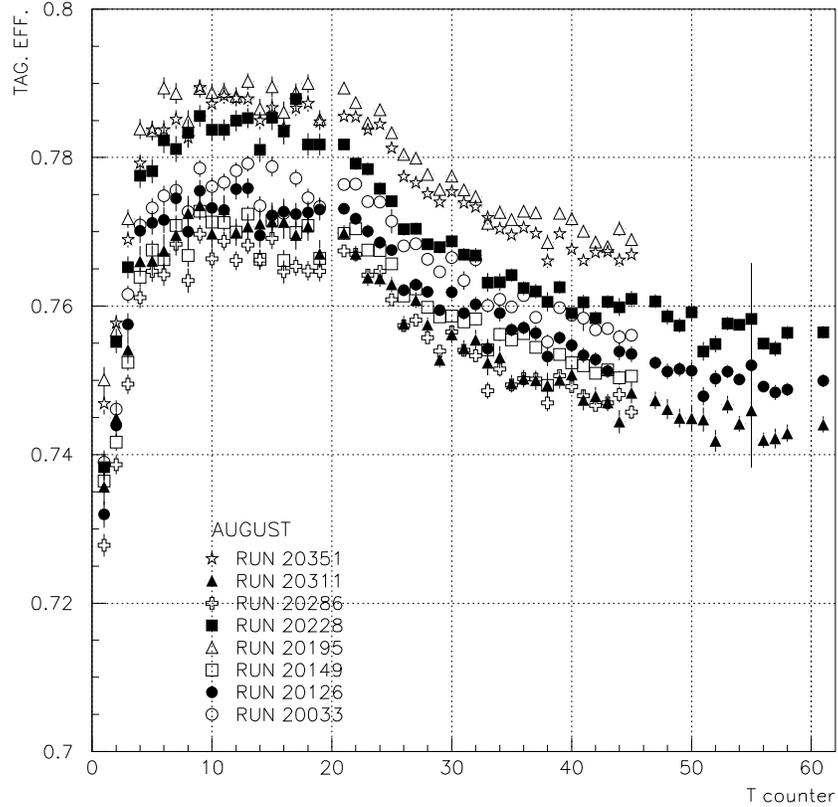,width=11cm}
 \caption{\small Behavior of the tagging efficiency for the tagger
spectrometer \mbox{T-counters} for different runs (indicated by the different
symbols). The tagging efficiency is defined as the number of tagged photons 
per energy interval divided by the total number of counts observed in the tagging 
counter defining that interval. The knowledge of this quantity is very important because it
directly affects the determination of the photo-reaction cross sections and  
strongly depends on the beam setup. }
\label{fig:tag-eff}
 \end{center}
\end{figure}

The overall tagging efficiency as a function of the 
tagger spectrometer \mbox{T-counter} is shown in 
Fig.~\ref{fig:tag-eff} for different runs.
As can be seen also from Fig.~\ref{fig:nline}
the higher numbered \mbox{T-counters}
correspond to high energy electrons
poorly deviated from the incident beam direction,
so that their energy release $\Delta E_0$ 
to the emitted photon is small.
According to the relation $\theta_\gamma \simeq m_e/\Delta E_0$ 
the photon emission angle is larger and
for this reason the collimation of the low energy photons is poor.
Thus, the loss in efficiency 
is due the larger fraction of photons cut out by the
collimator.
On the contrary, the dip in the efficiency in the very first part
of the tagging range is due to background coming from the electron beam.
 
\begin{figure}[htbp]
 \begin{center}
 \leavevmode
 \epsfig{file=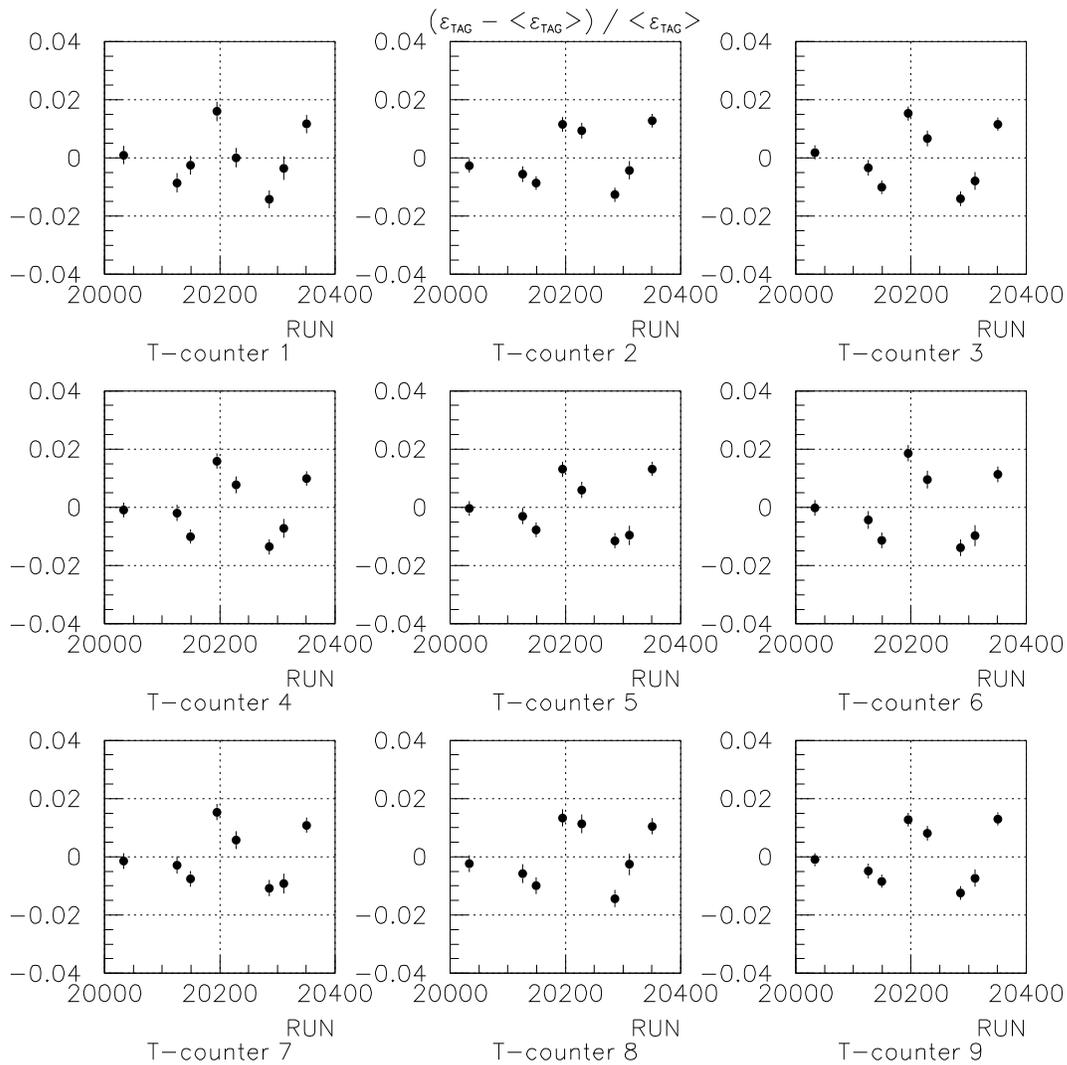,width=14cm,height=14cm}
 \caption{\small Variation of the tagging efficiency with respect to the 
mean value for some tagger spectrometer \mbox{T-counters}. 
As can be seen the variation is around 3-4\%.
}
\label{fig:var_tag_counts}
 \end{center}
\end{figure}

A more detailed example of the
stability of the tagging efficiency
can be seen in Fig~\ref{fig:var_tag_counts} where the variation of this 
quantity with respect to the mean value calculated for each run 
is shown for selected tagger counters. 
As it can be clearly seen, the tagging ratio is stable at level of $\simeq 3\div4\%$.

The normalization code have been used to process the normalization runs reported in Tab. \ref{tab:norm}.
\begin{table}[htbp]
\begin{center}
\begin{tabular}{|c|c|} \hline
August & September \\ \hline
20033 & 20612 \\ 
20126 & 20644 \\
20149 & 20697 \\ 
20195 & 20719 \\ 
20228 & 20720 \\ 
20286 & 20733 \\ 
20311 & 20754 \\ 
20351 & 20761 \\ 
      & 20762 \\ \hline 
\end{tabular}
\end{center}
\caption{\small Normalization runs acquired during the ``August'' and ``September'' 
run periods.}
\label{tab:norm}
\end{table}
Since some T and E-counters were out-of-work, the procedure have been skipped for some of them, more precisely:
T-counters: 20, 46, 59, 60 and E-counters: 66, 98, 101, 142, 214, 220, 222, 224, 239, 240, 285, 335, 337, 343, 380.

\clearpage
\subsection{Validation of the Photon Flux Normalization}
\indent
\par

Empty target runs have been used to
check the photon flux normalization results.
The collected statistics is not  
enough to perform a complete check 
over the whole incident photon energy range and
proton scattering angles.
Nevertheless the photon flux normalization
has been validated comparing
the distribution of the $z$ component of the interaction vertex
for proton events (normalized to the number of incident photons) 
between a full and empty target run.

\begin{figure}[htbp]
 \begin{center}
 \leavevmode
 \epsfig{file=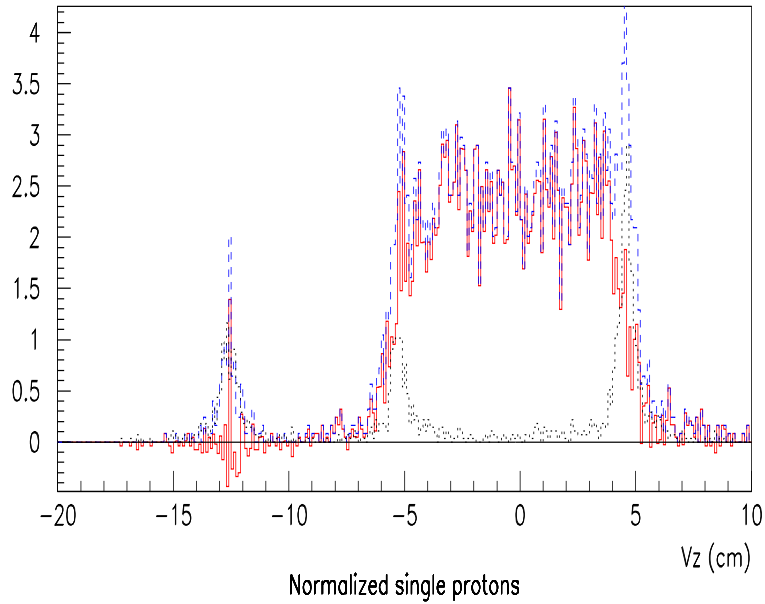,width=10cm,height=8cm}
 \caption{\small Distribution of the $z$-component of the interaction vertex 
for the proton events (normalized to the number of incident photons) between a full (blue line) and empty 
target (black line) run. The difference between the two distributions (red line) is seen to 
be less than 1\% in the interval  (\mbox{$-\rm{15} \leq z \leq -10$})~cm, confirming 
the validity of the photon normalization procedure.
}
\label{fig:empty}
 \end{center}
\end{figure}

The distributions are shown in Fig.~\ref{fig:empty} where
the empty target contribution (black line) exhibits two sharp peaks
at $z = -5$~cm and $z=5$~cm in addition to the one at $z=-12.5$~cm.
These peaks are due to the interaction of the photon beam with
the target cell entrance and exit windows.
The smaller peak at \mbox{$(-\rm{15} \leq z \leq -10)$}~cm 
is due to a structure holding the target cell into place.
The event distribution from the full target run, represented by the blue 
line of Fig.~\ref{fig:empty}, has contributions 
from events originating in 
the enclosure contains the liquid $D_2$. 
The difference between the full target and the 
empty target contributions is represented by the red line of Fig.~\ref{fig:empty}.
In order to validate the photon flux normalization result
the number of events originating 
in the interval (\mbox{$-\rm{15} \leq z \leq -10$})~cm has been
considered in the two cases.
If the photon normalization procedure is correct these two 
numbers must be very close.
In fact, the difference $N_{\rm{full}} - N_{\rm{empty}}$ between the 
number of events in the chosen $z$ interval is less then 1\% 
since \mbox{$N_{\rm{full}} = 10.86 \pm 0.93$} and \mbox{$N_{\rm{empty}} = 10.94 \pm 0.63$},
thus confirming the validity of the normalization procedure.

\clearpage
\section{Data Quality Check}
\indent
\par
In order to select runs collected with stable running conditions, 
a given number of data quality checks have been performed on the reconstructed 
files used for the data analysis.
The first quality checks have been performed during the ``cooking'' procedure where
several run-based parameters, as the overall number of particles, 
the number of detected $p$, $K^{\pm}$, and $\pi^\pm$  
normalized to the incident photon flux in each CLAS sector,
as well as the overall $z$ vertex distribution 
(average value and width) are required to be constant 
at the few \% level from run to run.

\begin{figure}[htbp]
\begin{center}
\leavevmode
\epsfig{file=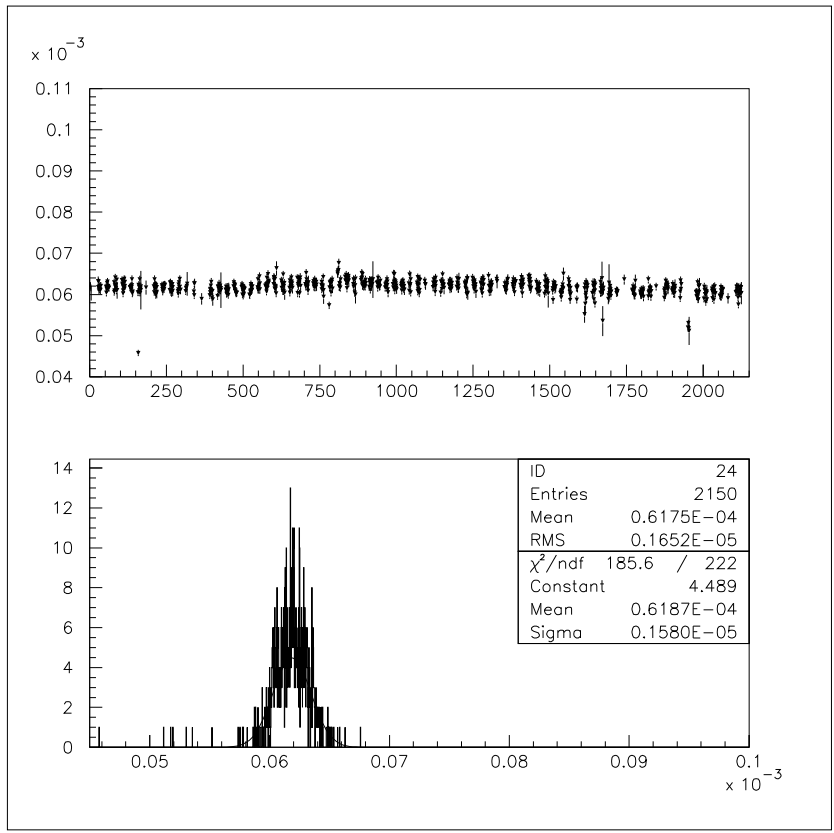,width=11cm}
\caption{\small Behavior of the normalized number 
of protons as a function of time for the ``August'' run period 
 characterized by the tagger spectrometer in the prescale ON configuration.
Top plot: for each run several files are analyzed (represented by the single points) 
and the normalized proton yield is seen to be constant in time.
Bottom Plot: Overall distribution of the normalized proton yield fitted with a 
Gaussian function. The ratio between the standard deviation $\sigma$ and
the mean $\mu$ is less than 3\%. 
}
\label{fig:dataq}
\end{center}
\end{figure}

An example of the stability relative to the normalized number 
of protons as a function of time is shown in Fig.~\ref{fig:dataq}
for the ``August'' run period characterized by the tagger
spectrometer in the prescale ON configuration.
For each run several files are analyzed and the average normalized
proton yield is seen to be constant in time at less than 3\% level.

After this first selection, a second more accurate quality check 
based on the deuteron two body photo-disintegration proton yield (normalized 
to the number of incident photons) is performed to
monitor the data quality for the present analysis.
Significant file-based quantities have been considered:
\begin{itemize}
\item{the number of events with at least one charged particle in the final state for each tagger \mbox{T-counter}, $N^i_{_T}$  
($T=1 \div 61$);}
\item{the number of photo-disintegration events per 100 MeV bin, $N^i_{E}$  ($E=1 \div 25$);}
\item{the number of photo-disintegration events per sector, $N^i_{S}$  ($S=1 \div 6$);}
\end{itemize}
where the index $i$ runs over the number of collected data files $N$
and each data file is considered a measurement of the above quantities.
In order to define a criterion for rejecting the data files (measurements) 
which yield results are not compatible with the statistical fluctuations of the sample,
the Chauvenet's criterion \cite{Chauvenet} has been applied.
This criterion is based on hypothesis that 
one result may be rejected if it shows an ``unusually large'' 
deviation from the average of the results obtained from the remaining $N$ measurements.
The meaning of unusually large is defined using a Gaussian 
distribution function to calculate the probability 
that such a result will occur: if this probability is larger than 
$1/N$, the result is not considered anomalous, since it can be due to 
a statical fluctuation of the data sample. 
On the other hand, if the probability is much smaller than $1/N$, 
then it is very unlikely that the deviation of the result from the average
is due to a fluctuation. 
In this case, the measurement is rejected and its unusually large deviation 
from the average of the remaining $N$ measurements is considered as originating
from a uncontrolled change in the experimental conditions
or as the result of mistake which was not recorded in the experimental logbook.
The ``rule of thumb'' used in the literature rejects a measurement 
if the probability of obtaining it is found to be less than $P_0=\frac{1}{2N}$.

In order to apply the Chauvenet's criterion the mean values
of $N^i_{T}$, $N^i_{E}$ and $N^i_{S}$ have been calculated
according to the following relations:
\begin{equation}
\label{eq:mean1}
\mu(A) =\frac{\sum_{i=1}^N\frac{ A^i}{\sigma^2(A^i)}}{\sum_{i=1}^N\frac{1}{\sigma^2(A^i)}}
\end{equation}\\
where $A^i=N^i_{T}$, $N^i_{E}$ or $N^i_{S}$ and $\sigma^2(A^i)$ is the related statistical error. 
The corresponding errors on the mean values are:
\begin{equation}
\label{eq:error1}
\sigma_\mu^2(A)=\frac{1}{\sum_{i=1}^N\frac{1}{\sigma^2(A^i)}}
\end{equation}\\
where the sum index $i$ runs over the total number of collected data files $N$.\\
For each of the quantities $A^i$, the variable $\Lambda$ can be defined as follows:
\begin{equation}
\label{eq:chis}
\Lambda^i(A)=\frac{\mu(A)-A^i}{\sigma(A^i)}
\end{equation}\\
Under the hypothesis that the variables $A^i$ follow a Gaussian distribution 
also the variable $\Lambda^i(A)$ is distributed according to a Gaussian 
curve with vanishing mean value and standard deviation equal to the unity.

The  Chauvenet's limiting probability $P_0$ defines an interval
of acceptability ($-\Delta_0$, $\Delta_0$) for $\Lambda^i$ such as:
\begin{equation}
\label{eq:delta}
P_0=\frac{1}{\sqrt{2\pi}} \int\limits_{-\Delta_0}^{\Delta_0}
{ \frac{1}{\sigma(\Lambda_i)} e^{\textstyle - \frac{1}{2} \left( \frac{\Lambda_i}{\sigma(\Lambda_i)} \right)^2 }}
\end{equation}\\
so that in the present analysis data files yielding  $A^i$ values 
not satisfying the Chauvenet's criterion have been discarded.

\begin{figure}[htbp]
\begin{center}
\leavevmode
\epsfig{file=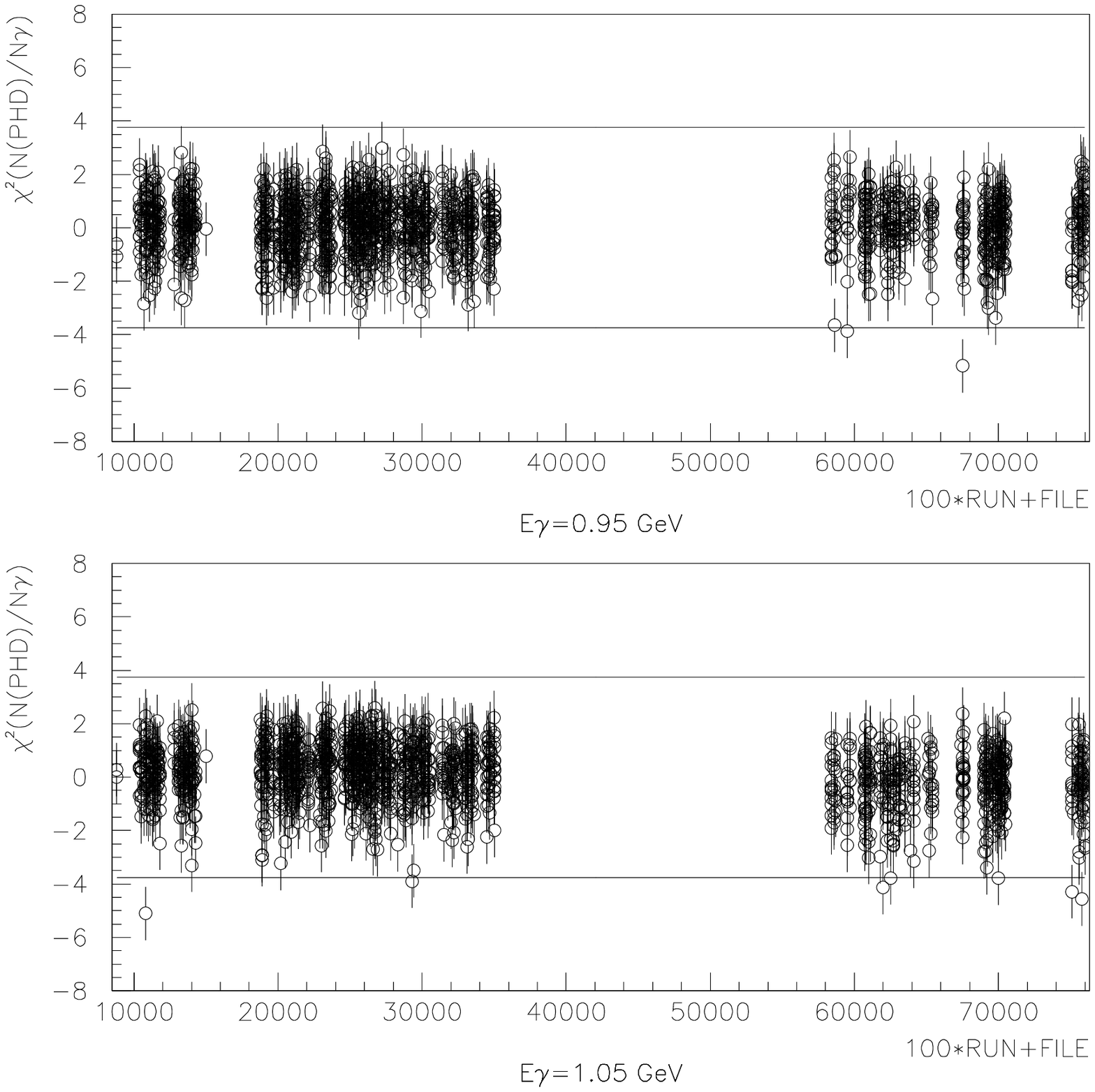,width=10cm}
\epsfig{file=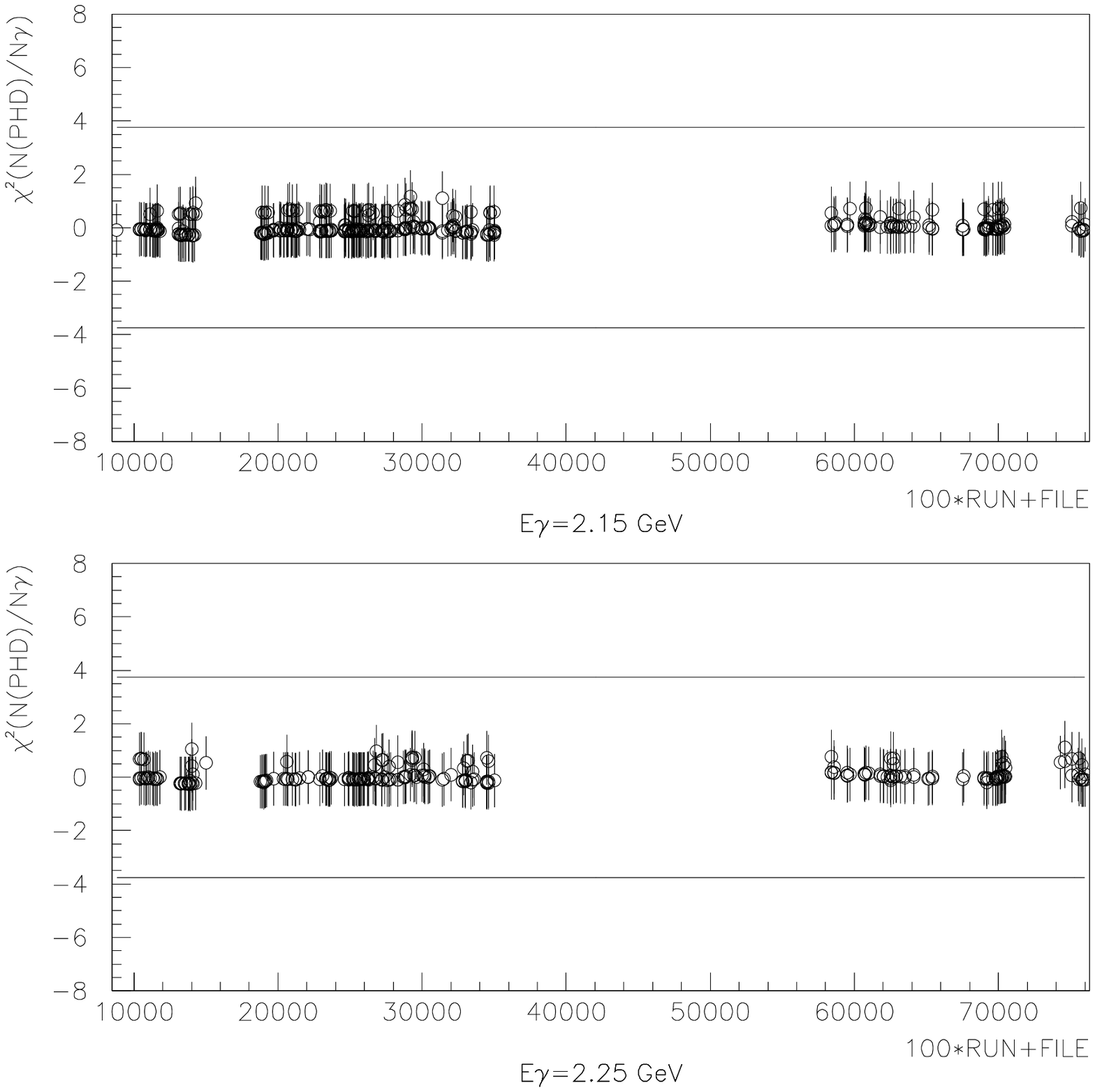,width=10cm}
\caption{\small  Distribution of the $\Lambda$ values calculated for 
the normalized photo-disintegration proton yield  $N^i_{E}$ 
at an incident photon energy of 0.95~-~1.05~GeV (top plots) and 2.15~-~2.25~GeV (bottom plots). 
The horizontal lines represent the fiducial limits derived from the Chauvenet's criterion.
The distribution is practically flat and most of the runs yield values of $\Lambda^i$ whitin
the limits of the fiducial interval \mbox{($-\Delta_0$, $\Delta_0$)} represented 
by the two horizontal lines. Only two files have been discarded from the overall dataset.}
\label{fig:chi1}
\end{center}
\end{figure}

An example of the $\Lambda$ distribution obtained for $A^i=N^i_{E}$ 
({\em i.e.} number of normalized photo-disintegration events) is shown in Fig.~\ref{fig:chi1} 
for two different bins of incident photon energies since this quantity 
represents the most critical yield to be monitored in the present
analysis. It can be seen that the distribution 
is practically flat and most of the data files yield values of  $\Lambda^i$ whitin
the limits of the fiducial interval \mbox{($-\Delta_0$, $\Delta_0$)} represented 
by the two horizontal lines.

Only two data files have been found to yield $\Lambda$ values out of the
confidence interval defined according to the criterion illustrated 
above and have been excluded from the current analysis.

Another important parameter to be investigated is the uniformity of 
the response of the six CLAS sectors since they can be regarded as six 
independent spectrometers. 
The uniformity has been monitored considering the ratio between 
the normalized photo-disintegration protons yields from 
sectors 1 to 5 over those from sector 6, as a function of time (or run). 
The choice of using CLAS sector 6 as a reference sector has 
been done since from the proton detection efficiency study,  
sector six has shown the best response 
(more details will be given in Sec.~\ref{sec:eff}).

\begin{figure}[htbp]
\begin{center}
\leavevmode
\epsfig{file=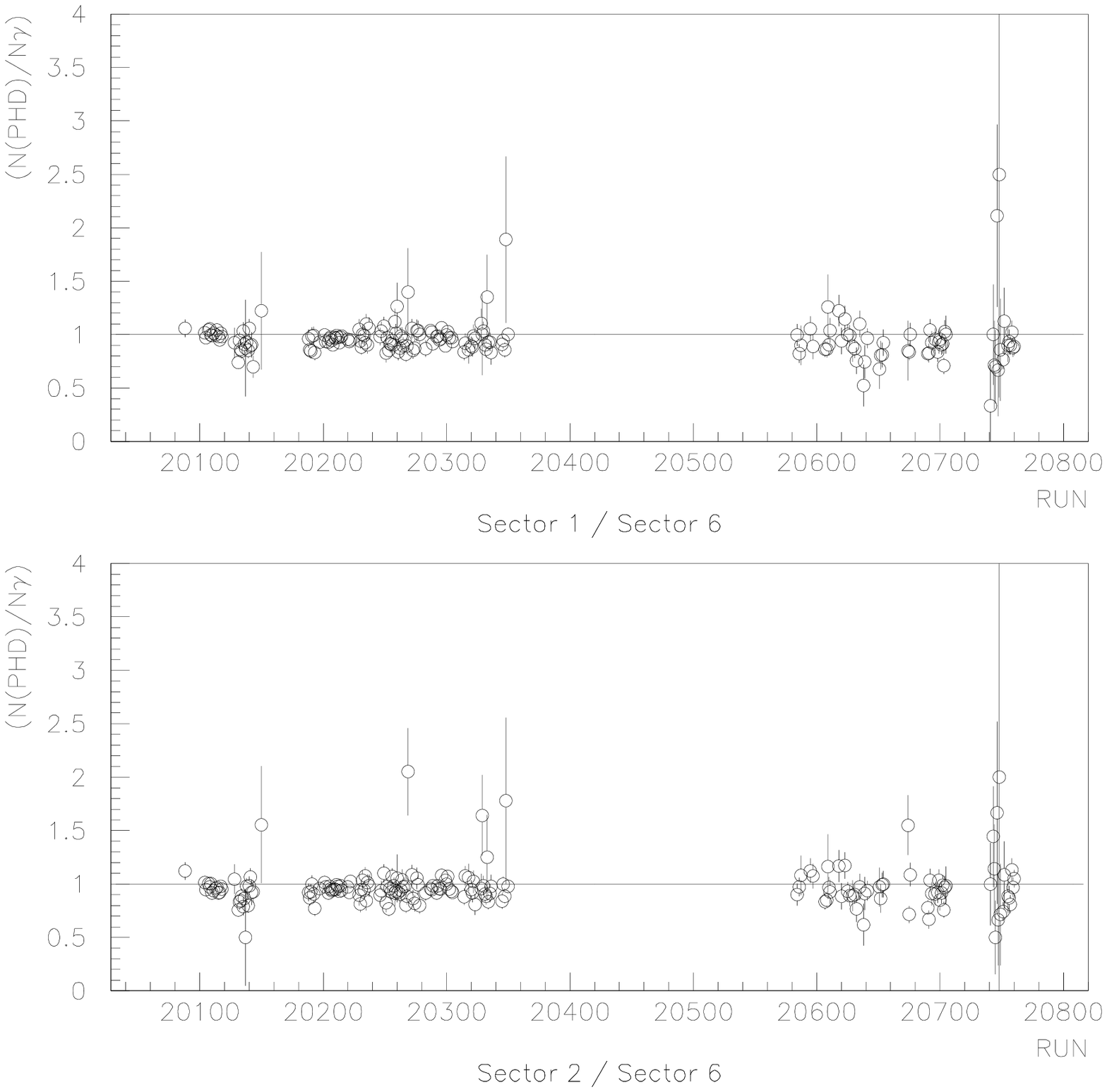,width=7.5cm}
\epsfig{file=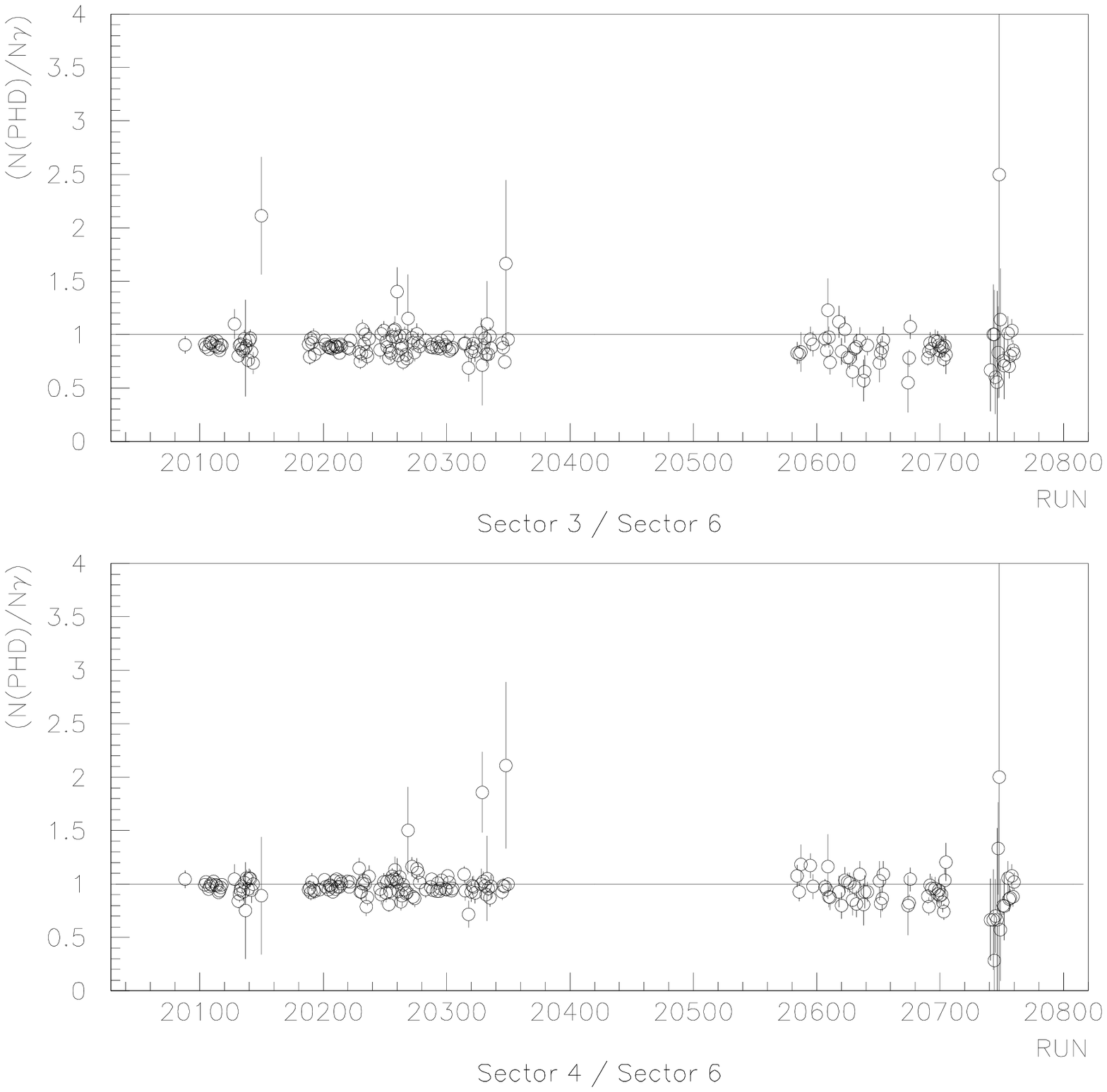,width=7.5cm}
\epsfig{file=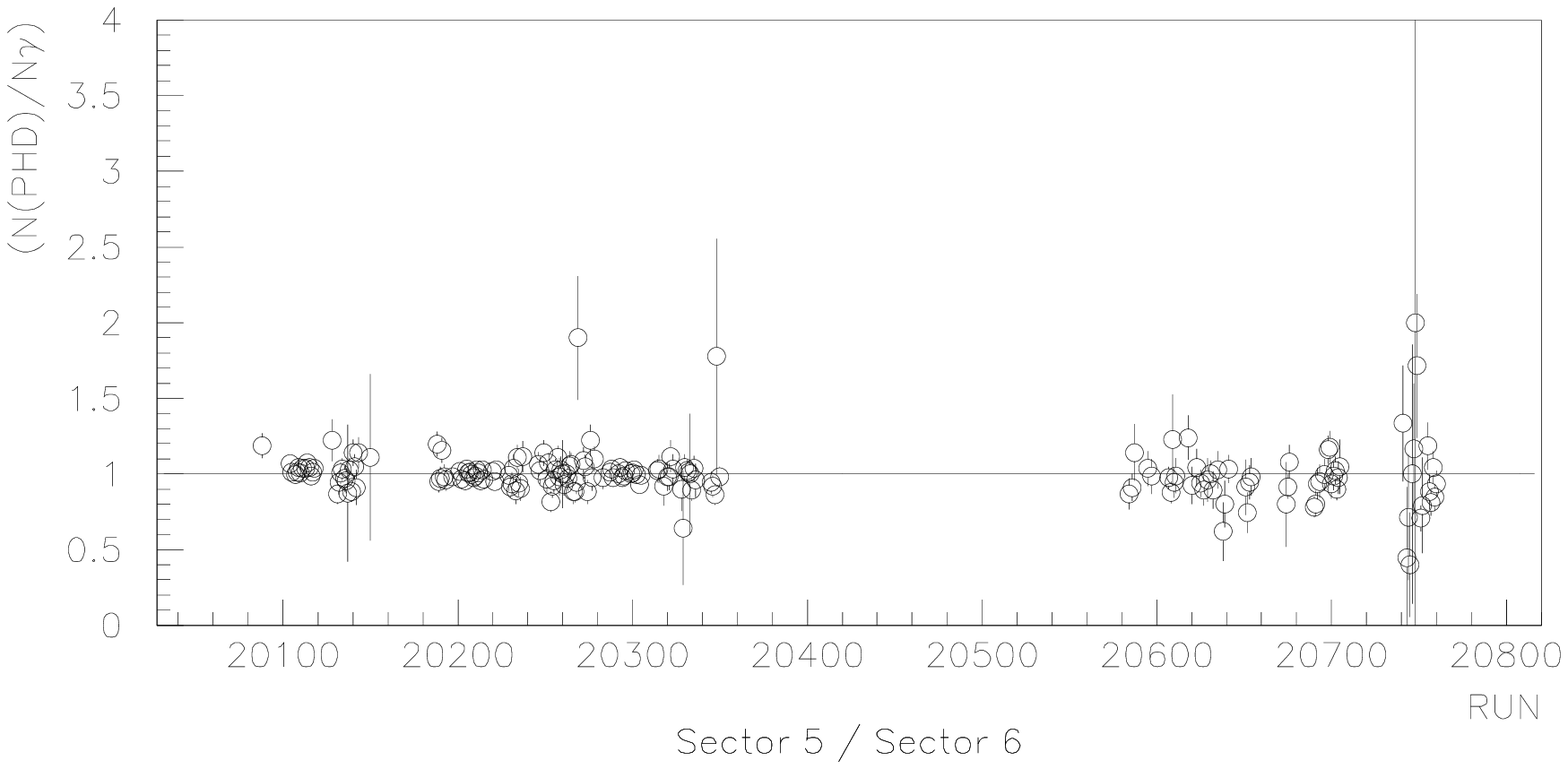,width=7.5cm}
\caption{ \small Ratio of the normalized photo-disintegration proton yields
from sectors 1 to 5 over sector 6 as a function of the run number. 
As it can be seen, all the distributions are flat and no time dependence on the
run period is shown. The average values of uniformity are reported in table \ref{tab:unif}.}
\label{fig:phd-sec1}
\end{center}
\end{figure}

As can be seen from Fig.~\ref{fig:phd-sec1},
the distributions of the uniformity for sectors 1 to 5 with 
respect to sector 6 are flat and no time dependence is evident. 
The values of the overall uniformity are reported in table \ref{tab:unif}.
The lower value of the uniformity for sector 3 can be understood
taking into account that a certain number time-of-flight paddles
were out of work, as will be shown in detail in Sec.~\ref{sec:eff}.
\begin{table}[htpb]
\begin{center}
\begin{tabular}{|c|c|} \hline
Sector & Uniformity \\ \hline
 S1/S6&  $ 0.97 $ \\
 S2/S6&  $ 0.97 $  \\
 S3/S6&  $ 0.91 $ \\
 S4/S6&  $ 0.99 $ \\
 S5/S6&  $ 1.01 $ \\ \hline
Average & $ 0.97 \pm 0.01$\\ \hline
\end{tabular}\\
\end{center}
\caption{\small Uniformity of the response of the CLAS sectors.}
\label{tab:unif}
\end{table}

In conclusion, data files not satisfying the data quality checks 
have been discarded and so forth excluded from the current analysis.
Following the procedures discussed above, about 7\% of the originally collected
data have been excluded. 

\clearpage
\subsection{Data Stability}
\indent
\par
\label{sec:stab}

The data files processed according to 
the quality checks described above 
have been additionally 
checked to make sure that
the experimental conditions 
in place during their acquisition
where stable.

Nevertheless, some experimental conditions 
must change since some runs have different requirements.
For example, the maximum incident energy $E_0$ of
the electrons may change as well as
the tagger prescale settings.

For this reason, the data stability
has been monitored using the total proton yield normalized to the incident photon flux
and the rate of accidental coincidences, 
with different tagger prescale settings
and different electron energies.

\begin{figure}[htbp]
\begin{center}
\leavevmode
\epsfig{file=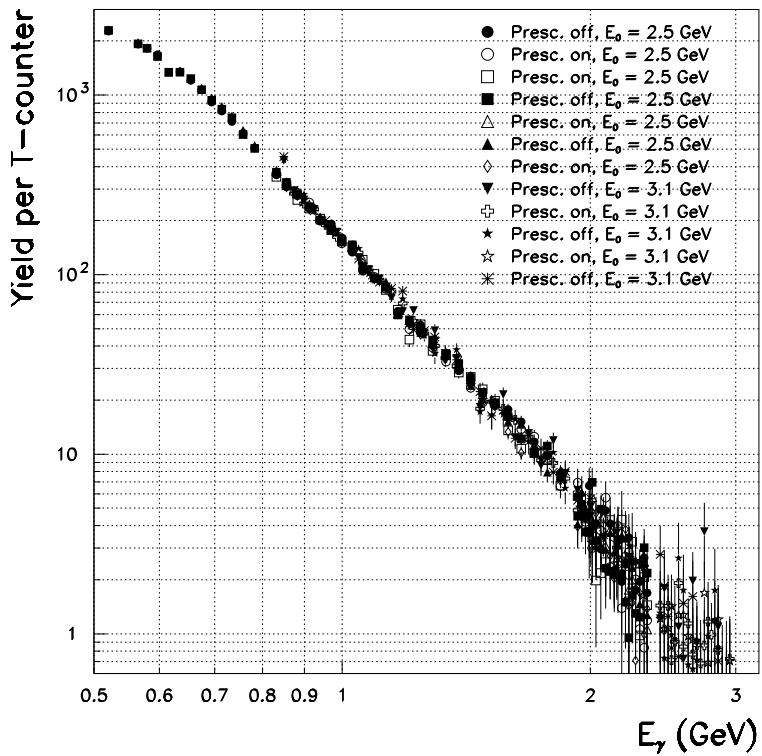,width=10cm}
\caption{\small Normalized total proton yield per \mbox{T-counter} as a function of incident photon energy: 
the different symbols refer to different running conditions 
(tagger prescale ON or OFF and incident electron energy $E_0$).
The behavior of the normalized proton yield per tagger \mbox{T-counter} does not 
show evident dependence on the experimental settings.
}
\label{fig:yield}
\end{center}
\end{figure}

The typical behavior of the normalized proton yield 
per tagger \mbox{T-counter} as a function of the incident photon energy is shown
in Fig.~\ref{fig:yield} for different prescale settings 
of the tagger spectrometer and for different initial electron
energies $E_0$. 
The overall normalized proton yield is 
proportional to the total photo-production cross section 
and is a decreasing function of the incident photon energy.
On the contrary, it does not show evident dependence 
on the experimental settings since  
the contributions of runs performed in different 
conditions (identified by the symbols shown in the legend of Fig.~\ref{fig:yield}) 
all have a similar behavior as a function of the incident photon energy.

For what concerns the accidental coincidences,
their contribution  affects the data taking 
since the total trigger for photo-production experiments in CLAS 
is given by the coincidence signals coming from the 
CLAS and the tagger spectrometer (Master OR).

The CLAS part of the trigger is formed when a charged particle, 
produced by an hadronic event in the target volume 
is detected in the near-by start counter scintillators, 
which seen from the target, 
cover the same solid angle as the time-of-flight system. 
The start counter signal has a time fluctuation of about $3\div4$~ns 
depending on production point of the particle in the target volume 
and impact position on the scintillators. 

At the end of its flight path through the CLAS, the charged
particle fires the time-of-flight scintillators 
so that a valid CLAS Level~1 trigger may be formed. 
The overall time fluctuation of the 
time-of-flight signal depends on several factors.
The velocity of the outgoing particle gives rise to a time fluctuation 
which can be estimated considering the time difference between 
a fast (high momentum) and slow (low momentum) pions, which is $\simeq 30$~ns. 
Then, an additional $\simeq 15$~ns fluctuation is introduced by 
the impact position of the particle in the TOF scintillator. 
In order to guarantee a suitable overlap with 
the start counter signal which is only~12 ns wide the width of 
the TOF signal is chosen to be 100~ns wide. 
\begin{figure}[htbp]
 \begin{center}
 \leavevmode
 \epsfig{file=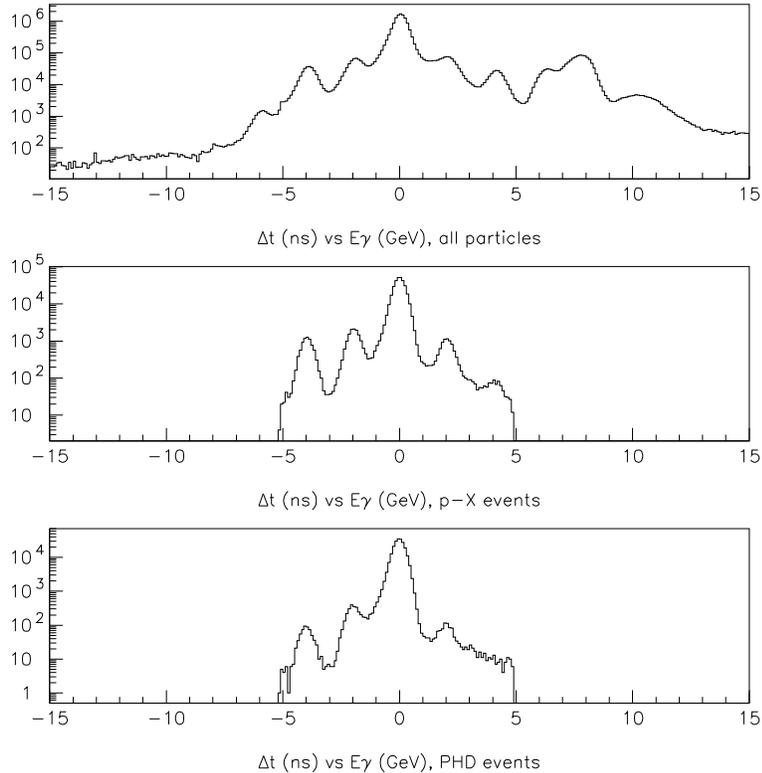,width=10cm}
 \caption{\small Distribution of the event time difference $\Delta t$ between 
the tagger time and the start counter time
for all outgoing charged particles (top plot), for proton events (middle plot),
and for photo-disintegration events (bottom plot).
}
\label{fig:accid}
 \end{center}
\end{figure}
On the tagger spectrometer side, each of the $7 \cdot 10^6$ $\gamma/$s 
flowing through the spectrometer can possibly produce a 
valid ``good photon'' trigger in one of the 121~$T$-counters
so that all the T-counters signals have to be aligned  
to produce a collective Master OR signal which is 15~ns wide.
In this arrangement accidental coincidences are possible if, 
for example, multiple hits in the tagger have timings 
that can all be associated with a ``good photon'' pattern. 

Another possibility is realized when a photon hits the target 
but its energy is not measured in the spectrometer being out of its tagging range 
(very low or extremely high energy photons).
Such photons may produce an hadronic event which is detected in CLAS 
while this event is accidentally in coincidence with a ``good photon'' tagger pattern 
which did not produce it.

An example of the time difference distribution $\Delta t$ between the tagger time 
and the start counter time is shown in Fig.~\ref{fig:accid}
for all outgoing charged particles (top plot), for proton events (middle plot),
and for photo-disintegration events (bottom plot).
The ratio $R$ of the number of events 
lying under time interval ($-1$,1)~ns to the average number 
of events under the two near by peaks gives 
the accidental coincidences level, that is:
\begin{equation}
R= \frac{\frac{1}{2} \left( N(-3,-1)+N(1,3) \right)}{N(-1,1)}
\end{equation}

\begin{figure}[htbp]
 \begin{center}
 \leavevmode
 \epsfig{file=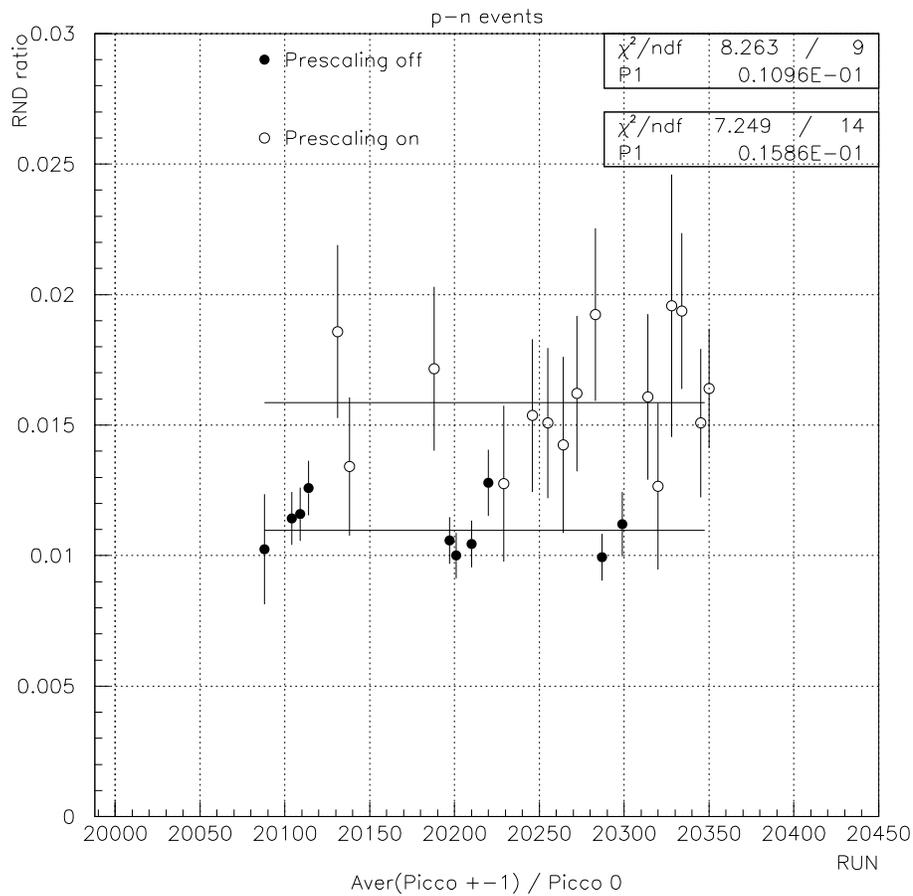,width=12cm}
 \caption{\small Level of accidental coincidences $R$ shown as a function of 
the run number. The relatively flat behavior has
two different levels: one is relative to 
the tagger prescale OFF configuration and has an average value around 1\%
while the other is relative to the tagger prescale ON configuration and is less then 2\%.
}
\label{fig:accid2a}
 \end{center}
\end{figure}
The values of $R$ as a function of the run number are shown 
in Fig.~\ref{fig:accid2a}.
It can be seen that the level of accidental coincidences has a
relatively flat behavior as a function of the run number (time).
On the contrary, two different levels are measured: one level is relative to 
the tagger prescale OFF configuration and has an average value around 1\%
while the other, relative to the tagger prescale ON configuration, is less then 2\%.

\newpage
\section{Momentum Corrections}
\indent
\par
\label{sec:pcorr}
The missing mass distributions from \mbox{$\gamma d \rightarrow p X$} 
events have been calculated over the whole range of incident 
photon energies and proton scattering angles and 
the resulting peak position has been found to be off with respect 
to the reference value given by the neutron rest mass.
This shift is not very large (of the order of 3\%, at maximum) nevertheless 
the distortion introduced in the missing mass distribution shape can
reduce the number of identified photo-disintegration events. 

A similar problem was found also in other CLAS experiments.
For example, in the elastic $ep$ scattering experiment 
the missing mass distribution peak for $e p \rightarrow e X$ events
is shifted by the order of few \% from the reference value of the 
proton rest mass~\cite{PITT}.
This effect could be due to small misalignments of the drift chambers
positions with respect to their nominal positions or to 
uncertainties in the magnetic field maps.
Another possibility concerns the knowledge of
the energy of the incoming photon which could be incorrect. 
Since it is not possible to disentangle between these two contributions,
the shift of the missing mass peak is usually corrected introducing
an empirical function $f(P_{0},\theta_{0},\phi_{0})$
depending on the measured proton \mbox{3-momentum} $P_0$:
\begin{figure}[htbp]
\begin{center}
\leavevmode
\epsfig{file=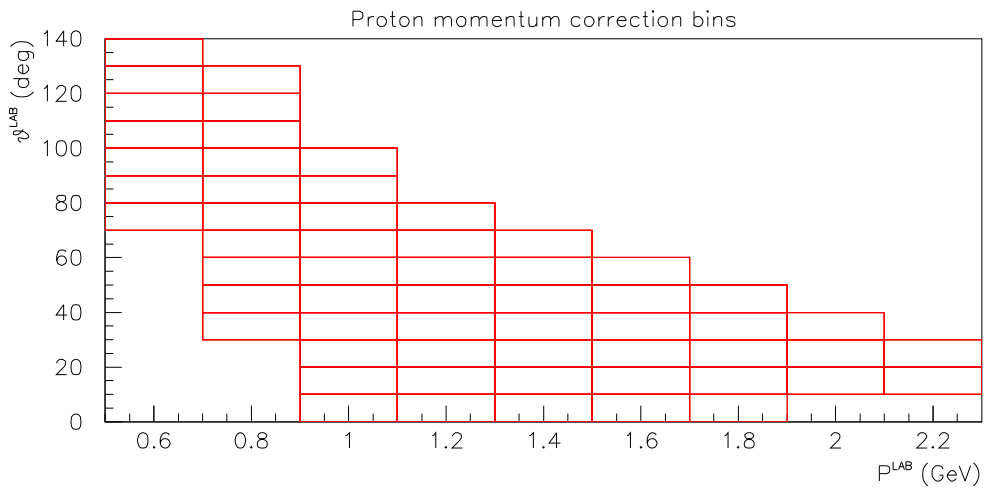,width=13cm}
\caption{\small  Region of the kinematic plane where the momentum corrections 
could be calculated using two body kinematics.}
\label{fig:pbins}
\end{center}
\end{figure}
\begin{equation}
P^\prime= P_0\, f(P_{0},\theta_{0},\phi_{0}) 
\label{eq:pfun}
\end{equation}\\
where $P^\prime$ is the corrected momentum and $\theta_{0}$ and $\phi_{0}$ 
are referred to measured proton angles in the LAB system.
Assuming that angles and photon energy are correctly measured,
the momentum correction function is factorized as follows:
\begin{equation}
P^{\prime} = P_{0}\,f_1(P^{\prime\prime}, \theta_{0})\, f_2(P_{0},\phi_{0})  
\label{eq:pfact}
\end{equation}\\
where $P^{\prime\prime}$, $f_1$, and $f_2$ are functions to be determined 
from the data according to the following procedure:
\begin{enumerate}
\item{first, the proton momentum range is divided in 200 MeV bins; }
\item{in each bin of proton momentum, photo-disintegration events
are selected from the  $\gamma d \rightarrow p X$ distributions.
In order to reduce background, a ($-2\sigma$, $2\sigma$) cut around the peak position
is applied, where $\sigma$ is the standard deviation obtained from 
a Gaussian fit of the missing mass distribution;}

\item{event by event the momentum expected on the base of two
body kinematics is calculated as a function of photon energy $E_\gamma$ and
the proton scattering angle $\theta_{0}$, which are assumed correctly measured.
An example of the region of the kinematic plane where this procedure has been applied
is shown in Fig.~\ref{fig:pbins}.
}

\item{Each bin of proton momentum is additionally 
divided in $5^\circ$ wide bins of the azimuthal angle $\phi$;}

\item{in each $\phi$-bin the ratio $R_\phi = P_\phi / P_{0}$
is calculated and its distribution is fitted using a Gaussian function
to obtain the average value $<R_\phi>$;
}

\item{the behavior of $<R_\phi>$ as a function of $\phi$
is reproduced using a $2^{rd}$ order polynomial ($f_2$ in Eq.~\ref{eq:pfact}).
An example is shown in Fig.~\ref{fig:pphi} for CLAS sector 1
and for three different values of proton momentum:
$P_0$:~0.55~GeV (top plot), 0.65~GeV (middle plot), and 0.75~GeV (bottom plot).
Then, a first correction is applied to the measured momentum $P_{0}$
as a function of $\phi$ in the form $P^{\prime\prime}=P_0\,f_2(P_{0},\phi_{0})$ }

\begin{figure}[htbp]
\begin{center}
\leavevmode
\epsfig{file=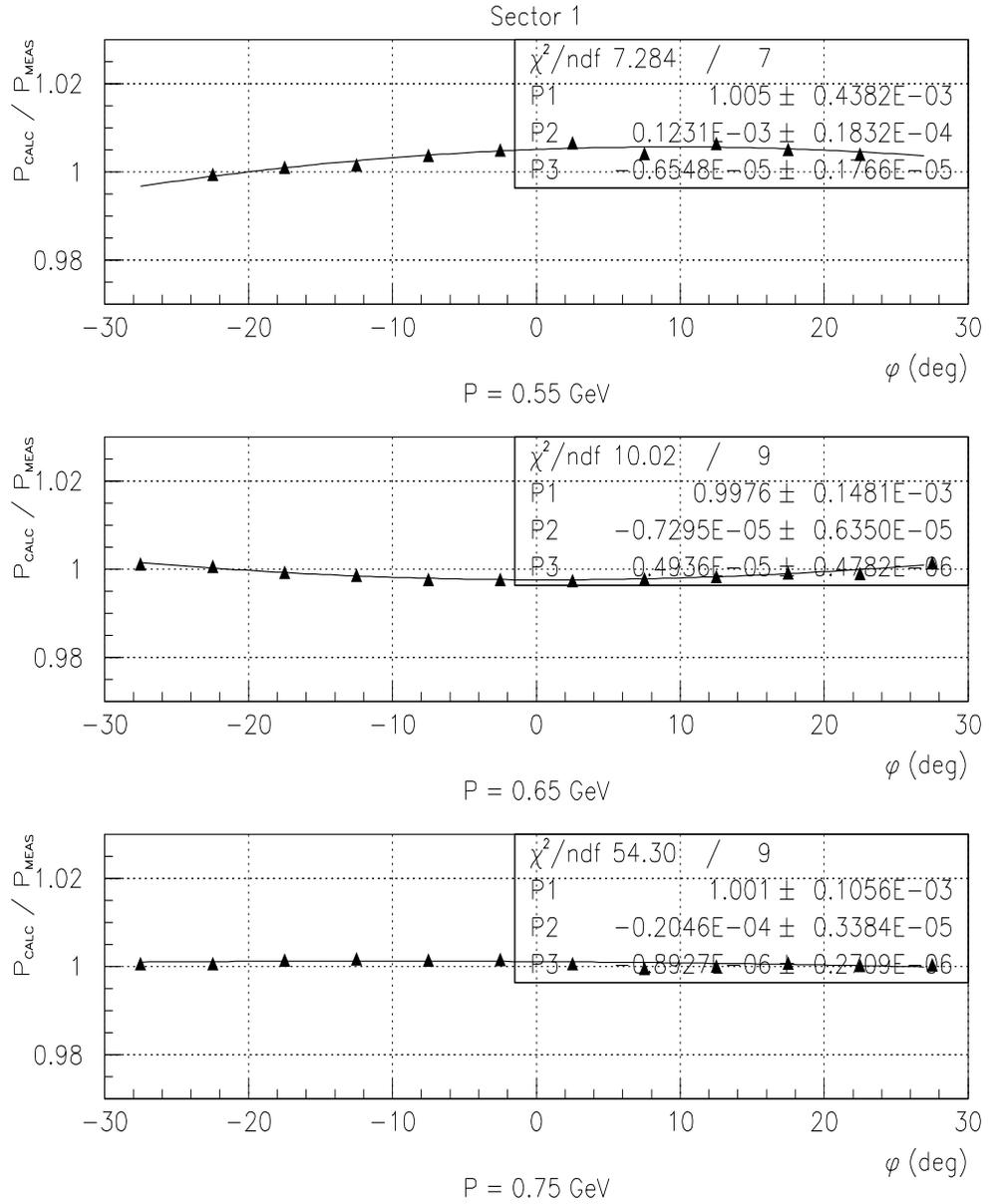,width=13cm,height=16cm}
\caption{\small Behavior of the ratio $<R_\phi>$ as a function of 
$\phi$ in CLAS sector 1 for three values of the  measured proton
momentum $P_0$:~0.55~GeV (top plot), 0.65~GeV (middle plot), and 0.75~GeV (bottom plot). 
In each $\phi$-bin the ratio $R_\phi = P_\phi / P_{0}$ is 
calculated and its distribution is fitted using a Gaussian function
to obtain the average value $<R_\phi>$.
}
\label{fig:pphi}
\end{center}
\end{figure}

\item{The procedure is repeated starting from point (4) and dividing
each bin of proton momentum in $10^\circ$ wide bins in 
the  scattering angle $\theta$. In this way, 
a second  correction factor is calculated and applied to the measured momentum $P_{0}$
as a function of $\theta$ ($f_1$ in Eq.~\ref{eq:pfact}). }
\end{enumerate}

\begin{figure}[htbp]
\begin{center}
\leavevmode
\epsfig{file=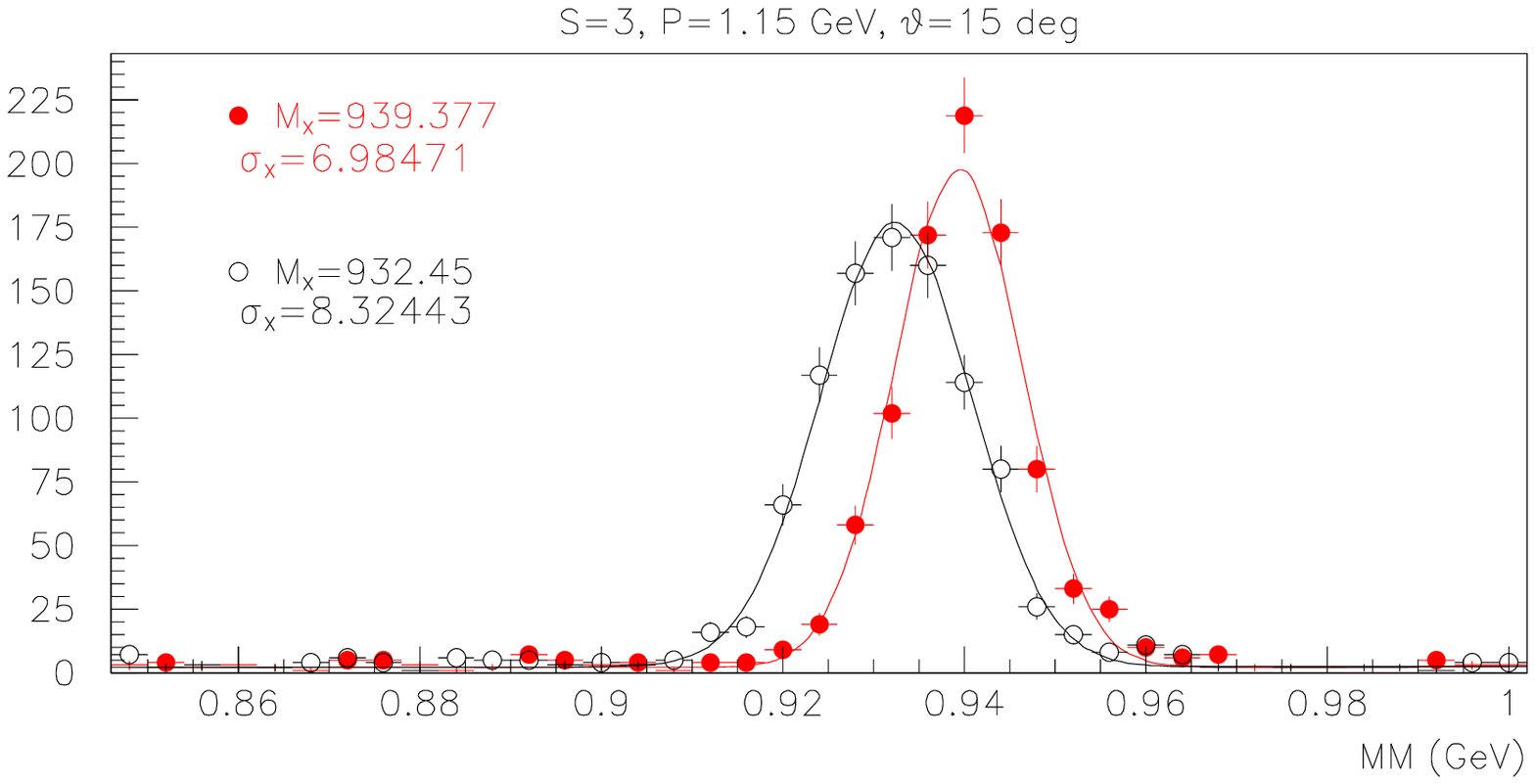,width=12cm}
\epsfig{file=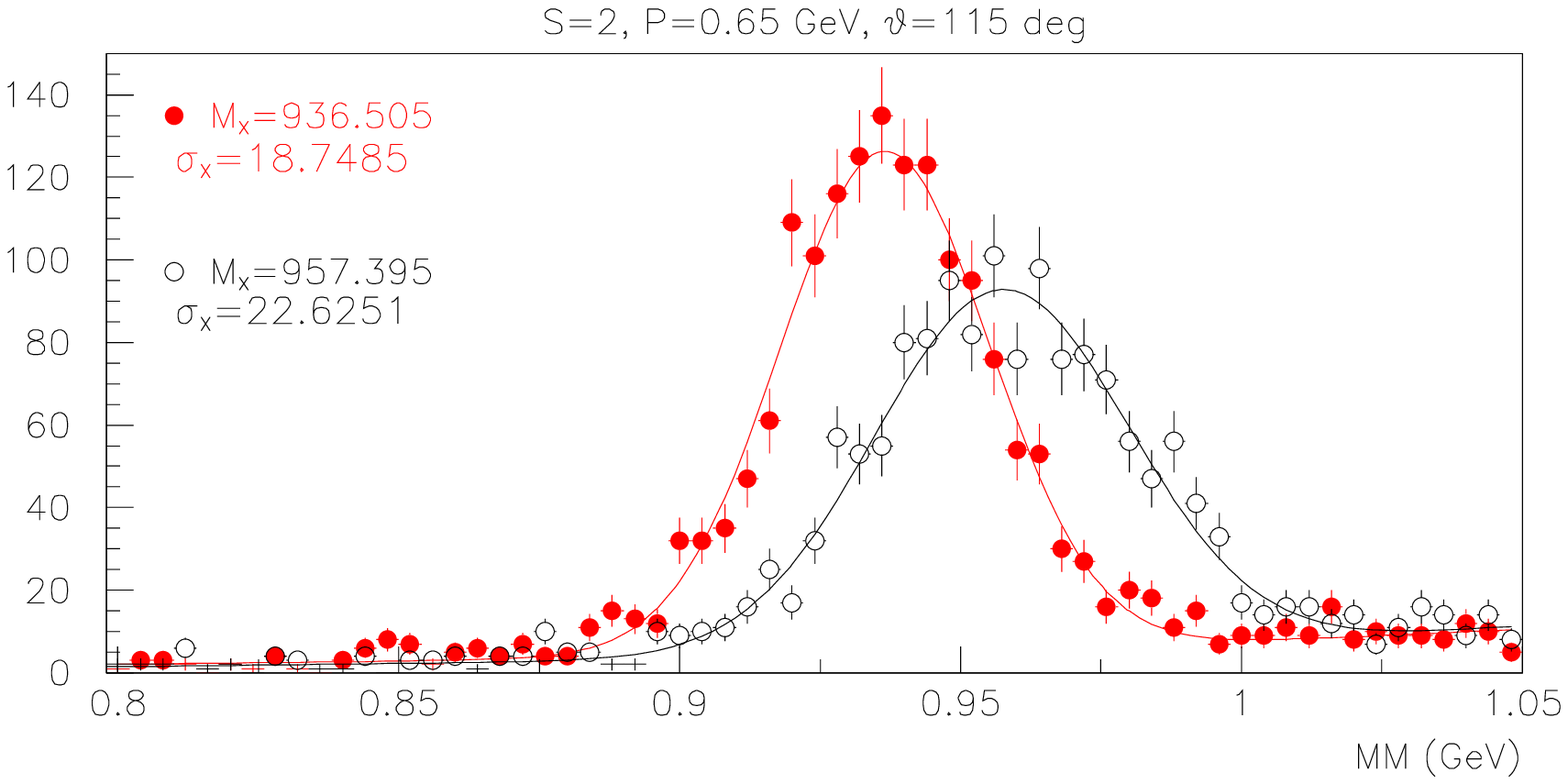,width=12cm}
\caption{\small  missing mass distributions for \mbox{$\gamma d \rightarrow pX$} events
for sector 2 (top plot) and sector 3 (bottom plot) 
for two different proton momenta and scattering angles: 
$P$=0.65~GeV, $\theta=115^\circ$ and $P$=1.15~GeV, $\theta=15^\circ$, respectively.
The missing mass distributions are sharper after the application of the correction procedure 
and their peak positions are closer to the neutron rest mass value.}
\label{fig:pcorr}
\end{center}
\end{figure}

An example of the effect of the momentum correction procedure is shown in 
\mbox{Fig.~\ref{fig:pcorr}} for CLAS sector 3 (top plot) and sector 2 (bottom plot) 
for two different proton momenta and scattering angles: 
$P$=0.65~GeV, $\theta=115^\circ$ and $P$=1.15~GeV, $\theta=15^\circ$, respectively.
The missing mass distributions are sharper after the correction procedure 
while the peak position (indicated by $M_X$) is closer to the neutron rest mass value.

\begin{figure}[htbp]
 \begin{center}
 \leavevmode
 \epsfig{file=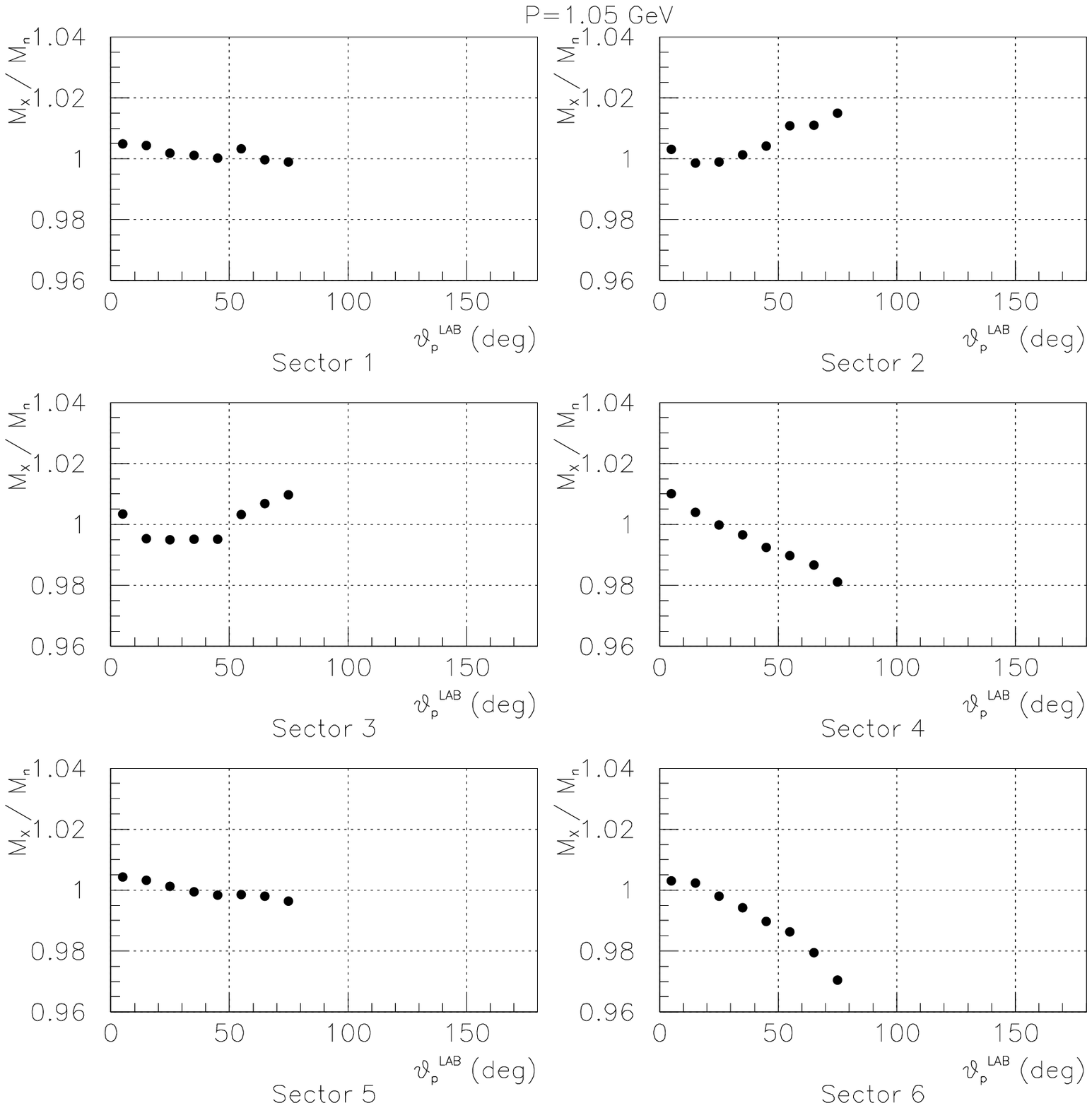,width=15cm}
 \caption{\small Before the momentum correction. Ratio $M_X/M_n$
(being $M_n$ the nominal value for the neutron rest mass) 
plotted against the proton scattering angle in the LAB system separately
for the six CLAS sectors and for the intermediate proton momentum of 1.05~GeV.
The deviation from the unity affects practically all six
CLAS sectors, and shows a dependence on the proton scattering angle.
The maximum deviation is of the order of 3\% in some sectors 
(specifically sector 4 and 6).
}
\label{fig:pcorr2}
 \end{center}
\end{figure}

\begin{figure}[htbp]
 \begin{center}
 \leavevmode
\epsfig{file=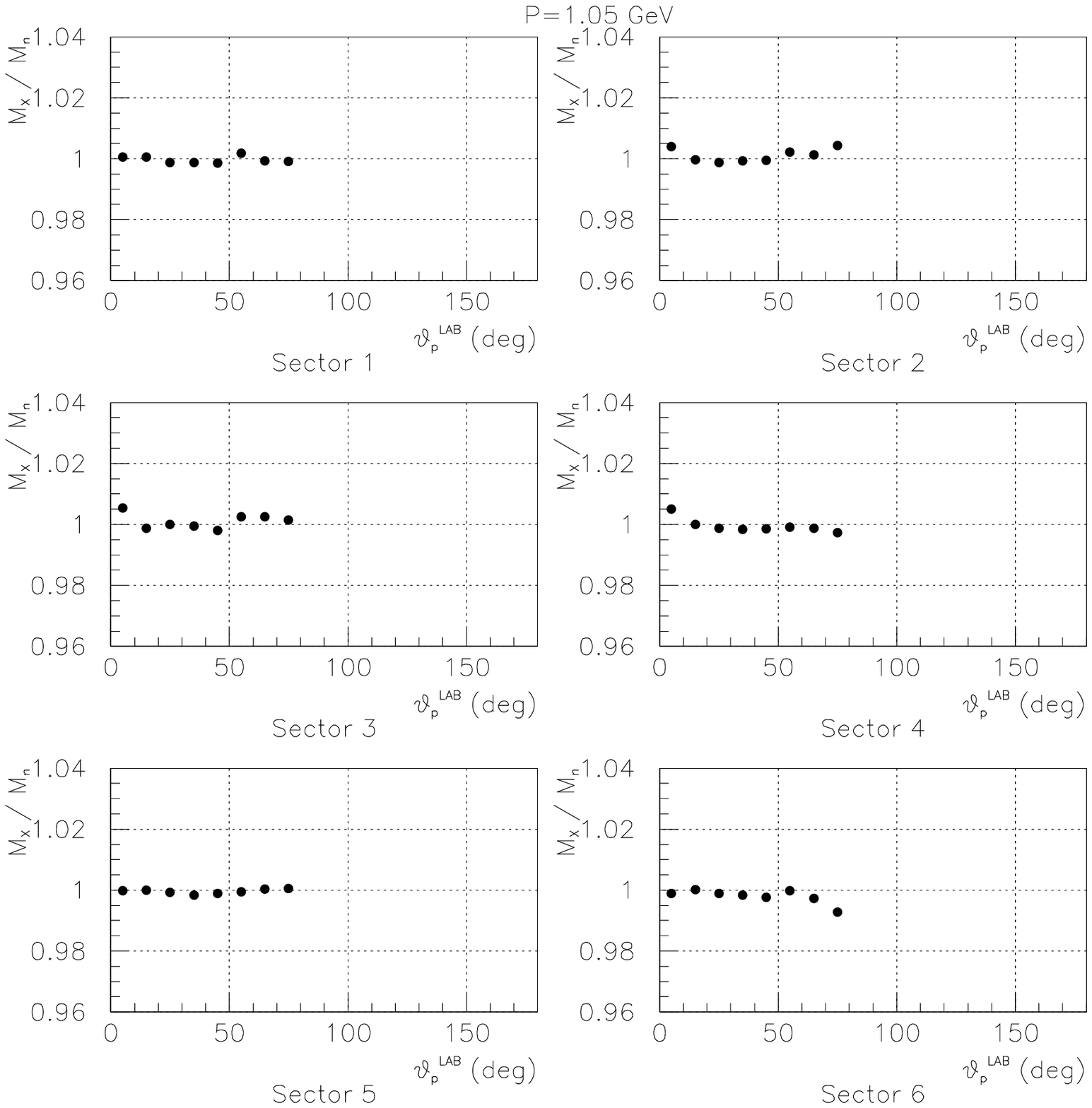,width=15cm}
 \caption{\small After the momentum correction. Ratio $M_X/M_n$
(being $M_n$ the nominal value for the neutron rest mass) 
plotted against the proton scattering angle in the LAB system separately
for the six CLAS sectors and for the intermediate proton momentum 1.05~GeV.
The situation is clearly improved compared to the one shown in Fig.~\ref{fig:pcorr2} 
since the values obtained for $M_X/M_n$ are of the order of 1 and the  
dependence on the proton scattering angle is removed by the correction.}
\label{fig:pcorr3}
 \end{center}
\end{figure}

In particular, Fig.~\ref{fig:pcorr2} shows the overall situation before the 
application of the momentum correction procedure.
As can be seen, the deviation from the unity of the ratio $M_X/M_n$,
(being $M_n$ the nominal value for the neutron rest mass)  
plotted against the proton scattering angle in the LAB system separately
for the six CLAS sectors and for the intermediate proton momentum of 1.05~GeV,
affects practically all sectors and shows 
a dependence on the proton scattering angle.
The maximum deviation is of the order of 3\% in some sectors 
(specifically sector 4 and 6).
The same quantity, after introducing the momentum correction procedure 
is shown in Fig.~\ref{fig:pcorr3}.
It can be seen that the deviation from the unity is greatly reduced in all
sectors together with the dependence of  $M_X/M_n$ on the proton scattering angle.

\begin{figure}[htbp]
 \begin{center}       
 \leavevmode
 \epsfig{file=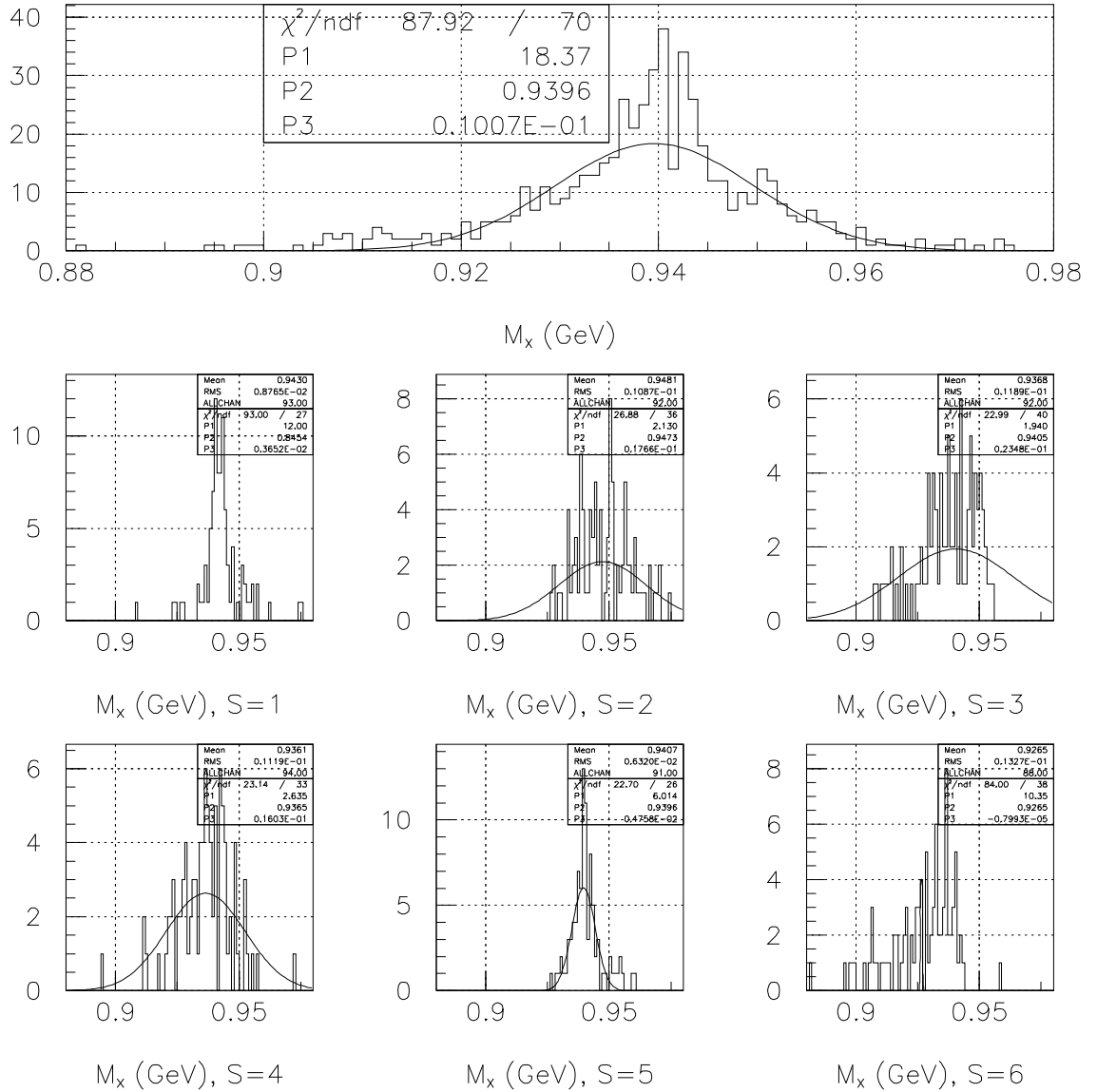,width=15cm}
 \caption{\small  Before the momentum correction.  Distributions of the $M_X$ values obtained from
the missing mass distributions for \mbox{$\gamma d \rightarrow pX$} events.
The top plot shows the overall $M_X$ distribution for the six CLAS sectors, while the 
remaining six plots show the contributions of the different sectors.
It is seen that, the $M_X$ distributions exhibit large widths which are different 
in the six CLAS sectors. }
\label{fig:pcorr4}
 \end{center}
\end{figure}

\begin{figure}[htbp]
 \begin{center}
 \leavevmode
\epsfig{file=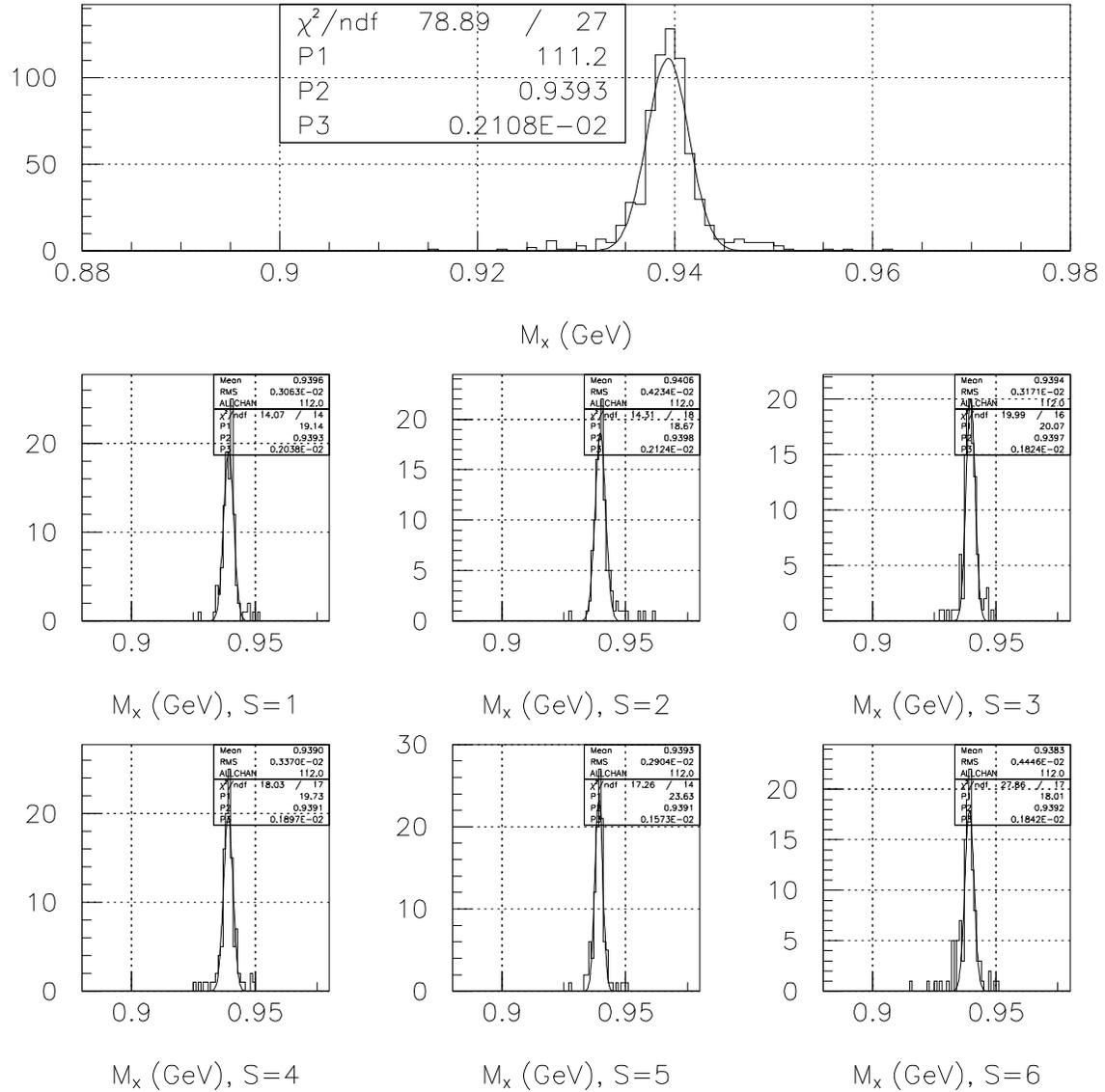,width=15cm}
 \caption{\small 
After the momentum correction. Distributions of the $M_X$ values obtained from
the missing mass distributions for \mbox{$\gamma d \rightarrow pX$} events.
The top plot shows the overall $M_X$ distribution for the six CLAS sectors, while the 
remaining six plots show the contributions of the different sectors.
After the application of the momentum correction procedure, the widths of the
$M_X$ distributions are reduced and are more uniform in the six sectors. 
}
\label{fig:pcorr5}
 \end{center}
\end{figure}

The distributions of the $M_X$ values obtained from
the missing mass distributions for \mbox{$\gamma d \rightarrow pX$} 
events before and after the application of the momentum correction 
procedure are shown in Figs.~\ref{fig:pcorr4} and \ref{fig:pcorr5} respectively.
The top plot shows the overall $M_X$ distribution for the six CLAS sectors, while the 
remaining six plots show the contributions of the different sectors.
As shown in Fig.~\ref{fig:pcorr4} before the momentum
correction is applied, the $M_X$ distributions
have large widths which are quite different in the six CLAS sectors. 
On the contrary, after the application of the momentum correction 
procedure, the widths of the $M_X$ distributions are reduced and
are uniform in the six sectors, as shown in 
Fig.~\ref{fig:pcorr5}.

\begin{figure}[htbp]
 \begin{center}
 \leavevmode
\epsfig{file=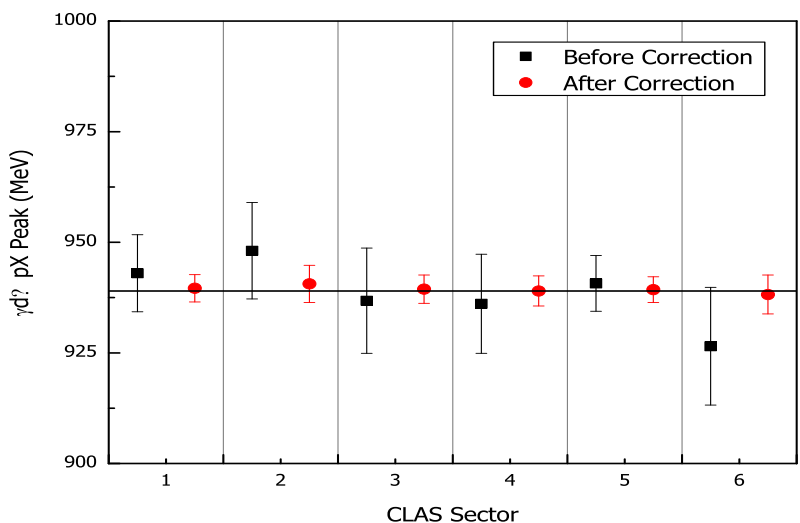,width=12cm,height=8cm}
 \caption{\small Position of the missing mass distribution peaks (associated with their
RMS) before (solid/black squares) and after the correction (solid/red dots).
The horizontal line represents the nominal value of the neutron rest mass.
}
\label{fig:pcorr-sum}
 \end{center}
\end{figure}

A summary of the improvement in the momentum reconstruction
introduced by the correction procedure is shown in Fig.~\ref{fig:pcorr-sum}.
The plot shows the position of the missing mass distribution peaks (associated with their
RMS) before (solid/black squares) and after  (solid/red dots) the correction
while the horizontal line represents the nominal value of the neutron rest mass.

For a more quantitative comparison, in Tab.~\ref{tab:pc} are given 
the numerical values of the missing mass distributions peak positions
and RMS in each sector before and after for the correction
together with the standard deviations obtained fitting the distributions after
the correction procedure.

\begin{table}[tbp]
\begin{center}
\begin{tabular}{|c|c|c|c|c|c|} \hline
Sector & Peak Before & RMS & Peak After & RMS & $\sigma$~fit \\ \hline

1&	943.0&	8.7&	939.6&	3.1	&2.1  \\
2&	948.1&	10.9&	940.6&	4.2	&2.1 \\
3&	936.8&	11.9	&939.4&	3.2	&1.8 \\
4&	936.1&	11.2&	939.0&	3.4	&1.9 \\
5&	940.7&	6.3&	939.3&	2.9	&1.6 \\
6&	926.5&	13.3&	938.2	&4.4	&1.8\\ \hline
ALL&	939.6&		&939.3	&	&2.1\\ \hline
\end{tabular}
\end{center}
\caption{\small Numerical values of the missing mass distributions peaks
and RMS before and after for the momentum correction procedure is applied 
in each sector, together with the standard deviations obtained 
fitting the corrected missing mass distributions.
}
\label{tab:pc}
\end{table}

\clearpage
\section{Proton Detection Efficiency}
\indent
\par
\label{sec:eff}

\subsection{Introduction}
\indent
\par
The extraction of the photo-disintegration differential cross section requires 
the knowledge of the number of \mbox{$\gamma d \rightarrow pn$} events produced 
in the selected bins of the outgoing proton momentum and scattering angle.
In order to count these events correctly the proton detection efficiency in CLAS
have to be evaluated. 

The CLAS proton detection efficiency cannot be extracted from the 
photo-disintegration data because it also requires the detection
of the photo-disintegration neutron in the CLAS.
This procedure is not convenient since:
\begin{itemize}
\item {neutrons are detected by the electromagnetic calorimeters,
having a much smaller angular coverage ($\theta_p^{\rm{LAB}} \leq 45^\circ$ or $70^\circ$ for two CLAS sectors)
than the time-of-flight scintillators which are used to detect protons; }
\item {the neutron detection efficiency of the calorimeters is less then 
50\% (see Ref.\cite{CNIM}).}
\end{itemize}

For the above reasons, to evaluate the single proton detection
efficiency a GEANT simulation of the CLAS detector has been used.

\subsection{GEANT Simulation}
\indent
\par
In order to simulate $\gamma d \rightarrow pn $ reaction in the CLAS, 
an appropriate number of events are produced using a simple 
Monte Carlo code. The events are generated uniformly in the CM system  
using an flat photon energy spectrum between 0.5 and 2.95~GeV 
and then Lorentz boosted in the LAB frame. 
Since the proton detection efficiency does not depend
on the production mechanism, no cross section or background 
information is taken into account by the event generator.

In order to have statistical errors smaller that those introduced from 
other quantities entering in the final cross section calculation, 
about $20 \cdot 10^6$ photo-disintegration events have been generated.
 
\subsection{Proton Detection Efficiency Evaluation using GSim}
\indent
\par
The proton detection efficiency calculated from the simulation 
is defined as the ratio between the number of generated events 
($N_{\gamma d \rightarrow pn}^{\rm{GEN}} $)
in a certain kinematic bin 
$( \Delta P_p^{\rm{LAB}},\Delta \theta_p^{\rm{LAB}},\Delta \phi)$ 
over the number of reconstructed events ($N_{\gamma d \rightarrow pn}^{\rm{REC}}$) 
in the same kinematic bin:
\begin{equation}
\epsilon_{\rm{GSIM}}  = \frac{N_{\gamma d \rightarrow pn}^{\rm{REC}}(\Delta P_p^{\rm{LAB}},\Delta \theta_p^{\rm{LAB}},\Delta \phi ) } 
{N_{\gamma d \rightarrow pn}^{\rm{GEN}}( \Delta P_p^{\rm{LAB}},\Delta \theta_p^{\rm{LAB}},\Delta \phi  ) }\ .
\label{eq:gsim-epsilon}
\end{equation}\\
A binning in proton momentum and polar scattering angle of 
$\Delta P_p^{\rm{LAB}}=100$ MeV and $\Delta \theta_p^{\rm{LAB}} =10^\circ$
is chosen to match the one used in the final cross-section calculation.
On the other hand, a binning of
$\Delta \phi=5^\circ$ is chosen to better investigate the 
azimuthal behavior of the CLAS proton detection efficiency.
In fact, according to the CLAS toroidal geometry, the reconstruction 
is worse on the borders of the six sectors, since these regions are adjacent
to the superconducting magnet coils.

The number of simulated photo-disintegration protons 
identified in CLAS (the numerator of Eq.~\ref{eq:gsim-epsilon})
by the particle ID is sensitive to the quality of momentum 
reconstruction. In order to discard protons whose momentum 
is not correctly reconstructed, 
the differences between the 
angles of generated and reconstructed protons
\begin{equation}
\Delta\theta= \theta_p^{\rm{GEN}}( P_p^{\rm{LAB}}) - \theta_p^{\rm{REC}}( P_p^{\rm{LAB}})\ , \   
\Delta\phi = \phi_p^{\rm{GEN}}( P_p^{\rm{LAB}}) - \phi_p^{\rm{REC}}( P_p^{\rm{LAB}})
\label{eq:gsim-diff}
\end{equation}
have been obtained and 
the standard deviations $\sigma_{\Delta\theta}$ and $\sigma_{\Delta\phi}$
of the distributions have been calculated from a Gaussian fit.

The momentum dependence of $\sigma_{\Delta\theta}$ and $\sigma_{\Delta\phi}$
has been fitted using a polynomial function in the range 
\mbox{($0.5 \leq P_p^{\rm{LAB}} \leq 3.2 $)~GeV} 
to obtain a variable cut used to discard reconstructed 
protons having polar and azimuthal angles differences outside the 
\mbox{($-3\sigma$, $3\sigma$)} interval, for $\theta$ and $\phi$ separately.

\begin{figure}[htbp]
\begin{center}
\leavevmode
\epsfig{file=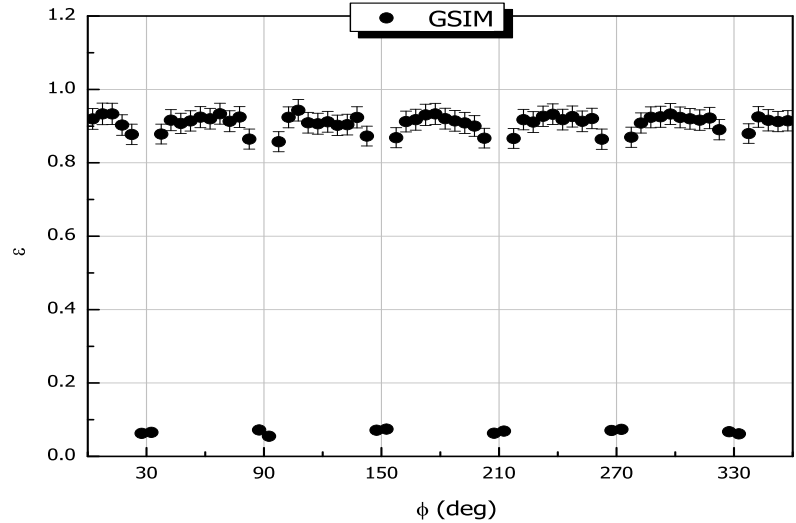,width=12cm,height=8cm}
\caption{\small The CLAS proton detection efficiency evaluated using GSim as a function 
of the azimuthal angle $\phi$ for $P_p^{\rm{LAB}}$=0.95~GeV, $\theta_p^{\rm{LAB}}=45^\circ$, and 
with a $\phi$ binning of $5^\circ$. The shapes of the six CLAS sectors are reflected by the
behavior of the proton detection efficiency and are clearly identified by the sudden 
decrease of the efficiency near the sector borders.} 
\label{fig:gsim-5deg}
\end{center}
\end{figure}

The result of the GSim evaluation of the proton detection efficiency in CLAS
is shown if Fig.~\ref{fig:gsim-5deg} as a function of the CLAS azimuthal 
angle $\phi$, for  \mbox{$P_p^{\rm{LAB}}$=0.95~GeV} and $\theta_p^{\rm{LAB}}=45^\circ$. 
Each CLAS sector has a $\phi$ extension of $60^\circ$ and it is sampled with 
a $\phi$ binning of $5^\circ$ using the procedure described above.
From Fig.~\ref{fig:gsim-5deg} it is clear that the proton detection
efficiency is almost constant (with an average value around 95\%) 
in the central regions of each CLAS sector while it drops vertically 
near the sector borders.

\begin{figure}[htbp]
\begin{center}
\leavevmode
\epsfig{file=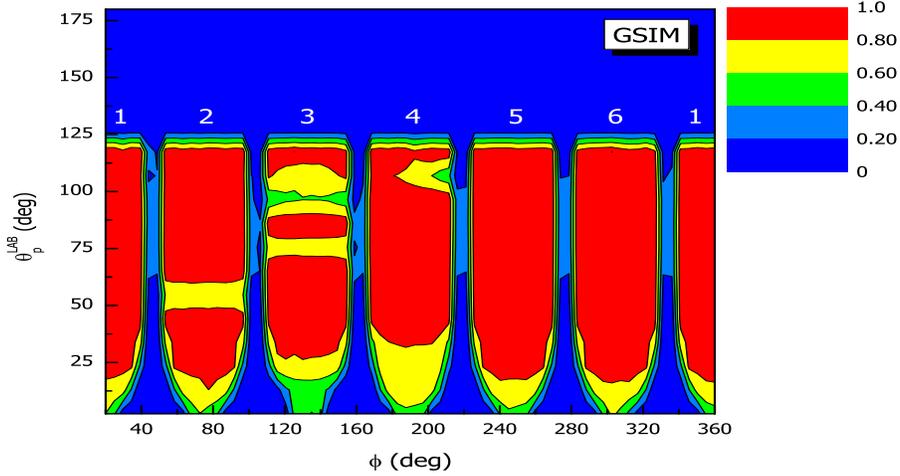,width=14cm,height=8cm}
\caption{\small 
Behavior of the proton detection efficiency obtained from GSim shown 
as a function of the azimuthal and polar angles (on the abscissa and on the ordinates, respectively)
of the generated protons for $P_p^{\rm{LAB}}$=0.95~GeV.
The quota represent the proton detection efficiency values and are shown
in different colors/grays according to the scale indicated on the right 
side of the plot. The CLAS sectors are numbered from left to right. 
In each sector, a region of practically constant high efficiency (close to $95\%$) 
is clearly identified.
The vertical low efficiency zones among sectors correspond 
the torus magnet coils. 
The horizontal gaps in the efficiency function (along $\theta$) correspond to 
to out of work time-of-flight paddles or drift chambers wires
taken into account in the simulation using the GPP tool.
}
\label{fig:gsim2}
\end{center}
\end{figure}

The overall behavior of the proton detection efficiency 
as obtained from the GSim evaluation is shown in the contour plot of Fig.~\ref{fig:gsim2}
as a function of the generated proton azimuthal and polar angles 
(running on the abscissa and on the ordinates, respectively)
for a momentum $P_p^{\rm{LAB}}$=0.95~GeV.
The quota represent the proton detection efficiency values
and are shown in different colors/grays according to the scale indicated
on the right side of the plot. 
The CLAS sectors are  numbered in the plot from left to right and 
in each sector, a  region of practically constant high efficiency 
(close to $95\%$) is clearly identified.
The vertical low efficiency zones among sectors correspond to the decrease
due to the presence of the torus magnet coils. 
The horizontal gaps in the efficiency function (along $\theta$) correspond to 
out of work time-of-flight paddles or drift chambers wires
introduced in the simulation by the GPP tool.

\subsection{Check of the GSim Evaluation Results}
\indent
\par
The proton detection efficiency in CLAS can also be measured from 
the g2 data using the \mbox{$\gamma d \rightarrow p \pi^- p$} reaction. 
The definition of the efficiency in this case is different 
from the one given in Eq.~\ref{eq:gsim-epsilon} for the simulated events
since the data does not convey any information on the ``generated'' particles.
For this reason, the correct definition for the proton detection efficiency in CLAS 
in this case is the following: 
\begin{equation}
\epsilon_{\rm{DATA}} \\
= \frac{N_{\gamma d \rightarrow p \pi^- p }^{\rm{PID}}( \Delta P_p^{\rm{LAB}}, \Delta \theta_p^{\rm{LAB}},\Delta \phi  ) } 
{N_{\gamma d \rightarrow p \pi^- X(p) }^{\rm{MM}}( \Delta P_p^{\rm{LAB}}, \Delta \theta_p^{\rm{LAB}},\Delta \phi ) - K( \Delta P_p^{\rm{LAB}}, \Delta \theta_p^{\rm{LAB}},\Delta \phi )       } \ .
\label{eq:data-epsilon}
\end{equation}\\
The quantity \mbox{${N_{\gamma d \rightarrow p \pi^- X(p) }^{\rm{MM}}}$}
in Eq.~\ref{eq:data-epsilon}
represents the number of protons identified searching for 
a $\pi^- p$ pair from the particle ID 
and then calculating the event missing mass according to the relation
$X^2(p) = (P_\gamma + P_d - P_p - P_{\pi^-})^2$, where $P_{i=\gamma,d,p,\pi^-}$  
indicates the particles 4-momenta.

On the contrary, $N_{\gamma d \rightarrow p \pi^- p }^{\rm{PID}}$ 
is the number of \mbox{$p \pi^- p$} events identified directly by the particle ID.
Clearly, in this last case, the particle ID identifies both protons, and
there is no reason to count one proton as ``missing'' 
and the other as ``identified'' so the calculation is repeated 
exchanging the proton labels.

The momenta of the identified particles 
in the final state have been compensated for the energy loss 
into the liquid $\rm{D}_2$ target and the start counter scintillators but
no momentum corrections such those described in Sec.~\ref{sec:pcorr} have
been applied in this case.

The bins of the kinematic variables 
$(\Delta P_p^{\rm{LAB}}, \Delta \theta_p^{\rm{LAB}},\Delta \phi)$
in Eq.~\ref{eq:data-epsilon} are always referred to the ``missing'' 
proton.
The binning in proton momentum and scattering angle 
is identical to the one used in the simulation case.
A different choice is done for the binning in the azimuthal angle $\phi$
chosen to be $\Delta \phi=10^\circ$
since the proton detection efficiency in CLAS 
measured from the $\gamma d \rightarrow p \pi^- p $ reaction
is mainly used to validate the simulation result 
and a full investigation of the CLAS proton detection capability 
in the $\phi$ coordinate is not necessary in this case.

Counting the number of protons identified by 
the event $\gamma d \rightarrow p \pi^- X$ 
(the first term in the denominator of Eq.~\ref{eq:data-epsilon}) 
requires the knowledge of the event missing mass distributions 
in each bin of proton momentum and angles.
These distributions have been calculated and fitted using 
a Gaussian curve and a straight line to identify the peak position 
(corresponding to the proton rest mass) and the background contribution
$K$ (second term in the denominator of Eq.~\ref{eq:data-epsilon}) as:
\begin{equation}
\label{eq:bkg-eff}
K(\Delta P_p^{\rm{LAB}}, \Delta \theta_p^{\rm{LAB}},\Delta \phi) = \int_{-3\sigma}^{3\sigma} (p_4+p_5M)\,dM
\end{equation}
where the fit parameters $p_4$, $p_5$ identify the straight line 
and $\sigma=p_3$ represents standard deviation of the Gaussian fitting 
function.
In order to count the number of protons associated with the 
exclusive event \mbox{$\gamma d \rightarrow p \pi^- X(p)$}
a ($-3\sigma$, $3\sigma$) cut around the peak of the  
missing mass distribution is applied.

\begin{figure}[htbp]
\begin{center}
\leavevmode
\epsfig{file=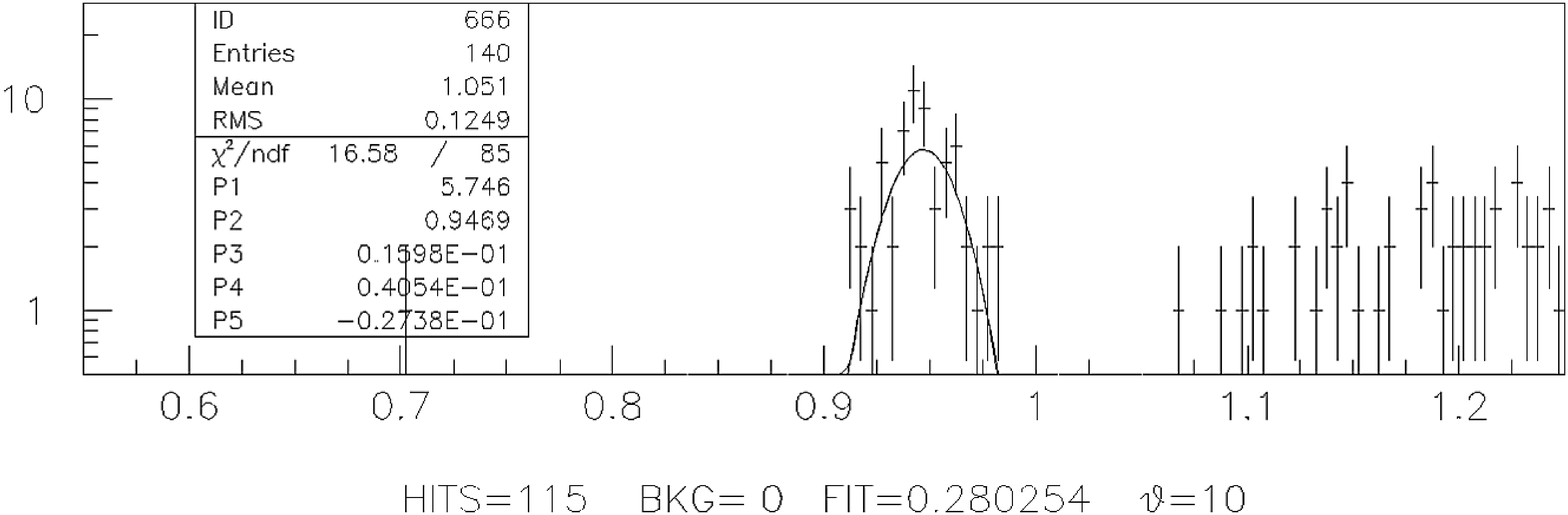,width=11cm,height=4cm}
\epsfig{file=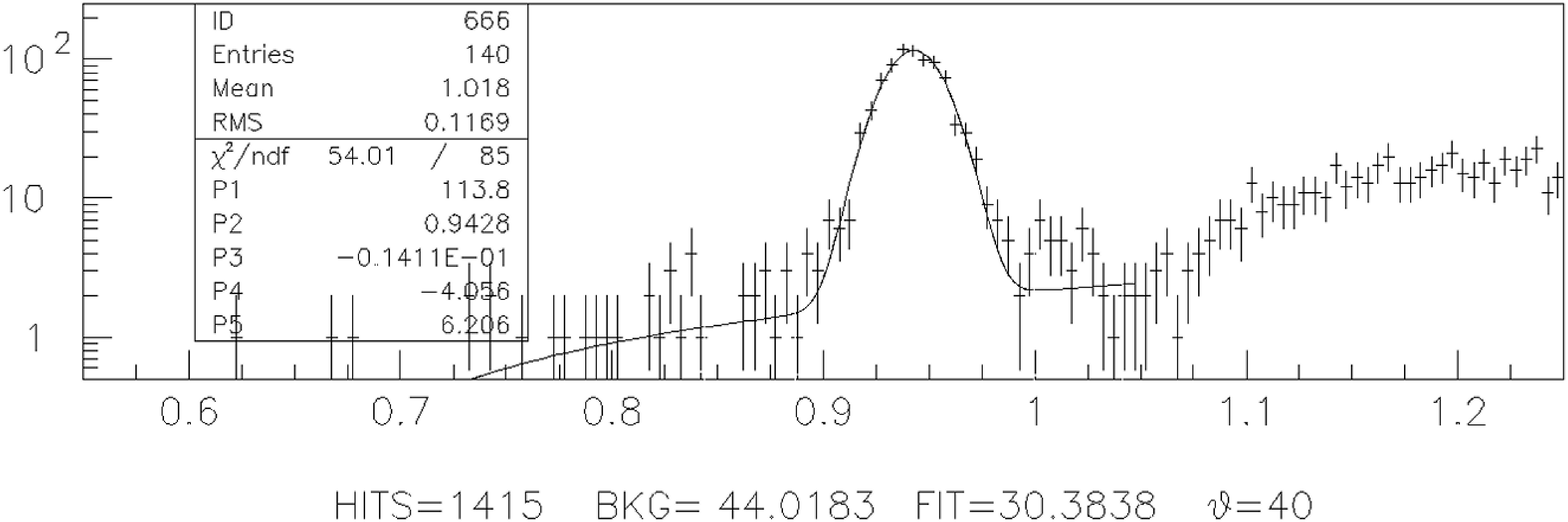,width=11cm,height=4cm}
\epsfig{file=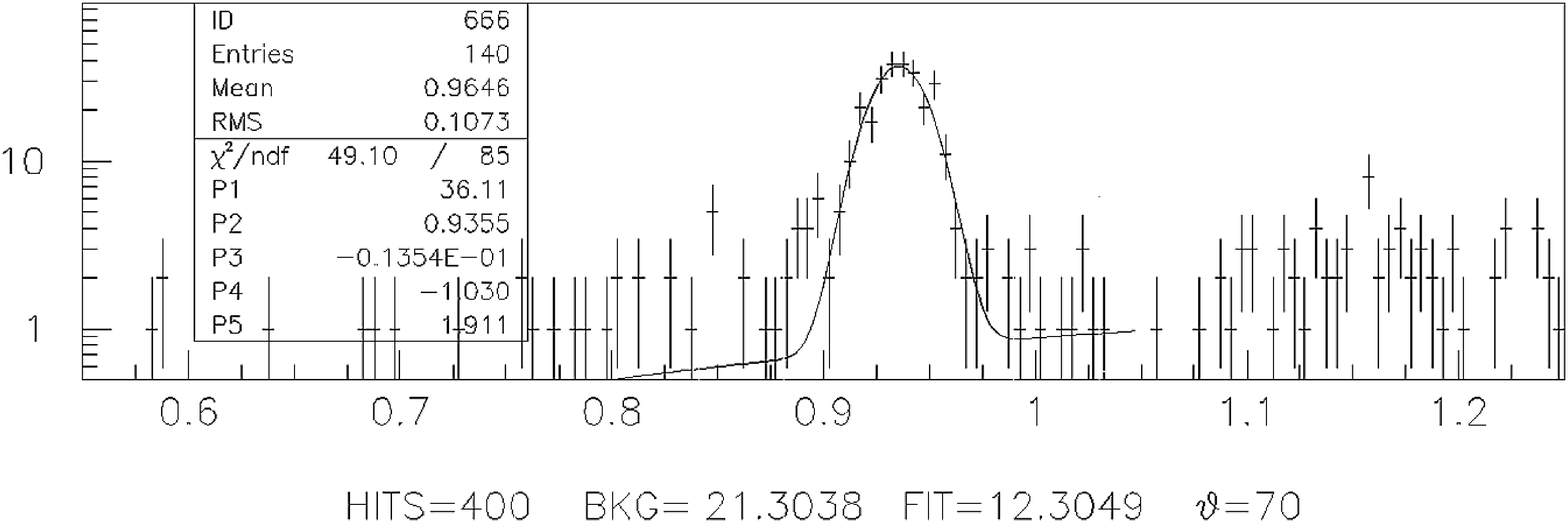,width=11cm,height=4cm}
\caption{ \small Missing mass distribution for the event
\mbox{$\gamma d \rightarrow p \pi^- X$}
at an intermediate proton momentum $P_p^{\rm{LAB}}=0.95$~GeV, 
an azimuthal angle $\phi=10^\circ$ (corresponding to the center of sector 1), 
and for three 
different proton scattering angles $\theta_p^{\rm{LAB}}$: 
$10^\circ$ (top), $40^\circ$ (middle), and $70^\circ$ (bottom).
The curves superimposed to the plots are Gaussian plus straight line
fitting functions calculated in the interval \mbox{($0.6-1.05$)~GeV}. 
The background contamination is evaluated according to Eq.~\ref{eq:bkg-eff}.}
\label{fig:p-p-pim}
\end{center} 
\end{figure}

Examples of the obtained \mbox{$\gamma d \rightarrow p \pi^- X$} 
missing mass distributions are shown in Fig.~\ref{fig:p-p-pim}
for an intermediate proton momentum $P_p^{\rm{LAB}}=0.95$~GeV, 
an azimuthal angle $\phi=10^\circ$, and for three 
different proton scattering angles $\theta_p^{\rm{LAB}}$: 
$10^\circ$ (top plot), $40^\circ$ (middle plot), and $70^\circ$ (bottom plot).
The shape of the distributions is well reproduced by the Gaussian
plus linear fitting function calculated in the interval \mbox{($0.6-1.05$)~GeV.}

As in the reconstruction of the simulated proton tracks,
the momentum determination is worse 
at the border of the six CLAS sectors since these regions are closer 
to the torus magnet coils.
In order to discard protons which momentum 
is not correctly reconstructed, 
the differences between the 
angles of ``identified'' and ``missing'' protons
\begin{equation}
\Delta\theta= \theta_p^{\rm{MM}}( P_p^{\rm{LAB}}) - \theta_p^{\rm{PID}}( P_p^{\rm{LAB}})\ , \   
\Delta\phi = \phi_p^{\rm{MM}}( P_p^{\rm{LAB}}) - \phi_p^{\rm{PID}}( P_p^{\rm{LAB}})
\label{eq:data-diff}
\end{equation}
have been obtained and 
the standard deviations $\sigma_{\Delta\theta}$ and $\sigma_{\Delta\phi}$
of the distributions have been calculated from a Gaussian fit.

The momentum dependence of $\sigma_{\Delta\theta}$ and $\sigma_{\Delta\phi}$
has been calculated using a polynomial fit in the range 
\mbox{($0.4 \leq P_p^{\rm{LAB}} \leq 2.3 $)~GeV} 
to obtain a variable cut used to discard reconstructed 
protons having polar and azimuthal angles differences outside the 
\mbox{($-3\sigma$, $3\sigma$)} interval, for $\theta$ and $\phi$ separately.

\begin{figure}[htbp]
\begin{center}
\leavevmode
\epsfig{file=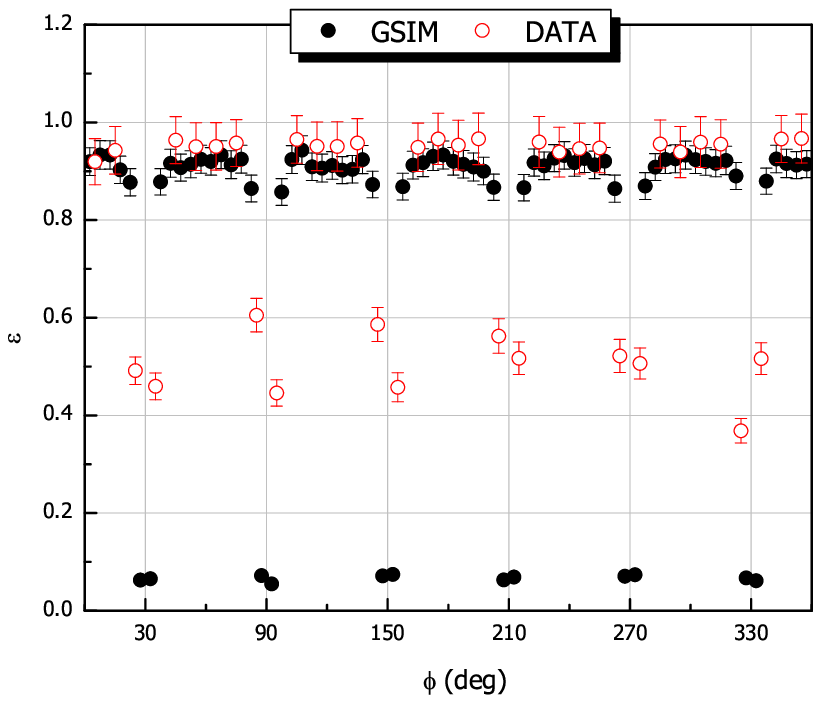,width=12cm,height=8cm}
\caption{\small CLAS proton detection efficiency in CLAS evaluated from GSIM 
(black/full dots) compared with the results measured 
from the \mbox{$\gamma d \rightarrow p \pi^- p$} reaction (red/open dots) for 
\mbox{$P_p^{\rm{LAB}}$=0.95~GeV} and \mbox{$\theta_p^{\rm{LAB}}=45^\circ$}. 
The resulting values for the proton detection efficiency are in good agreement.
Note the different $\phi$ binning. The datum corresponding 
to a sector border azimuthal angle $\phi_B$ obtained from  
the \mbox{$\gamma d \rightarrow p \pi^- p$} data
is roughly the average of the GSim results at $\phi_B-5^\circ$ and
$\phi_B+5^\circ$. 
}
\label{fig:data2}
\end{center}
\end{figure}

\begin{figure}[htbp]
\begin{center}
\leavevmode
\epsfig{file=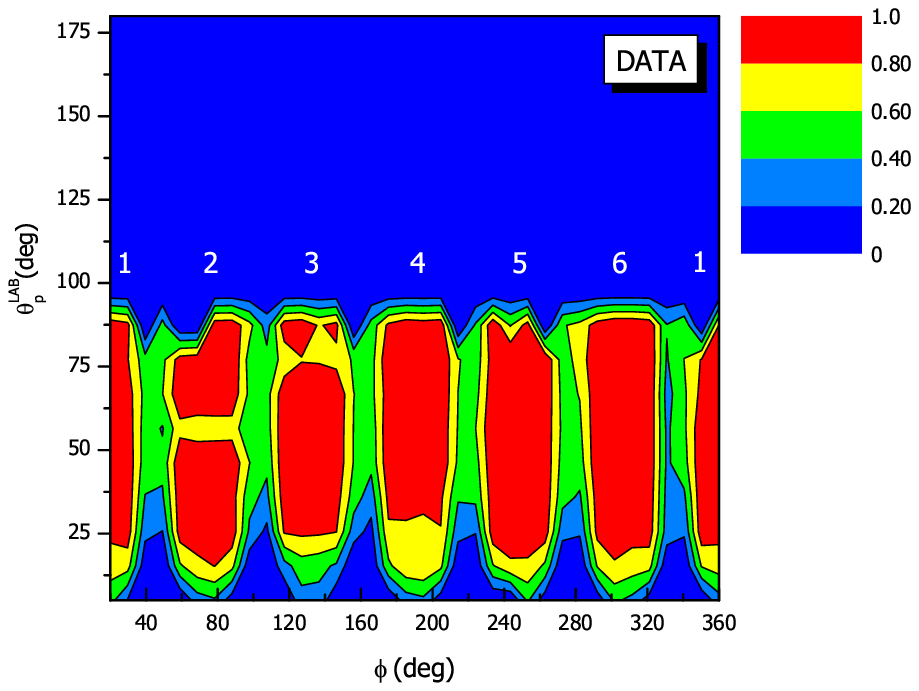,width=14cm,height=8cm}
\caption{\small Proton detection 
efficiency in CLAS measured from the \mbox{$\gamma d \rightarrow p \pi^- p$} reaction
shown as a function of the proton azimuthal and polar angles 
for  \mbox{$P_p^{\rm{LAB}}$=0.95~GeV}.
The quota represent the values of the proton detection efficiency 
in different colors/grays according to 
the scale indicated on the right side of the plot. 
The CLAS sectors are again numbered from left to right. 
The regions of constant high efficiency (close to $95\%$) can be clearly identified,
even if the sampling rate $\Delta \phi=10^\circ$ corresponds to a lower resolution.
The vertical low efficiency regions among sectors are due to the 
presence the CLAS torus magnet coils.
The horizontal gaps in the efficiency function (along $\theta$) correspond to 
to out of work time-of-flight paddles and drift-chamber wires.
}
\label{fig:data-kin}
\end{center}
\end{figure}

An example of the proton detection efficiency in CLAS measured from the 
\mbox{$\gamma d \rightarrow p \pi^- p$} reaction is shown in Fig.~\ref{fig:data2}
at the intermediate proton momentum $P_p^{\rm{LAB}}$=0.95~GeV 
and $\theta_p^{\rm{LAB}}=45^\circ$. 
The black/full dots represent the GSim result for the proton detection efficiency 
while the red/open dots are the current result 
from the \mbox{$\gamma d \rightarrow p \pi^- p$} data. 
As can be seen, the resulting efficiency values are in good agreement.
It should be noted the different  $\phi$  binning. The datum corresponding 
to a sector border azimuthal angle $\phi_B$ obtained from  
\mbox{$\gamma d \rightarrow p \pi^- p$} data
is roughly the average of the GSim results obtained at $\phi_B-5^\circ$ and
$\phi_B+5^\circ$.
The comparison between the results obtained using the two methods 
cannot be extended to all the momentum bins covered by the simulated  
\mbox{$\gamma d \rightarrow p n$} events since 
the phase space available to the \mbox{$\gamma d \rightarrow p \pi^- p$} reaction 
is smaller compared to the photo-disintegration case.
This is clearly seen in Fig.~\ref{fig:data-kin} where the proton detection 
efficiency, measured from the \mbox{$\gamma d \rightarrow p \pi^- p$} reaction,
is shown as a function of the proton azimuthal and polar angles 
(on the abscissa and on the ordinates, respectively) 
for \mbox{$P_p^{\rm{LAB}}$=0.95~GeV}.
The comparison of this result with Fig.~\ref{fig:gsim2}, evidently shows 
that in this case the proton is detected only up to \mbox{$\theta_p^{\rm{LAB}}=90^\circ$}
while in the photo-disintegration case, the angular coverage extends
up to \mbox{$\theta_p^{\rm{LAB}}=125^\circ$}, for this value of proton momentum.
The quota represent the values of the proton detection efficiency 
in different colors/grays according to 
the scale indicated on the right side of the plot. 
The CLAS sectors are numbered from left to right. 
The regions of constant high efficiency (close to $95\%$) can be clearly identified.
Note that the sampling rate \mbox{$\Delta \phi=10^\circ$} yields a lower resolution
with respect to the GSim result (Fig.~\ref{fig:gsim2}).
The vertical low efficiency regions among sectors are due to the 
presence of the torus magnet coils.
The horizontal gaps in the efficiency function (along $\theta$) correspond  
to out of work time-of-flight paddles and drift chambers wires.

The result on the proton detection efficiency in CLAS
measured from the  \mbox{$\gamma d \rightarrow p \pi^- p$} data
is used, in the present analysis, to validate the GSim result 
in the appropriate kinematic range.
In fact, the limitation in phase space coverage does not allow 
a full comparison between the final results nor the usage of the 
proton detection efficiency measured from the \mbox{$\gamma d \rightarrow p \pi^- p$ }
data in the calculation the final photo-disintegration cross section.

The information obtained 
from the \mbox{$\gamma d \rightarrow p \pi^- p$} data
is used to define the final timing and spatial resolutions 
smearing factors to be introduced in the simulation in order to match 
the characteristics of the simulated detector 
with those of the real apparatus. 
The comparison is done iteratively, running partial 
reconstructions on the simulated files 
until the required resolution smearing factors are determined.

After this procedure, spatial and timing resolutions 
smearing factors equal to 1 have been applied
to the simulation results for all proton scattering 
angles. An exception is done for very forward proton scattering
angles \mbox{$8^\circ \leq \theta_p^{\rm{LAB}} \leq 10^\circ$},
where a spatial resolution smearing factor equal to 2 has 
been used.

\begin{figure}[htbp]
\begin{center} 
 \leavevmode
\hspace*{-1.0cm} 
\epsfig{file=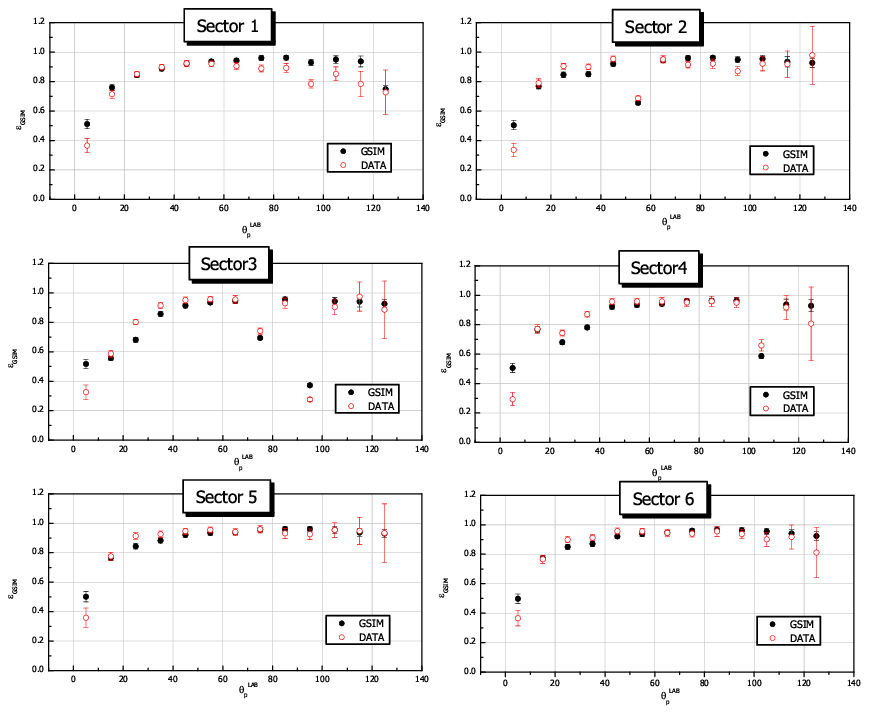,width=16cm,height=19cm}
\caption { \small 
Final comparison between the results obtained from the GSim evaluation 
of the proton detection efficiency using the \mbox{$\gamma d \rightarrow p n$} 
reaction (black/full dots) and the measurement 
from the data using the \mbox{$\gamma d \rightarrow p \pi^- p$} reaction
(red/empty dots).
The points represent the weighted averages of the detection efficiency 
over proton momenta in the interval \mbox{($0.5-1.0$)~GeV} 
for the sectors central region (defined by the interval \mbox{$\Delta \phi =\pm 10^\circ$}
assuming the sector azimuthal angle between $(-30^\circ,30^\circ)$) 
and are shown as a function of the proton scattering angle \mbox{$\theta_p^{\rm{LAB}}$}.
It is clearly seen that the GSim result is well matched 
by the \mbox{$\gamma d \rightarrow p \pi^- p$} result.
In addition, the actual detector status is well reproduced by the simulations: 
this is shown by the dips in the proton detection efficiency found
at the same scattering angle \mbox{$\theta_p^{\rm{LAB}}$} in both results. 
}
\label{fig:polar}
\end{center}
\end{figure}

A final comparison between the results obtained from the GSim evaluation 
of the proton detection efficiency using the \mbox{$\gamma d \rightarrow p n$} 
reaction (black/full dots) and the measurement 
from the data using the \mbox{$\gamma d \rightarrow p \pi^- p$} reaction
(red/empty dots) is shown Fig.~\ref{fig:polar} separately for the six CLAS sectors.
The points represent the weighted averages of the proton detection efficiency in CLAS 
over proton momenta in the interval \mbox{($0.5-1.0$)~GeV} 
for the sectors central region (defined by the interval \mbox{$\Delta \phi =\pm 10^\circ$}
assuming the sector azimuthal between $(-30^\circ,30^\circ)$) 
and are shown as a function of the proton scattering angle \mbox{$\theta_p^{\rm{LAB}}$}.
It is clearly seen how the GSim result is well matched 
by the \mbox{$\gamma d \rightarrow p \pi^- p$} result.
In addition, the actual detector status is well reproduced by the simulations: 
this is shown by the dips in the proton detection efficiency found
for the same scattering angle \mbox{$\theta_p^{\rm{LAB}}$} in both results (see, 
for instance, sector 3 for \mbox{$\theta_p^{\rm{LAB}}=75^\circ$} and $95^\circ$).

\begin{figure}[htbp]
\begin{center}
 \leavevmode
 \epsfig{file=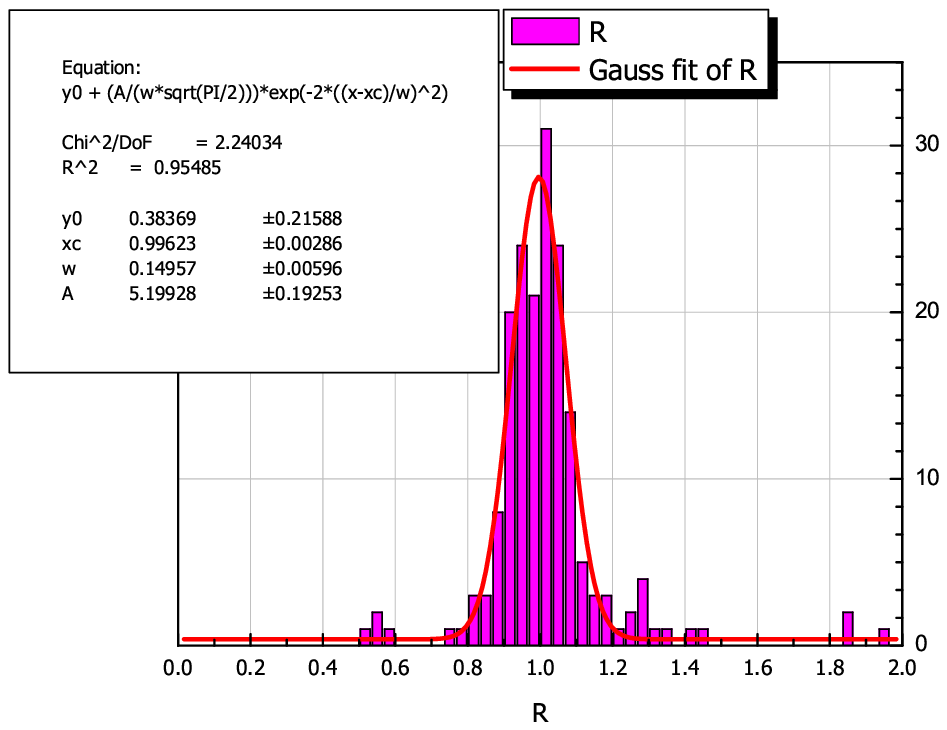,width=14cm,height=10cm}
  \caption{\small Distribution of the ratio \mbox{$R = \frac {\epsilon_{\rm{DATA}}} {\epsilon_{\rm{GSIM} }}$} 
between the proton detection efficiency measured from \mbox{$\gamma d \rightarrow p \pi^- p$}  
and simulated from \mbox{$\gamma d \rightarrow p n$} 
(defined only in the kinematic bins \mbox{$\Delta P_p^{\rm{LAB}}$}
and \mbox{$\Delta \theta_p^{\rm{LAB}}$} populated by both kind of data)
for the sectors central region (defined by the interval \mbox{$\Delta \phi =\pm 10^\circ$}
assuming the sector azimuthal angle between $(-30^\circ,30^\circ)$).
The parameters representing the Gaussian fit are shown in the legend. 
It can be seen that the mean value of the distribution is of the order of the unity while
the spread is about 8\%. 
}
\label{fig:ratio}
\end{center}
\end{figure}

In order to evaluate the uniformity between the two results, 
the distribution for the values of 
ratio \mbox{$R = \frac {\epsilon_{\rm{DATA}}} {\epsilon_{\rm{GSIM}}}$}
(defined only for the kinematic bins \mbox{$\Delta P_p^{\rm{LAB}}$} 
and \mbox{$\Delta \theta_p^{\rm{LAB}}$} populated by both kind of data)
between the efficiency calculated with the two techniques 
is shown in Fig.~\ref{fig:ratio} for proton momenta in the range  \mbox{($0.5-1.0$)~GeV}.
As can be seen from the legend, the mean value of the distribution 
is of the order of the unity while the spread is about 8\%.
The best agreement between the simulation and data is found for sector 6.
This finding can be used to proper normalize cross section data from other sectors.

\clearpage
\subsection{Fiducial Cuts and Detector Acceptance}
\indent
\par
The information obtained from 
the simulated \mbox{$\gamma d \rightarrow p n$} reaction  
in CLAS allows the determination of the so-called 
``fiducial'' regions in the azimuthal angle $\phi$.
In fact, a fiducial region can be defined, for each sector $S$ and each bin of proton momentum 
\mbox{$\Delta P_p^{\rm{LAB}}$} and scattering angle \mbox{$\Delta \theta_p^{\rm{LAB}}$}, 
as the $\phi$ section of the CLAS detector characterized by a constant efficiency.

The procedure used to identify the fiducial regions requires the 
calculation of a reference value for the proton detection efficiency 
(indicated as $\epsilon_0$) in each CLAS sector and for each bin of 
proton momentum and scattering angle.
This reference value is obtained averaging the contributions 
of the most central regions of each sector.
The corresponding indetermination associated to $\epsilon_0$ 
is indicated by $\delta \epsilon_0$.

Then, in each CLAS sector (for each bin of proton momentum 
and scattering angle) 
the proton detection efficiency, $\epsilon_i$, 
in the remaining $i^{\rm{th}}$ $\phi$-bin 
on the left and on the right of the central region is  investigated
and the values of $\epsilon_i$ whose error bars $\delta \epsilon_i$
do not overlap with the band defined 
by the relation \mbox{$\epsilon_0 \pm \delta \epsilon_0$} are rejected.

\begin{figure}[htpb]
\begin{center}
 \leavevmode
 \epsfig{file=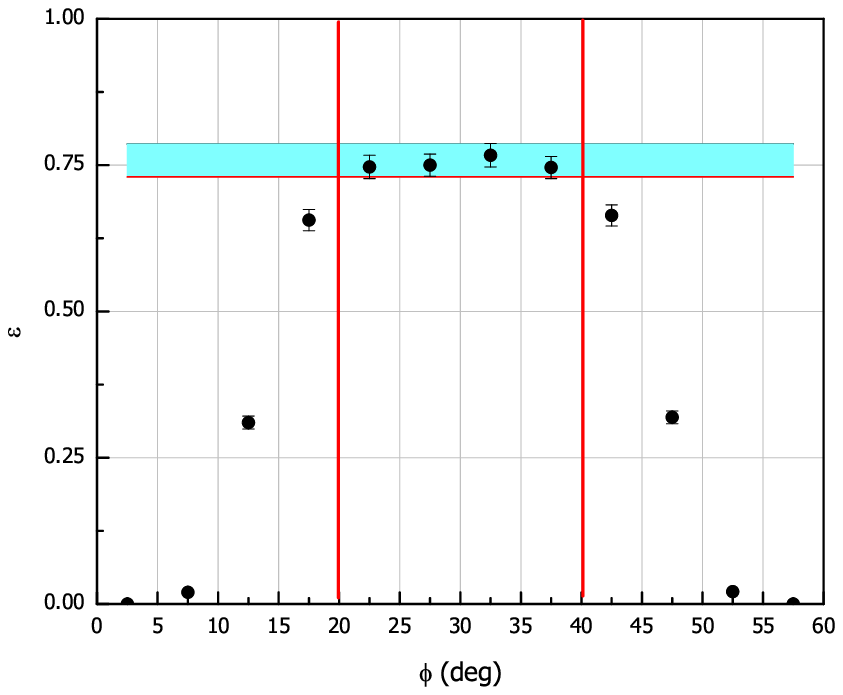,width=12cm,height=7cm}
\epsfig{file=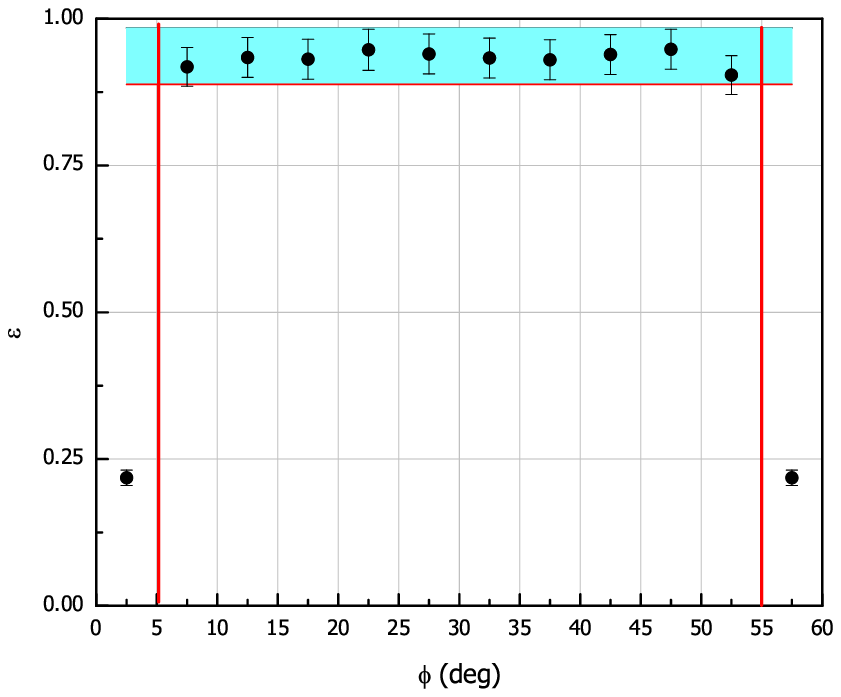,width=12cm,height=7cm}
 \caption{\small An example of fiducial cuts in CLAS sector 2 for 
\mbox{$P_p^{\rm{LAB}}$=1.15~GeV}. 
Upper panel: for \mbox{$\theta_p^{\rm{LAB}}=15^\circ$}. 
Lower panel: for \mbox{$\theta_p^{\rm{LAB}}=65^\circ$}.
The horizontal shaded area shows the allowed variation band \mbox{$\epsilon_0 \pm \delta \epsilon_0$}
while the vertical lines delimit the fiducial region according to the behavior of 
the proton detection efficiency as a function of the azimuthal angles $\phi$.}
\label{fig:fid-cuts}
\end{center}
\end{figure}

An example of this procedure is shown in Fig.~\ref{fig:fid-cuts}
for sector 2, for a proton momentum \mbox{$P_p^{\rm{LAB}}$=1.15~GeV} 
and the scattering angles \mbox{$\theta_p^{\rm{LAB}}=15^\circ$} (top plot) and
\mbox{$\theta_p^{\rm{LAB}}=65^\circ$} (bottom plot).

The horizontal shaded area corresponds to the allowed variation band
\mbox{$\epsilon_0 \pm \delta \epsilon_0$} while the vertical lines
mark the boundaries used to exclude efficiency data considered
out from the allowed variation band according to the criterion outlined above.
 It can be seen that the procedure retains only the contributions
from to the $\phi$ region where the efficiency behavior is constant.

\begin{figure}[htbp]
\begin{center}
 \leavevmode
 \epsfig{file=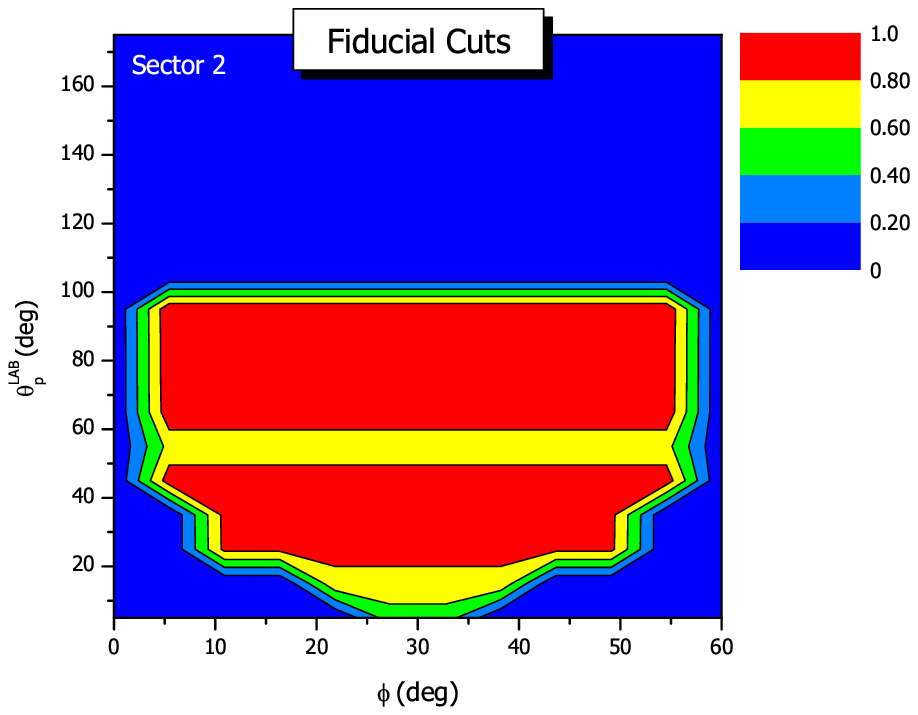,width=14cm,height=8cm}
 \caption{\small 
Fiducial cuts in the CLAS sector 2 for $P_p^{\rm{LAB}}$=1.15~GeV.
The boundaries of the fiducial $\phi$ regions (along the abscissa) 
in the sector are shown for each proton scattering angle (along the ordinates).
The mean values of the efficiency within the region are indicated by the color/gray
code shown in the legend. The dip found at \mbox{$\theta_p^{\rm{LAB}}=55^\circ$} is due to an
out of work time-of-flight paddle, as shown also in Fig.~\ref{fig:polar}
}
 \label{fig:eff-fid2}
\end{center}
\end{figure}

An example of the resulting fiducial cuts in the CLAS sector 2 
for $P_p^{\rm{LAB}}$=1.15~GeV is shown in the contour plot 
of Fig.~\ref{fig:eff-fid2}.
The boundaries of the fiducial $\phi$ regions (along the abscissa) 
are shown for each proton scattering angle (along the ordinates).
The mean values of the efficiency within the region are indicated by 
the color/gray code. 
The dip found at \mbox{$\theta_p^{\rm{LAB}}=55^\circ$} is due to an
out of work time-of-flight paddle, as shown also in Fig.~\ref{fig:polar}

The procedure outlined above is applied in each CLAS sector 
and for each bin of proton momentum and scattering angle 
in order to derive the full set of fiducial cuts. 
This is needed to properly weight the number of photo-disintegration
events entering in the final differential cross section calculation
since it will be expressed as a function 
of the outgoing proton momentum, scattering angle, and sector. 

For this reason, the $\phi$ information contained in the
original proton detection efficiency $\epsilon_{\rm{GSIM}}$ 
must be averaged over the CLAS sectors fiducial regions so to
obtain  the sector-based efficiency-acceptance function:
\begin{equation}
\eta_{\rm{GSIM}} \left( \Delta P_p^{\rm{LAB}},\Delta \theta_p^{\rm{LAB}}, S \right) 
= \frac{\sum_{i} w_i\, \epsilon_{\rm{GSIM}} \left ( \Delta P_p^{\rm{LAB}},\Delta \theta_p^{\rm{LAB}},\Delta \phi_i \right)}{\sum_i w_i}
\label{eq:gsim-sector}
\end{equation}
where 
\mbox{$1/w_i = (\delta \epsilon_{\rm{GSIM}} \left ( \Delta P_p^{\rm{LAB}},\Delta \theta_p^{\rm{LAB}},\Delta \phi_i \right))^2$}
is the weight assigned to each datum from the original efficiency function, 
$S$ stands for the CLAS sector, and the index $i$ runs over 
the $\phi$-bins of sector $S$ 
for the given proton momentum and scattering angle. 

\begin{figure}[htpb]
\begin{center}
\leavevmode
\epsfig{file=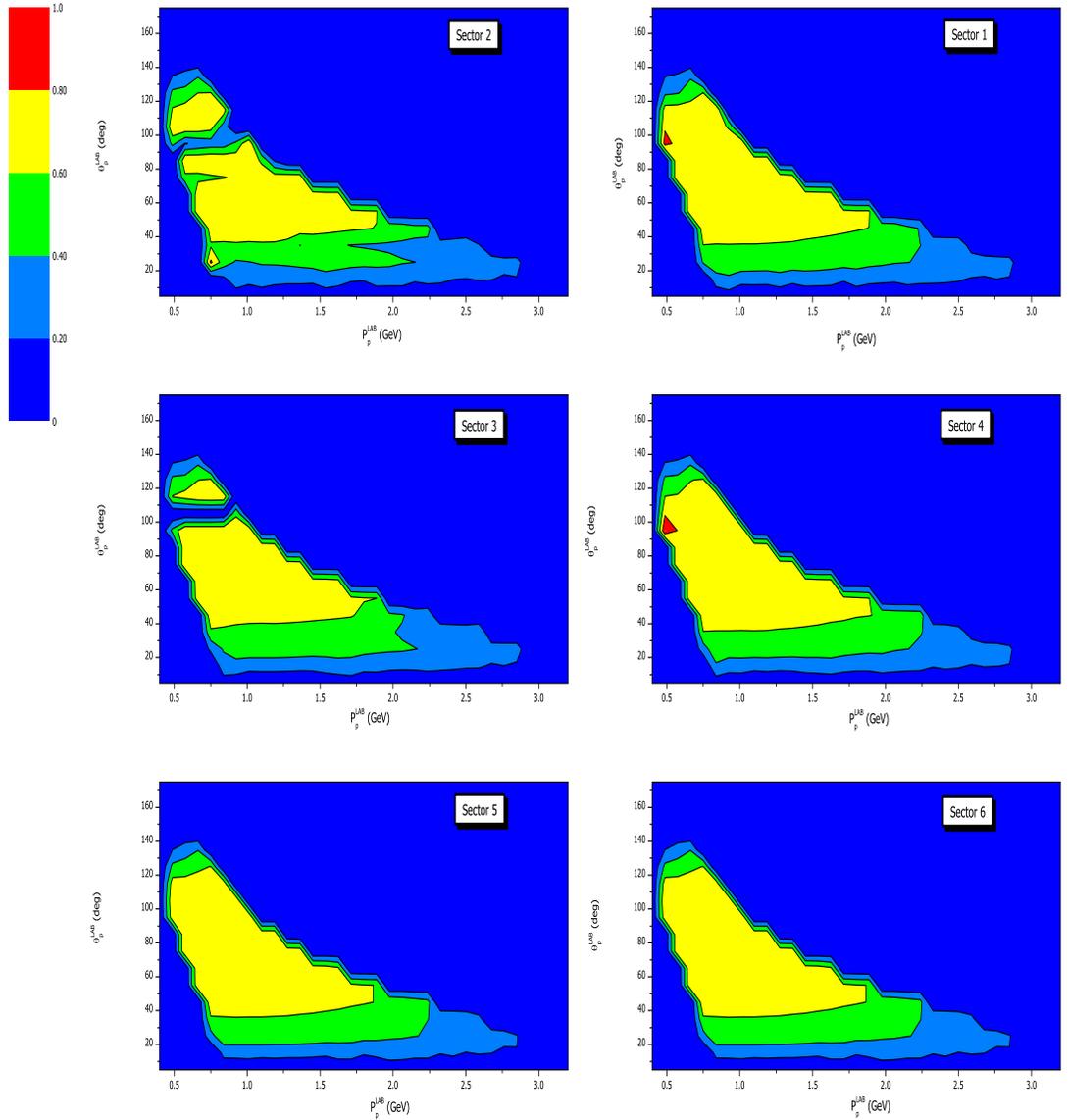,width=16cm,height=19cm}
\caption{\small 
Result of the calculation of the efficiency-acceptance function
for the six CLAS sectors.
The abscissa report the proton momentum while the ordinates run on the
proton scattering angle (both considered in the LAB system).
Quotas identify regions of constant values of the efficiency-acceptance 
functions (numeric values can be obtained from the color/gray scale 
shown on the left). As can be seen from  Fig.~\ref{fig:eff2dsect1}
the behavior of the six CLAS sectors is pretty uniform, except for
the holes introduced by out of work channels in some sectors.}
\label{fig:eff2dsect1}
\end{center}
\end{figure}

The result of this calculation for the efficiency-acceptance function
is shown in the contour plot of Fig.~\ref{fig:eff2dsect1} for the six CLAS sectors.
The abscissa report the proton momentum while the ordinates run on the
proton scattering angle (both considered in the LAB system).
Quotas identify regions of constant values of the efficiency-acceptance 
functions (numeric values can be obtained from the color/gray scale 
shown on the left). As can be seen from  Fig.~\ref{fig:eff2dsect1}
the behavior of the six CLAS sectors is pretty uniform, except for
the holes introduced by out of work channels in some sectors.

\newpage 
\section{Background Subtraction}
\indent
\par
\label{sec:bkg}
The determination of the number of ``true'' photo-disintegration events
requires the knowledge of the contamination from background events.
In fact, the missing mass distribution from $\gamma d \rightarrow pX$ events
usually exhibits a peak in correspondence to the neutron rest mass value 
sitting over a continuous background mainly due to photo-production
events originating from the target cell walls.

In the present analysis, the background contribution 
is evaluated from the data.
Only the events detected in the fiducial regions of the CLAS, 
defined using the fiducial cuts obtained in the previous section,
are taken into account. 
For each event, the missing mass is calculated and distributions are
obtained for each bin of proton scattering angle $\theta_p^{\rm{CM}}$ 
and incident photon energy $E_\gamma$.

\begin{figure}[htbp]
\begin{center}
\leavevmode
\epsfig{file=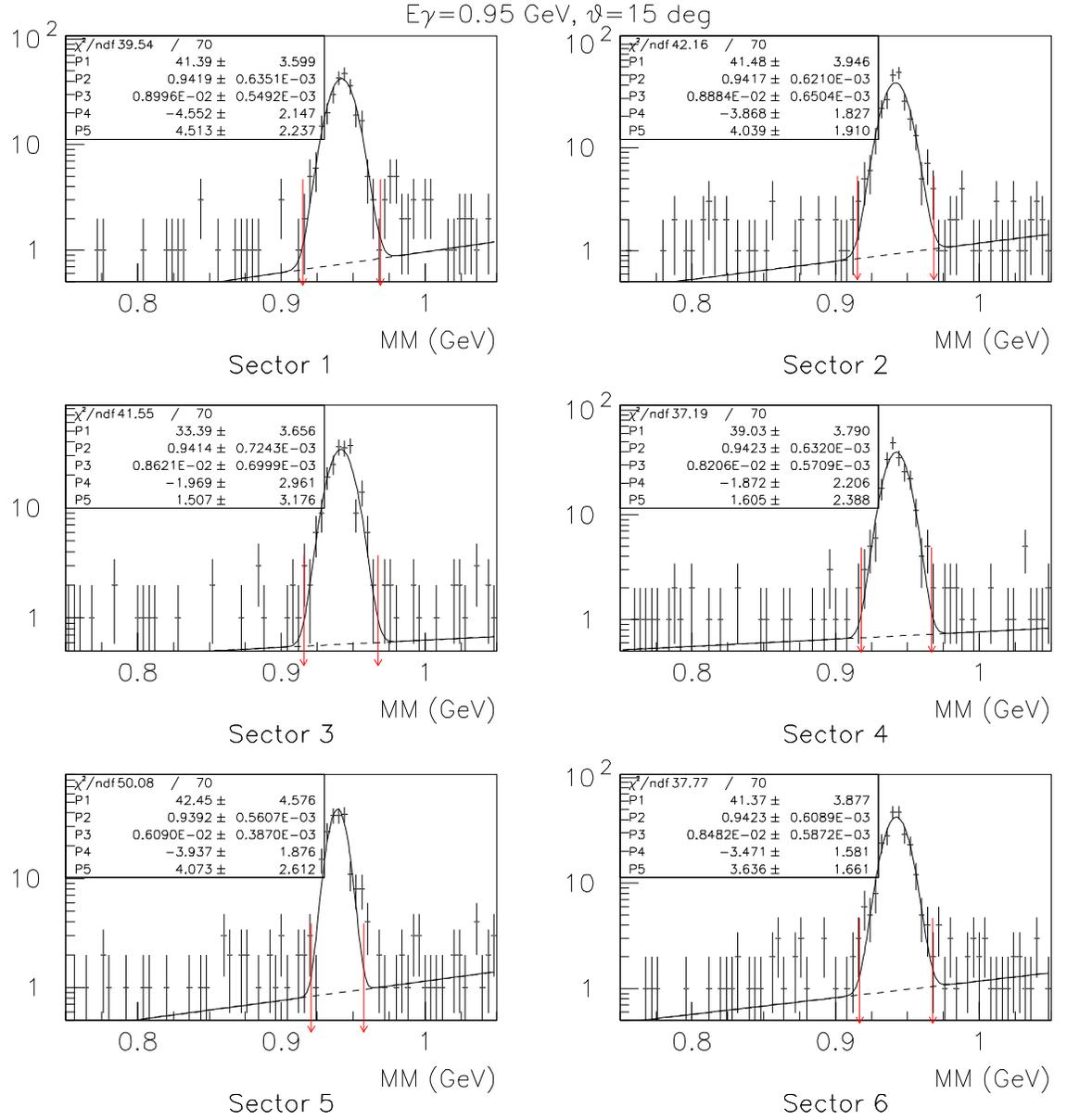,width=15cm}
\caption{\small Missing mass distributions for the six CLAS sectors, 
for the intermediate energy $E_\gamma$=0.95~GeV, and a forward scattering angle $\theta_p^{\rm{CM}}=15^\circ$:
the peak corresponding to the neutron rest mass is clearly identified.
Each distribution is fitted with a Gaussian function plus exponential curve to
take into account the continuous background.
The red/dotted arrows identify a $(-3\sigma,3\sigma)$ interval around the peak position,
being $\sigma$ the standard deviation of the Gaussian function.
}
\label{fig:bkg-mm}
\end{center}
\end{figure}

Examples of the obtained missing mass distributions are shown in Fig.
\ref{fig:bkg-mm} for the six CLAS sectors, 
for the intermediate energy $E_\gamma$=0.95~GeV, 
and a forward scattering angle $\theta_p^{\rm{CM}}=15^\circ$:
the peak corresponding to the neutron rest mass is clearly identified and
each distribution is fitted with a Gaussian function plus exponential curve to
take into account the continuous background.

\begin{figure}[htbp]
\begin{center}
\leavevmode
\epsfig{file=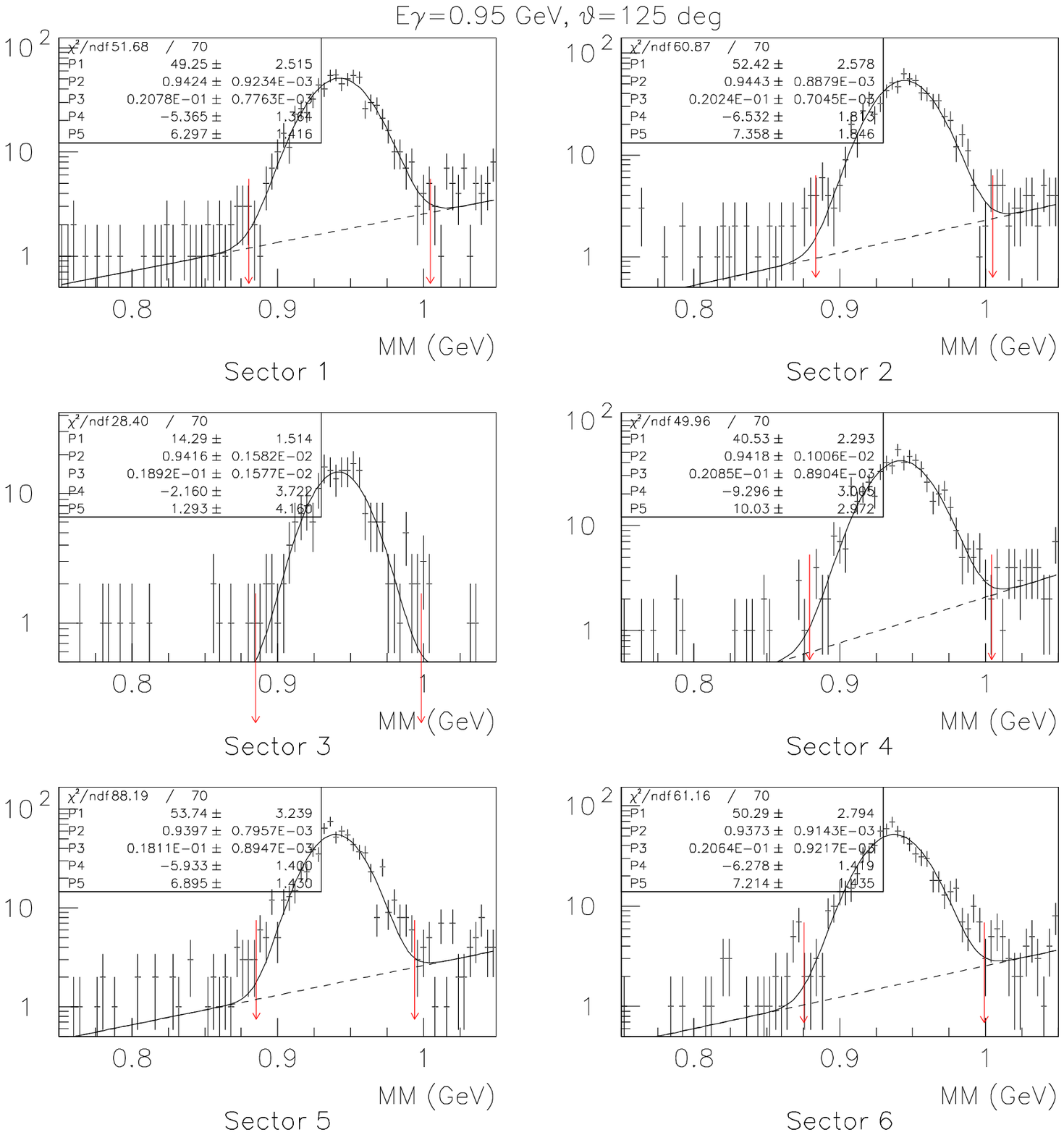,width=15cm}
\caption{\small Missing mass distributions for the six CLAS sectors for  $\theta_p^{\rm{CM}}=125^\circ$ and 
for an energy incident photon energy $E_\gamma$=0.95~GeV.
The distribution are wider and the neutron signal is less pronounced
since at backward scattering angles the resolution of the CLAS spectrometer is lower.
Nevertheless, the missing mass distributions are still well described by a Gaussian plus exponential function fit.
The red/dotted arrows identify a $(-3\sigma,3\sigma)$ interval around the peak position,
being $\sigma$ the standard deviation of the Gaussian function.}
\label{fig:bkg-mm1}
\end{center}
\end{figure}

At larger angles (for the same energy $E_\gamma$=0.95~GeV), 
the distribution are wider and the neutron signal is less pronounced, as shown in Fig.~\ref{fig:bkg-mm1}. 
Nevertheless, they are still well described by a Gaussian plus exponential function fit.

The contamination from background events under the missing mass distribution 
peak is calculated according to the relation:
\begin{equation}
\label{eq:bkg-data}
k(\theta_p^{\rm{CM}}, E_\gamma) = \frac{\int_{-3\sigma}^{3\sigma} e^{p_4+p_5M}dM}{N_{\rm{peak}}}
\end{equation}
where $\sigma$ is the standard deviation of the Gaussian function,
$p_4$ and $p_5$ are the parameters of the exponential function, 
and $N_{\rm{peak}}$ is the number of events under the 
($-3\sigma$, $3\sigma$) interval around the peak position. 

\begin{figure}[htbp]
\begin{center}
\leavevmode
\epsfig{file=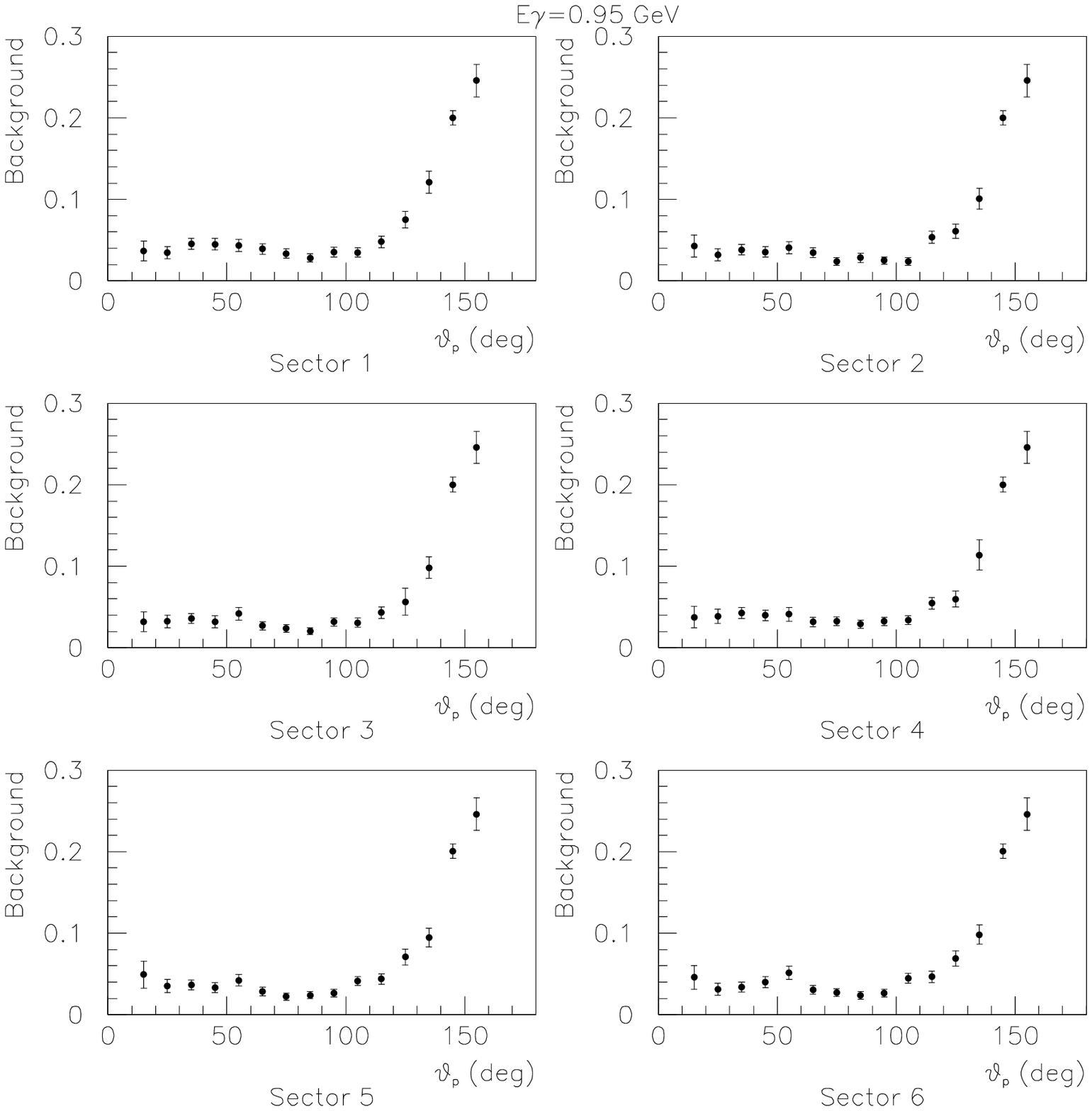,width=15cm}
\caption{\small The behavior of the background contamination $k$ as a function of
the proton scattering angle for the six CLAS sectors and for 
the intermediate incident photon energy $E_\gamma$=0.95~GeV.
The contribution is around 4\% and is relatively flat for scattering angles 
less than $90^\circ$.}
\label{fig:bkg-ang}
\end{center}
\end{figure}

The behavior of the background contamination $k$ as a function of
the proton scattering angle is shown in Fig.~\ref{fig:bkg-ang}
for the six CLAS sectors and for the intermediate incident photon energy 
$E_\gamma$=0.95~GeV.
The contribution is around 5\% and is relatively flat for scattering angles 
less than $90^\circ$. At larger angles the contamination is higher since the
missing mass distributions get wider 
since the momentum resolution of the detector decreases.

\begin{figure}[htbp]
\begin{center}
\leavevmode
\epsfig{file=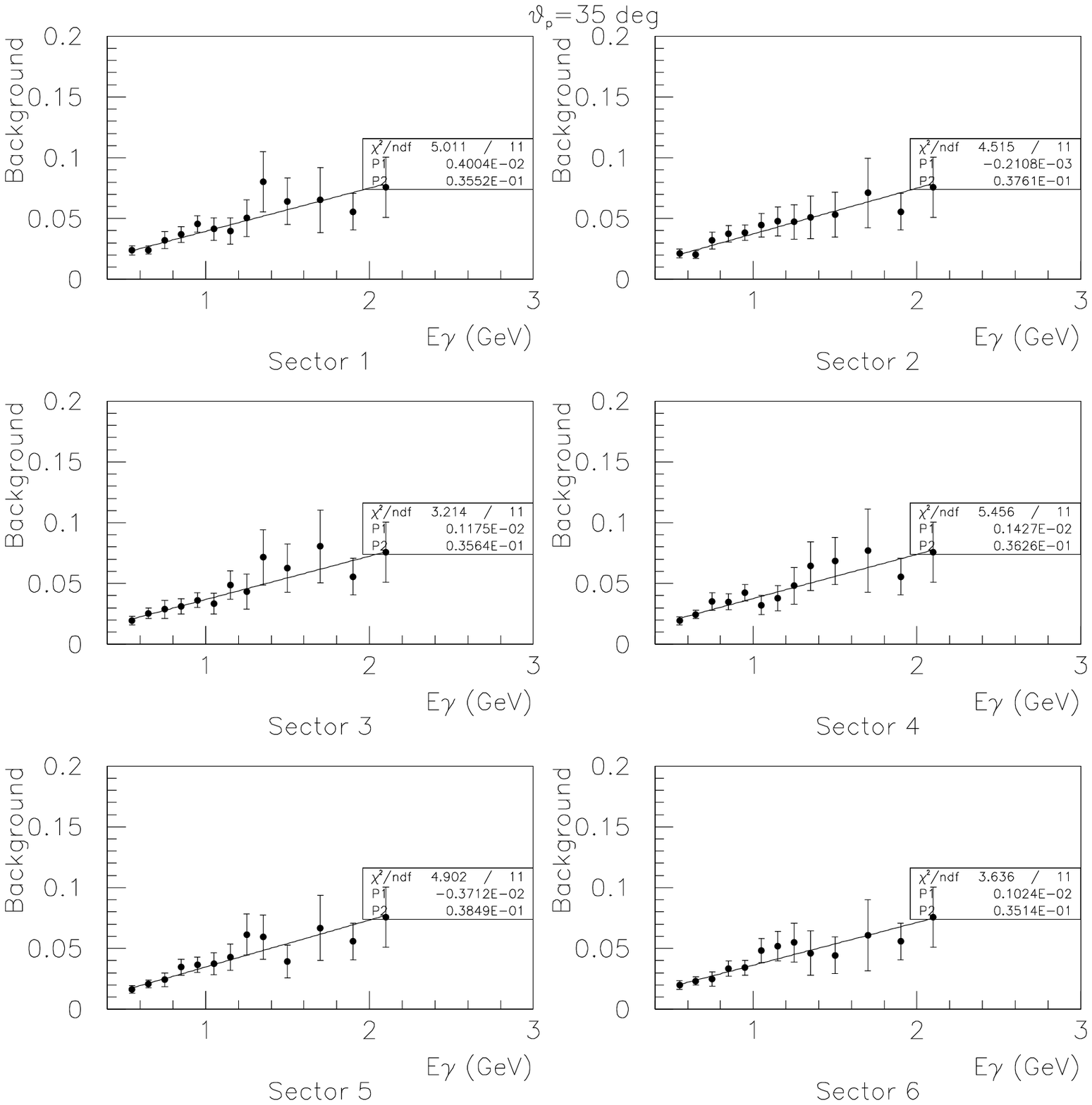,width=15cm}
\caption{\small  
Behavior of the background contamination $k$ as a function of the incoming
photon energy for the six CLAS sectors and for $\theta_p^{\rm{CM}}=35^\circ$.
The last three data points, corresponding to the energy interval
$1.8 \leq E_\gamma \leq 2.1$~GeV, are obtained averaging the contributions 
from all sectors and are identical in the plots shown. 
It is clearly seen that the data are well reproduced using a linear fit.
}
\label{fig:bkg-en}
\end{center}
\end{figure}

A similar situation occurs at high incident energies (more than
2.1~GeV) where the background subtraction cannot be directly performed
since the statistics is not enough to obtain 
a prominent neutron peak in the missing mass distributions.
To overcome this limitation, the behavior of the background 
contamination has been studied as a function of the incoming
photon energy and extrapolated up to 3~GeV using a linear fit. 
An example of the result is shown in Fig.~\ref{fig:bkg-en} 
for the six CLAS sectors and for $\theta_p^{\rm{CM}}=35^\circ$.
The last three data points, corresponding to the energy interval
$1.8 \leq E_\gamma \leq 2.1$~GeV, are obtained averaging the contributions coming
from all sectors since the statistics is not enough
to have a clear neutron signal in the missing mass spectra,
and are identical in the plots shown. 
It is clearly seen that the data are well reproduced using a linear fit.

\begin{figure}[htbp]
\begin{center}
\leavevmode
\epsfig{file=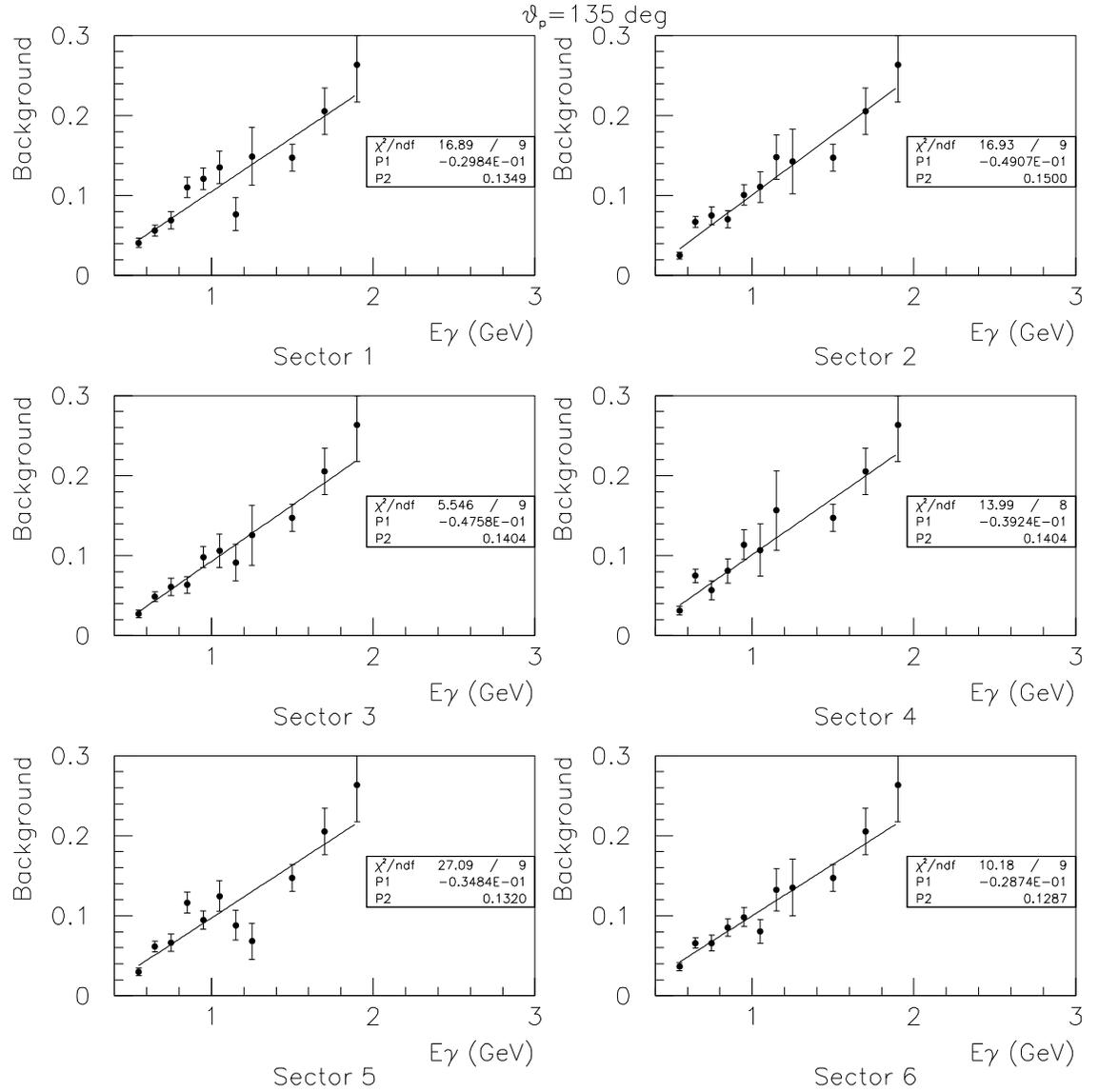,width=15cm}
\caption{\small 
Behavior of the background contamination $k$ as a function of the incoming
photon energy for the six CLAS sectors and for $\theta_p^{\rm{CM}}=135^\circ$.
The last three data points, corresponding to the energy interval
$1.8 \leq E_\gamma \leq 2.1$~GeV, are obtained averaging the contributions 
from all sectors and are identical in the plots shown. 
At this proton scattering angles the background contribution is larger
but it can still be extrapolated using a linear fit of the available points.
}
\label{fig:bkg-en1}
\end{center}
\end{figure}

At large proton scattering angles the background contamination $k$ is larger
but it can still be extrapolated using a linear fit of the available
points. This can be seen from Fig.~\ref{fig:bkg-en1} for the six
CLAS sectors and for $\theta_p^{\rm{CM}}=135^\circ$. 
In analogy to the situation described in Fig.~\ref{fig:bkg-en} the last three points 
are obtained averaging the contribution of all sectors.

\clearpage
\section{Evaluation of the Systematic Errors}
\indent
\par
Systematic errors may result from faulty
calibration of the experimental equipment or from
biased procedures used in the analysis of the data.
These errors must be estimated from reconsidering
the status of the experimental conditions and the 
techniques used to analyze the data.

In the present analysis, the main contributions to the 
overall systematic uncertainty come from 
the following procedures:

\begin{itemize}
\item{ determination of the number of incident photons;}
\item{ background subtraction;}
\item{ evaluation of the proton detection efficiency.}
\end{itemize}

The evaluation of the systematic error on the photon 
flux normalization, is done using the 
information on the tagging efficiency as 
described in Sec.~\ref{sec:norm}.
As a consequence, it can be seen that the systematic 
uncertainties are estimated to be around $3-4$\%.

The background subtraction 
procedure introduces a systematic error of the order of 10\%
when repeated with different cuts while 
the proton detection efficiency in CLAS has been 
evaluated using two different techniques 
(as shown in Sec.~\ref{sec:eff}).

The comparison between these results, 
in the full overlap of the kinematics 
of the two reactions introduces a systematic error 
of the order of 5\%.

The different
contributions are summarized in Tab.~\ref{tab:syst}.

\begin{table}[ht]
\begin{center}
\begin{tabular}{|c|c|} \hline
Procedure & systematic error \\ \hline
Photon flux normalization & 3-4\% \\
Background subtraction & 8-10\% \\
Proton detection efficiency & 5-6\%  \\ \hline
Overall  	&   10-12\% \\ \hline
\end{tabular}
\end{center}
\caption{\small Summary of the main contributions to the 
overall systematic uncertainty.}
\label{tab:syst}
\end{table}

It should be underlined anyhow that a more accurate 
evaluation of the systematic uncertainties
is still underway.

\clearpage
\section{Summary}
\indent
\par

In this Chapter the various phases 
of the data analysis chain are fully described. 
First, the CLAS event format and the reconstruction procedure 
are introduced since these structures have been used to access 
and process the g2 data to perform all subsequent analysis.

Then, the data calibration procedure is briefly described 
followed by the definition of the photo-disintegration event.
At this point, the photon flux normalization 
procedure has to be introduced since its needed to normalize 
the number of photo disintegration events used 
to calculate the final cross-section.

In order to obtain accurate results, 
data quality checks are performed using appropriate criteria 
and several corrections are introduced to improve the quality 
of the reconstruction.

The CLAS proton detection efficiency is evaluated both 
from a GEANT simulation of the detector and
from the data itself in order to verify the 
consistency between the results.
After this step, the fiducial regions of the detector are
defined.

Then, the background contamination is evaluated from the
data and the subtraction procedure is described.

Finally, an
estimation of the overall systematic error
has been given.

\setcounter{chapter} {3}     
\pagestyle{plain}
\chapter{Results}

\section{Introduction}
\indent
\par

This Chapter is devoted to the illustration of the final
results on the deuteron photo-disintegration cross section
obtained from the analysis of the full statistics of the data
collected with the CLAS detector during the g2 data production
period.

The CLAS data represent the first wide ranging survey of the 
deuteron photo-disintegration cross section in the few GeV
region ($0.55-3.0$~GeV) of incident photon energy.
The statistical uncertainties are of the order of $2 \div 5\%$
below 1.5~GeV, and of the order of 10\% for $E_\gamma$ aroud 2.5~GeV.

The CLAS experimental results are discussed and 
compared to the published data.
They give for the first time
the possibility to study the 
photo-disintegration cross section
as a function of the incident
photon energy over the full range of proton scattering angles,
thus expanding the view limited to four 
scattering angles (\mbox{$\theta_p^{\rm{CM}} = 36^\circ$},
$ 52^\circ$,  $ 69^\circ$, and  $89^\circ$) 
investigated in the past years (as discussed in the
first Chapter of the present thesis).

Then, the angular distributions of the cross 
sections, integrated over the CLAS 
azimuthal coordinate, 
are shown for incident photon energies 
from 0.5 to 3.0~GeV in 100~MeV bins.
These data provide for the first time, 
the cross section at very forward 
(\mbox{$\theta_p^{\rm{CM}} = 15^\circ$}) and
backward proton scattering angles (\mbox{$\theta_p^{\rm{CM}} = 155^\circ$})
allowing to test the available theoretical predictions 
in regions where the cross section behavior becomes 
steeper.

In addition, both the new CLAS results and the 
published angular distributions are compared to 
the non perturbative 
calculation of the Quark Gluon String Model (QGSM) 
since, as shown in the first Chapter of this
thesis, this model describes rather well the data.

\section{The Differential Cross Section}
\indent
\par

The photo-disintegration cross section is calculated 
weighting the number of protons (\mbox{$N_{\gamma d \rightarrow p n}$}) 
detected in the CLAS from photo-disintegration  
events in each bin of proton scattering angle and
incident photon energy, by the 
effective number of deuteron targets,
the absolute number of incident photons 
$N_\gamma(1-F)$, where $F$ is the photon beam attenuation
through the experimental target, 
the efficiency-acceptance 
function \mbox{$\eta_{\rm{GSIM}}$} (Sec.~\ref{sec:eff}) 
and taking into account the background contamination $k$ 
(as defined in Eq.~\ref{eq:bkg-data}), {\em i.e.}:
\begin{equation}
\frac{d\sigma}{d\Omega}(E_\gamma,\theta_p^{\rm{CM}})= 
\frac {A}{\rho x N_A}
\frac{1}{(1-F)\,N_\gamma \, \Delta \Omega }
\frac{N_{\gamma d \rightarrow pn}}{\eta_{\rm{GSIM}}} 
\left( 1 - k(E_\gamma,\theta_p^{\rm{CM}}) \right) \ ,
\label{eq:cross}
\end{equation}
where $A$ is the target molecular weight, $N_A$ is the Avogadro's number,
and $\rho$ and $x$ (as defined in Sec.~\ref{sec:vertex}) 
the target density and effective length, respectively.

The CLAS cross section data are fully reported 
in Sec.~\ref{sec:tabella} (at the end of the present Chapter) 
tabulated as a function of the incident 
photon energy (Tab.~\ref{tab:data}). 
For each energy bin,
the values of the differential cross section as $d\sigma/d\Omega$,
obtained summing the individual contributions of the six CLAS sectors, 
are reported as a function of the 
proton scattering angle in the CM system,
together with their statistical uncertainties.
The tables report also the values of $d\sigma/dt$ 
as a function of the proton scattering angle in the CM system,
with their associated statistical uncertainties.


\subsection{Uniformity of the Six CLAS Sectors}
\indent
\par

As stated in Chap. 2, 
the CLAS is composed of six independent 
sectors that can be considered as six similar
spectrometers. 
This allows for the investigation
of the uniformity of the 
final cross section results since
the measurements are performed 
independently in each spectrometer (sector),
and provides an estimate of the systematic uncertainty
due to the CLAS detector efficiency.

\begin{figure}[htbp]
\begin{center}
 \leavevmode
\vspace*{-1cm} 
\epsfig{file=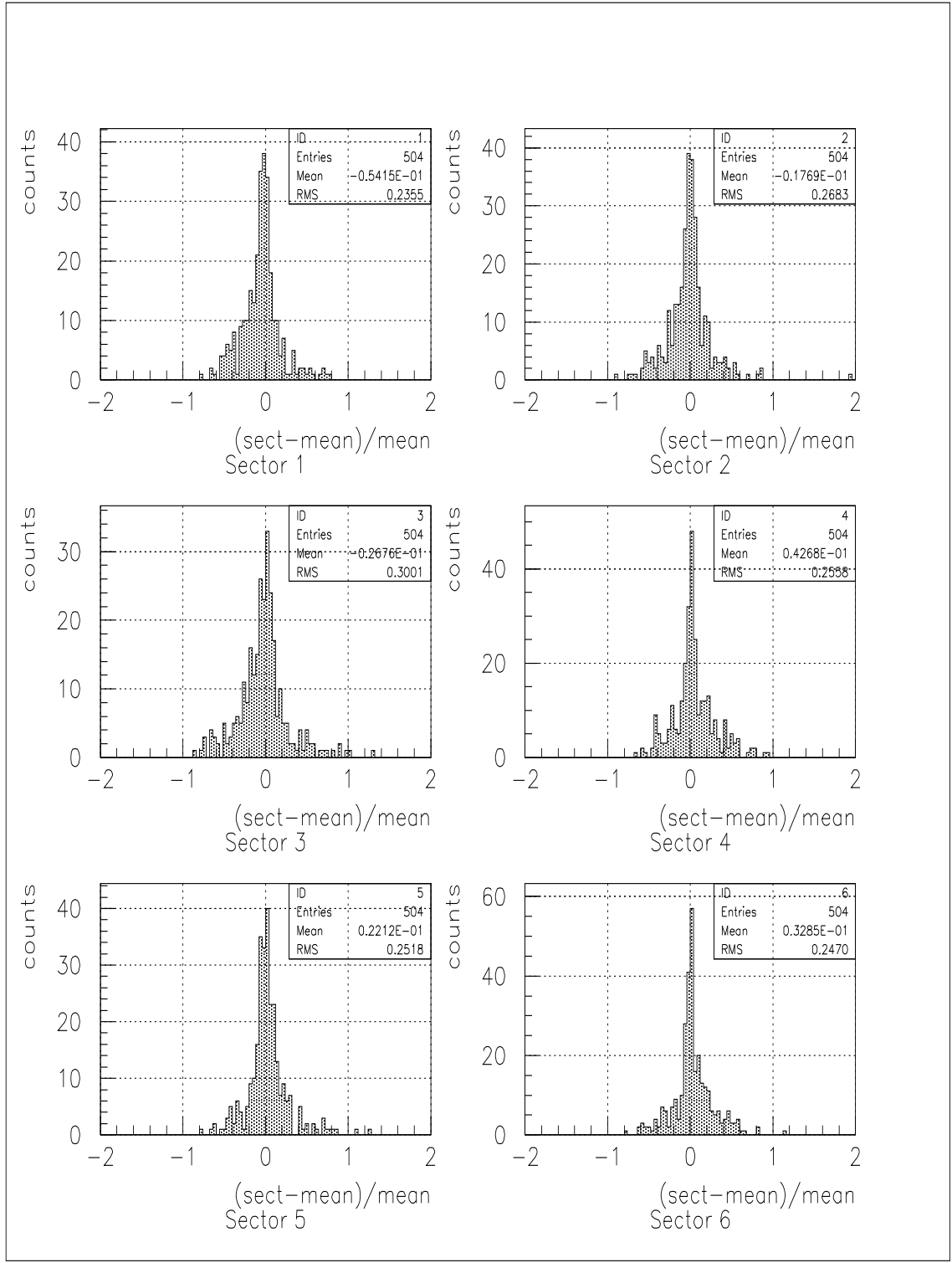,width=15cm,height=20cm}
 \caption{\small Distributions of the differences between the value of the 
cross section obtained in the individual sector and
its average over the six sectors, divided by the average itself, 
as defined in  Eq.~\ref{eq:syst} for each bin of proton scattering angle and
incident photon energy.
The result is shown separately for the six CLAS sectors 
(numbered from left to right and top to bottom).
In each panel the statistical parameters of the distributions are shown 
and it can be seen that a systematic shift 
of the order of $2\div5$\% is present in all sectors.}
\label{fig:xsect-st}
\end{center}
\end{figure}

The uniformity of the results 
has been evaluated considering the 
differences between the values of the 
cross section obtained 
in each bin of incident photon energy 
and proton scattering angle
for each sector $S$ (with $S=1\div6$)
and the related average over the six sectors, divided by the average itself, or:
\begin{equation}
V= \frac{\left(\frac{d\sigma}{d\Omega} \right)_S - \left < \frac{d\sigma}{d\Omega} \right>}
{ \left < \frac{d\sigma}{d\Omega} \right> }
\label{eq:syst}
\end{equation}\\
The resulting distributions for the values of $V$ 
are shown in Fig.~\ref{fig:xsect-st} separately for the six CLAS sectors
(numbered from left to right and top to bottom).
In each panel the statistical parameters of the distributions are shown 
and it can be seen that a systematic shift 
of the order of $2\div5$\% is present in all sectors.

\newpage
\subsection{The Energy Dependence of the Cross Section}
\indent
\par
The new CLAS data give for the first time
the opportunity to study the energy 
dependence of the deuteron photo disintegration 
differential cross section over the wide range
of proton scattering angles covered 
with the CLAS detector 
and with a very good statistical accuracy.

\begin{figure}[htbp]
\begin{center}
 \leavevmode
\epsfig{file=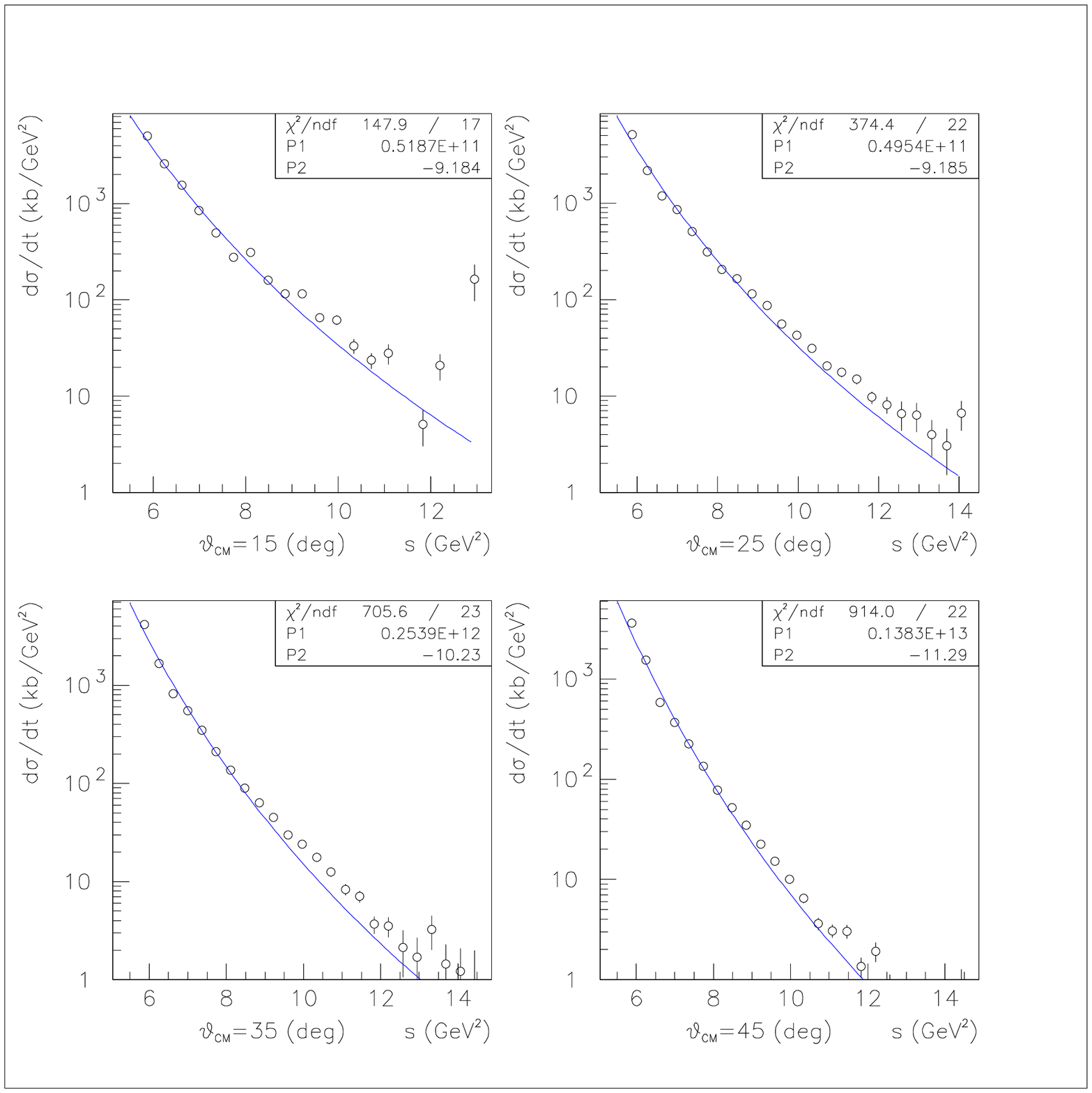,width=15cm,height=9cm}
\epsfig{file=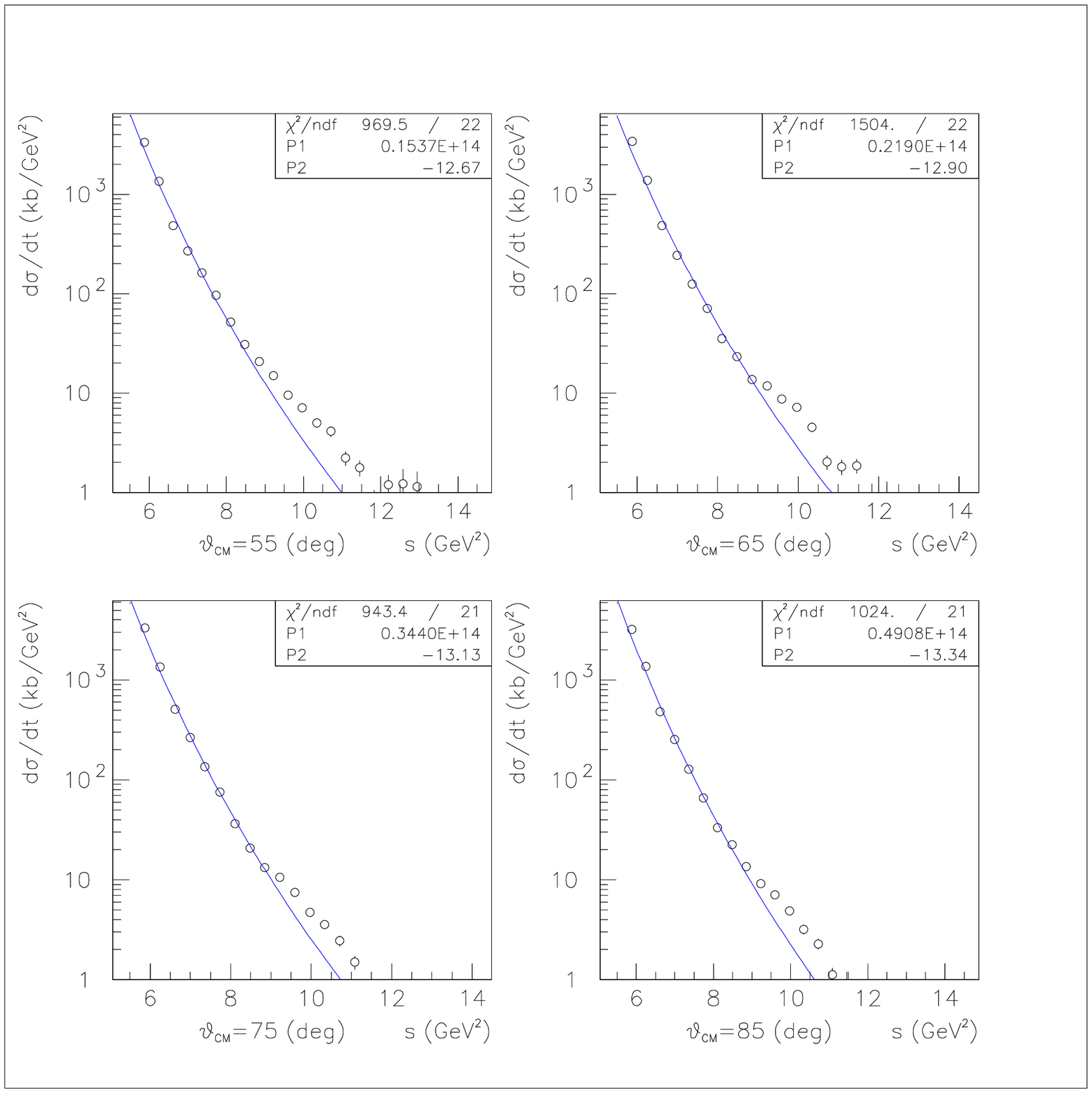,width=15cm,height=9cm}
\caption{\small 
The CLAS results on the differential cross section 
$\frac{d\sigma}{dt}$
as a function of the
total energy $s$ (where $t$ and $s$ are 
the momentum transfer to the proton and the total energy 
in the Mandelstam notation)
for proton scattering angles in the range
\mbox{$15^\circ \leq \theta_p^{\rm{CM}} \leq 85^\circ$}.
As can be seen,
the cross section decrease as a function of $s$ is
extremely steep and this behavior is enhanced as the
scattering angle approaches $90^\circ$.
The blue/solid line superimposed to the
experimental results represents 
a power law fitting function proportional to 
$s^n$.}
\label{fig:fit1}
\end{center}
\end{figure}

\begin{figure}[htbp]
\begin{center}
 \leavevmode
\epsfig{file=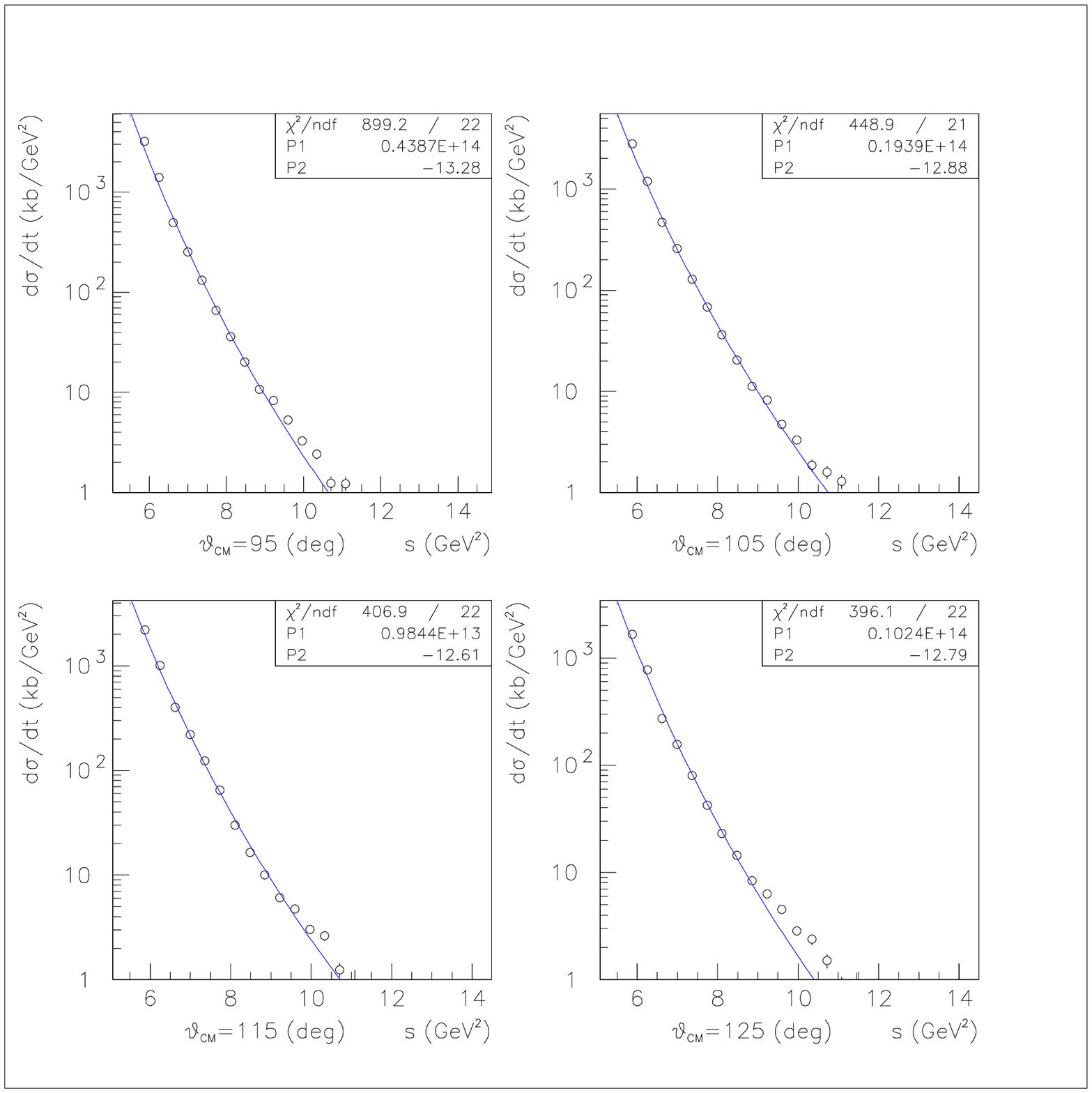,width=15cm,height=9cm}
\epsfig{file=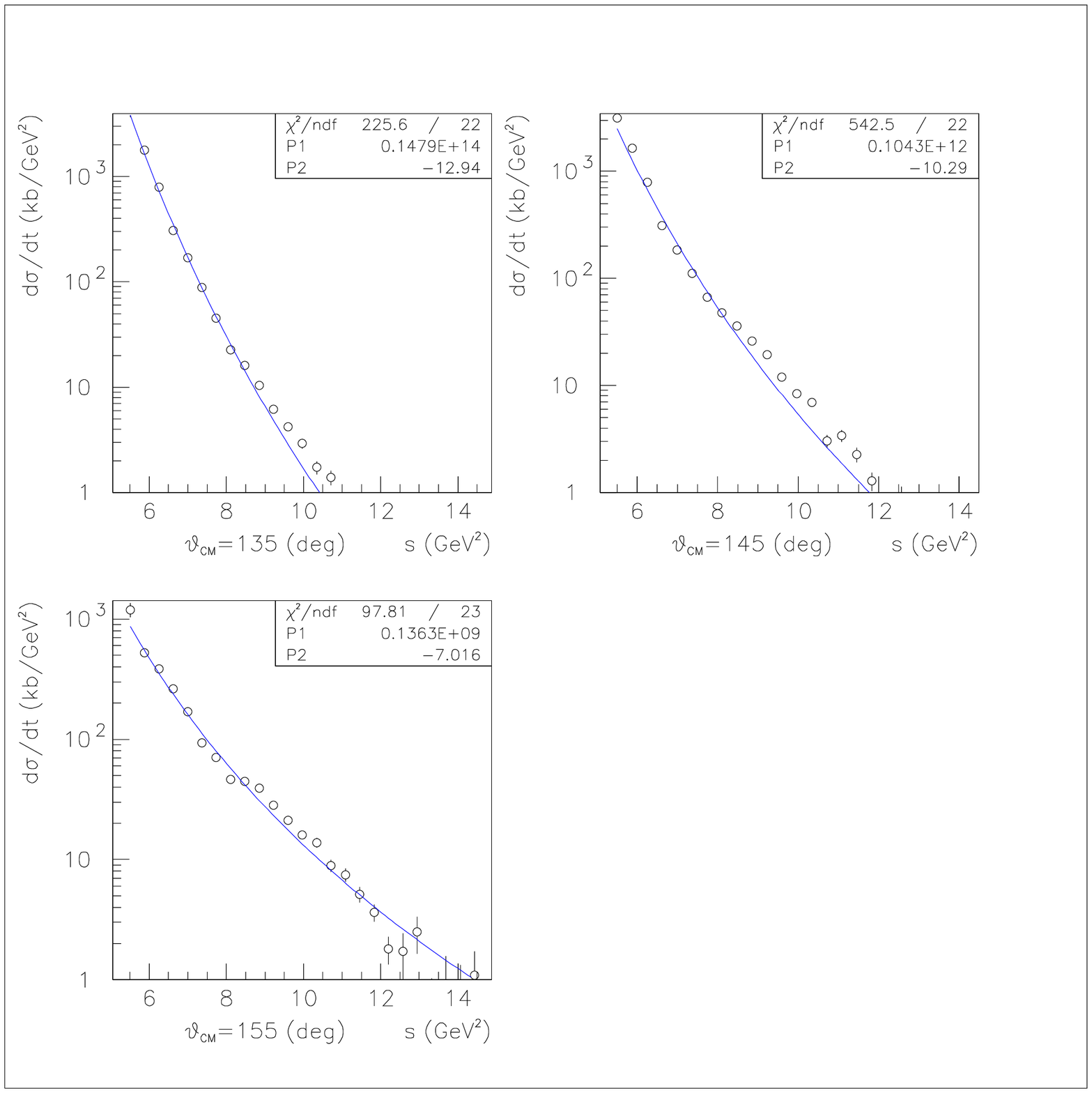,width=15cm,height=9cm}
\caption{\small 
The CLAS results on the differential cross section 
$\frac{d\sigma}{dt}$
as a function of the
total energy $s$ (where $t$ and $s$ are 
the momentum transfer to the proton and the total energy 
in the Mandelstam notation)
for proton scattering angles in the range
\mbox{$95^\circ \leq \theta_p^{\rm{CM}} \leq 155^\circ$}.
It can be seen that the cross section
tendency to a fast decrease as a function of $s$, slows down 
going toward backward scattering angles.
The blue/solid line superimposed to the
experimental results represents 
a power law fitting function proportional to 
$s^n$.
}
\label{fig:fit2}
\end{center}
\end{figure}

The results on the differential cross section 
$\frac{d\sigma}{dt}$
are presented in Fig.~\ref{fig:fit1} and Fig.~\ref{fig:fit2} 
(solid/red dots) as a function of the
total energy $s$ 
(here $t$, $u$, and $s$ are the momentum
transfer to the proton, to the neutron and
the total energy in the Mandelstam notation)
for proton scattering angles in the range
\mbox{$15^\circ \leq \theta_p^{\rm{CM}} \leq 155^\circ$}.
As can be seen from Fig.~\ref{fig:fit1} and Fig.~\ref{fig:fit2} 
the cross section decrease as a function of $s$ is
extremely steep and this behavior is enhanced for
scattering angles around $90^\circ$.
The blue/solid line superimposed to the
experimental results represents 
a power law fitting function proportional to 
$s^n$.

\begin{figure}[htbp]
\begin{center}
 \leavevmode
\epsfig{file=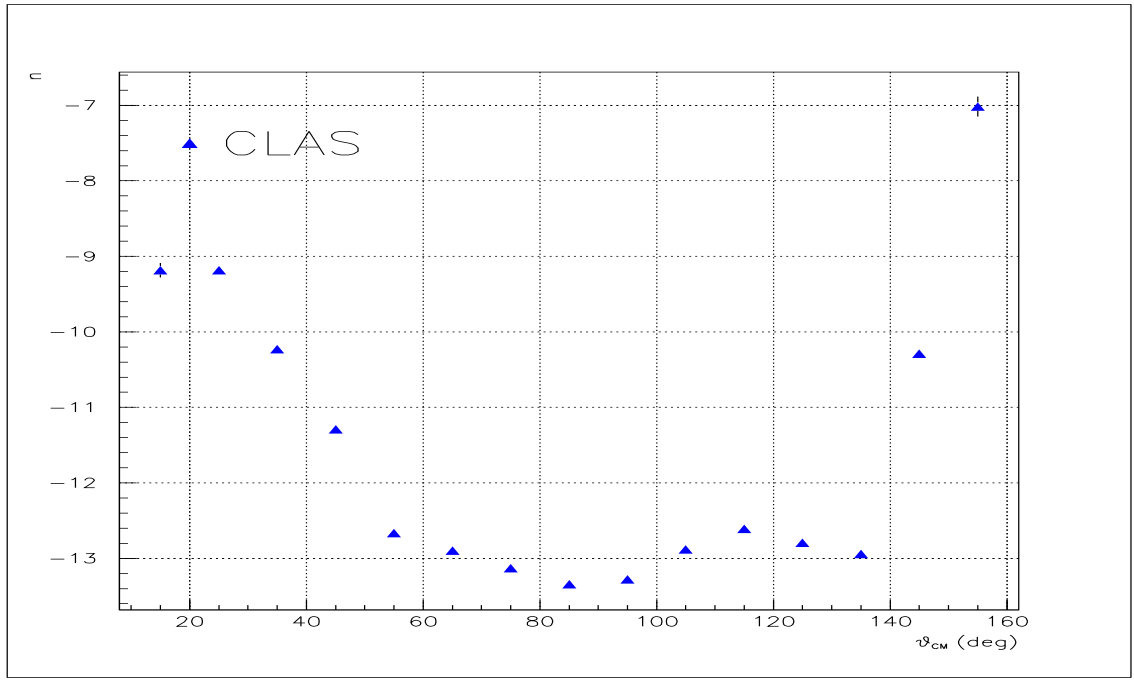,width=15cm,height=9cm}
\caption{\small  Exponents $n$ 
obtained by the angular survey on the cross section.
The data behavior  is of the type \mbox{$\frac{d\sigma}{dt} \simeq (\rm{const})s^n$}
as a function of the total energy $s$.
It can be seen that the cross section decrease
as a function of $s$ follows different exponents $n$ 
for different scattering angles.
At very forward and backward angles (corresponding to the dominance of the $t$ or 
$u$ channel, and where $u$ is the momentum transfer to the neutron) 
the exponent is of the order of $-9$ and $-7$ respectively. 
At more intermediate angles (around $90^\circ$),
where both the $t$-dominance and $u$-dominance are suppressed,
$n$ drops around $-13$.
A clear asymmetry exist between forward ($n=-9$) 
and backward angles ($n=-7$). The errors on the data points
are those given by MINUIT on the fitting parameter $n$ .
}
\label{fig:enne}
\end{center}
\end{figure}

The resulting exponents $n$ obtained 
from the fits of the cross section 
for the various angles
are plotted in Fig.~\ref{fig:enne}.
It can be seen that the cross section decrease
as a function of $s$ occurs according to different exponents
$n$ for the different scattering angles.
At very forward and backward angles,
corresponding to the dominance of the $t$ or 
$u$ channel, the exponent is of the order of $-9$ and $-7$ respectively.
At more intermediate angles (around $90^\circ$),
where both the $t$-dominance and $u$-dominance are suppressed,
$n$ drops around $-13$.
Moreover, one can also argue that
an asymmetry exist between forward ($n=-9$) and
backward angles ($n=-7$) and a local maximum can be identified 
for \mbox{$\theta_p^{\rm{CM}}=155^\circ$}.

\newpage
\subsection{Angular Distributions}
\indent
\par

\begin{figure}[htbp]
\begin{center}
 \leavevmode
\vspace*{-1.0cm} 
 \epsfig{file=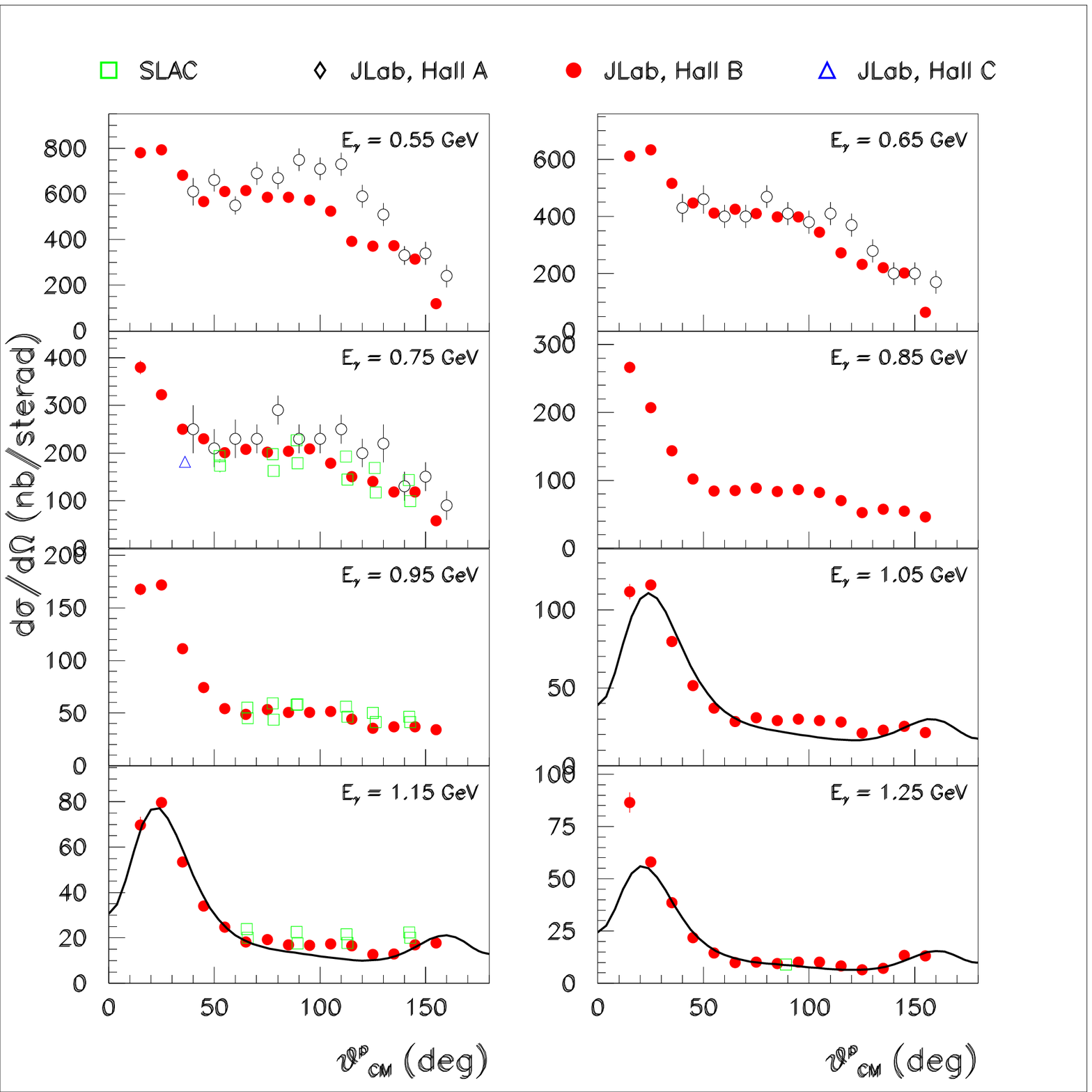,width=15cm,height=20cm}
 \caption{\small  Angular distributions of the 
deuteron photo-disintegration cross section measured by the 
CLAS (solid/red dots) in the incident photon energy range \mbox{$0.55-1.25$~GeV}.
It can be seen that the CLAS data cover a broad range in the proton scattering angle:
\mbox{$15^\circ \leq \theta_p^{\rm{CM}} \leq 155^\circ$}.
Results from Mainz~\cite{CRAW} (open dots), SLAC~\cite{FREE,BELZ} (open squares), 
JLab Hall A~\cite{SHU2} (open diamonds) and Hall C~\cite{BOCH,SHU1} (open triangles) are included 
and are generally in agreement with the CLAS data. 
The angular dependence of the photo-disintegration cross section shows a pronounced 
asymmetry between the forward and backward proton scattering angles.
For incident photon energies higher than 1~GeV,
the solid line represents the non perturbative calculation of the QGSM~\cite{gris} 
which well reproduces the data. In particular, the QGSM calculation
accounts for the forward to backward asymmetry shown by the data.
}
\label{fig:dist1}
\end{center}
\end{figure}

\begin{figure}[htbp]
\begin{center}
 \leavevmode
 \epsfig{file=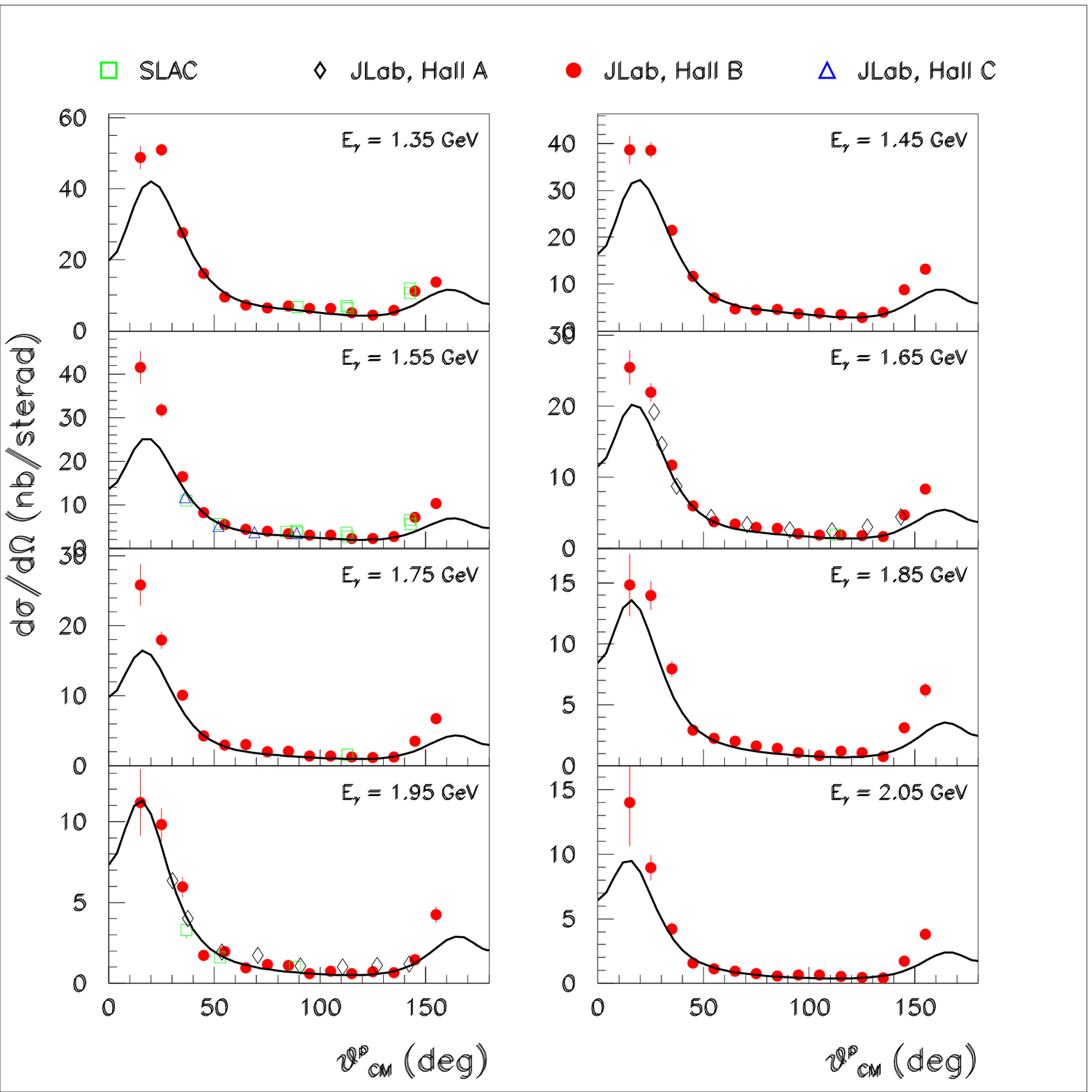,width=15cm,height=20cm}
 \caption{\small 
 Angular distributions of the deuteron photo-disintegration cross section
measured by the CLAS (solid/red dots) in the incident photon energy range \mbox{$1.35-2.05$~GeV}.
Results from already published experiments such as
SLAC~\cite{FREE,BELZ} (open squares), JLab Hall A~\cite{SHU2} (open diamonds) and Hall C~\cite{BOCH,SHU1} 
(open triangles) are included, where available.
The cross section asymmetry between the forward and backward angles 
persists in this photon energy range. The QGSM~\cite{gris} calculation 
(shown as a solid line) well reproduces the data behavior also in this energy range.
}
 \label{fig:dist2}
\end{center}
\end{figure}

\begin{figure}[htbp]
\begin{center}
 \leavevmode
 \epsfig{file=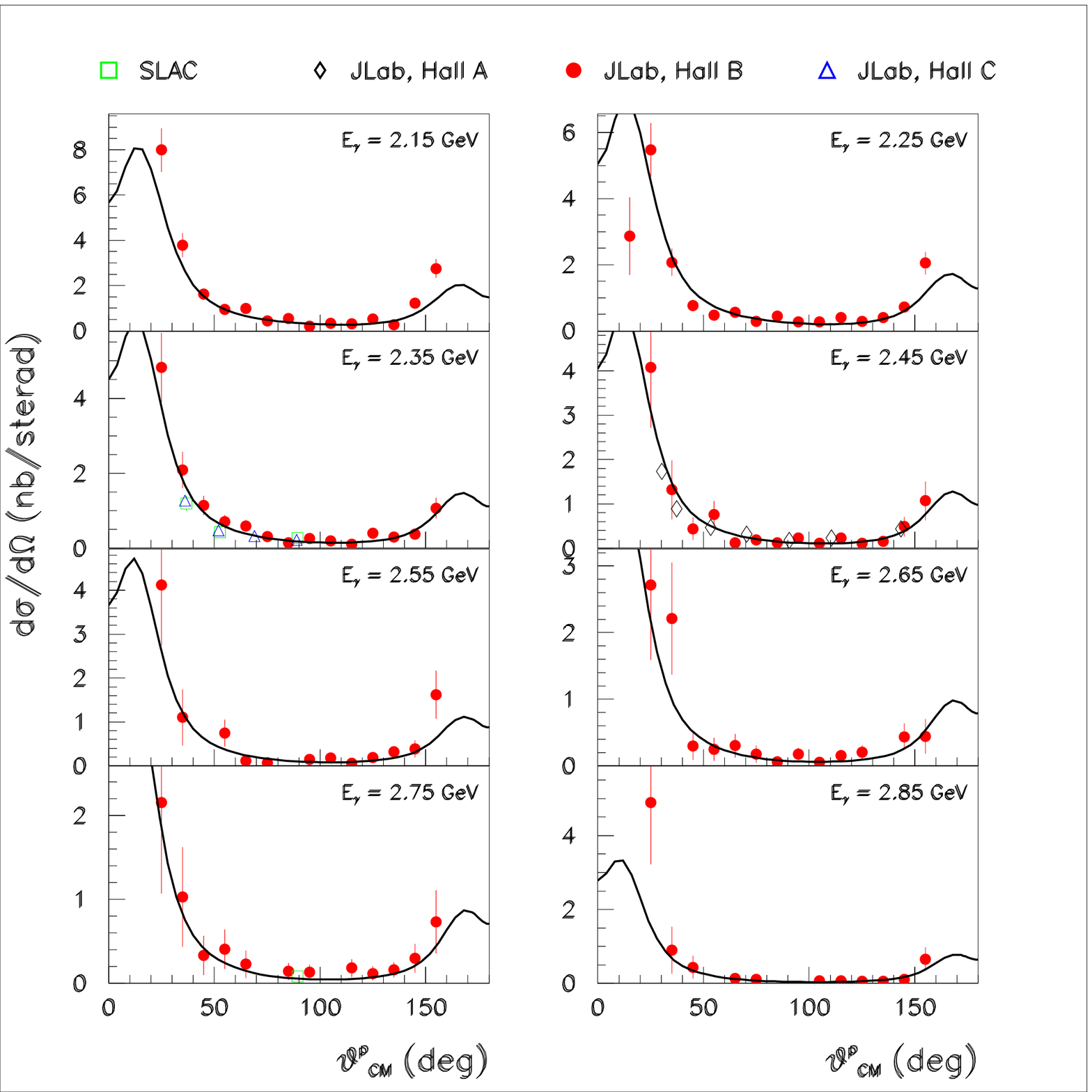,width=15cm,height=20cm}
 \caption{\small Angular distributions of the 
deuteron photo-disintegration cross section
measured by the CLAS (solid/red dots)
in the incident photon energy range \mbox{$2.15-2.85$~GeV}.
Results from SLAC~\cite{FREE,BELZ} (open squares), 
JLab Hall A~\cite{SHU2} (open diamonds) and Hall C~\cite{BOCH,SHU1} (open triangles) 
are included, where available.
The angular distributions are well reproduced by the QGSM~\cite{gris} 
calculation (shown as a solid line) which accounts for the data
asymmetry between forward and backward scattering directions.
}
\label{fig:dist3}
\end{center}
\end{figure}

The angular distributions derived using 
Eq.~\ref{eq:cross} 
are shown in Figs.~\ref{fig:dist1},  
Fig.~\ref{fig:dist2}, and Fig.~\ref{fig:dist3}
for incident photon energy bins of 100~MeV, 
in the range from 0.55 up to 2.9~GeV.
As seen from the figures,
the data cover a broad range in the proton scattering angle:
\mbox{$15^\circ \leq \theta_p^{\rm{CM}} \leq 155^\circ$}
allowing for the first time to investigate
the behavior of the cross section at 
very forward and backward scattering angles.

Results from published data are also included in the plots.
More specifically, results in the highest energy bins obtained by the 
Mainz/DAPHNE~\cite{CRAW} experiment overlap the CLAS data in the photon energy 
region \mbox{$0.55 \leq E_\gamma \leq 0.75$}~GeV.
At higher photon energies, results from the SLAC~\cite{FREE,BELZ} 
experiments are shown together with the recent data 
from the JLab Hall C~\cite{BOCH,SHU1} and Hall A~\cite{SHU2} collaborations.
As can be seen from Fig.~\ref{fig:dist1},
in the energy region below 1~GeV, the angular 
distributions still exhibit a trace of the excitation 
of baryon resonances in the second resonance region, 
which comprises $P_{11}(1440)$, $D_{13}(1520)$, and 
$S_{11}(1535)$. At \mbox{$E_\gamma = 0.55$~GeV}, 
the Mainz data are slightly higher 
then the CLAS data, mostly 
at intermediate scattering angles. 
This discrepancy is greatly reduced at forward and backward 
angles and also becomes less evident at higher photon energies
where the two datasets are well compatible 
to each other within the total statistical uncertainties.

At \mbox{$E_\gamma = 0.65$~GeV},
both the angular dependence and the absolute value
of the differential cross section of the CLAS data are
in good agreement with the Mainz result.  
Starting from \mbox{$E_\gamma = 0.75$~GeV}
the comparison 
can be extended also to the SLAC and JLab Hall C data.
It turns our that, in general, the CLAS results well 
agree with data from these experiments.

For incident photon energies higher than 
1~GeV the angular dependence of
the photo-disintegration cross section 
changes, and a more pronounced asymmetry emerges 
between the forward and backward
angles, as can be seen from 
the bottom plots of Fig.~\ref{fig:dist1}.

The angular distributions obtained in the energy range
$1.35 \leq E_\gamma \leq 2.85$~GeV are shown in Figs~\ref{fig:dist1} and \ref{fig:dist2}.
At \mbox{$E_\gamma =1.6$~GeV} and \mbox{$E_\gamma =1.9$~GeV} 
the CLAS results are compared with the angular 
distributions measured by the JLab Hall A 
collaboration which cover a few  scattering angles
in the range  \mbox{$26^\circ \leq \theta_p^{\rm{CM}} \leq 143^\circ$}.
The CLAS results confirms the Hall A data
and extends that result at the very 
forward and backward angular regions
where the cross section increases.
In general, the clear angular 
asymmetry shown by the CLAS data 
persists in the whole range of explored 
photon energies, as is confirmed by 
Fig.~\ref{fig:dist3} which shows the results 
up to \mbox{$E_\gamma =2.85$~GeV}.

At photon energies higher that 1~GeV, 
where the resonance contributions starts to fade,
the predictions of the QGSM~\cite{gris} are included in the 
plots. 
As shown in Figs~\ref{fig:dist1},~\ref{fig:dist2}, and 
Fig.~\ref{fig:dist3}, the angular distributions are 
well reproduced by the QGSM calculation 
(shown as a solid line).
This model accounts very well for the 
persistent forward to backward asymmetry seen 
in the data by invoking the interference of the iso-vector 
and iso-scalar components of the photon,
which is constructive at forward
angles and destructive in the backward direction.


\newpage
\subsection{The Cross Section Scaling}
\indent
\par

As fully discussed in the opening Chapter
of this thesis, the dependence of the differential 
cross section as a function of the incident 
photon energy has been often investigated 
in the past years to evidence the scaling behavior.

Such an effect is believed to show up
in the cross section data  
when the proton transverse momentum 
\begin{equation}
P^2_T=\frac{1}{2} E_\gamma M_D \sin^2{(\theta_p^{\rm{CM}})} \ ,
\label{eq:prot-mom}
\end{equation}
being $M_D$ the deuteron mass, is of the order of 1~(GeV/c)$^2$.
In this case, the pQCD assumption that soft and hard contributions
factorize may be realized.

In the case of deuteron photo-disintegration process, 
the cross section is expected to
scale with a unique simple power law of the square 
of the total energy $s$
\begin{equation}
\frac{d\sigma}{dt} = f\left(  \theta_p^{\rm{CM}}    \right)\,s^{-11} 
\label{eq:scal}
\end{equation}
according to the prediction
of the Constituent Counting Rules (CCR).
As can be seen from Eq.~\ref{eq:scal}
the scaling law does not depend, apart from a constant factor, 
on the proton scattering angle.

As seen from the results shown
in the previous paragraphs, 
the CCR prediction seems not in agreement,
at the photon energies of interest here, 
with the obtained values for the exponent
of the scaling law $n$ as shown in Fig.~\ref{fig:enne}.

This situation persists even at intermediate 
scattering angles (around $90^\circ$)
where momentum transfer to the outgoing
proton should be the highest possible
and the perturbative regime
is believed to be better established.

On the contrary, the present results 
indicate that pQCD is 
not plainly applicable to
the description of the photo-disintegration 
process in the region
of incident photon energy
explored here.



\begin{figure}[htbp]
\begin{center}
 \leavevmode
 \epsfig{file=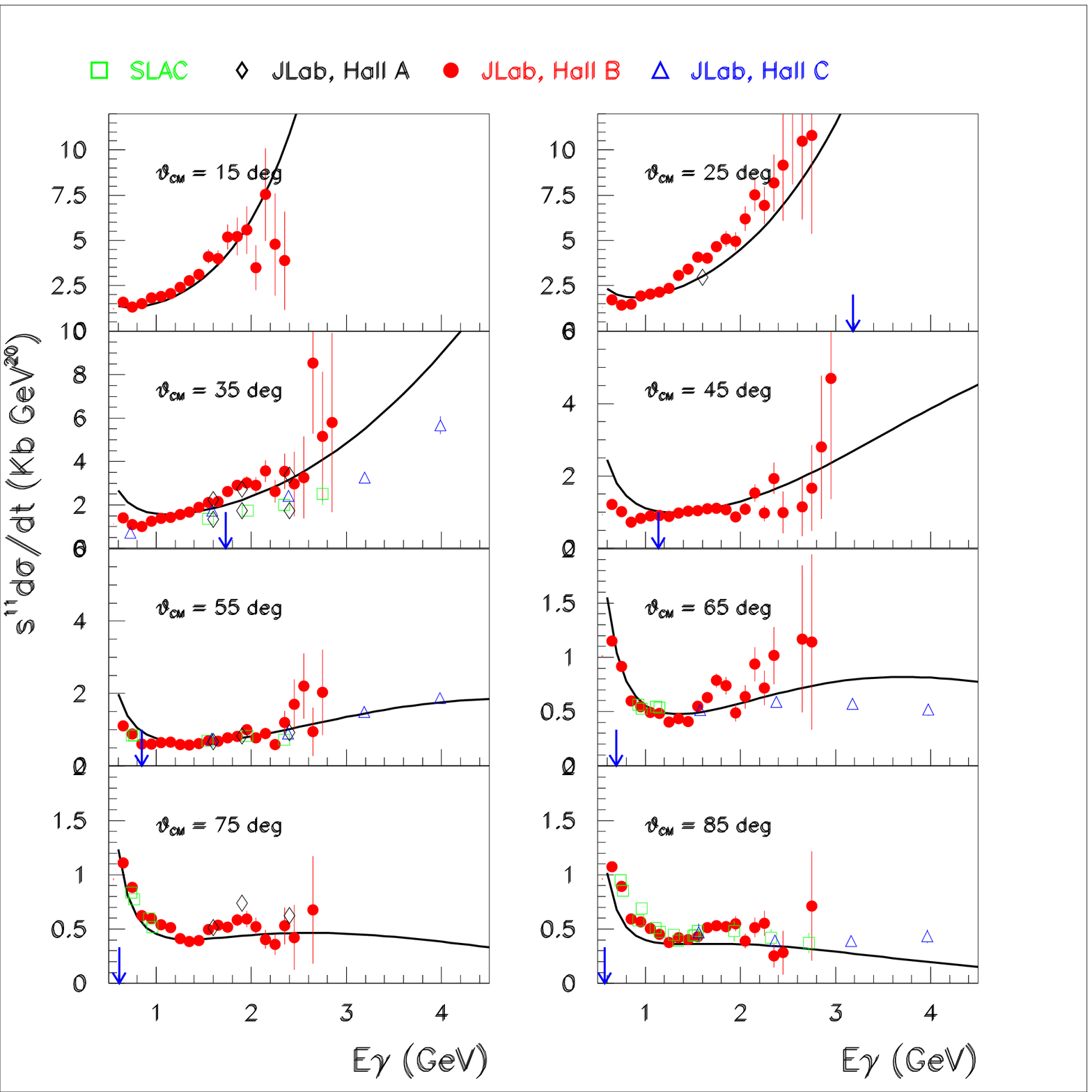,width=14cm,height=16cm}
 \caption{\small Deuteron photo-disintegration
differential cross section expressed as ${d\sigma}/{dt}$
multiplied by $s^{11}$ (being $t$ the momentum
transfer and $s$ the total energy in the Mandelstam notation)
in the angular range from \mbox{$\theta_p^{\rm{CM}} = 15^\circ$}
up to \mbox{$ \theta_p^{\rm{CM}} = 85^\circ$}.
The CLAS results are represented by the solid dots.
Also shown are previously available data:
SLAC~\cite{FREE,BELZ} data are represented by open squares, 
JLab Hall C~\cite{BOCH,SHU1} data by open triangles, and
JLab Hall A~\cite{SHU2} data by open diamonds.
The arrows in each panel, indicate the photon energies
where the onset of the $s^{-11}$ scaling is foreseen.
This limit correspond to proton transverse momenta of 1~GeV/c.
For the $\gamma d \rightarrow pn$ reaction the proton 
transverse momentum can be expressed
\mbox{$P^2_T=\frac{1}{2} E_\gamma M_D \sin^2{(\theta_p^{\rm{CM}})}$}.
According to the CCR,
a constant behavior of the cross section 
as a function of the incident photon energy is 
expected when the limit for $P_T$ is reached.
The solid line represents 
the QGSM~\cite{gris} non perturbative calculation 
for the cross section.
This model, as discussed in the first Chapter of this thesis,
gives a non perturbative description of the photo-disintegration
cross section assuming that the scattering amplitude at very high
energy is dominated by the exchange of three valence 
quarks in the $t$-channel.
}
\label{fig:dsdt1}
\end{center}
\end{figure}

\begin{figure}[htbp]
\begin{center}
 \leavevmode
 \epsfig{file=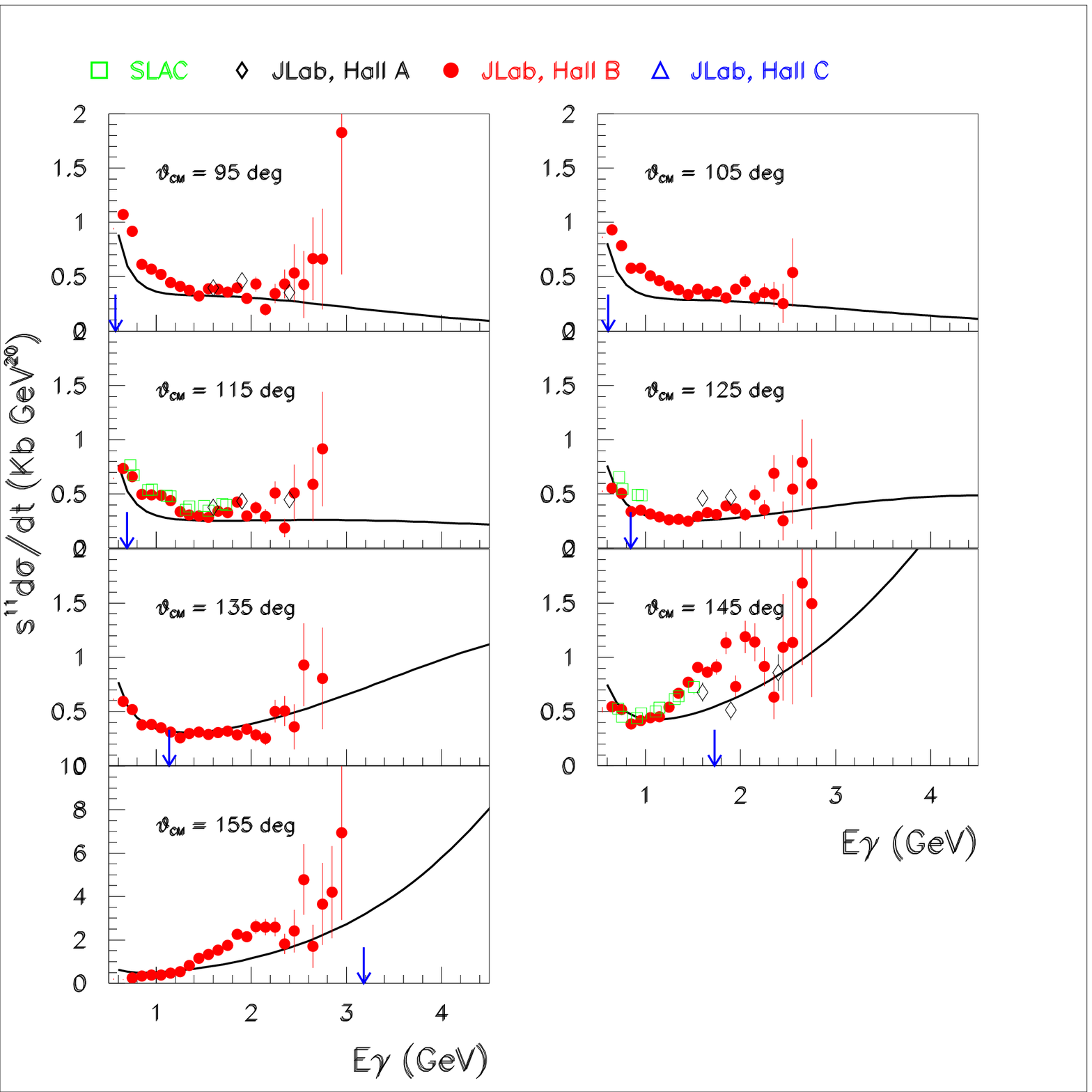,width=14cm,height=16cm}
 \caption{\small 
Deuteron photo-disintegration
differential cross section expressed as ${d\sigma}/{dt}$
multiplied by $s^{11}$ (being $t$ the momentum
transfer and $s$ the total energy in the Mandelstam notation)
in the angular range from \mbox{$\theta_p^{\rm{CM}} = 95^\circ$}
up to \mbox{$ \theta_p^{\rm{CM}} = 155^\circ$}.
The CLAS results are represented by the solid dots.
Also shown are previously available data:
SLAC~\cite{FREE,BELZ} data are represented by open squares, 
JLab Hall C~\cite{BOCH,SHU1} data by open triangles, and
JLab Hall A~\cite{SHU2} data by open diamonds.
The arrows in each panel, indicate the photon energies
where the onset of the $s^{-11}$ scaling is foreseen.
This limit correspond to proton transverse momenta of 1~GeV/c.
For the \mbox{$\gamma d \rightarrow pn$} reaction the proton 
transverse momentum can be expressed by
\mbox{$P^2_T=\frac{1}{2} E_\gamma M_D \sin^2{(\theta_p^{\rm{CM}})}$}.
According to the CCR,
a constant behavior of the cross section 
as a function of the incident photon energy is 
expected when the limit for $P_T$ is reached.
It can be seen that the asymptotic behavior for the cross section
is well established from \mbox{$\theta_p^{\rm{CM}} = 75^\circ$} 
up to \mbox{$\theta_p^{\rm{CM}} = 125^\circ$} 
while at backward angles, 
the cross section dependence 
on the photon energy follows 
powers of $s$ different from $-11$
as shown in Fig.~\ref{fig:enne}.
The solid line represents 
the QGSM~\cite{gris}, a non perturbative calculation 
for the cross section.
}
\label{fig:dsdt2}
\end{center}
\end{figure}

In order to better underline 
this result, the CLAS differential cross section data, 
$d\sigma/dt$, multiplied by the $s^{11}$ scaling
factor, are shown in Figs~\ref{fig:dsdt1} and~\ref{fig:dsdt2}
as a function of the 
incident photon energy
in the range $0.55-3.0$~GeV.

The $s^{11}$ factor has been used 
to evidence possible deviations
of the experimental data from 
the constant behavior
expected on the basis of plain pQCD.
In addition, it should be noticed
that the CLAS data cover 
the incident photon energy 
range in fine steps of 100~MeV 
thus allowing for an unprecedented precision
in sampling the cross section 
behavior in this energy region.

The panels of Fig.~\ref{fig:dsdt1} and Fig.~\ref{fig:dsdt2}
show, from left to right and from top to bottom 
the calculated values for 
$s^{11}\frac{d\sigma}{dt}$ in the angular range from 
$\theta_p^{\rm{CM}} = 15^\circ$
up to $ \theta_p^{\rm{CM}} = 155^\circ$.
The CLAS data are represented by the solid/red dots,
while the symbols relative to the data available in the literature  
are indicated in the legend found on the top of Fig.~\ref{fig:dsdt1}.

The arrows in each panel indicate the photon energies
where $P_T=$~1~GeV/c.
According to the CCR, a constant behavior of the $s^{11}$
scaled cross section as a function of 
the incident photon energy is expected 
beyond that limit.
As can be seen from Fig.~\ref{fig:dsdt1} and  Fig.~\ref{fig:dsdt2}
this situation is not perfectly realized by the 
experimental data.
It is clear that, for forward proton angles, 
\mbox{$\theta_p^{\rm{CM}} = 15^\circ\div 45^\circ$},
the data scale with a power
of the total energy $s$ lower then $-11$
according to the plot shown in Fig.~\ref{fig:enne}. 
The data show a trend compatible
with the CCR $s^{-11}$ scaling roughly in the range 
of  scattering angles \mbox{$\theta_p^{\rm{CM}} = 55^\circ \div 125^\circ$}. 
Again, at backward angles, 
the cross section dependence 
on the photon energy follows 
powers of $s$ quite lower than $-11$.

The observed deviation from the exact $n=-11$ prediction 
of plain pQCD for $P_T \geq 1$~GeV  
found for the exponent of the scaling power $n$
may indicate that pQCD is not fully applicable
in this energy region.

On the contrary, a non perturbative approach may be
more effective  in reproducing the behavior of the 
deuteron photo-disintegration cross section.
This done by the non perturbative QGSM 
calculation for the cross section, which is 
shown as a solid line in Figs.~\ref{fig:dsdt1} and~\ref{fig:dsdt2}.
Again, it is seen that a good agreement with the data 
is found for the QGSM predictions at all proton scattering angles.
\clearpage
\section{Summary and Conclusions}
\indent
\par

The aim of the present 
research project is the complete measurement of the deuteron 
photo-disintegration differential cross section 
in the few GeV region of incident photon energy.

This reaction is well suited for studying 
the interplay of nuclear and particle physics
in order to find out how and at what energy the transition 
from the hadronic picture of the deuteron, established in the reaction at 
low energies (below 1 GeV), to the quark-gluon picture, takes place.

With respect to this goal, 
the physical motivation for such an investigation 
has been discussed, 
showing that the already published 
data do not fill the gap of knowledge
on the physical mechanisms 
underlying the photo-disintegration process.

Different theoretical models have been
developed to describe the 
deuteron photo-disintegration 
cross section.
In particular, the QCD inspired models
have been reviewed since there is common 
agreement on the fact
that quark and gluons are more appropriate 
degrees of freedom 
for the description of the process  
in this energy region.
The theoretical questioning 
is not on the use of QCD but, if perturbative QCD
can be applied at this incident photon energies,
or, on the contrary, non perturbative QCD must 
be necessarily used.

In order to answer the above question,
the photo-disintegration cross section
needs to be known with great statistical
accuracy and over a broad angular range
so to investigate its behavior in the most
critical kinematic regions.

To this aim, the \mbox{$\gamma d \rightarrow pn$}
cross section has been measured  
with the CLAS large angle spectrometer at the Jefferson Lab Hall B,
over a wide range of proton scattering angles
(\mbox{$15^\circ \leq \theta_p^{\rm{CM}} \leq 155^\circ$}) 
using a photon beam
in the energy range from 0.55 up to 3.0~GeV.

This powerful investigation tool,
combining the unique characteristic of a nearly
$4\pi$ acceptance in
conjunction with a tagged photon beam of high energy and intensity
allowed the collection of a very high statistics and
high quality dataset during the end of year 1999.

The analysis of this large amount of data
constitutes the central part of this work
and has been fully reported in the third Chapter
of this thesis to illustrate the
procedures and techniques used to 
extract the differential cross section  data.

In the last part of the thesis,
the new cross section data
are discussed in order to contribute 
to the answering of the questions posed in the 
opening Chapter.

To this aim, the differential 
cross section is shown as a 
function of the total energy $s$.
The new CLAS data add 
significant information 
to the previous scenario,
allowing the study of  
the cross section 
decrease as $s^n$  
with an very good accuracy
and over a wide angular range.

In addition, the different exponents 
$n$ have been calculated over the complete
range of proton scattering angles
and an asymmetry is found between 
forward \mbox{($n=-9$)} and
backward angles \mbox{($n=-7$)}.

This result is reflected 
from the angular distributions
of the cross section.
The data measured at
very forward and backward angles
show that the photo-disintegration cross section 
exhibits a persistent asymmetry 
over the whole incident photon 
energy range.

For what concerns the 
theoretical interpretations, 
all cross section data (new and published) 
have been compared to 
the predictions of the available 
theoretical models based on QCD 
degrees of freedom.

The conclusion is a clear indication
that pQCD based models do not fully reproduce
the data behavior in this region of incident photon energies.
On the contrary, the non perturbative calculation
provided by the Quark Gluon String Model
has been able to well describe 
the overall data.

This result can be considered a 
further indication that a pQCD interpretation of the
deuteron photo-disintegration process may not be 
adequate in the few GeV region and a complete non 
perturbative theory is needed.

\newpage
\section{Tabulated Data}
\indent
\par
\label{sec:tabella}
\begin{table}[ht]
\begin{center}
\begin{tabular}{|c|c|c|c|c|c|} \hline
$E_\gamma$ &$\theta_p^{CM}$&$d\sigma/d\Omega$ & $\Delta(d\sigma/d\Omega)$ &$d\sigma/dt$ & $\Delta (d\sigma/dt)$\\ \hline
GeV     &    deg        &  nb/str       &    nb/str     &        kb/GeV$^2$    & kb/GeV$^2$ \\ \hline
0.55  &  15  &  780  &  23  &  7979  &  240   \\ 
 0.55  &  25  &  793  &  12  &  7958  &  122   \\ 
 0.55  &  35  &  682  &  8  &  6845  &  78   \\ 
 0.55  &  45  &  566  &  6  &  5679  &  61   \\ 
 0.55  &  55  &  610  &  6  &  6129  &  58   \\ 
 0.55  &  65  &  615  &  5  &  6174  &  52   \\ 
 0.55  &  75  &  585  &  5  &  5871  &  51   \\ 
 0.55  &  85  &  586  &  5  &  5884  &  49   \\ 
 0.55  &  95  &  574  &  5  &  5759  &  49   \\ 
 0.55  &  105  &  525  &  5  &  5271  &  48   \\ 
 0.55  &  115  &  393  &  5  &  3942  &  52   \\ 
 0.55  &  125  &  370  &  6  &  3207  &  46   \\ 
 0.55  &  135  &  372  &  5  &  3735  &  48   \\ 
 0.55  &  145  &  314  &  15  &  3154  &  146   \\ 
 0.55  &  155  &  119  &  16  &  1191  &  160   \\  \hline
 0.65  &  15  &  612  &  16  &  5022  &  131   \\ 
 0.65  &  25  &  633  &  9  &  5110  &  73   \\ 
 0.65  &  35  &  516  &  6  &  4161  &  45   \\ 
 0.65  &  45  &  447  &  5  &  3608  &  37   \\ 
 0.65  &  55  &  412  &  4  &  3322  &  31   \\ 
 0.65  &  65  &  426  &  4  &  3437  &  28   \\ 
 0.65  &  75  &  411  &  3  &  3319  &  27   \\ 
 0.65  &  85  &  399  &  3  &  3217  &  26   \\ 
 0.65  &  95  &  398  &  3  &  3211  &  27   \\ 
 0.65  &  105  &  345  &  3  &  2788  &  24   \\ 
 0.65  &  115  &  272  &  3  &  2198  &  22   \\ 
 0.65  &  125  &  233  &  3  &  1657  &  20   \\ 
 0.65  &  135  &  221  &  3  &  1780  &  23   \\ 
 0.65  &  145  &  203  &  4  &  1635  &  31   \\ 
 0.65  &  155  &  65  &  5  &  525  &  42   \\  \hline

\end{tabular}
\end{center}
\caption{ \small Differential cross section data.}
\label{tab:data}
\end{table}

\newpage
\begin{table}[tbp]
\begin{center}
\begin{tabular}{|c|c|c|c|c|c|} \hline
$E_\gamma$ &$\theta_p^{CM}$&$d\sigma/d\Omega$ & $\Delta(d\sigma/d\Omega)$ &$d\sigma/dt$ & $\Delta (d\sigma/dt)$\\ \hline
GeV     &    deg        &  nb/str       &    nb/str     &        kb/GeV$^2$    & kb/GeV$^2$ \\ \hline

 0.75  &  15  &  379  &  14  &  2586  &  94   \\ 
 0.75  &  25  &  322  &  8  &  2164  &  52   \\ 
 0.75  &  35  &  249  &  4  &  1674  &  28   \\ 
 0.75  &  45  &  230  &  4  &  1543  &  24   \\ 
 0.75  &  55  &  200  &  3  &  1345  &  20   \\ 
 0.75  &  65  &  208  &  3  &  1396  &  18   \\ 
 0.75  &  75  &  201  &  3  &  1352  &  17   \\ 
 0.75  &  85  &  203  &  2  &  1365  &  17   \\ 
 0.75  &  95  &  209  &  3  &  1403  &  17   \\ 
 0.75  &  105  &  178  &  2  &  1198  &  16   \\ 
 0.75  &  115  &  151  &  2  &  1010  &  14   \\ 
 0.75  &  125  &  140  &  4  &  772  &  13   \\ 
 0.75  &  135  &  118  &  2  &  794  &  15   \\ 
 0.75  &  145  &  118  &  3  &  792  &  21   \\ 
 0.75  &  155  &  57  &  5  &  385  &  32   \\  \hline
 0.85  &  15  &  266  &  8  &  1546  &  48   \\ 
 0.85  &  25  &  207  &  4  &  1186  &  25   \\ 
 0.85  &  35  &  144  &  2  &  823  &  13   \\ 
 0.85  &  45  &  102  &  2  &  581  &  9   \\ 
 0.85  &  55  &  84  &  1  &  484  &  8   \\ 
 0.85  &  65  &  85  &  1  &  486  &  6   \\ 
 0.85  &  75  &  89  &  1  &  507  &  6   \\ 
 0.85  &  85  &  84  &  1  &  480  &  6   \\ 
 0.85  &  95  &  87  &  1  &  495  &  6   \\ 
 0.85  &  105  &  82  &  1  &  469  &  6   \\ 
 0.85  &  115  &  70.3  &  1  &  402  &  6   \\ 
 0.85  &  125  &  52  &  1  &  273  &  5   \\ 
 0.85  &  135  &  58  &  1  &  307  &  6   \\ 
 0.85  &  145  &  54  &  1  &  311  &  7   \\ 
 0.85  &  155  &  46  &  3  &  263  &  15   \\  \hline

\end{tabular}
\end{center} 
\end{table}

\newpage
\begin{table}[tbp]
\begin{center}
\begin{tabular}{|c|c|c|c|c|c|} \hline
$E_\gamma$ &$\theta_p^{CM}$&$d\sigma/d\Omega$ & $\Delta(d\sigma/d\Omega)$ &$d\sigma/dt$ & $\Delta (d\sigma/dt)$\\ \hline
GeV     &    deg        &  nb/str       &    nb/str     &        kb/GeV$^2$    & kb/GeV$^2$ \\ \hline

 0.95  &  15  &  168  &  5  &  846  &  27   \\ 
 0.95  &  25  &  172  &  3  &  855  &  16   \\ 
 0.95  &  35  &  111  &  2  &  553  &  8   \\ 
 0.95  &  45  &  74  &  1  &  369  &  6   \\ 
 0.95  &  55  &  54.3  &  0.9  &  270  &  4   \\ 
 0.95  &  65  &  48.8  &  0.7  &  243  &  3   \\ 
 0.95  &  75  &  53.3  &  0.7  &  265  &  3   \\ 
 0.95  &  85  &  50.9  &  0.7  &  253  &  3   \\ 
 0.95  &  95  &  50.8  &  0.7  &  253  &  3   \\ 
 0.95  &  105  &  51.6  &  0.7  &  257  &  3   \\ 
 0.95  &  115  &  44.2  &  0.6  &  220  &  3   \\ 
 0.95  &  125  &  35.7  &  0.8  &  156  &  3   \\ 
 0.95  &  135  &  36.8  &  0.9  &  170  &  3   \\ 
 0.95  &  145  &  37.1  &  0.8  &  185  &  4   \\ 
 0.95  &  155  &  34  &  2  &  170  &  8   \\  \hline
 1.05  &  15  &  112  &  5  &  497  &  22   \\ 
 1.05  &  25  &  116  &  3  &  509  &  13   \\ 
 1.05  &  35  &  80  &  2  &  350  &  7   \\ 
 1.05  &  45  &  51  &  1  &  226  &  5   \\ 
 1.05  &  55  &  36.9  &  0.8  &  162  &  4   \\ 
 1.05  &  65  &  28.4  &  0.6  &  125  &  3   \\ 
 1.05  &  75  &  31  &  0.6  &  136  &  3   \\ 
 1.05  &  85  &  29.1  &  0.6  &  128  &  2   \\ 
 1.05  &  95  &  30  &  0.6  &  132  &  2   \\ 
 1.05  &  105  &  29.1  &  0.6  &  128  &  3   \\ 
 1.05  &  115  &  28.1  &  0.6  &  124  &  2   \\ 
 1.05  &  125  &  21  &  0.7  &  80  &  2   \\ 
 1.05  &  135  &  22.7  &  0.8  &  89  &  2   \\ 
 1.05  &  145  &  25.4  &  0.7  &  111  &  3   \\ 
 1.05  &  155  &  21  &  1  &  94  &  6   \\  \hline

\end{tabular}
\end{center} 
\end{table}

\newpage
\begin{table}[tbp]
\begin{center}
\begin{tabular}{|c|c|c|c|c|c|} \hline
$E_\gamma$ &$\theta_p^{CM}$&$d\sigma/d\Omega$ & $\Delta(d\sigma/d\Omega)$ &$d\sigma/dt$ & $\Delta (d\sigma/dt)$\\ \hline
GeV     &    deg        &  nb/str       &    nb/str     &        kb/GeV$^2$    & kb/GeV$^2$ \\ \hline
 
 1.15  &  15  &  70  &  4  &  277  &  14   \\ 
 1.15  &  25  &  80  &  2  &  313  &  9   \\ 
 1.15  &  35  &  53  &  1  &  210  &  5   \\ 
 1.15  &  45  &  34.1  &  0.8  &  134  &  3   \\ 
 1.15  &  55  &  24.7  &  0.6  &  97  &  3   \\ 
 1.15  &  65  &  18.1  &  0.5  &  71  &  2   \\ 
 1.15  &  75  &  19.2  &  0.4  &  75  &  2   \\ 
 1.15  &  85  &  16.9  &  0.4  &  66  &  2   \\ 
 1.15  &  95  &  16.7  &  0.4  &  66  &  2   \\ 
 1.15  &  105  &  17.4  &  0.4  &  68  &  2   \\ 
 1.15  &  115  &  16.5  &  0.4  &  65  &  2   \\ 
 1.15  &  125  &  12.8  &  0.5  &  42  &  1   \\ 
 1.15  &  135  &  12.8  &  0.5  &  45  &  2   \\ 
 1.15  &  145  &  17  &  0.6  &  67  &  2   \\ 
 1.15  &  155  &  18  &  1  &  70  &  4   \\  \hline
 1.25  &  15  &  86  &  5  &  311  &  17   \\ 
 1.25  &  25  &  58  &  2  &  206  &  7   \\ 
 1.25  &  35  &  39  &  1  &  137  &  4   \\ 
 1.25  &  45  &  21.8  &  0.7  &  78  &  2   \\ 
 1.25  &  55  &  14.5  &  0.5  &  52  &  2   \\ 
 1.25  &  65  &  10  &  0.4  &  35  &  1   \\ 
 1.25  &  75  &  10.2  &  0.3  &  36  &  1   \\ 
 1.25  &  85  &  9.3  &  0.3  &  33  &  1   \\ 
 1.25  &  95  &  10.1  &  0.3  &  36  &  1   \\ 
 1.25  &  105  &  10.2  &  0.3  &  36  &  1   \\ 
 1.25  &  115  &  8.4  &  0.3  &  30  &  1   \\ 
 1.25  &  125  &  6.5  &  0.3  &  23.1  &  1   \\ 
 1.25  &  135  &  7.2  &  0.4  &  23  &  1   \\ 
 1.25  &  145  &  13.4  &  0.5  &  48  &  2   \\ 
 1.25  &  155  &  13.1  &  0.8  &  46  &  3   \\  \hline

\end{tabular}
\end{center} 
\end{table}

\newpage
\begin{table}[tbp]
\begin{center}
\begin{tabular}{|c|c|c|c|c|c|} \hline
$E_\gamma$ &$\theta_p^{CM}$&$d\sigma/d\Omega$ & $\Delta(d\sigma/d\Omega)$ &$d\sigma/dt$ & $\Delta (d\sigma/dt)$\\ \hline
GeV     &    deg        &  nb/str       &    nb/str     &        kb/GeV$^2$    & kb/GeV$^2$ \\ \hline

 1.35  &  15  &  49  &  3  &  160  &  11   \\ 
 1.35  &  25  &  51  &  2  &  165  &  6   \\ 
 1.35  &  35  &  27.6  &  1  &  89  &  3   \\ 
 1.35  &  45  &  16.1  &  0.6  &  52  &  2   \\ 
 1.35  &  55  &  9.6  &  0.4  &  31  &  1   \\ 
 1.35  &  65  &  7.2  &  0.3  &  23  &  1   \\ 
 1.35  &  75  &  6.4  &  0.3  &  20.8  &  0.9   \\ 
 1.35  &  85  &  7  &  0.3  &  22.5  &  0.9   \\ 
 1.35  &  95  &  6.2  &  0.3  &  20.1  &  0.8   \\ 
 1.35  &  105  &  6.3  &  0.3  &  20.3  &  0.8   \\ 
 1.35  &  115  &  5.1  &  0.2  &  16.4  &  0.8   \\ 
 1.35  &  125  &  4.4  &  0.2  &  14.4  &  0.7   \\ 
 1.35  &  135  &  5.7  &  0.4  &  16.1  &  0.8   \\ 
 1.35  &  145  &  11.2  &  0.5  &  36  &  1   \\ 
 1.35  &  155  &  13.8  &  0.9  &  45  &  3   \\  \hline
 1.45  &  15  &  39  &  3  &  116  &  9   \\ 
 1.45  &  25  &  39  &  2  &  115  &  5   \\ 
 1.45  &  35  &  21.5  &  0.9  &  64  &  3   \\ 
 1.45  &  45  &  11.7  &  0.5  &  35  &  1   \\ 
 1.45  &  55  &  7  &  0.4  &  21  &  1   \\ 
 1.45  &  65  &  4.6  &  0.3  &  13.7  &  0.8   \\ 
 1.45  &  75  &  4.5  &  0.2  &  13.3  &  0.7   \\ 
 1.45  &  85  &  4.6  &  0.2  &  13.5  &  0.7   \\ 
 1.45  &  95  &  3.6  &  0.2  &  10.8  &  0.6   \\ 
 1.45  &  105  &  3.8  &  0.2  &  11.2  &  0.6   \\ 
 1.45  &  115  &  3.4  &  0.2  &  10.1  &  0.6   \\ 
 1.45  &  125  &  2.8  &  0.2  &  8.4  &  0.5   \\ 
 1.45  &  135  &  3.9  &  0.3  &  10.4  &  0.6   \\ 
 1.45  &  145  &  8.7  &  0.4  &  26  &  1   \\ 
 1.45  &  155  &  13.1  &  0.8  &  39  &  2   \\  \hline

\end{tabular}
\end{center} 
\end{table}

\newpage
\begin{table}[tbp]
\begin{center}
\begin{tabular}{|c|c|c|c|c|c|} \hline
$E_\gamma$ &$\theta_p^{CM}$&$d\sigma/d\Omega$ & $\Delta(d\sigma/d\Omega)$ &$d\sigma/dt$ & $\Delta (d\sigma/dt)$\\ \hline
GeV     &    deg        &  nb/str       &    nb/str     &        kb/GeV$^2$    & kb/GeV$^2$ \\ \hline

 1.55  &  15  &  42  &  4  &  115  &  10   \\ 
 1.55  &  25  &  32  &  2  &  87  &  4   \\ 
 1.55  &  35  &  16.4  &  0.8  &  45  &  2   \\ 
 1.55  &  45  &  8.2  &  0.4  &  22  &  1   \\ 
 1.55  &  55  &  5.5  &  0.3  &  14.9  &  0.9   \\ 
 1.55  &  65  &  4.3  &  0.3  &  11.8  &  0.7   \\ 
 1.55  &  75  &  3.9  &  0.2  &  10.6  &  0.6   \\ 
 1.55  &  85  &  3.3  &  0.2  &  9.2  &  0.6   \\ 
 1.55  &  95  &  3  &  0.2  &  8.3  &  0.5   \\ 
 1.55  &  105  &  3  &  0.2  &  8.2  &  0.5   \\ 
 1.55  &  115  &  2.2  &  0.2  &  6.1  &  0.4   \\ 
 1.55  &  125  &  2.3  &  0.2  &  6.3  &  0.5   \\ 
 1.55  &  135  &  2.7  &  0.3  &  6.2  &  0.5   \\ 
 1.55  &  145  &  7.1  &  0.4  &  19  &  1   \\ 
 1.55  &  155  &  10.3  &  0.7  &  28  &  2   \\  \hline
 1.65  &  15  &  25  &  2  &  65  &  6   \\ 
 1.65  &  25  &  22  &  1  &  56  &  3   \\ 
 1.65  &  35  &  11.7  &  0.7  &  30  &  2   \\ 
 1.65  &  45  &  6  &  0.4  &  15.2  &  1   \\ 
 1.65  &  55  &  3.7  &  0.3  &  9.5  &  0.7   \\ 
 1.65  &  65  &  3.4  &  0.3  &  8.7  &  0.6   \\ 
 1.65  &  75  &  2.9  &  0.2  &  7.5  &  0.5   \\ 
 1.65  &  85  &  2.8  &  0.2  &  7.1  &  0.5   \\ 
 1.65  &  95  &  2.1  &  0.2  &  5.3  &  0.4   \\ 
 1.65  &  105  &  1.8  &  0.1  &  4.7  &  0.4   \\ 
 1.65  &  115  &  1.9  &  0.1  &  4.8  &  0.4   \\ 
 1.65  &  125  &  1.8  &  0.1  &  4.5  &  0.4   \\ 
 1.65  &  135  &  1.7  &  0.2  &  4.2  &  0.4   \\ 
 1.65  &  145  &  4.7  &  0.3  &  12  &  0.8   \\ 
 1.65  &  155  &  8.3  &  0.6  &  21  &  2   \\  \hline

\end{tabular}
\end{center} 
\end{table}

\begin{table}[tbp]
\begin{center}
\begin{tabular}{|c|c|c|c|c|c|} \hline
$E_\gamma$ &$\theta_p^{CM}$&$d\sigma/d\Omega$ & $\Delta(d\sigma/d\Omega)$ &$d\sigma/dt$ & $\Delta (d\sigma/dt)$\\ \hline
GeV     &    deg        &  nb/str       &    nb/str     &        kb/GeV$^2$    & kb/GeV$^2$ \\ \hline

 1.75  &  15  &  26  &  3  &  62  &  7   \\ 
 1.75  &  25  &  18  &  1  &  43  &  3   \\ 
 1.75  &  35  &  10.1  &  0.7  &  24  &  2   \\ 
 1.75  &  45  &  4.2  &  0.3  &  10.1  &  0.8   \\ 
 1.75  &  55  &  3  &  0.3  &  7.1  &  0.6   \\ 
 1.75  &  65  &  3  &  0.2  &  7.2  &  0.6   \\ 
 1.75  &  75  &  2  &  0.2  &  4.7  &  0.4   \\ 
 1.75  &  85  &  2  &  0.2  &  4.9  &  0.4   \\ 
 1.75  &  95  &  1.4  &  0.1  &  3.3  &  0.3   \\ 
 1.75  &  105  &  1.4  &  0.1  &  3.3  &  0.3   \\ 
 1.75  &  115  &  1.3  &  0.1  &  3  &  0.3   \\ 
 1.75  &  125  &  1.2  &  0.1  &  2.9  &  0.3   \\ 
 1.75  &  135  &  1.2  &  0.1  &  2.9  &  0.3   \\ 
 1.75  &  145  &  3.5  &  0.3  &  8.3  &  0.6   \\ 
 1.75  &  155  &  6.8  &  0.6  &  16  &  1   \\  \hline
 1.85  &  15  &  15  &  3  &  33  &  6   \\ 
 1.85  &  25  &  14  &  1  &  31  &  3   \\ 
 1.85  &  35  &  8  &  0.7  &  18  &  1   \\ 
 1.85  &  45  &  2.9  &  0.3  &  6.5  &  0.7   \\ 
 1.85  &  55  &  2.2  &  0.2  &  5  &  0.5   \\ 
 1.85  &  65  &  2  &  0.2  &  4.5  &  0.5   \\ 
 1.85  &  75  &  1.6  &  0.2  &  3.6  &  0.4   \\ 
 1.85  &  85  &  1.4  &  0.1  &  3.2  &  0.3   \\ 
 1.85  &  95  &  1.1  &  0.1  &  2.4  &  0.3   \\ 
 1.85  &  105  &  0.8  &  0.1  &  1.9  &  0.2   \\ 
 1.85  &  115  &  1.2  &  0.1  &  2.6  &  0.3   \\ 
 1.85  &  125  &  1.1  &  0.1  &  2.4  &  0.3   \\ 
 1.85  &  135  &  0.8  &  0.1  &  1.7  &  0.2   \\ 
 1.85  &  145  &  3.1  &  0.3  &  6.9  &  0.6   \\ 
 1.85  &  155  &  6.2  &  0.6  &  14  &  1   \\  \hline

\end{tabular}
\end{center} 
\end{table}

\newpage
\begin{table}[tbp]
\begin{center}
\begin{tabular}{|c|c|c|c|c|c|} \hline
$E_\gamma$ &$\theta_p^{CM}$&$d\sigma/d\Omega$ & $\Delta(d\sigma/d\Omega)$ &$d\sigma/dt$ & $\Delta (d\sigma/dt)$\\ \hline
GeV     &    deg        &  nb/str       &    nb/str     &        kb/GeV$^2$    & kb/GeV$^2$ \\ \hline

 1.95  &  15  &  11  &  2  &  24  &  4   \\ 
 1.95  &  25  &  10  &  1  &  21  &  2   \\ 
 1.95  &  35  &  6  &  0.6  &  12  &  1   \\ 
 1.95  &  45  &  1.7  &  0.2  &  3.6  &  0.5   \\ 
 1.95  &  55  &  2  &  0.2  &  4.1  &  0.5   \\ 
 1.95  &  65  &  1  &  0.2  &  2  &  0.3   \\ 
 1.95  &  75  &  1.2  &  0.1  &  2.5  &  0.3   \\ 
 1.95  &  85  &  1.1  &  0.1  &  2.3  &  0.3   \\ 
 1.95  &  95  &  0.59  &  0.1  &  1.2  &  0.2   \\ 
 1.95  &  105  &  0.8  &  0.1  &  1.6  &  0.2   \\ 
 1.95  &  115  &  0.59  &  0.1  &  1.2  &  0.2   \\ 
 1.95  &  125  &  0.7  &  0.1  &  1.5  &  0.2   \\ 
 1.95  &  135  &  0.7  &  0.1  &  1.4  &  0.2   \\ 
 1.95  &  145  &  1.5  &  0.2  &  3  &  0.4   \\ 
 1.95  &  155  &  4.2  &  0.5  &  9  &  1   \\ 
 1.95  &  165  &  5  &  3  &  10  &  7   \\  \hline
 2.05  &  15  &  14  &  3  &  28  &  7   \\ 
 2.05  &  25  &  9  &  1  &  18  &  2   \\ 
 2.05  &  35  &  4.2  &  0.5  &  8  &  1   \\ 
 2.05  &  45  &  1.6  &  0.2  &  3.1  &  0.5   \\ 
 2.05  &  55  &  1.1  &  0.2  &  2.2  &  0.4   \\ 
 2.05  &  65  &  0.9  &  0.2  &  1.8  &  0.3   \\ 
 2.05  &  75  &  0.8  &  0.1  &  1.5  &  0.2   \\ 
 2.05  &  85  &  0.56  &  0.1  &  1.1  &  0.2   \\ 
 2.05  &  105  &  0.65  &  0.1  &  1.3  &  0.2   \\ 
 2.05  &  115  &  0.54  &  0.09  &  1.1  &  0.2   \\ 
 2.05  &  125  &  0.45  &  0.08  &  0.9  &  0.2   \\ 
 2.05  &  135  &  0.41  &  0.09  &  0.8  &  0.2   \\ 
 2.05  &  145  &  1.7  &  0.2  &  3.4  &  0.4   \\ 
 2.05  &  155  &  3.8  &  0.5  &  7.5  &  0.9   \\  \hline

\end{tabular}
\end{center} 
\end{table}

\begin{table}[tbp]
\begin{center}
\begin{tabular}{|c|c|c|c|c|c|} \hline
$E_\gamma$ &$\theta_p^{CM}$&$d\sigma/d\Omega$ & $\Delta(d\sigma/d\Omega)$ &$d\sigma/dt$ & $\Delta (d\sigma/dt)$\\ \hline
GeV     &    deg        &  nb/str       &    nb/str     &        kb/GeV$^2$    & kb/GeV$^2$ \\ \hline

 2.15  &  25  &  8  &  1  &  15  &  2   \\ 
 2.15  &  35  &  3.8  &  0.5  &  7.1  &  1   \\ 
 2.15  &  45  &  1.6  &  0.3  &  3  &  0.5   \\ 
 2.15  &  55  &  0.9  &  0.2  &  1.8  &  0.3   \\ 
 2.15  &  65  &  1  &  0.2  &  1.9  &  0.3   \\ 
 2.15  &  75  &  0.43  &  0.09  &  0.8  &  0.2   \\ 
 2.15  &  85  &  0.55  &  0.1  &  1  &  0.2   \\ 
 2.15  &  95  &  0.21  &  0.06  &  0.4  &  0.1   \\ 
 2.15  &  105  &  0.33  &  0.07  &  0.6  &  0.1   \\ 
 2.15  &  115  &  0.31  &  0.07  &  0.6  &  0.1   \\ 
 2.15  &  125  &  0.53  &  0.09  &  1  &  0.2   \\ 
 2.15  &  135  &  0.27  &  0.07  &  0.5  &  0.1   \\ 
 2.15  &  145  &  1.2  &  0.2  &  2.3  &  0.3   \\ 
 2.15  &  155  &  2.8  &  0.4  &  5.1  &  0.8   \\ 
 2.15  &  165  &  1  &  0.8  &  2  &  2   \\  \hline
 2.25  &  15  &  3  &  1  &  5  &  2   \\ 
 2.25  &  25  &  5.5  &  0.8  &  10  &  1   \\ 
 2.25  &  35  &  2.1  &  0.4  &  3.7  &  0.7   \\ 
 2.25  &  45  &  0.8  &  0.2  &  1.4  &  0.3   \\ 
 2.25  &  55  &  0.5  &  0.1  &  0.8  &  0.2   \\ 
 2.25  &  65  &  0.6  &  0.1  &  1  &  0.2   \\ 
 2.25  &  75  &  0.28  &  0.08  &  0.5  &  0.1   \\ 
 2.25  &  85  &  0.44  &  0.09  &  0.8  &  0.2   \\ 
 2.25  &  95  &  0.27  &  0.07  &  0.5  &  0.1   \\ 
 2.25  &  105  &  0.28  &  0.07  &  0.5  &  0.1   \\ 
 2.25  &  115  &  0.4  &  0.08  &  0.7  &  0.1   \\ 
 2.25  &  125  &  0.28  &  0.07  &  0.5  &  0.1   \\ 
 2.25  &  135  &  0.4  &  0.08  &  0.7  &  0.1   \\ 
 2.25  &  145  &  0.7  &  0.1  &  1.3  &  0.3   \\ 
 2.25  &  155  &  2  &  0.3  &  3.6  &  0.6   \\ 
 2.25  &  165  &  2  &  1  &  4  &  2   \\  \hline

\end{tabular}
\end{center} 
\end{table}

\newpage
\begin{table}[tb]
\begin{center}
\begin{tabular}{|c|c|c|c|c|c|} \hline
$E_\gamma$ &$\theta_p^{CM}$&$d\sigma/d\Omega$ & $\Delta(d\sigma/d\Omega)$ &$d\sigma/dt$ & $\Delta (d\sigma/dt)$\\ \hline
GeV     &    deg        &  nb/str       &    nb/str     &        kb/GeV$^2$    & kb/GeV$^2$ \\ \hline

 2.35  &  15  &  12  &  4  &  21  &  6   \\ 
 2.35  &  25  &  4.8  &  0.9  &  8  &  2   \\ 
 2.35  &  35  &  2.1  &  0.5  &  3.5  &  0.8   \\ 
 2.35  &  45  &  1.1  &  0.3  &  1.9  &  0.4   \\ 
 2.35  &  55  &  0.7  &  0.2  &  1.2  &  0.3   \\ 
 2.35  &  65  &  0.6  &  0.2  &  1  &  0.3   \\ 
 2.35  &  85  &  0.15  &  0.06  &  0.3  &  0.1   \\ 
 2.35  &  95  &  0.26  &  0.08  &  0.4  &  0.1   \\ 
 2.35  &  105  &  0.2  &  0.07  &  0.3  &  0.1   \\ 
 2.35  &  115  &  0.11  &  0.05  &  0.19  &  0.08   \\ 
 2.35  &  135  &  0.3  &  0.08  &  0.5  &  0.1   \\ 
 2.35  &  145  &  0.4  &  0.1  &  0.6  &  0.2   \\ 
 2.35  &  155  &  1.1  &  0.3  &  1.8  &  0.5   \\ 
 2.35  &  165  &  3  &  2  &  6  &  3   \\  \hline
 2.45  &  25  &  4  &  1  &  7  &  2   \\ 
 2.45  &  35  &  1.3  &  0.7  &  2  &  1   \\ 
 2.45  &  45  &  0.4  &  0.3  &  0.7  &  0.4   \\ 
 2.45  &  55  &  0.8  &  0.3  &  1.2  &  0.5   \\ 
 2.45  &  75  &  0.2  &  0.1  &  0.3  &  0.2   \\ 
 2.45  &  85  &  0.13  &  0.09  &  0.2  &  0.1   \\ 
 2.45  &  95  &  0.2  &  0.1  &  0.4  &  0.2   \\ 
 2.45  &  105  &  0.11  &  0.08  &  0.2  &  0.1   \\ 
 2.45  &  115  &  0.2  &  0.1  &  0.4  &  0.2   \\ 
 2.45  &  125  &  0.11  &  0.08  &  0.2  &  0.1   \\ 
 2.45  &  135  &  0.16  &  0.09  &  0.3  &  0.1   \\ 
 2.45  &  145  &  0.5  &  0.2  &  0.8  &  0.4   \\ 
 2.45  &  155  &  1.1  &  0.4  &  1.7  &  0.7   \\  \hline

\end{tabular}
\end{center} 
\end{table}

\newpage
\begin{table}[tbp]
\begin{center}
\begin{tabular}{|c|c|c|c|c|c|} \hline
$E_\gamma$ &$\theta_p^{CM}$&$d\sigma/d\Omega$ & $\Delta(d\sigma/d\Omega)$ &$d\sigma/dt$ & $\Delta (d\sigma/dt)$\\ \hline
GeV     &    deg        &  nb/str       &    nb/str     &        kb/GeV$^2$    & kb/GeV$^2$ \\ \hline

 2.55  &  15  &  106  &  44  &  165  &  68   \\ 
 2.55  &  25  &  4  &  1  &  6  &  2   \\ 
 2.55  &  35  &  1.1  &  0.6  &  1.7  &  1   \\ 
 2.55  &  55  &  0.7  &  0.3  &  1.1  &  0.5   \\ 
 2.55  &  105  &  0.2  &  0.1  &  0.3  &  0.2   \\ 
 2.55  &  125  &  0.2  &  0.1  &  0.3  &  0.2   \\ 
 2.55  &  135  &  0.3  &  0.1  &  0.5  &  0.2   \\ 
 2.55  &  145  &  0.4  &  0.2  &  0.6  &  0.3   \\ 
 2.55  &  155  &  1.6  &  0.5  &  2.5  &  0.8   \\  \hline
 2.65  &  25  &  3  &  1  &  4  &  2   \\ 
 2.65  &  35  &  2.2  &  0.8  &  3  &  1   \\ 
 2.65  &  45  &  0.3  &  0.2  &  0.4  &  0.3   \\ 
 2.65  &  65  &  0.3  &  0.2  &  0.4  &  0.3   \\ 
 2.65  &  75  &  0.2  &  0.1  &  0.3  &  0.2   \\ 
 2.65  &  95  &  0.17  &  0.1  &  0.3  &  0.1   \\ 
 2.65  &  115  &  0.15  &  0.09  &  0.2  &  0.1   \\ 
 2.65  &  125  &  0.2  &  0.1  &  0.3  &  0.2   \\ 
 2.65  &  145  &  0.4  &  0.2  &  0.6  &  0.3   \\ 
 2.65  &  155  &  0.4  &  0.3  &  0.6  &  0.4   \\  \hline

\end{tabular}
\end{center} 
\end{table}

\newpage
\begin{table}[tbp]
\begin{center}
\begin{tabular}{|c|c|c|c|c|c|} \hline
$E_\gamma$ &$\theta_p^{CM}$&$d\sigma/d\Omega$ & $\Delta(d\sigma/d\Omega)$ &$d\sigma/dt$ & $\Delta (d\sigma/dt)$\\ \hline
GeV     &    deg        &  nb/str       &    nb/str     &        kb/GeV$^2$    & kb/GeV$^2$ \\ \hline

 2.75  &  25  &  2  &  1  &  3  &  2   \\ 
 2.75  &  35  &  1.0  &  0.6  &  1.5  &  0.8   \\ 
 2.75  &  45  &  0.3  &  0.2  &  0.5  &  0.3   \\ 
 2.75  &  55  &  0.4  &  0.2  &  0.6  &  0.3   \\ 
 2.75  &  95  &  0.13  &  0.09  &  0.2  &  0.1   \\ 
 2.75  &  115  &  0.2  &  0.1  &  0.3  &  0.1   \\ 
 2.75  &  125  &  0.12  &  0.08  &  0.2  &  0.1   \\ 
 2.75  &  135  &  0.16  &  0.09  &  0.2  &  0.1   \\ 
 2.75  &  145  &  0.3  &  0.2  &  0.4  &  0.2   \\ 
 2.75  &  155  &  0.7  &  0.4  &  1  &  0.5   \\  \hline
 2.85  &  25  &  5  &  2  &  7  &  2   \\ 
 2.85  &  35  &  0.9  &  0.6  &  1.2  &  0.9   \\ 
 2.85  &  45  &  0.4  &  0.3  &  0.6  &  0.4   \\ 
 2.85  &  155  &  0.7  &  0.3  &  0.9  &  0.4   \\  \hline
 2.95  &  45  &  0.6  &  0.4  &  0.7  &  0.5   \\ 
 2.95  &  155  &  0.8  &  0.5  &  1.1  &  0.6   \\  \hline

\end{tabular}
\end{center} 
\end{table}

\clearemptydoublepage

\addcontentsline{toc}{chapter}{References}
\renewcommand{\baselinestretch}{1.0}
\markboth{}{}

\clearemptydoublepage


\begin{thebibliography}{99}
\small

\bibitem{HOLT} R.J. Holt, {\em Phys. Rev.} {\bf C41} 2400 (1990) 

\bibitem{PROS} P. Rossi, E. De Sanctis {\em et al.}, {\em Phys. Rev.} {\bf C40} 2412 (1989)

\bibitem{JENK} D.A. Jenkins  {\em et al.}, {\em Phys. Rev.} {\bf C50} 74 (1994)

\bibitem{ARE1} H. Aren\"ovel, M. Danos, and H.T. Williams, {\em Nucl. Phys.} {\bf A162} 12 (1971)

\bibitem{WILH} P. Wilhelm and H. Aren\"ovel, {\em Phys. Lett.} {\bf B318} 410 (1993)

\bibitem{LAGE} J.M. Laget, {\em Phys. Rep.} {\bf I69} 493 (1981)\\
J.M. Laget, {\em Can. J. Phys.} {\bf 62} 1046 (1984)\\ 
J.M. Laget, {\em Phys. Lett.} {\bf B 199} 493 (1987)

\bibitem{JAUS} W. Jaus, D. Bofinger and W.S. Woolcock, {\em Nucl. Phys.} {\bf A562} 477-500 (1993)

\bibitem{AREND} H.J. Arends {\em et al.}, {\em Nucl. Phys.} {\bf A412} 509 (1984)

\bibitem{EDSLEV} E. De Sanctis {\em et al.}, {\em Phys. Rev.} {\bf C34} 413 (1986);\\
P. Levi Sandri {\em et al.}, {\em Phys. Rev.} {\bf C39} 701 (1989)

\bibitem{MATT} MIT/BATES, http://filburt.mit.edu

\bibitem{WALL} P.A. Wallace {\em et al.}, {\em Nucl. Phys.} {\bf A532} 617 (1991)

\bibitem{LEGS} LEGS Data Release L1-3.0 (1994)

\bibitem{CRAW} R. Crawford {\em et al.}, {\em Nucl. Phys.} {\bf A603} 303-325 (1996)

\bibitem{BABA} K. Baba {\em et al.}, {\em Phys. Rev.} {\bf C48} 286 (1983)

\bibitem{CHIN} R. Ching and C. Schaerf, {\em Phys. Rev.} {\bf 141} 141 (1966)

\bibitem{DOUG} P. Dougan {\em et al.}, {\em Z. Phys. } {\bf A276} 55 (1976)

\bibitem{FREE} S.J. Freedman {\em et al.}, {\em Phys. Rev. } {\bf C48} 1864 (1993)

\bibitem{BELZ} J.E. Belz {\em et al.}, {\em Phys. Rev. Lett.} {\bf 74} 646 (1995)

\bibitem{HALLA} Jefferson Lab Experiments E89-01 (1989) and E99-008 (1999)

\bibitem{HALLB} Jefferson Lab Experiment E93-017 (1993)

\bibitem{HALLC} Jefferson Lab Experiment E89-012 (1989) and E96-003 (1996)

\bibitem{BRO1} S.J. Brodsky, G.R. Farrar, {\em Phys. Rev. Lett.} {\bf 31} 1153 (1998)

\bibitem{GILM} R. Gilman and F. Gross, {\em J. Phys. G.}, Nucl. Part. Phys. {\bf 28} R37 (2002)\\
R. Gilman and F. Gross, Review Article WM-01-113, JLAB-PHY-01-25 (2001)

\bibitem{BOCH} C. Bochna {\em et al.}, {\em Phys. Rev. Lett.} {\bf 81} 4576 (1998)

\bibitem{SHU1} E.C. Shulte {\it et al.}, {\em Phys. Rev. Lett.} {\bf 87} 102302-1 (2001)

\bibitem{SHU2} E.C. Shulte {\em et al.}, {\em Phys. Rev. } {\bf C} 0422011 (2002)

\bibitem{RNA} 
S.L. Brodsky, J.R. Hiller, Phys. Rev. {\bf C28} 475 (1983)

\bibitem{RNAEL} 
S.J. Brodsky, B. T. Chertok, Phys. Rev. {\bf D14} 3003  (1976)\\
S.J. Brodsky, B. T. Chertok, Phys. Rev. Lett. {\bf 37} 269 (1976)\\
S.J. Brodsky, C-R. Ji, G. P. Lepage, Phys. Rev. Lett. {\bf 51}  83 (1983)\\
L.C. Alexa {\it et al.},  Phys. Rev. Lett. {\bf 82}  1374 (1999)\\
D. Abbott {\it et al.},  Phys. Rev. Lett. {\bf 82}  1879 (1999)\\

\bibitem{HRM} 
L.L. Frankfurt {\it et al.}, Phys. Rev. Lett. {\bf 84}  3045 (2000)\\
L.L. Frankfurt {\it et al.}, Nucl. Phys. {\bf A663\&664}  349 (2000)

\bibitem{AMEC} 
A.E.L. Dieperink, S. I. Nagorny, Phys. Lett. {\bf B456}  9 (1999)

\bibitem{gris} V.Yu. Grishina {\it et al.}, {\em Eur. Phys. J.} {\bf A10} 355 (2001) 

\bibitem{LEV} M. Levin, {\em Sov. Phys. Usp.} {\bf 16} 600 (1973)

\bibitem{CLOS} F. Close and N. Isgur, Phys. Lett. {\bf B509} 81-86 (2001)

\bibitem{FEDRO} 
M. Mirazita, Proc. of {\em XL Internal Winter Meeting on Nuclear
Physics}, Bormio, Jan. 21-26, 2002, hep/ph 0206213 \\
F. Ronchetti, Proc. of {\em Quark and Nuclear Physics 2002}, \"Julich, 
Germany, Jun. 9-14 2002 in printing by Eur. Phys. J. {\bf A}  \\
F. Ronchetti, Proc. of {\em IV Conference on Quark Confinement and the Hadron Spectrum}, 
Gargnano, Italy, Sept. 10-14 2002, in printing by World. Sci. Pub.  \\
P. Rossi, Proc. of {\em Electron-Nucleus Scattering VII}, 
Isola d'Elba, Italy Jun. 24-28 2002 

\bibitem{SARG} M.M. Sargisan, Private Communication (2003)

\bibitem{hhc} G.P. Lepage, S.J. Brodsky, {\em Phys. Rev.} {\bf D22}  2157 (1980)\\
S.J. Brodsky, G.P. Lepage, Phys. Rev. {\bf D24}  2848 (1981)

\bibitem{IOF} B.L. Ioffe and A.V. Smilga, {\em Nucl. Phys.} {\bf B216} 373 (1983)

\bibitem{ISGU} N. Isgur and C.H. Llewellyn Smith,  {\em Phys. Lett.}  {\bf B217} 535 (1989)\\
N. Isgur and C.H. Llewellyn Smith, {\em Phys. Rev. Lett.}   {\bf 52 } 1080 (1984)

\bibitem{RADI} A.V. Radyushkin, {\em Nucl. Phys.} {\bf A 532}, 141 (1991)

\bibitem{AZNA} I.G. Aznaurian, A.S. Bagdasarian, S.V. Esaibegian, and N.L. Ter-Isaakian,
{\em Sov. J. Nucl. Phys.} {\bf 55} 1099 (1992)

\bibitem{pol} K. Wijesooriya {\it et al.}, {\em Phys. Rev. Lett.} {\bf 86}  2975 (2001), \\
and references therein.

\bibitem{qnp-vera} V.Yu. Grishina {\it et al.}, Proc. of {\em Quark and Nuclear Physics 2002}, \"Julich, 
Germany, June 9-14 2002 in printing by {\em Eur. Phys. J.} {\bf A}\\

\bibitem{hhcprot} D.G. Grabb {\it et al.}, {\em Phys. Rev. Lett.} {\bf 65}  3241 (1990)

\bibitem{landmecprot} T. Gousset {\it et al.}, {\em Phys. Rev.} {\bf D53}  1202 (1996)\\
C. Carlson, M. Chachkhunashvili, {\em Phys. Rev.} {\bf D45}  2555 (1992)

\bibitem{landmecgam} A. Afanasev {\it et al.}, {\em Phys. Rev.} {\bf D61}  034014 (2000)

\bibitem{qnp-vera12}
I.P. Auer {\it et al.}, {\em Phys. Rev. Lett.} {\bf 62} 2649 (1989)

\bibitem{adam} F. Adamian {\it et al.}, {\em Eur. Phys. J.} {\bf A8} 423 (2000)



\bibitem{CRE} {
M. Creutz,  {\em Phys. Rev.} {\bf D21} 2308 (1980)\\ 
M. Creutz,  {\em Phys. Rev. Lett.} {\bf 45} 313 (1980)\\ 
M. Creutz,  {\em Quarks, Gluons, and Lattice}, Cambridge U.P. (1983)}

\bibitem{SHIF} M.A. Shifman, A.I. Vainstein, and V.I. Zakharov,\\
{\em Nucl. Phys.} {\bf B147} 385, 448, 519  (1979) 

\bibitem{THOO} G. T'Hooft, {\em Nucl. Phys.} {\bf B72} 461 (1974)

\bibitem{VEN} { 
G. Veneziano, {\em Phys. Lett.} {\bf B52} 220 (1974)\\
G. Veneziano, {\em Nucl. Phys.} {\bf B117} 519 (1976)}

\bibitem{CIA} M. Ciafalloni, G. Marchesini, and G. Veneziano, {\em Nucl. Phys.} {\bf B98} 472 (1975)

\bibitem{CAS1} A. Casher, J. Kogut, and L. Susskind,  {\em Phys. Rev.} {\bf D10} 732 (1974)

\bibitem{CAS2} A. Casher, H. Neuberger, and S. Nussinov,  {\em Phys. Rev.} {\bf D20} 179 (1979)

\bibitem{ART} X. Artru and G. Menneiser, {\em Nucl. Phys.} {\bf B70} 93 (1974)

\bibitem{AND1} B. Andersson, G. Gustafson, and C. Peterson, {\em Phys. Lett.} {\bf B71} 337 (1977)

\bibitem{AND2} B. Andersson, G. Gustafson, and C. Peterson, {\em Z. Phys.} {\bf C1} 105 (1979)

\bibitem{KA1} A.B. Kaidalov, {\em JTEP Lett.} {\bf 32} 474 (1980)\\
A.B. Kaidalov, {\em Sov. J. Nucl. Phys.} {\bf 33} 733 (1981)

\bibitem{K93} L.A. Kondratyuk, E. De Sanctis {\em et al.}, , {\em Phys. Rev.} {\bf C48} 2491 (1993)

\bibitem{KA3} A.B. Kaidalov, {\em Phys. Lett.} {\bf B116} 459 (1982)\\
A.B. Kaidalov and K. Ter-Martirosyan, {\em Phys. Lett.} {\bf B117} 247 (1982)

\bibitem{KA5} A.B. Kaidalov, {\em Z. Phys.} {\bf C12} 63 (1982)

\bibitem{KA6} A.B. Kaidalov and P. Volkovitsky, {\em Sov. J. Nucl. Phys.} {\bf 35} 720, 909 (1982)

\bibitem{KA7} A.B. Kaidalov and K. Ter-Martirosyan, {\em Sov. J. Nucl. Phys.} {\bf 39} 979 (1984)\\
A.B. Kaidalov and K. Ter-Martirosyan, {\em Sov. J. Nucl. Phys.} {\bf 40} 135 (1984)

\bibitem{ions} E.E. Zabrodin {\it et al.}, {\em Phys. Lett.} {\bf B508} 184 (2001)

\bibitem{CHE} G.F. Chew and C. Ronzenzweig, {\em Phys. Rep.} {\bf 41} 26 (1978)

\bibitem{KA9} A.B. Kaidalov {\em Surv. High En. Phys.} {\bf 13} 265 (1999)

\bibitem{CAP}{
A. Capella {\em et al.}, {\em Z. Phys.} {\bf C3} 329 (1980) \\
A. Capella and J. Tran Thanh Van, {\em Phys. Lett.} {\bf B114} 450 (1982) \\
A. Capella {\em et al.}, {\em Phys. Rep.} {\bf 326} 225 (1994)
}

\bibitem{COLL} P.D.B. Collins and E.J. Squires, {\em Regge Poles in Particle Physics}, \\
Springer Tracts in Modern Physics {\bf 45} (1970)

\bibitem{BJO} J. Bjorken, SLAC Report, PUB-95-6949 (1995) 

\bibitem{SEGR} E. Segr\'e, {\em Nuclei e Particelle}, Zanichelli, (1966)  

\bibitem{LEON} M. Leon,  {\em Particle Physics: an Introduction}, Academic Press (1973)

\bibitem{BORN} K.D. Born {\em et al.}, {\em Phys. Rev} {\bf D40} 1653 (1989)

\bibitem{LYU1} V.A. Lyubimov, {\em Sov. J. Nucl. Phys. Usp.} {\bf 20} 691 (1977)

\bibitem{BRA} A. Brandt and the UA8 Collaboration, {\em Nucl. Phys.} {\bf B514} 3 (1998)

\bibitem{INO} A.E. Inopin, hep/th 0012248 

\bibitem{CHI} Z. Chikovani, L. Jenkowsky and F. Paccanoni, {\em Mod. Phys. Lett.} {\bf A6} 1409 (1991) 

\bibitem{KA10} A.B. Kaidalov, {\em Sov. Phys. Usp.} {\bf 14} 600 (1972)

\bibitem{KA11} A.B. Kaidalov, {\em Sov. Phys. Usp.} {\bf 53} 872 (1991)

\bibitem{KKG} {
C. Guaraldo, A.B. Kaidalov, L.A. Kondratyuk, Y.S. Golubeva,\\
{\em Yad. Fiz} {\bf 59} 1896 (1996); {\em Phys. At. Nuclei} {\bf 59} 1832 (1996)
}

\bibitem{ALLA} {
J.V. Allaby {\em et al.}, {\em Phys. Lett.} {\bf B29} 198 (1969) \\
U. Amaldi {\em et al.}, {\em Lett. Nuovo Cimento} {\bf 4} 121 (1972) \\
H.L. Anderson {\em et al.}, {\em Phys. Rev.} {\bf D3} 1536 (1971) - {\bf D9} 580 (1974) 
}

\bibitem{BAG} {  
C. Bagalin {\em et al.}, {\em Nucl. Phys.} {\bf B37}, 639 (1972) \\
A.J. Pawliki {\em et al.}, {\em Phys. Rev. Lett.} {\bf 31} 665 (1988) \\ 
C. Evangelista {\em et al.}, {\em Nucl. Phys.} {\bf B131} 54 (1977) 
}

\bibitem{BIZZ}{ 
R. Bizzarri {\em et al.}, {\em Lett. Nuovo Cimento} {\bf 2} 431 (1969); \\
G.A. Smith, in {\em The Elementary Structure of Matter}\\ 
ed. by J.M. Richard {\em et al.}, (Springer-Verlag, 219, Berlin (1988); \\
J. Reidleberger {\em et al.}, {\em Phys. Rev.} {\bf C40} 2717 (1989); \\
M.P. Bussa, in {\em Proceedings of the Second Biennial Conference LEAP92}\\
ed. by C. Guaraldo, Courmayeur (1992);\\
A. Zenoni and F. Iazzi, {\em Nucl. Phys.} {\bf A558} (1993) 
}

\bibitem{BBG} M.M. Brisdudova, L. Burakovsky, and T. Goldamn, {\em Phys. Rev.} {\bf D} 61 054013 (2000)

\bibitem{KAKO} A.B. Kaidalov, L.A. Kondratyuk, {\em et al.}, {\em Phys. od At. Nuclei} {\bf 63} 1395, 
1409 (2000)

\bibitem{ARNO} R.G. Arnold {\em et al.}, {\em Phys. Rev. Lett.} {\bf 57} 174 (1986)

\bibitem{ARMS} T. Armstrong {\em et al.}, {\em Phys. Rev. Lett.} {\bf 70} 1212 (1993)

\bibitem{BISE} D. Bisello {\em et al.}, {\em Nucl. Phys. } B {\bf 411} 3 (1994)

\bibitem{BEBE} C.J. Bebek {\em et al.}, {\em Phys. Rev.} D {\bf 13} 25 (1976)

\bibitem{BISE1} D. Bisello {\em et al.},  {\em Phys. Lett.} B {\bf 220} 321 (1989)

\bibitem{MILA} J. Milana, S. Nussinov, and M.G. Olsson, {\em Phys. Rev. Lett.} {\bf 71} 2533 (1993)

\bibitem{KOBZ} I. Kobzarev {\em et al.}, {\em Sov. J. Nucl. Phys.} {\bf 45} 330 (1987)

\bibitem{HITO} H. Ito, {\em Progr. Theor. Phys.} {\bf 84} 94 (1990)

\bibitem{N6} 
B.D. Anderson {\em et al.}, to be submitted to {\em  Nucl. Instr. \& Meth. } {\bf A}

\bibitem{N7}
O.K. Baker {\em et al.} {\em  Nucl. Instr. \& Meth. }{\bf A367} 92 (1995)

\bibitem{N37}
D.I. Sober {\em et al.}, {\em  Nucl. Instr. \& Meth. }{\bf A440} 263 (2000)

\bibitem{ADON}
N. Bianchi {\em et al.}, {\em  Nucl. Instr. \& Meth. }{\bf A311} 172 (1992)\\
N. Bianchi {\em et al.}, {\em  Nucl. Instr. \& Meth. }{\bf A317} 434 (1992)

\bibitem{TIER} 
T. Auger, PhD thesis, DAPNIA/SPhN-99-01T, (1999) 

\bibitem{N8}
M.D. Mestayer {\em et al.}, {\em  Nucl. Instr. \& Meth. }{\bf A449} 81 (2000)

\bibitem{N9}
D.S. Carman {\em et al.}, {\em  Nucl. Instr. \& Meth. }{\bf A419} 315 (1998)

\bibitem{N10} 
L.M. Qin {\em et al.}, {\em  Nucl. Instr. \& Meth. }{\bf A411} 265 (1998)

\bibitem{N11} 
G. Adams {\em et al.}, {\em  Nucl. Instr. \& Meth. }{\bf A465} 414 (2001) 

\bibitem{N12} 
E.S. Smith {\em et al.}, {\em  Nucl. Instr. \& Meth. }{\bf A432} 265 (1999)

\bibitem{N13} 
M. Amarian {\em et al.}, {\em  Nucl. Instr. \& Meth. }A{\bf 460} 239  (2001)

\bibitem{N14}
M. Anghinolfi {\em et al.}, {\em  Nucl. Instr. \& Meth. }{\bf A447}, 424 (2000)

\bibitem{N15}
S. Taylor {\em et al.}, {\em  Nucl. Instr. \& Meth. }A{\bf 462} 484 (2001)


\bibitem{N18} 
P. Rossi {\em et al.}, {\em  Nucl. Instr. \& Meth. }{\bf A381}  32 (1996)

\bibitem{N20} 
M. Ripani {\em et al.}, {\em  Nucl. Instr. \& Meth. }{\bf A406} 403 (1998)

\bibitem{N21} 
P. Rossi {\em et al.}, JLab CLAS Note 2001-005, (2001)

\bibitem{N24} 
G.P. Heath, {\em  Nucl. Instr. \& Meth. } {\bf A278} 431 (1989)

\bibitem{N25}
D.C. Doughty Jr. {\em et al.}, {\em IEEE Trans. Nucl. Sci.} {\bf NS-39} 241 (1992)

\bibitem{N30}
E. Jastrzembski  {\em et al.}, Proc. of the {\em IEEE NPSS Real Time Conference} 538 (1999)

\bibitem{DD}
C. Witzig, {\em The DD system}, BNL Note, {\bf 510C} (1995)

\bibitem{N31}
CODA: CEBAF On Line Data Acquisition Manual, JLab Internal Report (1997)

\bibitem{N32}
V.H. Gyurjyan, http://www.jlab.org/$\sim$gurjyan/dosumentations.htm\\
S. Barrow, CODA 2.0, JLab Internal Report (1997)

\bibitem{BOS}
V. Blobel {\em et al.}, The BOS System for the CLAS Detector, JLab Internal Report (1995) 

\bibitem{ELOSS}
E. Pasyuk, JLab Internal Report (2000)

\bibitem{NORM}
E. Pasyuk and J. Ball, JLab Internal Report (2000) \\
J. Ball, http://www.jlab.org/$\sim$jimball

\bibitem{ATT}
R. Schumacher, JLab CLAS Note, 2001-010 (2001)

\bibitem{Chauvenet}
H.D. Young, {\em Statistical Treatment of Experimental Data},\\
McGraw-Hill Book Company, NY (1962)

\bibitem{PITT}
K.Y. Kim {\em et al.}, JLab CLAS Note, 2001-18 (2001)

\bibitem{CNIM}
B. Mecking and the CLAS Collaboration, {\em The CLAS Detector},\\
in printing by {\em Nucl. Instr. \& Meth.} (2003)

\bibitem{SOB}
D. Sober, http://www.jlab.org/$\sim$sober/misc.html





%













\end{thebibliography}
\end{document}